%% file: ociamthesismain.tex
\documentclass[11pt]{ociamthesis}

\usepackage{amssymb}
\usepackage{hyperref}
\usepackage{graphicx}
\usepackage{caption}
\usepackage{commath}
\usepackage{wrapfig}
\usepackage{geometry}
\usepackage{amsmath}
\usepackage{braket}
\usepackage{subcaption}
\usepackage[framed,numbered,autolinebreaks,useliterate]{mcode}
\usepackage{algorithm}
\usepackage{mathtools}
\usepackage{multirow}
\usepackage{mdframed}

\usepackage{fancyhdr}
\usepackage[english]{babel}
\usepackage{amsthm}

\theoremstyle{definition}
\newtheorem{definition}{Definition}[section]

\theoremstyle{definition}
\newtheorem*{definition*}{Definition}

\theoremstyle{remark}
\newtheorem*{remark}{Remark}

\theoremstyle{definition}
\newtheorem{example}{Example}[section]

\theoremstyle{definition}
\newtheorem{theorem}{Theorem}[section]

\theoremstyle{definition}
\newtheorem*{theorem*}{Theorem}

\usepackage{nomencl}
\makenomenclature

\renewcommand{\nompreamble}{This list describes symbols and notation that will be used throughout the paper.}

\title{Chemical Integration of ODEs\\[1ex] using Idealized Abstract Solutions}

\author{Su Hyeong Lee}

\begin{document}

\baselineskip=18pt plus1pt

\setcounter{secnumdepth}{3}
\setcounter{tocdepth}{3}

\cfoot{}
\maketitle     
\cfoot{}
\clearpage
\cfoot{\thepage}

\begin{romanpages}          
\tableofcontents            
\listoffigures              

\include{notationandsymbols}  
\end{romanpages}            

\include{chapter1}
\include{chapter2}
\include{chapter3}

\include{chapter4}
\include{chapter5}
\include{chapter6}
\include{chapter7}
\include{conclusions}

\appendix
\include{appendix1}

\include{appendix2}
\include{appendix3}

\include{appendix4}
\include{appendix4andhalf}
\include{appendix5}
\include{appendix6}

\addcontentsline{toc}{chapter}{Bibliography}

\bibliography{sample}
\bibliographystyle{unsrt}

\end{document}

%% file: notationandsymbols.tex
\nomenclature{$\mathbf{X}$}{State vector $(\mathrm{X}_1, \dots, \mathrm{X}_N)$, for $N$ total specimen}

\nomenclature{$\mathrm{X}_i$}{Biochemical specimen of interest, for indices $i \in \mathbb{Z}_{\ge 1}$. Conventionally, we interchangeably use the same notation to denote molecule counts of specimen. Exceptions to this notation are in $\Psi_{QSST}$ and $\Psi_{Kow}$, where $\mathrm{Y}$ is used, and in Quasi-Steady State Approximations, where fast specimen is denoted by $\mathrm{Z}$}

\nomenclature{$x_i$}{Concentration of specimen $\mathrm{X}_i$, interchangeably expressed $[\mathrm{X}_i]$ by convention}

\nomenclature{$\boldsymbol{\nu}$}{Stoichiometry matrix with columns $\boldsymbol{\nu}_j$ for $j \in \{1,\dots,N\}$}

\nomenclature{$\boldsymbol{\nu}_j$}{State change vector $(\nu_{1j}, \dots, \nu_{Nj})^\top$ for reaction $j$, for $N$ total specimen}

\nomenclature{$\nu_{ij}^-$}{Stoichiometric coefficients of the reactant $\mathrm{X}_i$ of reaction channel $j$}

\nomenclature{$\nu_{ij}^+$}{Stoichiometric coefficients of the product $\mathrm{X}_i$ of reaction channel $j$}

\nomenclature{$k_i$}{The rate coefficient of reaction channel $i$}

\nomenclature{$\boldsymbol{k}$}{Rate vector $(k_1,\dots,k_M)$, for $M$ total reaction channels}

\nomenclature{$\varnothing$}{Represents untracked quantity, such as saturated specimen whose concentration may be considered unchanging}

\nomenclature{$V$}{Volume of the reacting solution, to which rates of the reaction channels must be scaled accordingly. Unless noted otherwise, we take $V = 100$ throughout the paper}

\nomenclature{QSST}{Quasi-Steady State Approximation as described in \cite{TomiInversePaper}}

\nomenclature{QSSA}{Quasi Steady State Approximation as described in \cite{Wilhelm}}

\nomenclature{$F^\mathcal{X}$}{The set of elements of the form $\sum_{\substack{f \in F \\ s \in \mathcal{X}}} fs$, where $F$ is a set and $\mathcal{X}$ is identified as the basis elements; in the coordinate representation, we view elements of $\mathbb{F}^\mathcal{X}$ as a column vector. For example, $\mathbb{R}^{s_1,s_2}_{\ge 0} = \{ r_1s_1 + r_2s_2 : r_i \in \mathbb{R}_{\ge 0} \} = \{(r_1,r_2)^\top : r_i \in \mathbb{R}_{\ge 0}\}$ }

\nomenclature{$F_{\ge a}$}{The set of elements $b \in F$ such that $b \ge a$, where $a \in F$ and $F$ is a poset with non-strict partial ordering $\ge$. For example, $\mathbb{Z}_{\ge 0} = \{b \in \mathbb{Z}: b \ge 0 \}$}

\nomenclature{$x \downarrow 0$}{The limit as $x$ tends to $0$ from above }

\nomenclature{$\sqcup$}{Represents disjoint union. For a quick definition, consider the two sets $F_1,F_2$; then $F_1 \sqcup F_2 = \{(f_1,f_2):f_1 \in F_1, f_2 \in F_2 \}$. This induces the canonical identification $F_1 \subset F_1 \sqcup F_2$ by viewing $F_1=\{(f_1,f):f_1 \in F_1 \}$ for any fixed identity element $f \in F_2$}

\nomenclature{$\subset$}{Represents non-strict inclusion (i.e. a subset)}

\printnomenclature

%% file: chapter1.tex
\chapter{Introduction}\label{oldchapter1}

Broadly speaking, problems in chemical reaction network theory may be classified under two distinct categories. The first category is \textit{direct problems}, where a chemical reaction network is explicitly proposed and its properties investigated through deterministic analysis and stochastic simulation. The reaction network is selected through a combination of intuition, biological design principles, and data from laboratory experiments. This induces the second category of \textit{inverse problems}, where a set of desirable properties are provided (such as time series realizations), and compatible reaction networks are constructed. It has been shown that unique inversion is impossible in deterministic settings, due to differing reaction networks translating into identical ODE systems describing their dynamics~\cite{easySparsePaper}. However, a general inversion framework for designing \textit{any} reaction network compatible with \textit{any} desired property of interest has yet to be presented. 

In a paper by Plesa et al~\cite{TomiInversePaper}, an inverse problem framework for generating a chemical reaction network undergoing a supercritical homoclinic bifurcation is presented. In a later publication~\cite{TomiStatInferencePaper}, a similar framework is employed to construct two dimensional reaction systems displaying limit cycle bifurcations, which are subsequently posed as test systems for statistical inference. In particular, it is left as an open question as to whether or not Discrete Fourier Transforms~\cite{FFTBookTheory} and Autocorrelation Functions~\cite{AutoCorrelationBookTheory} will fail to distinguish the topological shift taking place in systems displaying a bifurcation sharply at the bifurcation point, which are two modern methods commonly used to differentiate between random systems with similar stochastic dynamics~\cite{AutoCorrAndDFT}.

In this paper, we seek to establish a small step toward answering the ambitious question: \textit{Given an arbitrary ODE system, does there exist a chemical reaction network that encapsulates its dynamics?} We do so by proposing an inversion framework to non-uniquely invert a large class of ODEs into chemical reaction networks closely approximating their dynamical properties.  

In Chapters~\ref{oldchapter2} and~\ref{oldchapter3}, we develop critical theory and code which form the foundation of this work. The notations used to describe chemical reaction network theory in the literature are rather involved, thus we aim to provide an intuitive interpretation of the formal notation by frequently introducing examples to motivate the formalisms. Even in rudimentary systems, careful application of the developed inversion strategies demonstrate a rich range of non-trivial behaviours that assert important implications about the strengths and weaknesses of each inversion technique. We consequently cycle through multiple test systems designed to simplify the analysis in Chapters~\ref{oldchapter4} and~\ref{oldchapter5}. We conclude a detailed framework for the chemical integration of ordinary differential equations based on our observations. 

Our work culminates in Chapter~\ref{oldchapter6}, where the utility of the framework is verified by chemically simulating ODE systems displaying oscillatory and chaotic dynamics. Specifically, we simulate the shifted pendulum, the R\"{o}ssler Attractor, and the Lorenz Attractor\footnote{The chemical simulations of the two chaotic attractors via Gillespie~\cite{GillespieFirstandDirectReactions} are available in video format in: https://youtu.be/CCplQMosFLI} after transcribing deterministic dynamics onto a chemical reaction network. The framework is further applied to simulate an ODE system undergoing a Hopf bifurcation, and a relevant time series analysis is provided. This subsequently confirms the inadequacy of proposed statistical methods~\cite{AutoCorrAndDFT} in distinguishing deterministic topological differences induced in ODE systems within the vicinity of their bifurcation using individual time series realizations, thereby validating the concerns raised in the previous work~\cite{TomiStatInferencePaper}. Through these experiments, we confirm the capabilities of the developed framework in capturing the dynamics of meaningful ODE systems.

The proposed inversion algorithm has the potential to mass generate high-dimensional chemical reaction networks approximating a very wide range of exotic\footnote{There is a degree of subjectiveness in which reactions are considered exotic. We follow the conventions of Plesa et al~\cite{TomiInversePaper}, where exotic systems demonstrate multistability, oscillations or chaos. A system is called regular if its trajectories converge to a globally steady equilibrium.} dynamics, as well as instantaneously producing multiple novel test systems for biological, numerical, and statistical analysis. A historical introduction to the Law of Mass Action and further motivations are included in Appendix~\ref{appendixA}. 

%% file: chapter2.tex
\chapter{Background Theory and Notation}\label{oldchapter2}

The formal notation and background theory are inspired by various sources throughout the literature~\cite{TomiStatInferencePaper,implicitTauLeap,originaltauleap,KowalskiWeak}, and in particular we follow~\cite{TomiInversePaper} quite closely, from which most of the notation is motivated with major modifications. We work through very simple example systems to present an intuitive understanding of the formal definitions. 

\begin{mdframed}
\begin{definition}[Reaction Network]\label{reactionnetwork}

A triple $\{\mathcal{X}, \mathcal{V}, \mathcal{R}\}$ is called a \textit{reaction network} or a \textit{network} if:

(A) $\mathcal{X} = \{\mathrm{X}_1, \dots, \mathrm{X}_N\}$ has finite cardinality, whose elements are called \textit{specimen} of the reaction network, 

(B) $\mathcal{V} \subset \mathbb{Z}_{\ge 0}^{\mathcal{X}} = \{\sum_{i=1}^N b_i \mathrm{X_i}: b_i\in \mathbb{Z}, b_i \ge 0\}$ has finite cardinality (see Appendix~\ref{chapter2appendix}), whose elements $\boldsymbol{\nu}_{j}^{\pm} = (\nu_{1j}^{\pm},\dots,$ $\nu_{Nj}^{\pm} )^\top$ are called the stoichiometric coefficients. We impose that $\forall \mathrm{X}_i \in \mathcal{X}$, $\exists \boldsymbol{\nu}_{j}^{\pm} \in \mathcal{V}$ such that the $i$-th component of vector $\boldsymbol{\nu}_{j}^{\pm}$ is non-zero, that is, all specimen are relevant to the network (see below),

(C) The \textit{reaction set} $\mathcal{R} \subset \mathcal{V} \times \mathcal{V}$ is a binary relation whose elements $\left(\boldsymbol{\nu}^-_{j}, \boldsymbol{\nu}^+_{j^\prime}\right)$ are written $\boldsymbol{\nu}^-_{j} \rightarrow \boldsymbol{\nu}^+_{j^\prime}$ and called \textit{reaction} or \textit{reaction channel}. $\boldsymbol{\nu}^-_{j}$ is called \textit{stoichiometric coefficients of reactants}, and $\boldsymbol{\nu}^+_{j^\prime}$ \textit{stoichiometric coefficients of products}. The \textit{order} of a reaction $\boldsymbol{\nu}^-_{j} \rightarrow \boldsymbol{\nu}^+_{j^\prime}$ is defined as $\norm{\boldsymbol{\nu}^-_{j}}_{\ell^1} = \sum_{i=1}^N \nu_{ij}^-$. $\mathcal{R}$ satisfies:

(i) $\forall \boldsymbol{\nu}_{j}^{\pm} \in \mathcal{V},$  $(\boldsymbol{\nu}_{j}^{\pm} \rightarrow \boldsymbol{\nu}_{j}^{\pm}) \notin \mathcal{R}$,

(ii) $\forall \boldsymbol{\nu}^-_{j} \in \mathcal{V}$, $\exists \boldsymbol{\nu}^+_{j^\prime} \in \mathcal{V}$ such that $\left(\boldsymbol{\nu}^-_{j} \rightarrow \boldsymbol{\nu}^+_{j^\prime}\right) \in \mathcal{R}$ or $\left(\boldsymbol{\nu}^+_{j^\prime} \rightarrow \boldsymbol{\nu}^-_{j}\right) \in \mathcal{R}$,

(D) There may exist a \textit{null specimen} $\varnothing \in \mathcal{X}$ which represents an untracked quantity, such as a chemical whose supply is continuously replenished, hence not of interest to the network.
\end{definition}
\end{mdframed}

\begin{remark}
Intuitively speaking, condition (C)(i) imposes that a reaction channel may not react to produce its reactants unchanged. (C)(ii) may be interpreted as having all stoichiometric coefficients represent a set of either reactant or product specimen, and requiring that there exists a corresponding reaction channel within the network that renders the coefficients relevant. Furthermore, conditions (A)--(D) ensure that $\mathcal{R}$ alone is sufficient to deduce $\mathcal{X}$, $\mathcal{V}$, thus the reaction network is often abbreviated as $\mathcal{R}$. In an abuse of notation for clarity, we often enumerate the elements of $\mathcal{R}$ and replace them by positive integers $\mathcal{R} = \{1, \dots, |\mathcal{R}|\}$. 
\end{remark}

\begin{remark}
Again for clarity, we have made two exceptions to denoting specimen by $\mathrm{X}$. External specimen induced by Quasi-Steady State Transformations and Kowalski Transformations (Sections~\ref{QSSA},~\ref{ForcedVariable}) are denoted by $\mathrm{Y}$, and fast specimen introduced by Quasi-Steady State Approximations (Section~\ref{quasisteadystatetransformation}) are denoted $\mathrm{Z}$.
\end{remark}

\noindent For example, consider $\mathcal{R} = \{(0,0)^\top \rightarrow (1,0)^\top, (1,0)^\top \rightarrow (1,1)^\top\} = \{1,2 \}$,
\begin{equation}\label{veryeasysystem0}
    r_1: \quad \varnothing \stackrel{ }{\longrightarrow} \mathrm{X}_1 , \quad r_2 : \quad \mathrm{X}_1 \stackrel{ }{\longrightarrow} \mathrm{X}_1 + \mathrm{X}_2.
\end{equation}
From $\mathcal{R}$, we determine that $\mathcal{X} = \{\mathrm{X}_1,\mathrm{X}_2\}$ with the null specimen $\varnothing$ additionally added. Clearly, $\mathcal{V} = \{(0,0)^\top, (1,0)^\top, (1,1)^\top\}$. Now assume that the reactions $r_1, r_2$ possess reaction coefficients $k_1, k_2$. The Law of Mass Action (Appendix~\ref{appendixA}) gives the dynamics of this system as
\begin{equation}\label{veryeasysystem}
\begin{aligned}
&\Dot{x}_1 =k_1, \\
&\dot{x}_2=k_2 x_1,
\end{aligned}
\end{equation}
where $x_i := [\mathrm{X}_i]$ represent specimen concentrations. We wish to formalize the concept of such biochemically realizable ODE systems, which we call \textit{(mass-action) kinetic systems}. 

\begin{mdframed}
\begin{definition}[Reaction Rate Equations and Kinetic Systems]
Let $\mathcal{R}$ be a network. The \textit{rate vector} is defined by
\begin{equation}
\mathbf{k}:=\left(k_1,\dots,k_{|\mathcal{R}|}\right) \end{equation}
where $k_i \in \mathbb{R}$ are called the \textit{rate coefficients} or \textit{rate constants}. Given reaction $r_j : \boldsymbol{\nu}_j^- \xrightarrow{} \boldsymbol{\nu}_j^+$ indexed by $j \in \mathcal{R}$, we define the \textit{rate representation} $\kappa :  \mathcal{R} \times \mathbb{R}^{\mathcal{X}} \to \mathbb{R}^{\mathcal{X}}$ as
\begin{equation}\label{raterepresentation}
    \kappa(j,\mathbf{x}) = k_j \left( \boldsymbol{\nu}_j^+-\boldsymbol{\nu}_j^-\right) \mathbf{x}^{\boldsymbol{\nu}_j^-}, \quad\text{where} \quad \mathbf{x}^{(\nu_{1j}^-,\dots,\nu_{Nj}^-)^\top} := \prod_{i = 1}^N x_i^{\nu_{ij}^-}.
\end{equation}
As is standard, we follow the convention that $0^0 = 1$. 

A pair $(\mathcal{R},\mathbf{k})$ is called a \textit{(mass-action) kinetic system} if $k_j > 0$, $\forall j \in  \mathcal{R}$. $\mathbf{x}$ is interpreted as the concentrations of chemical specimen, and the concentration dynamics are determined by the \textit{Reaction Rate Equations (RREs)}
\begin{equation}\label{kineticsystem}
    \dot{\mathbf{x}}=\mathcal{K}(\mathbf{x} ; \mathcal{R}) := \sum_{j \in \mathcal{R}} \kappa (j,\mathbf{x}) = \sum_{j \in \mathcal{R}} k_j \left( \boldsymbol{\nu}_j^+-\boldsymbol{\nu}_j^-\right) \mathbf{x}^{\boldsymbol{\nu}_j^-}.
\end{equation}
Following established naming convention, we call $\boldsymbol{\nu}_j := \boldsymbol{\nu}_j^+-\boldsymbol{\nu}_j^-$ the \textit{state change vector} of reaction $j$ and $\boldsymbol{\nu}$ the  \textit{Stoichiometry matrix} with $j$-th column $\boldsymbol{\nu}_j$, whose $(i,j)$-th entry is denoted $\nu_{ij}$. 
\end{definition}
\end{mdframed}
In particular, note that the RREs~\eqref{kineticsystem} express~\eqref{veryeasysystem} in the case~\eqref{veryeasysystem0}. Furthermore, it should be emphasized that we abstractly allow rate coefficients to be negative. 

\section{Cross-Negative Terms and Kinetic Transformations}\label{crossnegativesection}

Denote the set of all $|\mathcal{X}|$-dimensional polynomials with degree less than or equal to $m$ as
\begin{equation}
    \mathbb{P}_{m}\left(\mathbb{R}^{\mathcal{X}} ; \mathbb{R}^{\mathcal{X}}\right) := \left\{\mathcal{P}(\mathbf{x}): \mathbb{R}^{\mathcal{X}} \to \mathbb{R}^{\mathcal{X}},\  \operatorname{deg}(\mathcal{P}(\mathbf{x})) \leq m \right\}.
\end{equation}
In particular, an $|\mathcal{X}|$-dimensional polynomial ODE system of degree $m$ may be written 
\begin{equation}\label{polynomialrighthandside}
    \Dot{\mathbf{x}} = \mathcal{P}(\mathbf{x}),
\end{equation}
where $\mathcal{P}(\mathbf{x}) \in \mathbb{P}_{m}\left(\mathbb{R}^{\mathcal{X}} ; \mathbb{R}^{\mathcal{X}}\right)$. If the polynomial right hand side can be identified as RREs of a mass-action kinetic network, it is called a \textit{kinetic function}, and if no corresponding mass-action kinetic network exists, a \textit{non-kinetic function}. ODE systems defined by non-kinetic functions depict trajectories which cannot be realized in chemical reaction networks. For example, consider the simple system
\begin{equation}\label{easysystem}
\begin{aligned}
&\Dot{x_1}=-1 \\
&\Dot{x_2}=k_2 x_1.
\end{aligned}
\end{equation}
The culprit responsible for non-kineticness is the term $-1$, which drives the concentration of specimen $\mathrm{X}_1$ negative even when $x_1 = 0$. Such dynamics cannot be realized in biochemical systems as concentrations are necessarily non-negative. We now categorically define terms responsible for non-kineticness in a polynomial ODE system, called \textit{cross-negative terms}.

\begin{mdframed}
\begin{definition}[Cross-Negative Terms]\label{crossnegative}
For the polynomial right hand side $\mathcal{P}_s(\mathbf{x})$ of the $s$-coordinate of~\eqref{polynomialrighthandside}, i.e. the right hand side of $\Dot{x_s}$, consider the restricted polynomial \\ $\mathcal{P}_s(x_1,\dots,x_{s-1},0,x_{s+1},\dots, x_n)$ with $x_s = 0$ and all other input variables untouched. All terms in the original equation $\mathcal{P}_s(\mathbf{x})$ which possess negative coefficients in the restricted polynomial are called \textit{cross-negative terms}.
\end{definition}
\end{mdframed}

\noindent Due to being a nuisance in the inversion of polynomial systems into chemical reaction networks, methods have been proposed in the literature to rid cross-negative terms while preserving dynamics (see Sections~\ref{QSSA},~\ref{ForcedVariable}). Once cross-negative terms have been removed, we may perform a \textit{canonical inversion} into a network whose RREs recover the ODE system.

\begin{mdframed}
\begin{definition}[Canonical Inversion]
 Consider the kinetic system 
\begin{equation}\label{morekineticsystem}\tag{\ref{kineticsystem}}
        \dot{\mathbf{x}}=\mathcal{K}(\mathbf{x} ; \mathcal{R}) = \sum_{j \in \mathcal{R}} k_j \left( \boldsymbol{\nu}_j^+-\boldsymbol{\nu}_j^-\right) \mathbf{x}^{\boldsymbol{\nu}_j^-} = \sum_{j \in \mathcal{R}} k_j \left(\begin{array}{c}
\nu_{1j} \\
\vdots \\
\nu_{Nj}
\end{array}\right) \mathbf{x}^{\boldsymbol{\nu}_j^-},
\end{equation}
where $N = |\mathcal{X}|$ and $\boldsymbol{\nu}$ is the Stoichiometry matrix. Every term $k_j \nu_{ij}x_1^{\nu_{1j}^-}\dots x_N^{\nu_{Nj}^-}$ that appears in the right hand side for $\nu_{ij} \neq 0$ can be \textit{canonically inverted} into the reaction
\begin{equation}\label{canonicalinversion}
    r_{ij}: \quad \nu_{1j}^- \mathrm{X}_1 + \dots + \nu_{Nj}^- \mathrm{X}_N \stackrel{k_j \norm{\nu_{ij}}}{\longrightarrow} \nu_{1j}^- \mathrm{X}_1 + \dots + \left(\nu_{ij}^- + \operatorname{sign}(\nu_{ij})\right) \mathrm{X}_i + \dots + \nu_{Nj}^- \mathrm{X}_N,
\end{equation}
which produces a network of at most $|\mathcal{X}\times\mathcal{R}|$ reactions.
\end{definition}
\end{mdframed}

\noindent Finally, within our framework of allowing rate coefficients to be negative, the system~\eqref{easysystem} may be non-chemically realized (by imitating the derivation of RREs for kinetic systems) as
\begin{equation}
    r_1: \quad \varnothing \stackrel{-1}{\longrightarrow} \mathrm{X}_1 , \quad r_2 : \quad \mathrm{X}_1 \stackrel{k_2}{\longrightarrow} \mathrm{X}_1 + \mathrm{X}_2.
\end{equation}
We will call such an abstract inversion, which has no biochemical interpretation, a \textit{non-kinetic inversion}. 

Note that every cross-negative term may be inverted in this manner. If $\nu_{ij}$ corresponds to a cross-negative term in~\eqref{morekineticsystem}, we flip the sign of $\nu_{ij}$ and perform the canonical inversion, and then multiply the resulting rate coefficient with $-1$. 

\begin{mdframed}
\begin{definition}[Non-kinetic Canonical Inversion]\label{noncanonicalinversiondefinition}
Consider an $|\mathcal{X}|$-\\dimensional polynomial ODE system 
\begin{equation}\label{properlyreferencedsystem}
        \dot{\mathbf{x}}=\mathcal{P}(\mathbf{x}) = \sum_{j = 1}^{M} k_j \left(\begin{array}{c}
\nu_{1j} \\
\vdots \\
\nu_{Nj}
\end{array}\right) \mathbf{x}^{\boldsymbol{\nu}_j^{ -}},\quad \text{for } N = |\mathcal{X}|,
\end{equation}
where $k_j \in \mathbb{R}$, $\nu_{ij} \in \mathbb{Z}$, $\boldsymbol{\nu}_{j}^{ - } \in \mathbb{Z}_{\ge 0}^\mathcal{X}$, and $M$ is sufficiently large. Assume the existence of a cross-negative term. All kinetic terms are inverted canonically into the reaction~\eqref{canonicalinversion}, and all cross-negative terms\footnote{Thereby implying $\boldsymbol{\nu}_{ij} \neq 0$ by definition.} are inverted abstractly into the reaction 
\begin{equation}\label{nonkineticcanonicalinversion}
    r_{ij}: \quad \nu_{1j}^{-} \mathrm{X}_1 + \dots + \nu_{Nj}^{-} \mathrm{X}_N \stackrel{k_j \nu_{ij}}{\longrightarrow} \nu_{1j}^{-} \mathrm{X}_1 + \dots + \left(\nu_{ij}^{-} + 1\right) \mathrm{X}_i + \dots + \nu_{Nj}^{-} \mathrm{X}_N.
\end{equation}
The resulting chemically non-realizable network with negative reaction coefficients is called a \textit{non-kinetic canonical inversion}.
\end{definition}
\end{mdframed}
In particular, every polynomial ODE system induces a reaction network $\mathcal{R}$ paired with a rate representation $\kappa$ as in~\eqref{raterepresentation}, via a canonical inversion or a non-kinetic canonical inversion. The non-kinetically inverted network must have a negative rate coefficient.

\section{Quasi-Steady State Transformation (QSST)}\label{QSSA}
Now, we present a kineticization technique described by Plesa et al~\cite{TomiInversePaper}.

\begin{mdframed}
\begin{definition}[Quasi-Steady State Transformation]\label{QSSAnewdefinition}
Consider an $|\mathcal{X}|$- \\ dimensional polynomial ODE system
\begin{equation}\tag{\ref{polynomialrighthandside}}
    \Dot{\mathbf{x}} = \mathcal{P}(\mathbf{x}),
\end{equation}
and its (possibly non-kinetic) canonical inversion network $\mathcal{R}$. Given any specimen $\mathrm{X}_s \in \mathcal{X}$, disjointly partition $\mathcal{R} = \mathcal{R}_1^s \cup \mathcal{R}_2^s$ such that $r \in \mathcal{R}_1^s$ are kinetic and $r \in \mathcal{R}_2^s$ are non-kinetic. We rewrite the system~\eqref{polynomialrighthandside} as
\begin{equation}\label{QSSAoriginalequation}
\begin{aligned}
\Dot{x_{s}} &=\sum_{j \in \mathcal{R}^s_{1}} k_j\nu_{s j} \mathbf{x}^{\boldsymbol{\nu}_j^-}-\sum_{j^\prime \in \mathcal{R}^s_{2}} \norm{k_{j^\prime}\nu_{s j^\prime}} \mathbf{x}^{\boldsymbol{\nu}_{j^\prime}^-}, \quad \text {for } \mathrm{X}_s \in \mathcal{X}, \\
x_{s}&\left(t_{0}\right) \ge 0, \quad t_0 \text{ initial time},
\end{aligned}
\end{equation}
where $\mathbf{k}$ is the rate vector, $\boldsymbol{\nu}$ is the Stoichiometry matrix, and $\boldsymbol{\nu}_j^-$ are the stoichiometric coefficients of reactants of channel $j$. Disjointly partition $\mathcal{X} = \mathcal{X}_1 \cup \mathcal{X}_2$ where $\mathcal{X}_1$ contains kinetic species (i.e. $\mathrm{X}_s \in \mathcal{X}$ such that the right hand side of $\dot{x_s}$ is kinetic) and $\mathcal{X}_2$ non-kinetic species, respectively. We further enforce\footnote{Doing so allows for dynamics to be preserved in the transformation. If $x_s = 0$,~\eqref{QSSAIntroducedSystem} incites a fast blow-up in $y_s$ as $\mu \downarrow 0$.} the initial condition $x_s(t_0) > 0$ for $\mathrm{X}_s \in \mathcal{X}_2$. Then, the \textit{degenerate system} is given by 
\begin{equation}
\begin{aligned}
\Dot{x_{s}} =\sum_{j \in \mathcal{R}^s_{1}} k_j\nu_{s j} \mathbf{x}^{\boldsymbol{\nu}_j^-}-\sum_{j^\prime \in \mathcal{R}^s_{2}} \norm{k_{j^\prime}\nu_{s j^\prime}} \mathbf{x}^{\boldsymbol{\nu}_{j^\prime}^-}, & \quad \text {for } \mathrm{X}_s \in \mathcal{X}_{1}, \\
\Dot{x_{s}}=\sum_{j \in \mathcal{R}^s_{1}} k_j\nu_{s j} \mathbf{x}^{\boldsymbol{\nu}_j^-} -\omega_{s}^{-1} x_{s} p_{s}(\mathbf{x}) y_{s} \left(\sum_{j^\prime \in \mathcal{R}^s_{2}} \norm{k_{j^\prime}\nu_{s j^\prime}} \mathbf{x}^{\boldsymbol{\nu}_{j^\prime}^-}\right) , & \quad \text {for } \mathrm{X}_s \in \mathcal{X}_{2},
\end{aligned}
\end{equation}
which satisfies all aforementioned initial conditions, with an \textit{adjoined system} for $\mathrm{X}_s \in \mathcal{X}_2$ given by
\begin{equation}\label{QSSAIntroducedSystem}
    \begin{aligned}
    \mu \Dot{y_{s}} &=\omega_{s}-x_{s} p_{s}(\mathbf{x}) y_{s}, \\
    y_{s}\left(t_{0}\right) &\geq 0, \quad t_0 \text{ initial time.}
    \end{aligned}
\end{equation}
Here $\mu, \omega_s \in \mathbb{R}_{>0}$, $y_s(t_0)$ may be \textit{any} non-negative value, and the polynomial $p(\mathbf{x})$ with $s$-coordinate $p_s(\mathbf{x})$ must map the non-negative set $\mathbb{R}_{\ge 0}^\mathcal{X}$ into the positive cone $\mathbb{R}_{>0}^{\mathcal{X}_2}$. The complete system composed of the degenerate and adjoined systems is called the \textit{general system}. Then, the map\footnote{Note the usage of disjoint unions and the induced canonical identification $\mathcal{X} \subset \mathcal{X} \sqcup \mathcal{X}_2$. These concepts are concisely summarized in the List of Symbols and Notation page (v) for readers who are unfamiliar.} $\Psi_{QSST} :  \mathbb{P}_{m}\left(\mathbb{R}^{\mathcal{X}} ; \mathbb{R}^{\mathcal{X}}\right) \rightarrow$ $\mathbb{P}_{m^\prime}\left(\mathbb{R}^{\mathcal{X}\sqcup\mathcal{X}_{2}} ; \mathbb{R}^{\mathcal{X}\sqcup\mathcal{X}_{2}}\right)$ which maps the right hand side of~\eqref{polynomialrighthandside} to the right hand side of the general system is called the \textit{Quasi-Steady State Transformation}. 
\end{definition}
\end{mdframed}
We note that the general system is kinetic, that is, $\Psi_{QSST}(\mathcal{P}(\mathbf{x}))$ may be canonically inverted. The power of the Quasi-Steady State Transformation comes from the following miraculous theorem:

\begin{theorem}\label{QSSTtheorem}
The general system induced by $\Psi_{QSST}$ is asymptotically equivalent to~\eqref{polynomialrighthandside} in the limit $\mu \downarrow 0$, given the agreement of the initial conditions posed in Definition~\ref{QSSAnewdefinition}.
\end{theorem}
\noindent See Appendix~\ref{ProofImportant} for an extended proof, which follows the concise proof given in~\cite{TomiInversePaper}. The essence is the application of Tikhonov’s theorem~\cite{QSSATomiTheoryTichnoff} (Theorem~\ref{TikhonovTheoreminAppendix}) to recover dynamics of the original ODE system in the limiting case. In Chapters~\ref{oldchapter5} and~\ref{oldchapter6},
$\Psi_{QSST}$ will become our prime kineticization strategy with $\mu = 10^{-6}$ and $\omega_s, p_s(\mathbf{x}) \equiv 1$ unless stated otherwise. Though not explicitly mentioned in Definition~\ref{QSSAnewdefinition}, we emphasize that for any $\mathrm{X}_s \in \mathcal{X}_2$, observing $x_s(t_1) = 0$ for $t_1 \ge t_0$ immediately breaks the dynamical equivalence, as the original ODE system may be reinitialized at $t=t_1$ and propagated onward, violating a core assumption of $\Psi_{QSST}$.

\section{Kowalski Transformation}\label{ForcedVariable}
Kowalski~\cite{KowalskiWeak} provides another method of eliminating cross-negative terms in polynomial ODE systems, which is analytically simpler. 

\begin{mdframed}
\begin{definition}[Kowalski Transformation]\label{kowalskidefinition}
Consider an identical setup to Definition~\ref{QSSAnewdefinition}, wherein an arbitrary polynomial ODE system may be rewritten:
\begin{equation}\label{systemtokowalskitransform}
\begin{aligned}
\Dot{x_{s}} &=\sum_{j \in \mathcal{R}^s_{1}} k_j\nu_{s j} \mathbf{x}^{\boldsymbol{\nu}_j^-}-\sum_{j^\prime \in \mathcal{R}^s_{2}} \norm{k_{j^\prime}\nu_{s j^\prime}} \mathbf{x}^{\boldsymbol{\nu}_{j^\prime}^-}, \quad \text{for } \mathrm{X}_s \in \mathcal{X}, \\
x_{s}\left(t_{0}\right) &\ge 0, \quad \text{for } \mathrm{X}_s \in \mathcal{X}_1, \quad \text{and }  x_{s}\left(t_{0}\right) > 0, \quad \text{for } \mathrm{X}_s \in \mathcal{X}_2.
\end{aligned}
\end{equation}
Note in particular that we have carried over the definitions of $\mathcal{X}_1$, $\mathcal{X}_2$ and imposed analogous initial conditions. We introduce the variables $y_s = 1/x_s$ for $\mathrm{X}_s \in \mathcal{X}_2$, which gives the \textit{degenerate system}  
\begin{equation}\label{Kowalskidegenerate}
\begin{aligned}
\Dot{x_{s}} =\sum_{j \in \mathcal{R}^s_{1}} k_j\nu_{s j} \mathbf{x}^{\boldsymbol{\nu}_j^-}-\sum_{j^\prime \in \mathcal{R}^s_{2}} \norm{k_{j^\prime}\nu_{s j^\prime}} \mathbf{x}^{\boldsymbol{\nu}_{j^\prime}^-}, & \quad \text {for } \mathrm{X}_s \in \mathcal{X}_{1}, \\
\Dot{x_{s}}=\sum_{j \in \mathcal{R}^s_{1}} k_j\nu_{s j} \mathbf{x}^{\boldsymbol{\nu}_j^-} - x_{s} y_{s} \left(\sum_{j^\prime \in \mathcal{R}^s_{2}} \norm{k_{j^\prime}\nu_{s j^\prime}} \mathbf{x}^{\boldsymbol{\nu}_{j^\prime}^-}\right) , & \quad \text {for } \mathrm{X}_s \in \mathcal{X}_{2},
\end{aligned}
\end{equation}
which satisfies the initial condition given in~\eqref{systemtokowalskitransform}. The \textit{adjoined system} to $\mathrm{X}_s \in \mathcal{X}_2$ is given by
\begin{equation}\label{Kowalskiadjoined}
    \begin{aligned}
    \Dot{y_{s}} = -\frac{1}{x_s^2}\cdot \frac{\mathrm{d} x_{s}}{\mathrm{d} t} &= -y_s^2\left(\sum_{j \in \mathcal{R}^s_{1}} k_j\nu_{s j} \mathbf{x}^{\boldsymbol{\nu}_j^-}-\sum_{j^\prime \in \mathcal{R}^s_{2}} \norm{k_{j^\prime}\nu_{s j^\prime}} \mathbf{x}^{\boldsymbol{\nu}_{j^\prime}^-}\right), \\
    y_{s}\left(t_{0}\right) &= \frac{1}{x_{s}\left(t_{0}\right)}, \quad t_0 \text{ initial time.}
    \end{aligned}
\end{equation}
The complete system composed of the degenerate and adjoined systems is called the \textit{general system}. In particular, note that the general system is mass-action kinetic and may be canonically inverted. The map $\Psi_{Kow} :  \mathbb{P}_{m}\left(\mathbb{R}^{\mathcal{X}} ; \mathbb{R}^{\mathcal{X}}\right) \rightarrow$ $\mathbb{P}_{m^\prime}\left(\mathbb{R}^{\mathcal{X} \sqcup \mathcal{X}_{2}} ; \mathbb{R}^{\mathcal{X}\sqcup \mathcal{X}_{2}}\right)$ which maps the right hand side of~\eqref{systemtokowalskitransform} to the right hand side of the general system is called the \textit{Kowalski Transformation}. 
\end{definition}
\end{mdframed}

We end this section with a warning. Although $\Psi_{Kow}$ preserves deterministic dynamics within the solution manifold $y_s(t) = 1 / x_s(t)$, there is no guarantee that the preservation will translate to stochastic realizations in which noise dislocates trajectories from the intended manifold. Without stability in the newly established dimensions $y_s$ or any sort of asymptotic preservation, there is no expectation that the topology of the transformed system will match that of the original system near fixed points, despite $\Psi_{Kow}$ allowing for a canonical inversion of the general system. Although deterministically valid, it remains an open question if $\Psi_{Kow}$ is useful for synthetically implementing ODE systems within wet lab settings.

%% file: chapter3.tex



%% file: chapter4.tex
\chapter{Polynomialization, Bimolecularization, and Inversion}\label{oldchapter3}

Kerner~\cite{KernerSleightofHand} has shown that very general nonlinear ODE systems may be reduced to so-called Riccati systems, whose right hand side is of polynomial form which has been quadraticized. That is, even non-autonomous ODE systems comprising of almost ``all cases which arise in practice'' 
\begin{equation}\label{verygeneralnonautonomoussystem}
    \Dot{x_{i}} = f_{i}\left(x_{1}, x_{2}, \ldots, x_{N}, t\right), \quad i\in \{1,2, \ldots, N\},
\end{equation}
may be represented by the Riccati system
\begin{equation}\label{Riccati}
    \Dot{\eta_{i}}=A^{i}+\left(\sum_{\alpha \in \{1,\dots,N^\prime \}} B^i_{\alpha} \eta_{\alpha}\right) +\left(\sum_{\alpha, \beta \in \{1,\dots,N^\prime \}} C^i_{\alpha \beta} \eta_{\alpha} \eta_{\beta}\right), \quad i \in \{1,2, \ldots, N^\prime\},
\end{equation}
where $N^\prime\ge N$, $f_i : \mathbb{R}^{N+1} \to \mathbb{R}$, $A^{i},B^i_{\alpha}, C^i_{\alpha \beta} \in \mathbb{R}$. The first step in the reduction to Riccati form is to identify the time variable $t$ as a new variable $x_{N+1}$, whose dynamics are determined by $\dot{x}_{N+1} = 1$. This allows~\eqref{verygeneralnonautonomoussystem} to be subsumed by
\begin{equation}\label{verygeneralautonomoussystem}
    \Dot{x_{i}}=f_{i}\left(x_{1}, x_{2}, \ldots, x_{N}, x_{N+1}\right), \quad i\in \{1,2, \ldots, N+1\},
\end{equation}
which implies that consideration of ODE systems in the autonomous form is sufficient for the chemical simulation of arbitrary ODE systems.

\section{Kerner Polynomialization}\label{quadraticization}
Kerner does not provide a formal algorithm for reducing the system~\eqref{verygeneralautonomoussystem} and instead resorts to explaining the methodology via examples. It is indeed difficult to define a rigorous algorithm for his procedure, which essentially relies on recursively identifying differentiable functions as new variables in the ODE system. Differentiating the introduced variables may yield functions previously unobserved within the system, and may further increase its nonlinearity. It is heuristically assumed that for most ODE systems observed in practice, this iterative process of ``differentiating-away~\cite{KernerSleightofHand}'' will eventually terminate.

This procedure works for systems composed of elementary functions such as exponential, hyperbolic, and trigonometric functions, as well as elliptic and Bessel functions, which arise frequently in mechanical models of complex processes.

\begin{example}
 Consider the system
\begin{equation}\label{naturalvariablesample}
    \begin{aligned}
    &\Dot{x_1} = \cos(x_1)\cos(x_2),\\
    &\Dot{x_2} = \exp(x_1) \exp(x_2),
    \end{aligned}
\end{equation}
where we introduce the variables 
\begin{equation}
    x_3 = \cos(x_1),\quad x_4 = \cos(x_2), \quad x_5 = \exp (x_1), \quad x_6 = \exp (x_2).
\end{equation}
Substitution gives
\begin{equation}
    \begin{aligned}
    &\Dot{x_1} = x_3x_4,\quad\Dot{x_2} = x_5x_6,\quad\Dot{x_3} = -\sin(x_1)\Dot{x_1} = -\sin(x_1) x_3x_4 , \\ 
    &\Dot{x_4} = - \sin(x_2) x_5x_6, \quad \Dot{x_5} = \exp(x_1)\Dot{x_1} = x_3x_4x_5, \quad \Dot{x_6} =  x_5x_6^2 .
    \end{aligned}
\end{equation}
Further introducing the variables $x_7 = \sin(x_1)$, $x_8 = \sin(x_2)$ gives the third degree polynomial system
\begin{equation}
    \begin{aligned}
    \Dot{x_1} &= x_3x_4,\quad\Dot{x_2} = x_5x_6,\quad \Dot{x_3} = -x_3x_4x_7 , \quad \Dot{x_4} = - x_5x_6x_8,\\ 
     &\Dot{x_5} = x_3x_4x_5, \quad \Dot{x_6} =  x_5x_6^2, \quad \Dot{x_7} = x_3^2x_4, \quad \Dot{x_8} =  x_4x_5x_6.
    \end{aligned}
\end{equation}
\end{example}

\section{General Quadraticization Algorithm}\label{generalquadraticizationalgorithm}

Once the system has been reduced to polynomial form of degree $m$ with $N$-many variables $x_1, \dots, x_N$, we rewrite the system as
\begin{equation}\label{generalpolynomialsystem}
    \dot{x}_{h}=\sum_{\ell \in \{0, \dots, m\}} \Lambda_h^\ell :=\Lambda_h^0 + \sum_{\ell \in \{1, \dots, m\}}\left( \sum_{\substack{c_j \in \{1, \dots, N\}, \\ \forall j \in \{1, \dots, \ell \}}}  \Phi^{\ell,h}_{c_1 \dots c_\ell} \left(\prod_{j \in \{1, \dots, \ell \} } x_{c_j} \right) \right), \quad \Phi^{\ell,h}_{c_1 \dots c_\ell} \in \mathbb{R}, 
\end{equation}
for $h \in \{1, \dots, N\}.$ This describes an arbitrary multivariate polynomial system whose right hand side has degree at most $m$, where the summation inside the brackets is taken over all vectors $(c_{1}, \dots, c_{\ell})^\top$ whose every entry is an element of the set $\{1,\dots,N \}$. Noting that elements of the polynomial ring $\mathbb{R}[x]$ are commutative, we identify $\Phi^{\ell,h}_{c_1 \dots c_\ell}  = \Phi^{\ell,h}_{c_1^\prime \dots c_\ell^\prime} $ if and only if $\{c_1,\dots ,c_\ell\} = \{c_1^\prime,\dots ,c_\ell^\prime\}$ with no ordering, where repeated elements are tracked in the set. That is, for $\ell = 3$ and $N = 2$, we have that $\Phi^{3,h}_{1,1,2} = \Phi^{3,h}_{2,1,1}$ as both $\{1,1,2\}$ and $\{2,1,1\}$ have two $1$'s and one $2$. Here, commas have been used in the indices of $\Phi$ to emphasize that $c_{1}, c_2, c_3 \in \{1,2\}$. 

For any $c_{j^\prime}, c_{j^{\prime\prime}} \in \{1, \dots, N\}$, we will now define $\eta_{c_{j^\prime}c_{j^{\prime\prime}}}:=x_{c_{j^\prime}}x_{c_{j^{\prime \prime}}}$ and differentiate away the monomials of degree greater than $2$ in an iterative manner. This implies that in every iteration, we reduce the degree of the first two variable multiples $x_{c_1} x_{c_2}$ of each summand of $\Lambda^\ell_h$ for $\ell \ge 2$ by $1$, given that $\Phi^{\ell,h}_{c_1 \dots c_\ell}$ is non-zero. To take an example for $\ell = 2$, we have
\begin{equation}
    \Lambda_h^2 =  \sum_{\substack{c_j \in \{1, \dots, N\}, \\\forall j \in \{1, 2 \}}}  \Phi^{2,h}_{c_1  c_2}  x_{c_1} x_{c_2} = \sum_{\substack{c_j \in \{1, \dots, N\}, \\ \forall j \in \{1, 2 \}}}  \Phi^{2,h}_{c_1  c_2}  \eta_{c_1 c_2},
\end{equation}
and for $\ell = m$, we have
\begin{equation}
    \Lambda_h^m = \sum_{\substack{c_j \in \{1, \dots, N\}, \\ \forall j \in \{1, \dots, m \}}}  \Phi^{m,h}_{c_1 \dots c_m}  x_{c_1}\dots x_{c_m} = \sum_{\substack{c_j \in \{1, \dots, N\}, \\ \forall j \in \{1, \dots, m \}}}  \Phi^{m,h}_{c_1 \dots c_m}  \eta_{c_1 c_2}x_{c_3}\dots x_{c_m}.
\end{equation}
The highest degree term in $\Lambda^{m}_h$ is $\eta_{c_1 c_2}x_{c_3}\dots x_{c_m}$, which is of degree $m-1$. To avoid confusion, it may be noted that the ordering of an $\ell$-degree term into $x_{c_1}\dots x_{c_\ell}$ need not be specified to prove the termination of this quadraticization algorithm. It may also be noted that if $\Phi^{\ell,h}_{c_1 \dots c_\ell}$ is zero for some choice of the indices (in particular for $\ell = m$) it is possible for the non-linearity to increase temporarily after an iteration, but we will now see that this also does not impact the termination of the algorithm in a finite number of iterations.

Differentiating $\eta_{c_{j^\prime} c_{j^{\prime \prime}}}$ gives
\begin{equation}\label{importantconsideration}
    \dot{\eta}_{c_{j^\prime} c_{j^{\prime \prime}}} = \dot{x}_{c_{j^\prime}}x_{c_{j^{\prime \prime}}} + x_{c_{j^\prime}}\dot{x}_{c_{j^{\prime \prime}}}.
\end{equation}
By symmetry with respect to indices $j^\prime$, $j^{\prime \prime}$, we consider the term $\dot{x}_{c_{j^\prime}}x_{c_{j^{\prime \prime}}}$ without loss of generality:
\begin{equation}
    \dot{x}_{c_{j^\prime}}x_{c_{j^{\prime \prime}}} =  \left(\Phi^{0,j^\prime} + \sum_{\substack{c_1 \in \{1, \dots, N\}}}  \Phi^{1,j^\prime}_{c_1}  x_{c_1}  +\dots +  \sum_{\substack{c_j \in \{1, \dots, N\}, \\ \forall j \in \{1, \dots, m \}}}  \Phi^{m,j^\prime}_{c_1 \dots c_m}  x_{c_1}\dots x_{c_m}\right)x_{c_{j^{\prime \prime}}}.
\end{equation}
In each term of the right hand side of $ \dot{x}_{c_{j^\prime}}$, we identify the first two $x$-variables from the left as $\eta$-variables:
\begin{equation}
    \left(\Phi^{0,j^\prime} + \sum_{\substack{c_1 \in \{1, \dots, N\}}}  \Phi^{1,j^\prime}_{c_1}  x_{c_1} + \dots + \sum_{\substack{c_j \in \{1, \dots, N\}, \\ \forall j \in \{1, \dots, m \}}}  \Phi^{m,j^\prime}_{c_1 \dots c_m} \eta_{c_1,c_2} x_{c_3}\dots x_{c_m}\right)x_{c_{j^{\prime \prime}}}.
\end{equation}
Despite commutativity, we strictly right-multiply the terms inside the brackets by $x_{c_{j^{\prime \prime}}}$ and identify the first two $x$-variables from the right as $\eta$-variables, that is,
\begin{equation}
    \Phi^{0,j^\prime} x_{c_{j^{\prime \prime}}} + \sum_{\substack{c_1 \in \{1, \dots, N\}}}  \Phi^{1,j^\prime}_{c_1}  x_{c_1} x_{c_{j^{\prime \prime}}} +\dots +  \sum_{\substack{c_j \in \{1, \dots, N\}, \\ \forall j \in \{1, \dots, m \}}}  \Phi^{m,j^\prime}_{c_1 \dots c_m} \eta_{c_1,c_2} x_{c_3}\dots x_{c_m}x_{c_{j^{\prime \prime}}}
\end{equation}
\begin{equation}\label{complicatedequation}
    = \Phi^{0,j^\prime} x_{c_{j^{\prime \prime}}} + \sum_{\substack{c_1 \in \{1, \dots, N\}}}  \Phi^{1,j^\prime}_{c_1}  \eta_{c_1 c_{j^{\prime \prime}}} +\dots +  \sum_{\substack{c_j \in \{1, \dots, N\}, \\ \forall j \in \{1, \dots, m \}}}  \Phi^{m,j^\prime}_{c_1 \dots c_m} \eta_{c_1,c_2} x_{c_3}\dots x_{c_{m-1}}\eta_{c_m c_{j^{\prime \prime}}}.
\end{equation}
The highest degree term has the variables $\eta_{c_1,c_2} x_{c_3}\dots x_{c_{m-1}}\eta_{c_m c_{j^{\prime \prime}}}$, which is of degree $m-1$. We remark that when the summand $x_{c_{j^\prime}}\dot{x}_{c_{j^{\prime \prime}}}$ is considered in equation~\eqref{importantconsideration}, the algorithm proceeds in reverse direction; that is, we strictly left-multiply by $x_{c_{j^\prime}}$ to expand the brackets and introduce the variable $\eta_{c_{j^\prime} c_1}$. This process is illustrated in the sample cubic~\eqref{translatedvariable1}.

As~\eqref{complicatedequation} is again of polynomial form, we may repeat this process, reducing the degree of the highest degree term by $1$ at each iteration. The algorithm terminates once all terms in the right hand side have been quadraticized, resulting in the Riccati system~\eqref{Riccati}. Note that the zeroth iteration starts off with $N$-many variables $x_1, \dots, x_N$, and the first iteration introduces at most ${}_{N+1} C_2$-many variables $\eta_{j^\prime j^{\prime \prime}}$. 

As each iteration decreases the degree of the polynomial system by $1$, we may cycle through $m-2$ iterations at maximum until complete quadraticization. Therefore, a loose upper bound for the total number of variables $N_{\mathrm{itr}} \ge N^\prime$ (see~\eqref{Riccati}) is given by the recursive formula
\begin{equation}\label{weakupperbound}
    N_{\mathrm{itr}} = N_{(\mathrm{itr} - 1)} + {}_{N_{(\mathrm{itr} - 1)}+1} C_2, \quad N_0 = N, \quad \text{for} \quad \mathrm{itr} \in \{1, \dots, m-2 \},
\end{equation}
where $\mathrm{itr}$ is the iteration number. 

Finally, we must verify that the resulting system is capable of subsuming the dynamics of the original system. This is done by imposing the initial conditions
\begin{equation}\label{icthing}
    \eta_{c_{j^\prime}c_{j^{\prime\prime}}}=x_{c_{j^\prime}}x_{c_{j^{\prime \prime}}} \quad \text{at } t = t_0,
\end{equation}
at every iteration. Then, differentiation of $\eta_{c_{j^\prime}c_{j^{\prime\prime}}}-x_{c_{j^\prime}}x_{c_{j^{\prime \prime}}}$ with respect to $t$ shows that the identity~\eqref{icthing} is maintained for all times $t \ge t_0$. A step-by-step application of the algorithm on a simple cubic is given as Example~\ref{easyexamplequadraticize}.

Note that there is no guarantee that this algorithm will provide the optimal quadraticization, neither in the number of iterations required nor in the number of variables in the quadraticized system. However, it is entirely constructive and proves the existence of a quadraticization algorithm. Furthermore, we imposed an ordering on variable multiples (e.g. $x_{c_1}\dots x_{c_m}$) despite commutativity. This is a technicality used to simplify the proof of algorithm termination, and need not be enforced during implementation.

From a biochemical standpoint, reactions resulting from a collision of more than two molecules are very rare. Occasionally reactions with three reactants \cite{Wilhelm} are described, but they are more accurately approximations of a hidden bimolecular network. Therefore, ODE systems describing concentration dynamics of reported reaction networks in biochemical applications are at most quadratic, and quadraticization is a powerful tool which assists us in reducing any polynomial ODE system into a form realizable by mass-action kinetics for use in Synthetic Biology. 

However, quadraticization may form additional cross-negative terms during the degree reduction process, in which case kineticization techniques described in Chapter~\ref{oldchapter2} must be used to rid the problematic terms and the quadraticization repeated. Kowalski Transformations (Section~\ref{ForcedVariable}) may increase the degree of the system by $2$, and Quasi-Steady State Transformations (Section~\ref{QSSA}) may increase the degree by an amount proportional to the degree of polynomial $p_s(\mathbf{x})$, complicating the bimolecularization of high-dimensional networks. It is therefore of interest to see if a purely bimolecular reaction network approximating the true dynamics can always be designed. For this purpose, we suggest the \textit{Quasi-Steady State Approximation (QSSA)} as a suitable alternative if quadraticization into a kinetic system fails. 

\section{Quasi-Steady State Approximation (QSSA)}\label{quasisteadystatetransformation}

Wilhelm~\cite{Wilhelm} proposes a general strategy for deriving bimolecular approximations from multi-molecular reactions. We demonstrate the technique on trimolecular reactions, but the generalization to higher orders is fairly straightforward. Interested readers are encouraged to view the original work.

We consider a single trimolecular reaction
\begin{equation}\label{confusinggeneralform}
r_1^\prime: \quad \nu_{11}^{\prime-} \mathrm{X}_{1}+\nu_{21}^{\prime-} \mathrm{X}_{2}+\nu_{31}^{\prime-} \mathrm{X}_{3} \stackrel{k_1^\prime}{\rightarrow} \nu_{11}^{\prime+} \mathrm{X}_{1}+\nu_{21}^{\prime+} \mathrm{X}_{2}+\nu_{31}^{\prime+} \mathrm{X}_{3}+\sum_{\ell \in I} \nu_{\ell 1}^{\prime+} \mathrm{X}_{\ell},
\end{equation}
whose deterministic counterpart is given by the RREs~\eqref{kineticsystem}: 
\begin{equation}\label{deterministictrimolecularsystem}
\begin{aligned}
&\Dot{x_1}=\left(\nu_{11}^{\prime+}-\nu_{11}^{\prime-}\right) k_1^\prime x_{1}^{\nu_{11}^{\prime-}} x_{2}^{\nu_{21}^{\prime-}} x_{3}^{\nu_{31}^{\prime-}}, \\
&\Dot{x_2}=\left(\nu_{21}^{\prime+}-\nu_{21}^{\prime-}\right) k_1^\prime x_{1}^{\nu_{11}^{\prime-}} x_{2}^{\nu_{21}^{\prime-}} x_{3}^{\nu_{31}^{\prime-}}, \\
&\Dot{x_3}=\left(\nu_{31}^{\prime+}-\nu_{31}^{\prime-}\right) k_1^\prime x_{1}^{\nu_{11}^{\prime-}} x_{2}^{\nu_{21}^{\prime-}} x_{3}^{\nu_{31}^{\prime-}}, \\
&\Dot{x_\ell}=\nu_{\ell1}^{\prime+} k_1^\prime x_{1}^{\nu_{11}^{\prime-}} x_{2}^{\nu_{21}^{\prime-}} x_{3}^{\nu_{31}^{\prime-}},\quad \ell \in I.
\end{aligned}
\end{equation}
We show that the dynamics dictated by~\eqref{deterministictrimolecularsystem} may be approximated in the limit $k_2 \to \infty$ via the bimolecular reaction network
\begin{equation}\label{thisintermediateequation}
\begin{aligned}
&r_1: \quad \mathrm{X}_{i}+\mathrm{X}_{j} \stackrel{k_1}{\rightarrow} \mathrm{Z}, \quad r_2: \quad \mathrm{Z} \stackrel{k_2}{\rightarrow} \mathrm{X}_{i}+\mathrm{X}_{j}, \\
&r_3: \quad \mathrm{X}_{k}+ \mathrm{Z} \stackrel{k_{3}}{\rightarrow} \nu_{i3}^+ \mathrm{X}_{i}+\nu_{j3}^+ \mathrm{X}_{j}+\nu_{k3}^+ \mathrm{X}_{k}+\sum_{\ell \in I} \nu_{\ell 3}^+ \mathrm{X}_{\ell}+\nu_{z3}^+ \mathrm{Z},
\end{aligned}
\end{equation}
where the rate coefficients are chosen to induce the fast relaxation of intermediary specimen $\mathrm{Z}$ to negligible quantities. The indices $i,j,k$ need not be distinct and are chosen such that $\mathrm{X}_i,\mathrm{X}_j,\mathrm{X}_k$ represent all reactants of $r_1^\prime$. Upon introducing $\varepsilon :=1 / k_{2}$, the deterministic dynamics of the network are given by the ODE system:
\begin{equation}\label{properintermediateequation}
\begin{aligned}
\varepsilon \Dot{x_i} &=z+\varepsilon\left(\nu_{i3}^+ k_{3} x_{k} z-k_{1} x_{i} x_{j}\right) , \\
\varepsilon \Dot{x_j} &=z+\varepsilon\left(\nu_{j3}^+ k_{3} x_{k} z-k_{1} x_{i} x_{j}\right) ,\\
\Dot{x_k} &=\left(\nu_{k3}^+ -1\right) k_{3} x_{k} z ,\\
\varepsilon \Dot{z} &=-z+\varepsilon\left(\left(\nu_{z3}^+-1\right) k_{3} x_{k} z+k_{1} x_{i} x_{j}\right) ,\\
\Dot{x_\ell} &=\nu_{\ell 3}^+ k_{3} x_{k} z, \quad \ell \in I.
\end{aligned}
\end{equation}
We introduce abstract specimen $\Sigma_1 = \mathrm{X}_i + \mathrm{Z}$, $\Sigma_2 = \mathrm{X}_j + \mathrm{Z}$ as an estimate to shadow the molecular counts of specimen $\mathrm{X}_i, \mathrm{X}_j,$ respectively. Performing a change of coordinates $\sigma_{1} = x_{i}+z$, $ \sigma_{2} = x_{j}+z$ to~\eqref{properintermediateequation} for elimination of $x_i, x_j$ gives
\begin{equation}\label{formattingissue}
    \begin{aligned}
\Dot{\sigma_1} &= \Dot{x_i} + \Dot{z} =\left(\nu_{i3}^++\nu_{z3}^+-1\right) k_{3} x_{k} z, \\
\Dot{\sigma_2} &= \Dot{x_j} + \Dot{z} =\left(\nu_{j3}^++\nu_{z3}^+-1\right) k_{3} x_{k} z,\\
    \varepsilon \Dot{z}&=\varepsilon k_{1}\left(\sigma_{1}  -z\right)\left(\sigma_{2}-z\right)-z+\varepsilon\left(\nu_{z3}^+-1\right) k_{3} x_{k} z, \\
\Dot{x_k}&=\left(\nu_{k3}^+-1\right) k_{3}  x_{k} z , \quad \Dot{x_\ell} = \nu_{\ell 3}^+ k_{3} x_{k} z, \quad \ell \in I,
    \end{aligned}
\end{equation}
where Wilhelm et al~\cite{Wilhelm,WilhelmAdvancedSchneider} justifies the use of the matched asymptotic expansion
\begin{equation}
    z=A\left(\sigma_{1}, \sigma_{2}, x_{k}, x_{\ell}\right)+\varepsilon B\left(\sigma_{1}, \sigma_{2}, x_{k}, x_{\ell}\right)+\mathcal{O}\left(\varepsilon^{2}\right).
\end{equation}
Comparing powers of $\mathcal{O}(\epsilon^0)$ gives $A = 0$. Moving to the next order,
\begin{equation}
    \mathcal{O}(\varepsilon): \quad   \Dot{A} = k_1\left(\sigma_1 - A \right)\left(\sigma_2 - A \right) - B + \left(\nu_{z3}^+-1 \right)k_3 x_k A,
\end{equation}
which gives 
\begin{equation}
    0 = k_1\sigma_1 \sigma_2 - B \iff B =  k_1\sigma_1 \sigma_2.
\end{equation}
Therefore for small $\varepsilon$, we may approximate the fast variable $z$ in first order as:
\begin{equation}
    z = 0 + \varepsilon k_1 \sigma_1 \sigma_2 + \mathcal{O}(\varepsilon^2) \approx \frac{k_{1}}{k_{2}} \sigma_{1} \sigma_{2}.
\end{equation}
Substitution into~\eqref{formattingissue} gives
\begin{equation}\label{goodapprox}
  \begin{aligned}
\Dot{\sigma_1} &=\left(\nu_{i3}^++\nu_{z3}^+-1\right) \frac{k_{1} k_{3}}{k_{2}} x_{k} \sigma_{1} \sigma_{2}, \\
\Dot{\sigma_2} &=\left(\nu_{j3}^++\nu_{z3}^+-1\right) \frac{k_{1} k_{3}}{k_{2}} x_{k} \sigma_{1} \sigma_{2}, \\
\Dot{x_k} &=\left(\nu_{k3}^+-1\right) \frac{k_{1} k_{3}}{k_{2}} x_{k} \sigma_{1} \sigma_{2}, \\
\Dot{x_\ell} &=\nu_{\ell3}^+ \frac{k_{1} k_{3}}{k_{2}} x_{k} \sigma_{1} \sigma_{2}, \quad \ell \in I.
\end{aligned}  
\end{equation}
Identifying $\sigma_1, \sigma_2$ as approximations of $x_i,x_j,$~\eqref{deterministictrimolecularsystem} and~\eqref{goodapprox} have the same functional form. Any choice of stoichiometric coefficients and rate coefficients to make the two equations identical, and which also guarantee $z \downarrow 0$, yield valid bimolecular approximations. A step-by-step demonstration of the strategy in bimolecularizing the reaction  
\begin{equation}
    r_1:\quad \mathrm{X}_1 + 2\mathrm{X}_2 \stackrel{k}{\rightarrow} 3\mathrm{X}_2 + \sum_{\ell \in \{3,\dots,10\}} \nu_{\ell 1} \mathrm{X}_{\ell}
\end{equation}
is provided in Appendix~\ref{newexample}.

\section{Polynomialization and AutoGillespie}\label{AutoGillespieSection}

\begin{mdframed}
(\textbf{AutoGillespie Algorithm Summary})

\noindent\rule{16.3cm}{0.5pt}
\vspace{1pt}

\noindent \textbf{(A1):} Take an arbitrary polynomial $\mathcal{P}(\mathbf{x})$ as input, which we interpret as the right hand side of the ODE system~\eqref{polynomialrighthandside}.

\vspace{10pt}
\noindent \textbf{(A2):} Apply the kineticization techniques $\Psi_{Kow}$ or $\Psi_{QSST}$ to remove cross-negative terms and derive the corresponding general system.

\vspace{10pt}
\noindent \textbf{(A3):} Canonically invert the general system into a chemical reaction network, which we feed into the Gillespie Algorithm~\cite{GillespieFirstandDirectReactions} for stochastic simulation.
\end{mdframed}

In this section, we propose Taylor expansions as a general polynomialization strategy which may also be used as a remedy when Kerner Polynomialization fails to terminate or becomes intractably complex. When all arbitrary ODE systems arising in practice are considered, cross-negative terms will invariably appear in series expansions. It is furthermore difficult to estimate \textit{a priori} how high orders of expansion must be for sufficient encapsulation of the dynamics of the original system, thus an algorithm capable of autonomously stochastically simulating an arbitrary polynomial ODE system after inversion into a chemical reaction network must be developed. 

The key to designing such an algorithm is the detection and tracking of cross-negative terms in input polynomials, which enables the painless application of nonlinear transformations for translation into a mass-action kinetic system. The development of this algorithm is very technical, lengthy, and (for some readers) possibly tedious. We have therefore elected to provide a verbose conversational pseudocode in Appendix~\ref{GillespieAutoGillespie}. Readers interested in reproducing the plots in Chapter~\ref{oldchapter6} are strongly recommended to program their own versions of the algorithm\footnote{Alternatively, a sample AutoGillespie code is freely available to the public at the following link on Github: https://github.com/leesh-1/AutoGillespie} after reading the relevant portion of this paper, as many stochastic simulations we consider are ghastly to implement manually by hand. Only a very brief and simplified summary of the algorithm, called \textit{AutoGillespie Algorithm}, has been given in this section as \textbf{(A1-A3)} for the convenience of the reader.

As we increase the order of Taylor expansions, the canonically inverted network grows vastly high dimensional. The version of AutoGillespie summarized here, which does not attempt any tricks with the Gillespie Algorithm, must siphon through a large number of reaction channels and the computational cost incurred confines simulations to early timescales, refusing to propagate beyond several non-dimensional seconds for the systems considered in Chapter~\ref{oldchapter6}. Therefore, it is desirable to seek a modified version of the algorithm that produces long-time stochastic trajectories even when the network has an excessive number of reaction channels. 

This may be accomplished by altering the structure of the well-established Gillespie Algorithm. Within the canonical inversion, any reaction channel adds or removes precisely one specimen from the reactants, which we take pains to track during the development of the AutoGillespie Algorithm. In other words, for specimen $\mathrm{X}_s\in \mathcal{X}$, all terms that appear within the kineticized right hand side of $\Dot{x}_s$ are canonically inverted to a reaction channel that either adds or removes one $\mathrm{X}_s$ molecule. We may exploit this feature to formalize any $|\mathcal{X}|$-dimensional polynomial into a reaction network with at most $4|\mathcal{X}|$ pseudo-reaction channels. This process is laid out in Appendix~\ref{GillespieAutoGillespie}.

In this approach, increasing the order of expansion of a multivariate polynomial drastically from $5$ to $20$ yields no substantive increase in run time in the Gillespie simulation, although the symbolic computation of the expansion takes longer. Given that that the computing platform has enough resources to perform a symbolic computation for series expansions of high order, the Gillespie Algorithm is almost guaranteed to produce long time trajectories. This modified AutoGillespie Algorithm is exclusively used in Chapter~\ref{oldchapter6}. 

\subsection{Optimal Polynomial Selection Strategy}\label{polynomialoptimizationmethod}

Taylor expansions attempt to derive the best fit polynomial to a given function by using local information obtainable at the point of expansion. In a similar vein, we aim to capture the coefficients $\Phi^{\ell,h}_{c_1 \dots c_\ell}$ in the  polynomial structure~\eqref{generalpolynomialsystem} that produces the best fit with the right hand side of the original system. We may minimize the error between the polynomial structure and the original system over some distance metric, such as the $\ell_2$-norm. Alternatively, we may consider weighted norms, designed to add emphasis to critical points in the original system so that the optimized polynomial structure will locally demonstrate topological similarities near points of interest. Such an approach is amenable to a theoretical analysis, though evaluation of whichever distance metric is used may grow intractable for higher order polynomial structures without the support of a symbolic computing platform.

A non-constructive numerical approach is given by adapting an unconstrained optimization algorithm to minimize some quantity representing a close fit between the polynomial structure~\eqref{generalpolynomialsystem} and the original system. For this purpose, we use the gradient-free Nelder-Mead Simplex Method of Lagarias et al~\cite{fminsearchtheory}, readily implemented via the \mcode{fminsearch} package~\cite{fminsearch} in \mcode{Matlab}.

An advantage of the numerical approach is that the minimization may be done over a norm-free quantity $L$, representing the deviance between the polynomial structure and the desired properties of the original system. It is possible to partition the domain of the original system into `essential' and `non-essential' segments, and minimize $L$ over unions of specific segments while neglecting the rest, or as in the analytical approach, add emphasis (weights) to certain segments to enhance the fit. We may further impose that cross-negative terms cannot be entertained in the polynomial structure. This results in a constrained optimization problem, which may be resolved via an unconstrained optimization algorithm by applying a very harsh penalty once cross-negative terms appear\footnote{We have set the norm-free quantity $L = 10^{18}$ if cross-negative terms are detected in the polynomial structure.}. A disadvantage of the numerical approach is that the optimization algorithm may fail to converge to a local minimum, or the local minimum may be woefully insufficient to approximate the given input.

\section{Closing Remarks}
We make a few closing remarks before testing the applicability of the proposed techniques in producing appropriate chemical systems in the coming chapters.

Firstly, we have been unable to locate a serious consideration of series expansions in the chemical inversion of ODE systems within the literature. The author suspects that the rationale may be that previously known methods (e.g. Kerner Polynomialization and $\Psi_{Kow}$) are proven to fully preserve deterministic dynamics, thus there is no point in using approximate expansions (which provide \textit{approximate} dynamics and does not even aim for full preservation of information) for chemical inversion if even precise methods available do not succeed in preserving deterministic expectations in the chemical simulation. However, deterministic dynamics are known to manifest differently in stochastic settings\footnote{Examples of this are provided in Appendix~\ref{appendixA}, via a simulation of the VKBL circadian oscillator.}, and it may very well turn out that series expansions are more optimal for chemical inversion. We assert that it is improper for series expansions not to be considered in the chemical integration of arbitrary ODEs in molecular computing~\cite{moleculecomputegeneralexplain}.

Secondly, an implementation of most of the techniques introduced in Chapters~\ref{oldchapter2} and~\ref{oldchapter3} in the stochastic setting are virtually non-existent, perhaps with the exception of $\Psi_{QSST}$~\cite{TomiInversePaper}. Nearly all previous studies appear to have been carried out deterministically and the sufficient translation of ideal dynamics to stochastic settings intrinsically assumed. This powerful assumption is in dire need of proper validation, and by no means should polynomial inversion strategies into chemical reaction networks be considered a resolved problem.

As we move on from theoretical discussions to stochastic simulations, we are in completely uncharted territory. Our simulations reveal a wide array of entirely unexpected dynamics and mysteries, some of which we pose as open problems to be further explicated by future research. Based on the results of our simulations, we will formulate--and also test--a coherent inversion framework for the successful chemical integration of arbitrary ODE systems. We now cycle through multiple test systems designed to simplify the analysis, in order to elucidate the strengths and weaknesses of each inversion technique. 

%% file: chapter5.tex
\chapter{Selection of Polynomialization Strategy}\label{oldchapter4}

The test system is chosen from elementary functions that recur in practice, and of those, the exponential function for its simplicity. The system is designed to be a one-equilibrium model with regular dynamics: 
\begin{equation}\label{model1}
    \Dot{x_{1}} = 1 - x_{1}\exp{(x_{1})}.\tag{Model 1}
\end{equation}
The unique stable steady state is given by $x^* \approx 0.5671$. We monitor the effects of Kerner Polynomialization~\eqref{model2} and series expansions~\eqref{model3} on~\ref{model1} to determine which strategy should be recommended for the inversion framework. 

Kerner Polynomialization introduces $x_2 = \exp{(x_{1})}$ and imposes the constraint $x_2(0) = \exp{(x_1(0))}$ for initial time $t_0 = 0$, giving
\begin{equation}\label{model2}
    \begin{aligned}
&\Dot{x_1}=1-x_1x_2,\\
&\Dot{x_2}=x_2-x_1x_2^2.
\end{aligned}\tag{Model 2}
\end{equation}
We contrast its performance against the Taylor series expansion of arbitrary order,
\begin{equation}\label{model3}
   \Dot{x_1}= 1 - x_1-x_1^2-x_1^3/2+ \dots \tag{Model 3}
\end{equation}
It is pertinent to address a concern of studying canonical inversions of truncated series expansions, which produces networks with highly multi-molecular reactions. It should be emphasized that such networks will not be found naturally, and is therefore unlikely to be a candidate network for complex biochemical mechanisms. However, we have decided to include their analysis as an intermediary step towards being able to implement synthetic biochemical integrators of arbitrary ODE systems via molecular computing~\cite{moleculecomputegeneralexplain}. To generate candidate bimolecular networks realizable naturally or in laboratories, it is apt to further apply quadraticization techniques  or Quasi-Steady State Approximations (Sections~\ref{generalquadraticizationalgorithm},~\ref{quasisteadystatetransformation}) for order reduction.

To realize mass-action kinetic systems in laboratory settings as synthetic biochemical processes, such as to compute roots of nonlinear equations~\cite{primitivecomputationCRN,molecularRootcomputation} (or in our case, to solve differential equations by measuring chemical concentrations), we must perform our analysis under the assumption that the rate coefficients of reaction channels as well as measured solution concentrations have been contaminated. Trivial analytical and computational progress may be made for \ref{model2} assuming perturbations in initial solution concentrations, and is included in Appendix~\ref{initialconcentration}. It is verified that contamination in initial data stabilizes the steady state to an incorrect value with the deviance proportional (percentagewise) to the perturbation in $x_2(0)$.

\section{Analysis of~\ref{model2}}
To study the effects of stochastic contamination in rate coefficients, \ref{model2} is formalized as
\begin{equation}\label{formalizedmodel2}
    \begin{aligned}
&\Dot{x_1}=k_1-k_2x_1x_2, \\
&\Dot{x_2}=k_3x_2-k_4x_1x_2^2,
\end{aligned}
\end{equation}
where $k_i = 1$, $i \in \{ 1,2,3,4\}$. We take for granted that realistic perturbations of rate coefficients are at the very least less than $100 \%$ of their absolute value. Elementary algebra gives that \eqref{formalizedmodel2} has a steady state if and only if
\begin{equation}\label{steadystateconditionmodel2}
    \frac{k_1}{k_2} = \frac{k_3}{k_4} \iff \frac{k_3}{k_1} = \frac{k_4}{k_2}.
\end{equation}
Any realistic perturbation which does not satisfy this identity removes the equilibrium of~\ref{model2}. 

\begin{figure}[h!]
  \begin{subfigure}[b]{0.5\textwidth}
    \includegraphics[width=\textwidth]{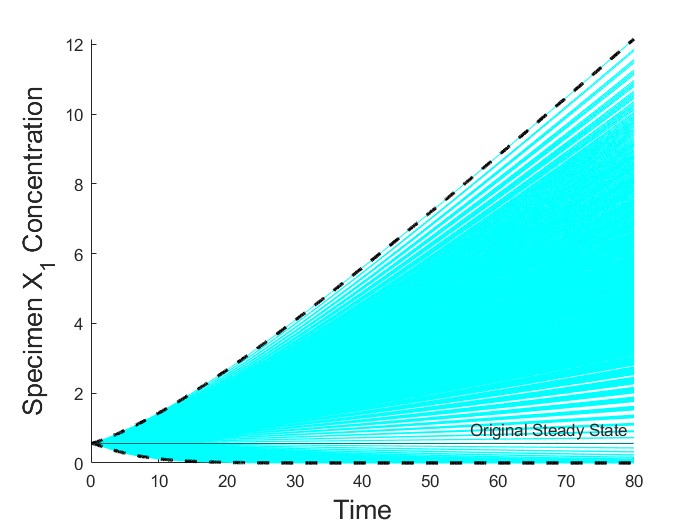}
    \caption{ }
  \end{subfigure}
  \hfill
  \begin{subfigure}[b]{0.5\textwidth}
    \includegraphics[width=\textwidth]{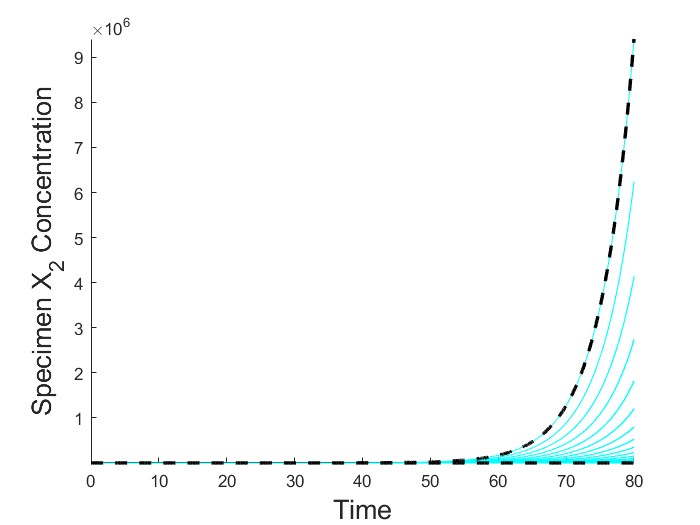}
    \caption{ }
  \end{subfigure}
    \begin{subfigure}[b]{0.5\textwidth}
    \includegraphics[width=\textwidth]{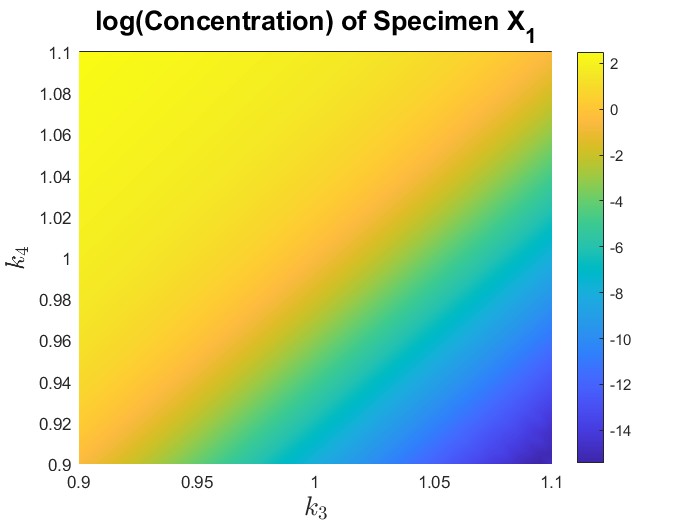}
    \caption{ }
  \end{subfigure}
  \hfill
  \begin{subfigure}[b]{0.5\textwidth}
    \includegraphics[width=\textwidth]{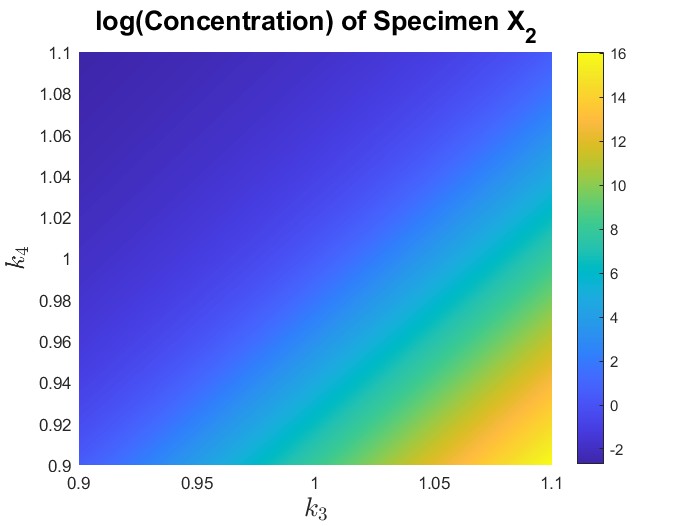}
    \caption{ }
  \end{subfigure}
  \caption[Explosion-decay effect of perturbations on~\ref{model2}]{More information about plot generation is included in Appendix~\ref{FigureGeneration}. A $10 \%$ perturbation is applied to $k_3$,$k_4$ for $k_1=k_2=1$ and the dynamics of $\mathrm{X}_1$,$\mathrm{X}_2$ are studied using $400$ deterministic trajectories numerically solving~\eqref{formalizedmodel2}. Interpolation is used to generate the pseudo-colour plots (c-d) at $t = 80$ and all trajectories are plotted in (a-b). The boundaries are plotted with dashed black lines and identified as the maximal perturbation $k_3 = 1.1$ and minimal perturbation $k_4 = 0.9$, and vice versa with the indices switched. }
  \label{dynamicsofthefirstmodel}
\end{figure}

In Figure~\ref{dynamicsofthefirstmodel}, most perturbations completely destabilize the trajectory and lead to uncontrolled blow-ups in one of the specimen as expected. For a meticulous analysis, we observe from (c) that perturbations applied to produce $k_3 \ge k_4$ pull trajectories of $x_1$ below the unperturbed steady state. Such trajectories cannot cross the $x_1$-axis (setting $x_1 = 0$ in~\eqref{formalizedmodel2} gives $\dot{x_1} = k_1 > 0$), which we observe in (a) where $x_1$-trajectories press closer to the horizontal axis without being driven negative. In (c-d), we observe an inverse correlation between the concentration dynamics of the two specimen: $x_1$ diverges as $x_2$ decays and vice versa. The only exception is when~\eqref{steadystateconditionmodel2} is satisfied, where both specimen stabilize to their predicted steady state value. Surprisingly, our regular model is displaying a phenomenon depicted in Synthetic Biology as \textit{Nonexistent Equilibrium Catastrophe (NEC)}~\cite{TomiCatastrophePaper}, which has not previously been reported nor investigated in the context of polynomialization.

\subsection{Nonexistent Equilibrium Catastrophe (NEC)}\label{NECAppendix}
Figure~\ref{dynamicsofthefirstmodel} showcases the \textit{Nonexistent Equilibrium Catastrophe (NEC)} phenomenon, expounded in~\cite{TomiCatastrophePaper} under the name \textit{Negative Equilibrium Catastrophe}. This catastrophe is observed within poorly designed molecular controllers in Synthetic Biology, whose goal is to successfully manipulate the dynamics of high-dimensional intracellular networks while preserving stability. A number of factors complicate the design. For example, intracellular networks can be mercilessly complex, frustrating all efforts toward any sort of exact analytical progress. Furthermore, many components of the network may be unknown, such as the initial specimen concentrations or precise reaction channel structures.

In practice, this may result in an inappropriately embedded molecular controller network forcing all critical points of the controlled system to be negative where they cannot be biochemically realized or even destroying them outright, resulting in an unintended blow-up of select specimen within the cell instead of stabilization. Such modified biochemical networks can have catastrophic consequences, leading to a lethal build-up of unnecessary species detrimental to the proper functioning of a cell. 

\subsection{Deterministic Escape Detection}
When NEC is detected, we may attempt to mollify the impact of the catastrophe by operating the network only for short time intervals, depriving NEC of the opportunity to take effect. For this purpose, we have computed deterministic \textit{escape times}, where initial concentrations of the specimen are set to their unperturbed steady states, and the time elapsed until trajectories evolved under perturbed reaction coefficients have deviated sufficiently from their intended (unperturbed) steady states are recorded. Escape times may act as an early indication of the suitability of any proposed reaction network to be synthesized in the laboratory.

A source of optimism for Kerner Polynomialization applied to~\ref{model1} is that escape times explode superexponentially, whose power laws are summarized in Table~\ref{table0}. This indicates that under controlled perturbations, the system enters a pseudo-steady state which competently elongates the time of validity of biochemically simulated results, indicating that we need not reject the polynomialization~\ref{model2} solely due to the existence of NEC.

\begin{figure}[h!]
    \includegraphics[width=.33\textwidth]{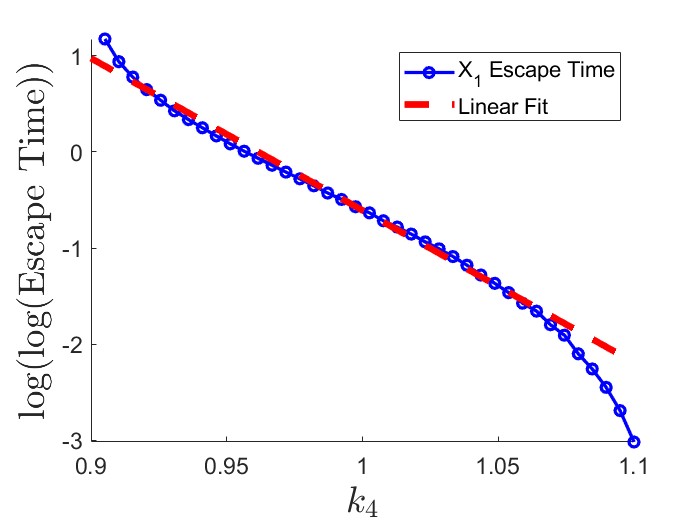}\hfill
    \includegraphics[width=.33\textwidth]{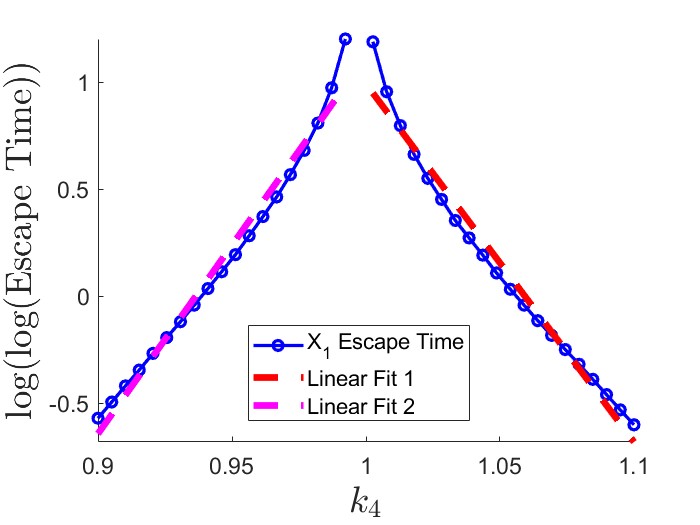}\hfill
    \includegraphics[width=.33\textwidth]{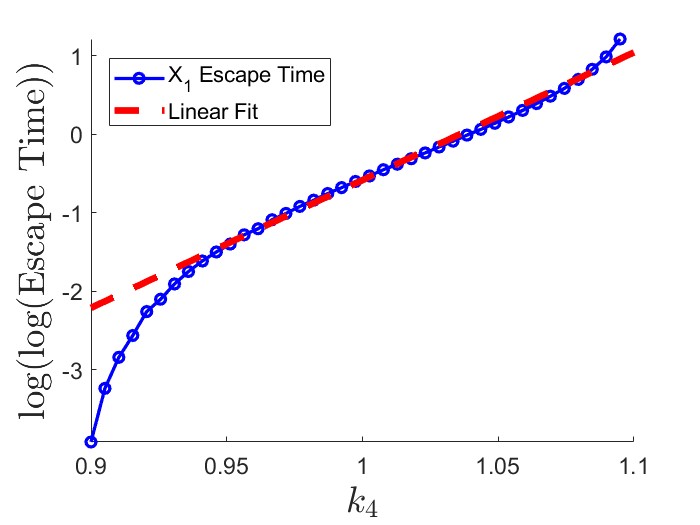}\hfill
    \caption[Superexponential explosion of escape times for~\ref{model2}]{$k_3 = 0.9,1,1.1,$ from left to right, where vertical axis plots strongly damped specimen $\mathrm{X}_1$ escape time data passed twice through $\log$. Data have been segmented and portions removed before linear fitting to achieve best fits. Specimen $\mathrm{X}_2$ displays analogous trends (plots not included), and Table~\ref{table0} gives their superpower-law values. Data is drawn from Figure~\ref{escapetimeoriginal}.}\label{superexponentialexplosion}
\end{figure}

\begin{figure}[h]
  \begin{subfigure}[b]{0.5\textwidth}
    \includegraphics[width=\textwidth]{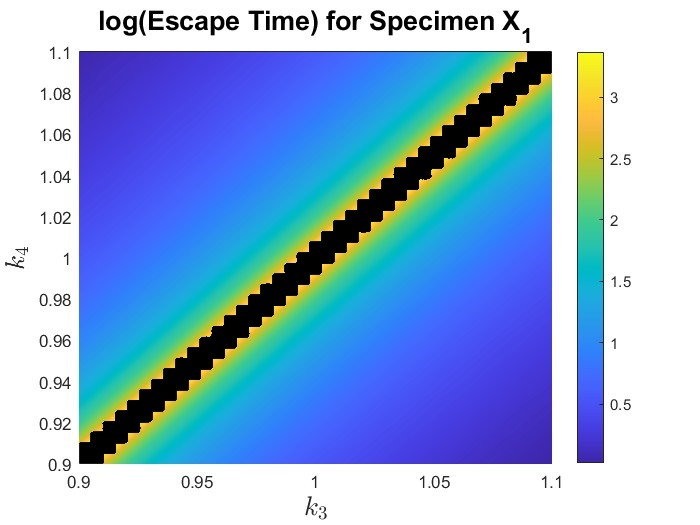}
  \end{subfigure}
  \hfill
  \begin{subfigure}[b]{0.5\textwidth}
    \includegraphics[width=\textwidth]{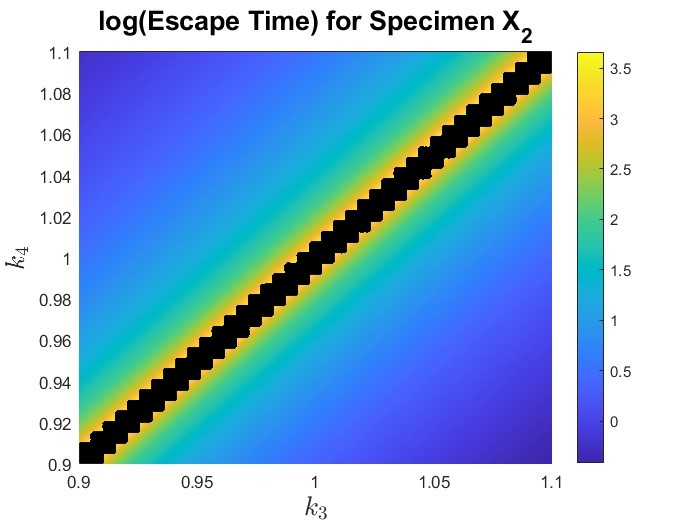}
  \end{subfigure}
  \caption[Escape times for specimen $\mathrm{X}_1$ and $\mathrm{X}_2$ on a $\log$ scale]{Escape times for specimen $\mathrm{X}_1$ and $\mathrm{X}_2$ on a $\log$ scale. Regions in which escape is not observed is drawn in black, which agrees with the steady-state condition~\eqref{steadystateconditionmodel2}.}
  \label{escapetimeoriginal}
\end{figure}

\begin{table}[h!]
    \centering
    \resizebox{\textwidth}{!}{%
\begin{tabular}{ |p{3cm}||p{3cm}|p{3cm}|p{3cm}|p{3cm}| }
 \hline
 \multicolumn{5}{|c|}{Superexponential Power-Law for Figure \ref{superexponentialexplosion}} \\
 \hline
 Type & $k_3 = 0.9$  & $k_3 = 1$  & $k_3 = 1$  & $k_3 = 1.1$\\
 \hline
 Specimen $\mathrm{X}_1$ & $-15.84$ & $17.52 $ & $-16.86$ & $16.28$\\
$\mathrm{X}_1$ Confidence & $(-16.51, -15.17)$ & $(16.29, 18.75)$ & $(-18.18, -15.54)$ & $(15.68, 16.88)$ \\
Specimen $\mathrm{X}_2$ & $-18.54$ & $21.77$  & $-18.08$ & $21.03$ \\
$\mathrm{X}_2$ Confidence & $(-19.49, -17.59)$ & $(20.98, 22.56)$ & $(-18.99, -17.17)$ & $(20.21, 21.84)$ \\
\hline
\end{tabular}}
\caption{The fitted values of superexponential power-laws $\varrho$ of escape times in Figure~\ref{superexponentialexplosion} are given, along with $95\%$ confidence intervals. The fit is performed against the equation $x_2 \propto \exp{(\exp(\varrho x_1))}$. }
\label{table0}
\end{table}

\subsection{Inversion into Chemical Reaction Network}
Being mass-action kinetic,~\ref{model2} may be realized as a chemical reaction network through at least two distinct formulations. Non-uniqueness can arise from interpreting the $x_1x_2^2$ term as the reaction $\mathrm{X}_1 + 2\mathrm{X}_2 \xrightarrow[]{} \mathrm{X}_1+\mathrm{X}_2$ or $\mathrm{X}_1 + 2\mathrm{X}_2 \xrightarrow[]{} \mathrm{X}_1$. We study two deterministically equivalent networks, Formulation $1$ being~\eqref{Formulation1} and Formulation $2$ being~\eqref{Formulation2}. Unless noted otherwise, all reacting solutions are of volume $V = 100$ in Chapter~\ref{oldchapter4}, which scales up the equilibrium molecule counts and intuitively puts any measured differences within our simulations to perspective.

\begin{figure}[h]
  \begin{subfigure}[b]{0.245\textwidth}
    \includegraphics[width=\textwidth]{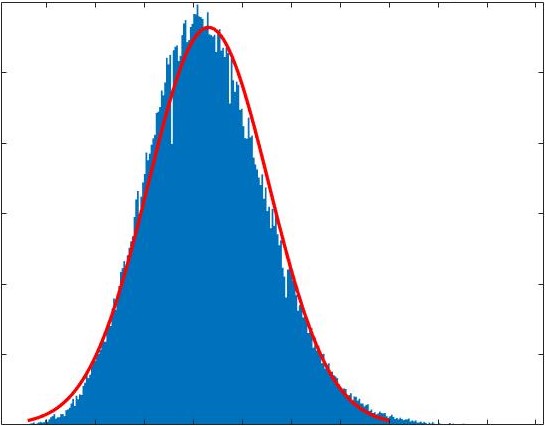}
  \end{subfigure}
  \begin{subfigure}[b]{0.245\textwidth}
    \includegraphics[width=\textwidth]{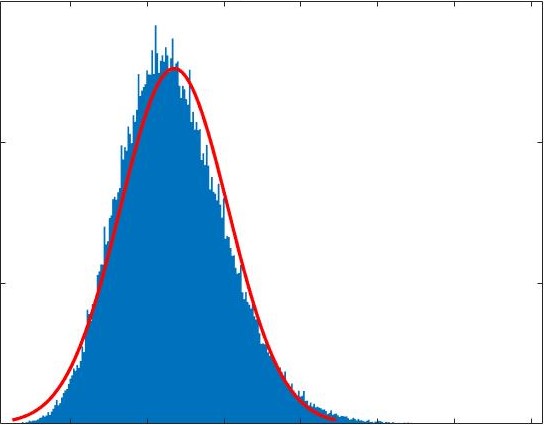}
  \end{subfigure}
    \begin{subfigure}[b]{0.245\textwidth}
    \includegraphics[width=\textwidth]{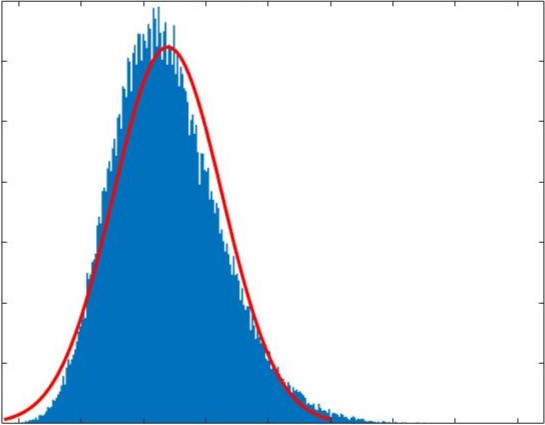}
  \end{subfigure}
    \begin{subfigure}[b]{0.245\textwidth}
    \includegraphics[width=\textwidth]{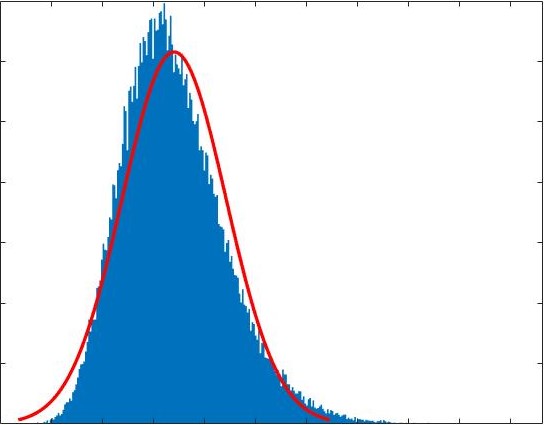}
  \end{subfigure}
  \caption[Distribution for specimen $\mathrm{X}_2$ in~\eqref{Formulation1} measured at $t = 10,20,30,40$]{Sampled distribution for specimen $\mathrm{X}_2$ in~\eqref{Formulation1} measured at $t = 10,20,30,40$, for volume $V = 1500$. Horizontal axis represents molecule counts, and vertical axis represents obtained probability mass function over $10^5$ realizations. The Gaussian fit, performed using a pre-existing \mcode{Matlab} package~\cite{gausianproper}, skews progressively with time and is consistent with data shown in Figure~\ref{whyisthisplotincluded}, where averaged $\mathrm{X}_2$-trajectories depict a continually increasing upward bias.}
  \label{distributioninvestigationplot}
\end{figure}

\begin{figure}[h!]
  \begin{subfigure}[b]{0.5\textwidth}
    \includegraphics[width=\textwidth]{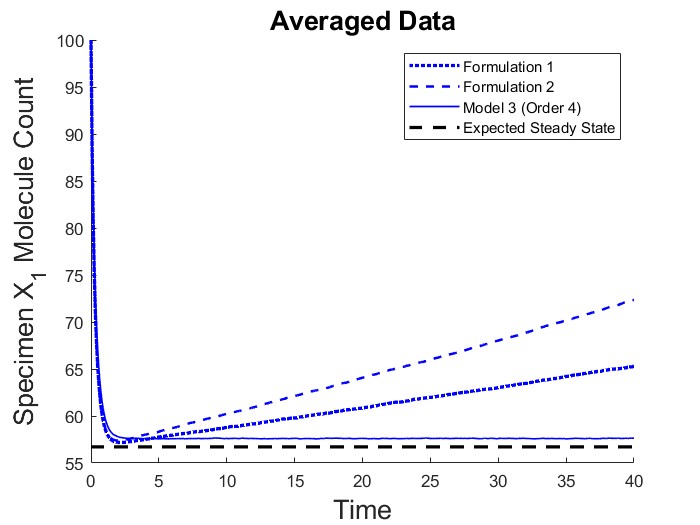}
    \caption{ }
  \end{subfigure}
  \hfill
  \begin{subfigure}[b]{0.5\textwidth}
    \includegraphics[width=\textwidth]{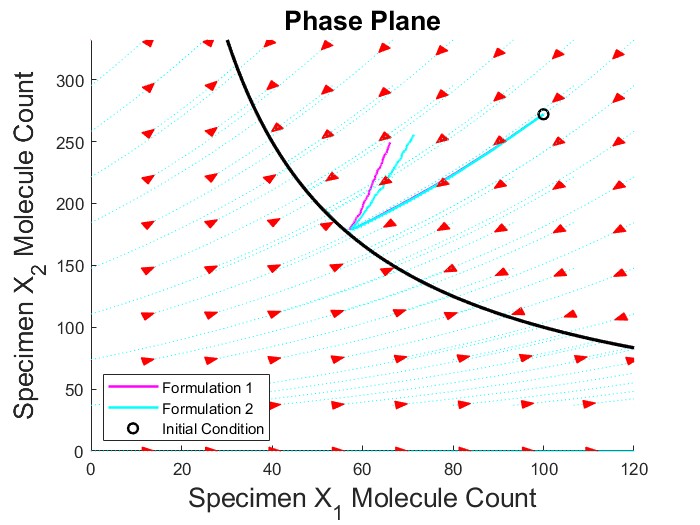}
    \caption{ }
\end{subfigure}
  \caption[Trajectory comparison for~\ref{model2} and~\ref{model3}]{Averaged stochastic trajectories in (a) show the upward tilt for specimen $\mathrm{X}_1$ for Formulations $1$~\eqref{Formulation1} and $2$~\eqref{Formulation2}. The fourth order Taylor network (canonically inverted cubic Taylor expansion,~\ref{model3}) performs optimally over repeated sampling, evidenced by its close proximity to the deterministic root of~\ref{model1} at $V = 100$, denoted as ``Expected Steady State''. (b) plots the trajectories of the two formulations in the phase plane, where the deviance is more clearly emphasized. Red arrows point in direction of flow, and deterministic paths are traced in cyan. The steady state wall $x_2 = 1/x_1 \iff \mathrm{X}_2 = V^2/\mathrm{X}_1$ of~\ref{model2} is drawn in black.}
  \label{onlythemostsnecessary}
\end{figure}

Stochastic simulations are a good indicator of the behaviours of synthetic implementations, and Figure~\ref{onlythemostsnecessary} shows that trajectories of~\eqref{Formulation1} and~\eqref{Formulation2} quickly present an upward bias. However, the correct steady state (root of~\ref{model2}) is inferrable by measuring the point at which the averaged stochastic paths are deflected from the steady state wall. This variation (upward bias) appears to correlate with a skew in the sampled histograms of specimen molecule counts which worsens with time (Figure~\ref{distributioninvestigationplot}). As the system is propagated, exploding solutions are detected as outliers which tilt the averaged trajectories upward. Relevant plots are included in Appendix~\ref{importantbitinappendix}.

\begin{minipage}{.45\textwidth}
\begin{equation}\label{Formulation1}
    \begin{aligned}
&r_{1}: \quad \varnothing \stackrel{k_{2}}{\longrightarrow} \mathrm{X}_1,\\
&r_{2}: \quad \mathrm{X}_1+\mathrm{X}_2 \stackrel{k_{2}}{\longrightarrow} \mathrm{X}_2, \\
&r_{3}: \quad \mathrm{X}_2 \stackrel{k_{3}}{\longrightarrow} 2\mathrm{X}_2,\\
&r_{4}: \quad \mathrm{X}_1+2\mathrm{X}_2 \stackrel{k_{4}}{\longrightarrow} \mathrm{X}_1+\mathrm{X}_2, \\ 
& \mathbf{k} = [1,1,1,1],
\end{aligned}
\end{equation}
\end{minipage}
\begin{minipage}{.45\textwidth}
\begin{equation}\label{Formulation2}
    \begin{aligned}
&r_{1}: \quad \varnothing \stackrel{k_{2}}{\longrightarrow} \mathrm{X}_1,\\
&r_{2}: \quad \mathrm{X}_1+\mathrm{X}_2 \stackrel{k_{2}}{\longrightarrow} \mathrm{X}_2,\\
&r_{3}: \quad \mathrm{X}_2 \stackrel{k_{3}}{\longrightarrow} 2\mathrm{X}_2,\\
&r_{4}: \quad \mathrm{X}_1+2\mathrm{X}_2 \stackrel{k_{4}}{\longrightarrow} \mathrm{X}_1, \\
&\mathbf{k} = \left[1,1,1,\frac{1}{2}\right].
\end{aligned}
\end{equation}
\end{minipage}

\vspace{2pt}

We hypothesize based on our simulations (expounded further in Appendix~\ref{importantbitinappendix}) that in biochemical systems in which NEC is predicted to continually grow specimen concentrations \textit{a priori}, a tendency for blow-ups will also be detected in repeated samplings of stochastic realizations, which are further exaggerated in small-volume environments. Stochasticity has been reported to blur the topological boundaries near bifurcation points~\cite{TomiStatInferencePaper}, making this a very natural hypothesis. We will later discover that~\ref{model3} expanded to the third degree displays excellent convergence and variance properties which are preserved in its bimolecularization (Section~\ref{selectionofbimolecularstrait}, Figure~\ref{summarizedplot} (d),(f)), demonstrating the potential of polynomial approximations as a satisfactory alternative to Kerner Polynomialization.

\section{Analysis of~\ref{model3}}\label{proofofconcept}

\begin{figure}[h!]
  \begin{minipage}[c]{0.44\textwidth}
    \includegraphics[width=\textwidth]{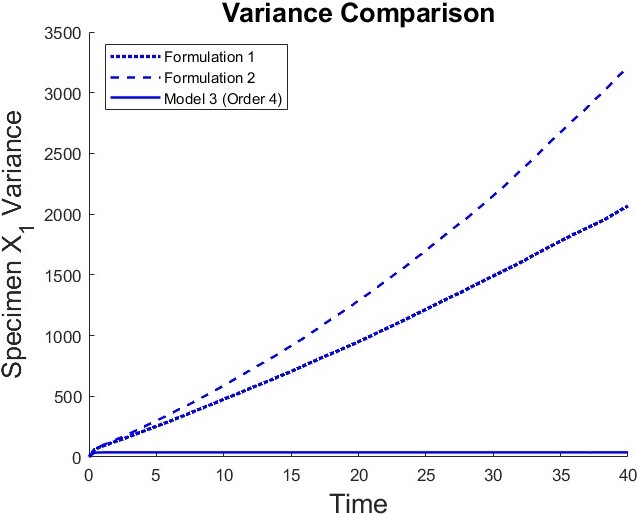}
  \end{minipage}\hfill
  \begin{minipage}[c]{0.55\textwidth}
  \caption[Variance of Formulations $1,2$ and Taylor network]{Further analysis of the data averaged in Figure~\ref{onlythemostsnecessary} (a) shows that the fourth order Taylor network demonstrates well-regulated variance, which enables smaller sampling for accurate aggregate data. Increasing volume $V$ leaves negligible qualitative distinctions. Figure~\ref{taylorinvestigation} (c) verifies that the variance is surprisingly preserved as the order of expansion is greatly diversified.}
  \label{excellentvariance}
  \end{minipage}
\end{figure}

We contrast~\ref{model2} with the dynamics of~\ref{model3} expanded to varying orders. The variance for the two formulations~\eqref{Formulation1} and~\eqref{Formulation2} obtained via Kerner Polynomialization shows an approximate linear increase in Figure~\ref{excellentvariance}. In contrast, the canonically inverted truncated Taylor expansions (called \textit{Taylor networks}) demonstrate sustained outstanding variance with no perceptible change in distribution as time progresses (Figure~\ref{taylorinvestigation} (c) or~\ref{skewinx} in Appendix).

This suggests another advantage series expansions may possess--in biochemical settings where Monte Carlo methods are costly to implement, far fewer sampling may be sufficient for accurate averaging when synthesizing a Taylor network (e.g. induced from~\ref{model3}) which is discovered to display tightly regulated variance, instead of synthesizing~\ref{model2} for example. Note that for synthetic implementations it must be verified that the variance is maintained after bimolecularization\footnote{A further discussion is included in Section~\ref{selectionofbimolecularstrait}, where bimolecularization strategies are studied. See Figure~\ref{summarizedplot} (f), for how the variance of the Taylor network of order $4$ varies under bimolecularization. In all bimolecularizations we consider, the molecule count variance is at maximum $80$--far below the exploding variance (near $10^3$ at $t=20$) observed for the formulations~\eqref{Formulation1} and~\eqref{Formulation2}  in Figure~\ref{excellentvariance}. Identical reactor volumes $V=100$ are taken  for Figures~\ref{excellentvariance},~\ref{summarizedplot}.}, where for the current test system we will later observe excellent variance regulation even after bimolecularization. Of course, there is no guarantee that this phenomenon will more generally translate to arbitrary ODE systems, however it does provide a proof of principle to the hypothesis that \textit{approximate} Taylor networks can possess qualitatively distinct dynamics that may be more desirable than \textit{exact} networks~\eqref{Formulation1},~\eqref{Formulation2} whose RREs~\eqref{model2} fully preserve deterministic dynamics of the original model~\eqref{model1}.

In Figure~\ref{taylorinvestigation}, a $10\%$ random perturbation to the rates of all reaction channels is assumed in the Taylor networks induced by~\ref{model3}. (a) computes the deterministic real roots of the Taylor expansions, revealing the existence of artificially generated negative roots\footnote{In some contexts within the literature, such critical points may be referred to as \textit{artefacts}.}. Induced by a polynomial approximation, Taylor networks suffer from all the usual caveats that come with higher order polynomial-based models. Though artefacts are inevitably generated in the polynomial approximation process, we expect them to grow less relevant as the accuracy of the approximation is enhanced, typically by expanding to higher orders. However, a disadvantage of Taylor networks may arise from the \textit{multiple root issue}, where for some polynomials of large degree, minute numerical instabilities in their coefficients may lead to massive perturbations in their computed roots (see Wilkinson polynomials~\cite{WilkinsonPolynomialProperReference}), which can generate incorrect steady states.

\begin{figure}[h!]
  \begin{subfigure}[b]{0.325\textwidth}
    \includegraphics[width=\textwidth]{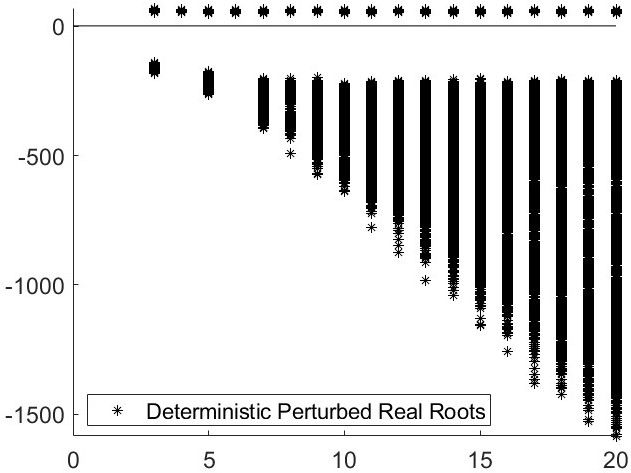}
    \caption{ }
  \end{subfigure}
  \hfill
    \begin{subfigure}[b]{0.325\textwidth}
    \includegraphics[width=\textwidth]{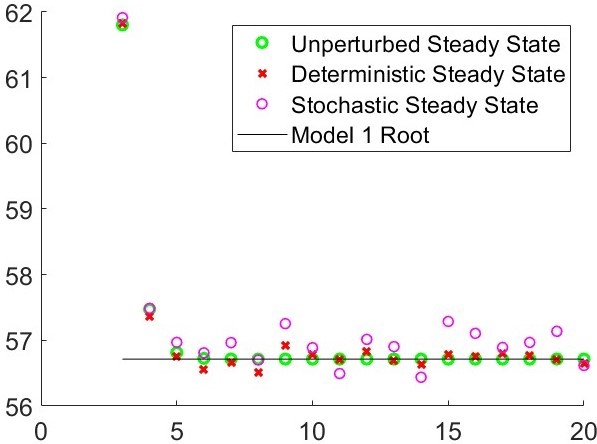}
    \caption{ }
  \end{subfigure}
  \hfill
  \begin{subfigure}[b]{0.325\textwidth}
    \includegraphics[width=\textwidth]{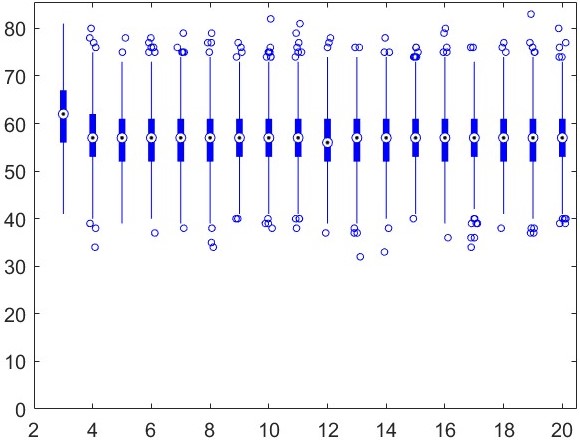}
    \caption{ }
  \end{subfigure}
  \caption[Taylor networks under random perturbations in reaction channels]{A random $10\%$ perturbation has been made in every reaction channel of the Taylor network, obtained by expanding from the $3$rd to the $20$th order (horizontal axis). Averages are taken over $10^3$ realizations, and we plot molecule counts in the vertical axis. ``Unperturbed Steady State'' in (b) refers to the deterministic positive root of~\ref{model3}.}
  \label{taylorinvestigation}
\end{figure} 

Figure~\ref{taylorinvestigation} (b) averages over deterministic and stochastic positive roots of \ref{model3} whose every coefficient has been randomly perturbed by $10\%$, or otherwise left unperturbed. The root of the simulated canonical inversion, denoted ``Stochastic Steady State'', is considered to be the end position of the $\mathrm{X}_1$-trajectory after  sufficient time has elapsed. We further analyze the data in (c), which gives the box plot of the stochastic roots in (b). \ref{model3} displays outstanding variance from expansions as low as of third order. Outliers are depicted as hollow balls.

In summary, we have established that series expansions can act as an attractive alternative to Kerner Polynomialization. Particularly,~\ref{model2} possesses exploding solutions and rapidly increasing variance, whereas~\ref{model3} suffers from neither of the deficiencies. The advantages of Kerner Polynomialization in precisely preserving deterministic dynamics may not translate well to the stochastic setting; as shown in this chapter, polynomial approximations using series expansions can be more suitable for synthetic implementation. However, it is difficult to assess the general superiority of one method over the other which must hold for all alternative ODE systems--series expansions also suffer from disadvantages which may be emphasized for different models. We therefore do not select one strategy over the other and utilize both polynomialization techniques in our inversion framework.

\subsection{On Optimal Polynomial Selection}\label{optimalpolyselection}
Before ending the chapter on polynomialization strategies, we briefly remark on polynomial approximations to~\ref{model1}. So far, Taylor expansions have essentially been used as a proxy for the optimum polynomial approximation. We propose an open question regarding Taylor networks: \textit{Which point of expansion is the most suitable to represent the ODE system for the purposes of inversion into a chemical reaction network?} We heuristically observe that expansion around critical points preserves the local topology, but often destroys any limit cycles we wish to approximate by the expansion; we further remark that for symmetric ODE systems, expanding around symmetry-breaking points often performs worse in containing exotic dynamics manifested by the original model in the polynomial approximation, a suspected phenomenon for which we have yet to provide a suitable explanation.

\begin{figure}[h]
  \begin{subfigure}[b]{0.5\textwidth}
    \includegraphics[width=\textwidth]{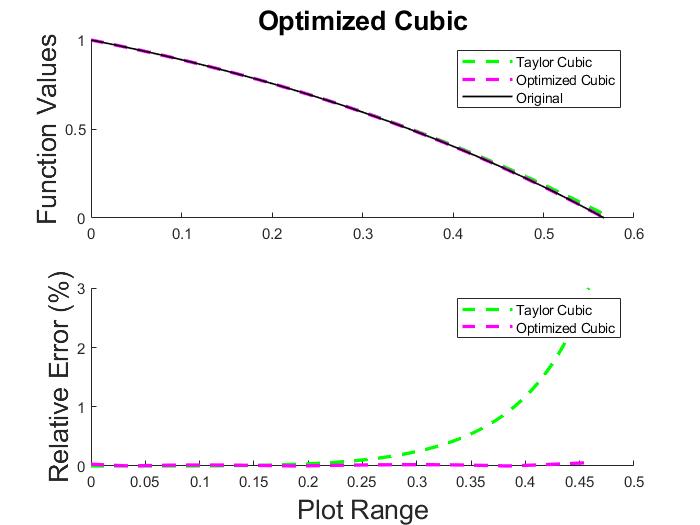}
    \caption{Expansion to order $4$ (degree $3$)}
  \end{subfigure}
  \hfill
  \begin{subfigure}[b]{0.5\textwidth}
    \includegraphics[width=\textwidth]{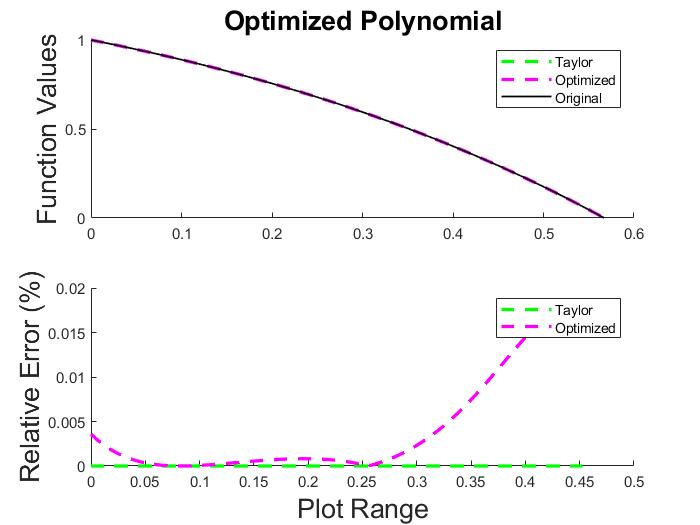}
    \caption{Expansion to order $11$ (degree $10$)}
  \end{subfigure}
  \caption[Performance of the optimization procedure]{Though showing superior approximation for low orders, the optimization procedure is quickly superseded by Taylor expansions for higher orders. The relative errors in the functional value of optimum polynomials computed using the optimization algorithm and series expansions, relative to the original right hand side of~\ref{model1}, are given in the bottom row. }
  \label{optimalselect}
\end{figure}

Using the method articulated in Section~\ref{polynomialoptimizationmethod}, we now carry out a rudimentary investigation of a polynomial structure (right hand side of~\eqref{model5}) with good fit
\begin{equation}\label{model5}
\Dot{x_1} = \sum_{\ell = 0}^{m} a_{\ell} x_1^{\ell},
\end{equation}
by optimizing over the coefficients using the Nelder-Mead Simplex Method~\cite{fminsearchtheory}, implemented via~\cite{fminsearch}. Imposing that there are no cross-negative terms in one-dimension corresponds to asserting that $a_0 \ge 0$. We minimize over the numeric $\ell_2$-norm, with data inputs taken over $99$ equidistant intervals representing the plot range\footnote{That is, we sum the square of the distance, representing deviance, between the target function and the polynomial structure~\eqref{model5} to compute the cost $L$. The deviance is summed over $100$ data points equally spaced apart in the domain in which we wish to approximate the target function. For Figure~\ref{optimalselect}, the domain was taken to be $[0,x^*] \approx [0,0.5671]$ for optimization.}.

Figure~\ref{optimalselect} (a) compares the performance of the algorithm within cubic structures, where the optimum polynomials are computed to be
\begin{gather}
    - 0.7320x_1^3 - 0.9101x_1^2 - 1.0116x_1 + 1.0003, \quad\text{root}\approx 0.5672, \tag{Optimized Cubic} \\ - 0.5x_1^3 - x_1^2 - x_1 + 1, \quad\text{root}\approx 0.5747.\tag{Taylor Cubic}
\end{gather}
The expected critical point is $x^* \approx 0.5671$, which is much closer to the root of the optimized cubic. However, their positions are quickly reversed; (b) shows that for expansions of the $11$th order, the Taylor polynomial appears to supersede the optimized polynomial. This is undesirable because the polynomial structure~\eqref{model5} has been designed to encompass all Taylor expansions of~\ref{model1}.

\begin{figure}[h]
  \begin{subfigure}[b]{0.5\textwidth}
    \includegraphics[width=\textwidth]{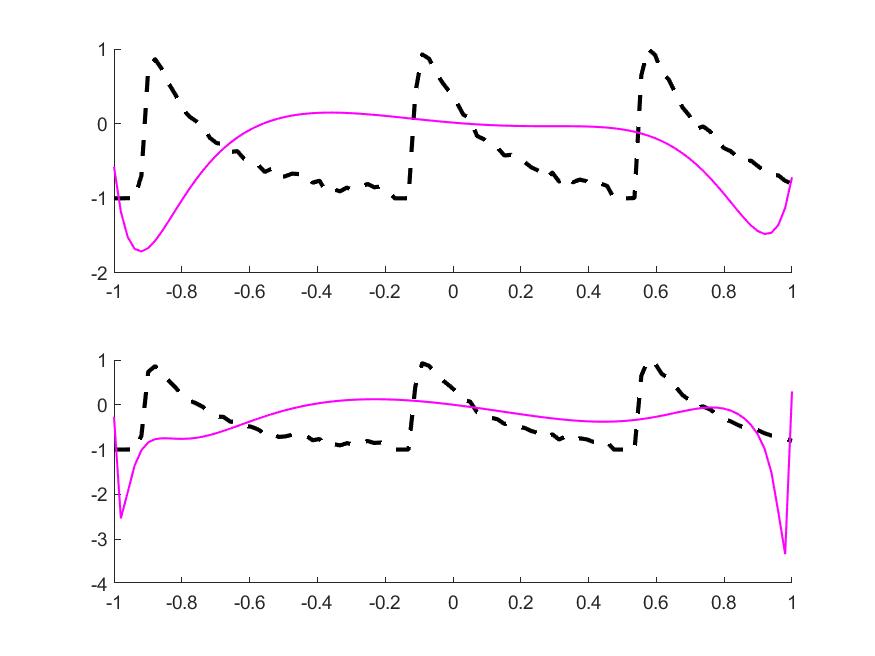}
  \end{subfigure}
  \hfill
  \begin{subfigure}[b]{0.5\textwidth}
    \includegraphics[width=\textwidth]{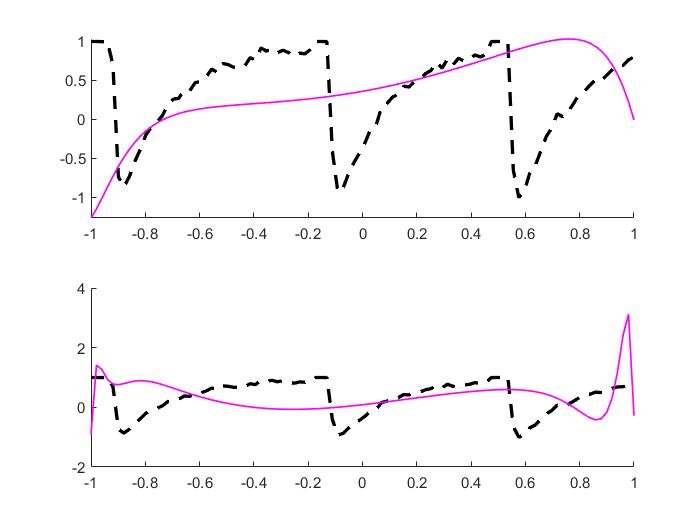}
  \end{subfigure}
  \caption[Comparison of polynomial structures in the optimization algorithm]{Polynomial structures of orders $21, 102$ have been utilized for the optimization process, for top, bottom, respectively. The dashed lines represent stochastically simulated Repressor $\mathrm{R}$ data, whereas the solid lines give the optimum polynomial found by the optimization procedure. We see that increasing orders does not necessarily translate to enhanced precision. Horizontal axis was time and vertical axis the concentration of $\mathrm{R}$ before affine transformation (see discussion).
  }
  \label{circadianoptimalselect}
\end{figure}

A consistent theme we observe is that higher orders of the polynomial structure do not necessarily translate to better performance of the optimization algorithm. Case in point, in Figure~\ref{circadianoptimalselect} we have purely for experimental purposes attempted to find a polynomial representation of the dynamics of the VKBL circadian Repressor $\mathrm{R}$, further expounded in Appendix~\ref{appendixA} as a supplement example to Chapter~\ref{oldchapter1}. Interested readers are encouraged to view the relevant section. Affine transformations have been taken on the domain and target space to restrict both ranges to the interval $[-1,1]$ before optimization, and the same procedure repeated after reflection of the oscillator data over the horizontal axis. As is evident in the plots, it is difficult to deduce any additional benefits the order $102$ structure procures over the order $21$ structure. Numerically, the cost of deviance from the input data is computed to be $L \approx 3208$ for all plots in Figure~\ref{circadianoptimalselect}. In contrast, optimizing over a polynomial structure of order $502$ incurs the cost $L \approx 4973410$.

It is slightly distressing that Lagrange interpolation~\cite{properlagrangeinterpolation} to the $13$th degree seems to provide a better fit (plot not included) to the oscillator. It should however be emphasized that the resulting polynomial yielded multiple coefficients to the order $\mathcal{O}(10^{23})$, thus its credibility is highly suspect due to problems arising from numerical precision. We expect that even fairly aged deep-learning approaches such as multilayer perceptrons~\cite{MultilayerPerceptronProperBook} will enhance the performance of the optimization strategy, although the training process will take much longer\footnote{Unlike the illustrated approach which requires no training at all and produces results within tens of minutes. One may also find splines to be a better approach to approximating apparently non-smooth data as in Figure~\ref{circadianoptimalselect}, but piecewise polynomials have no kinetic interpretation.}. See Appendix~\ref{appendixE}, where the idea is briefly discussed as a possible extension.

%% file: chapter6.tex
\chapter{Selection of Kineticization and Bimolecularization Strategy}\label{oldchapter5}

In Chapter~\ref{oldchapter4}, we have established Taylor expansions (and more generally polynomial approximations obtained through any suitable method) as an attractive alternative to Kerner Polynomialization, which fully preserves deterministic dynamics but may perform worse under stochastic settings. We now investigate kineticization and bimolecularization strategies, the latter especially critical if synthetic implementations are desired. 

We contrast two kineticization strategies, the Quasi-Steady State Transformation $\Psi_{QSST}$ (Section~\ref{QSSA}) and the Kowalski Transformation $\Psi_{Kow}$ (Section~\ref{ForcedVariable}). The test system is chosen to be non-kinetic linear decay, expressed in the first dimension of~\eqref{easysystem},
\begin{equation}\label{lineardecay}
    \Dot{x_1}=-1\quad \implies \quad r_1: \quad \varnothing \stackrel{-1}{\longrightarrow} \mathrm{X}_1.
\end{equation}
$\Psi_{Kow}$ introduces the variable $y_1 = 1/x_1$, under which the system~\eqref{lineardecay} transforms as~\eqref{thing} and is canonically inverted to the network~\eqref{network}:

\begin{minipage}{.45\textwidth}
\begin{equation}\label{thing}
    \begin{aligned}
    &\Dot{x_1}=-x_1y_1, \\
    &\Dot{y_1}=y_1^2,
    \end{aligned}
\end{equation}
\end{minipage}
\begin{minipage}{.45\textwidth}
\begin{equation}\label{network}
\begin{aligned}
    &r_1^\prime: \quad \mathrm{X}_1 + \mathrm{Y}_1 \stackrel{1}{\longrightarrow} \mathrm{Y}_1 ,\\
    &r_2^\prime : \quad 2\mathrm{Y}_1 \stackrel{1}{\longrightarrow} 3\mathrm{Y}_1.
\end{aligned}
\end{equation}
\end{minipage}

\vspace{2pt}

\noindent In contrast, applying $\Psi_{QSST}$ to~\eqref{lineardecay} yields 

\begin{minipage}{.45\textwidth}
\begin{equation}\label{QSSAthing}
    \begin{aligned}
    &\Dot{x_1}=-\omega_{x_1}^{-1} x_1 p_{1}(x_1,y_1)y_1, \\
    &\mu \Dot{y_1}=1-x_1p_{1}(x_1,y_1)y_1,
    \end{aligned}
\end{equation}
\end{minipage}
\begin{minipage}{.45\textwidth}
\begin{equation}\label{QSSAnetwork}
\begin{aligned}
    &r_1^{\prime\prime}: \quad \mathrm{X}_1 + \mathrm{Y}_1 \stackrel{1}{\longrightarrow} \mathrm{Y}_1 ,\\
    &r_2^{\prime\prime}: \quad \varnothing \stackrel{1/\mu}{\longrightarrow} \mathrm{Y}_1, \\
    &r_3^{\prime\prime}: \quad \mathrm{X}_1 + \mathrm{Y}_1 \stackrel{1/\mu}{\longrightarrow} \mathrm{X}_1,
\end{aligned}
\end{equation}
\end{minipage}

\vspace{2pt}

\begin{figure}[h!]
  \begin{subfigure}[b]{0.5\textwidth}
    \includegraphics[width=\textwidth]{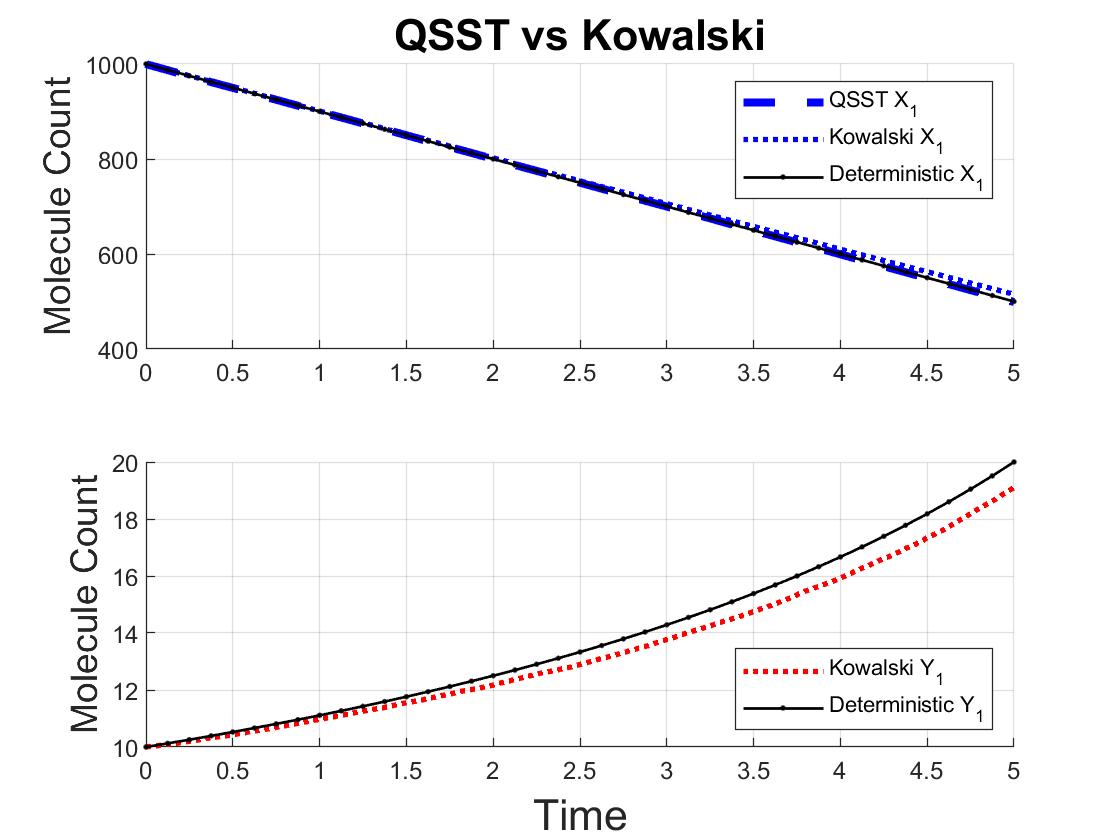}
    \caption{ }
  \end{subfigure}
  \hfill
  \begin{subfigure}[b]{0.5\textwidth}
    \includegraphics[width=\textwidth]{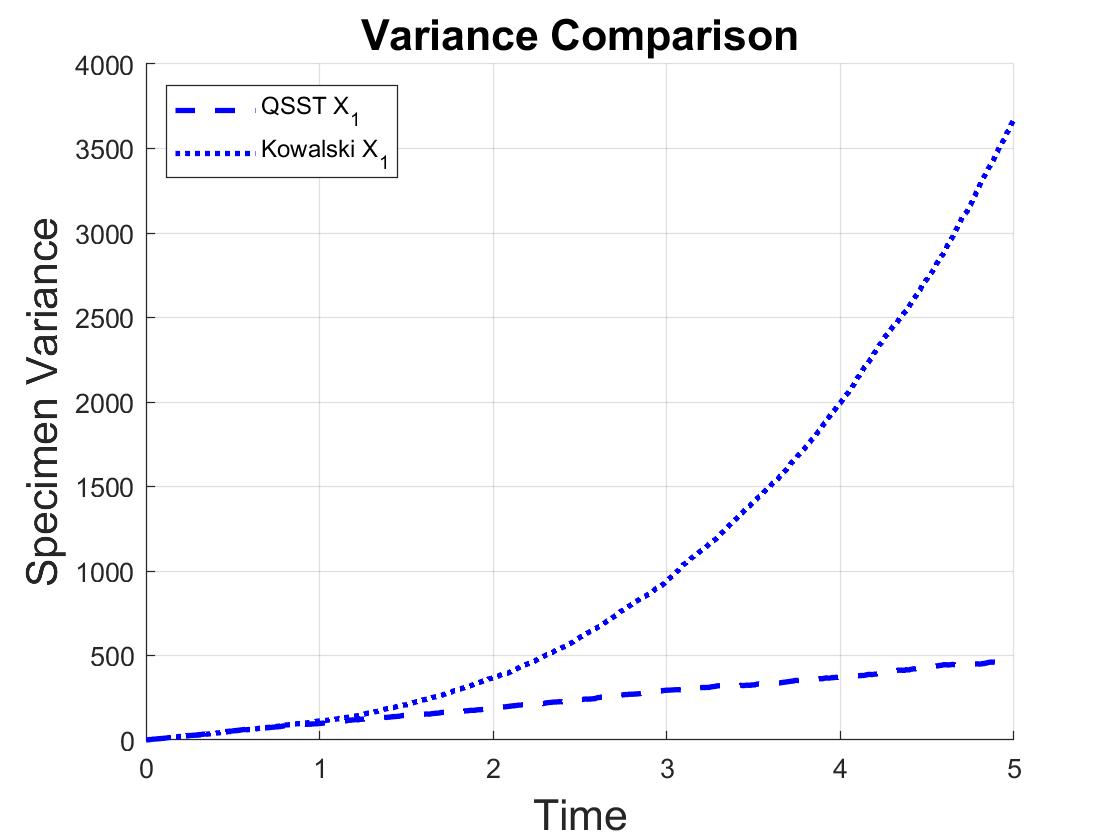}
    \caption{ }
  \end{subfigure}
  \caption[Demonstration of validity of kineticization techniques.]{Demonstration of  kineticization techniques $\Psi_{Kow}$, $\Psi_{QSST}$.}
  \label{QSSAplot}
\end{figure}

\begin{figure}[h!]
  \begin{subfigure}[b]{0.5\textwidth}
    \includegraphics[width=\textwidth]{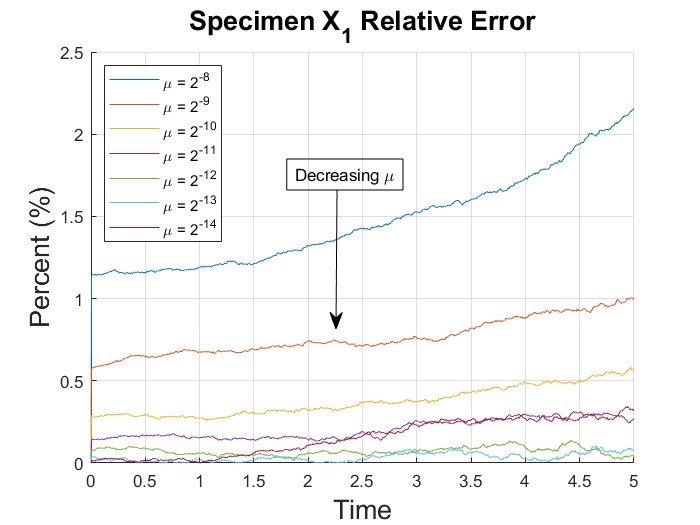}
    \caption{ }
  \end{subfigure}
  \hfill
  \begin{subfigure}[b]{0.5\textwidth}
    \includegraphics[width=\textwidth]{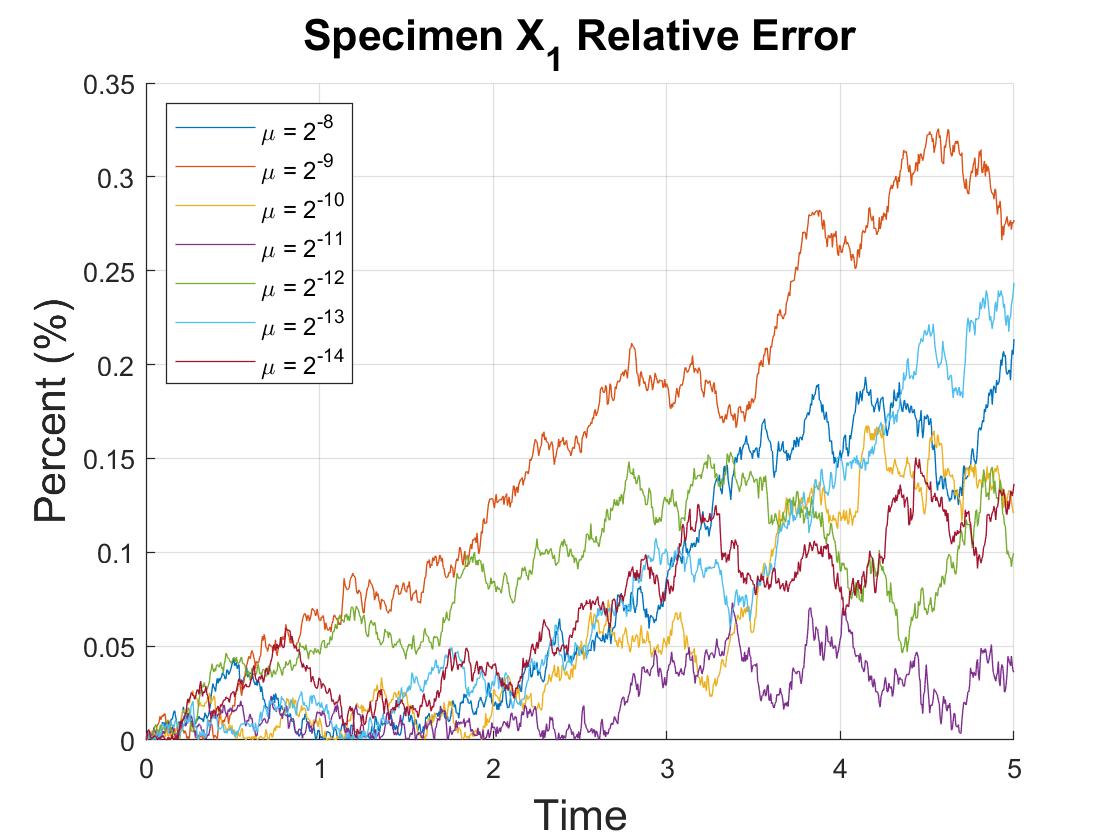}
    \caption{ }
  \end{subfigure}
  \caption[A comparison of varying $\mu$ in $\Psi_{QSST}$]{In (a), we take $\mu = 2^{-8}, \dots, 2^{-14}$ for $\mathrm{Y}_1(0) =30V$ for $\Psi_{QSST}$, where the averaged stochastic trajectories approach the deterministic trajectory as $\mu \downarrow 0$. In (b), we have initialized the specimen $\mathrm{Y}_1$ to negligible amounts, where no substantive difference is observed for all values of $\mu$. Relative errors are taken with respect to the deterministic solution in Figure~\ref{QSSAplot} (a).}
  \label{QSSAComparison}
\end{figure}

\noindent where we have used $p_{1}(x_1,y_1) \equiv 1$, $\omega_{x_1} =1$ for the inversion~\eqref{QSSAnetwork}. Linear decay~\eqref{lineardecay} predicts a decrease in molecule count given by $\mathrm{X}_1(t) = \mathrm{X}_1(0)-Vt$ for reactor volume $V= 100$, which $\Psi_{Kow},\Psi_{QSST}$ correctly approximate in Figure~\ref{QSSAplot} (a). However in (b), we see an explosive increase in variance for $\Psi_{Kow}$ which is unobserved for $\Psi_{QSST}$.

\begin{figure}[h!]
  \begin{subfigure}[b]{0.5\textwidth}
    \includegraphics[width=\textwidth]{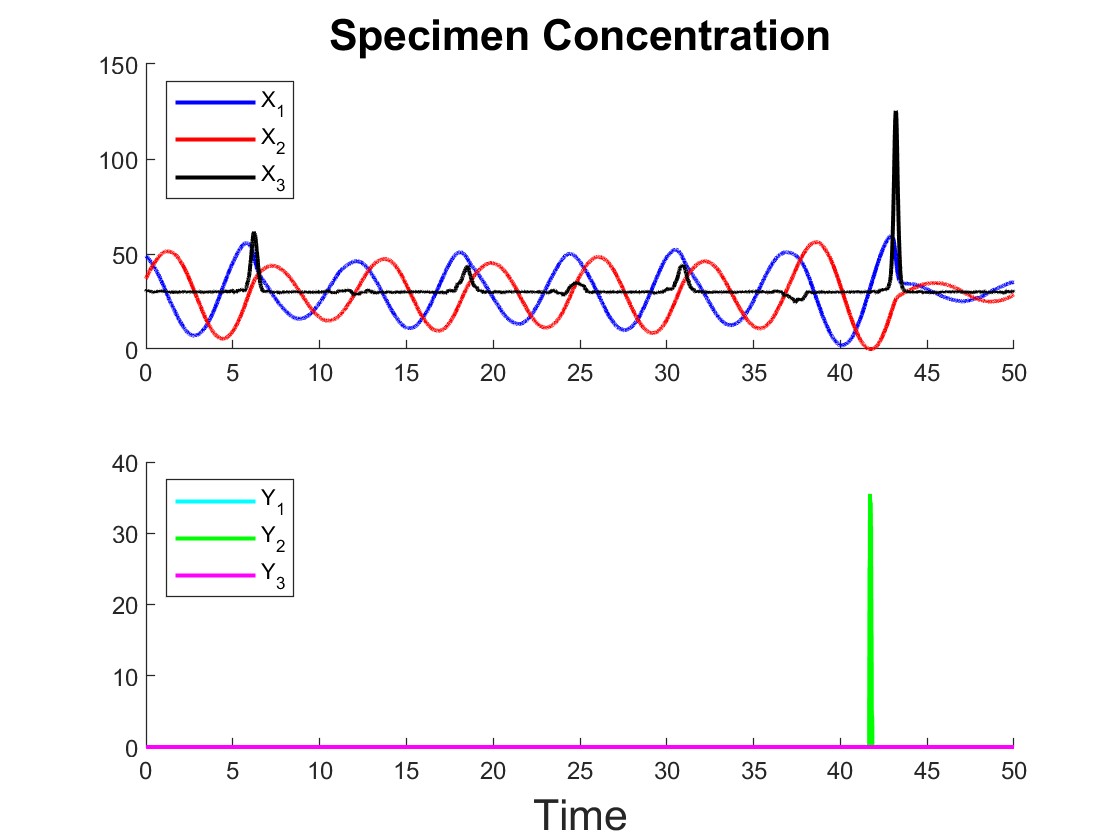}
    \caption{ }
  \end{subfigure}
  \hfill
  \begin{subfigure}[b]{0.5\textwidth}
    \includegraphics[width=\textwidth]{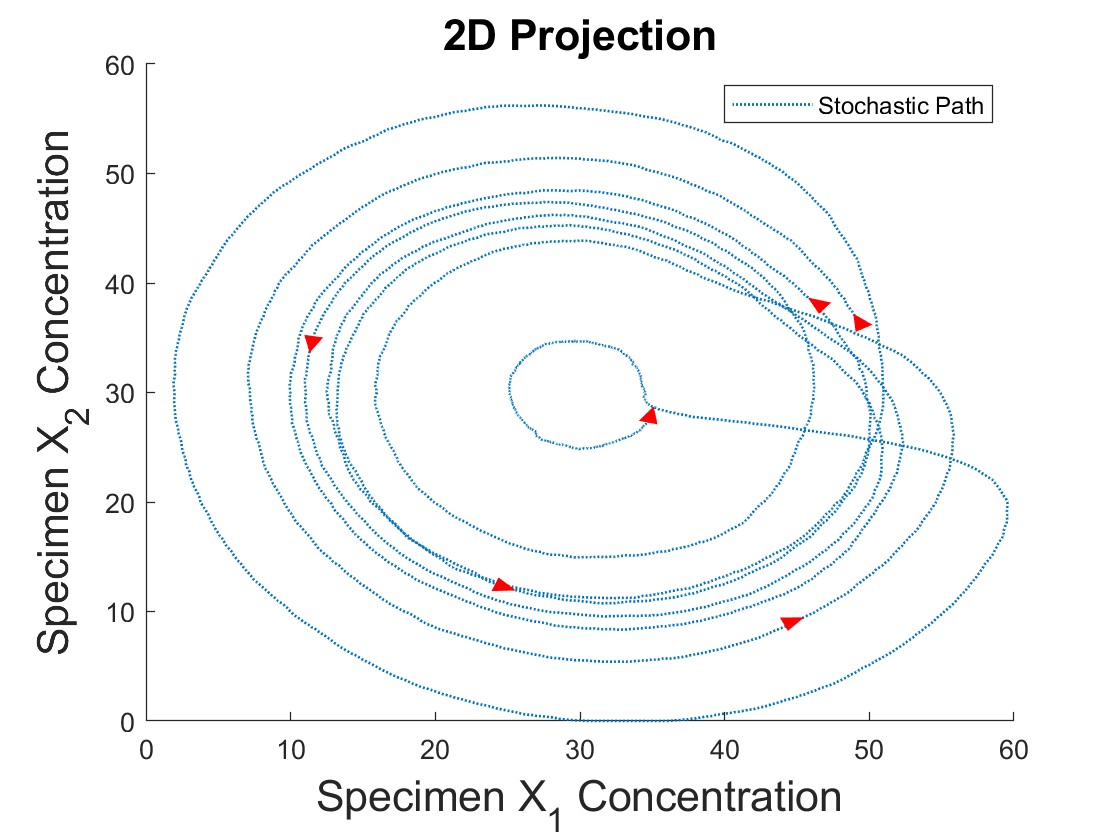}
    \caption{ }
  \end{subfigure}
  \hfill
  \begin{subfigure}[b]{0.5\textwidth}
    \includegraphics[width=\textwidth]{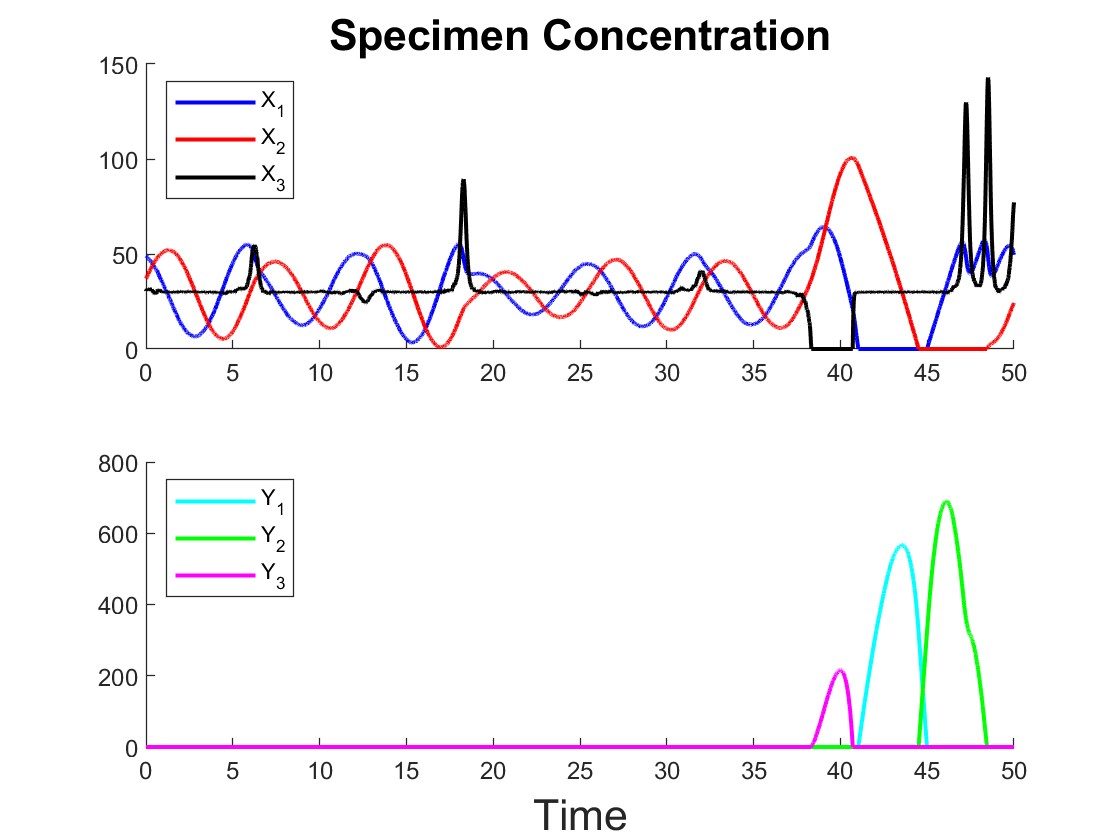}
    \caption{ }
  \end{subfigure}
  \hfill
  \begin{subfigure}[b]{0.5\textwidth}
    \includegraphics[width=\textwidth]{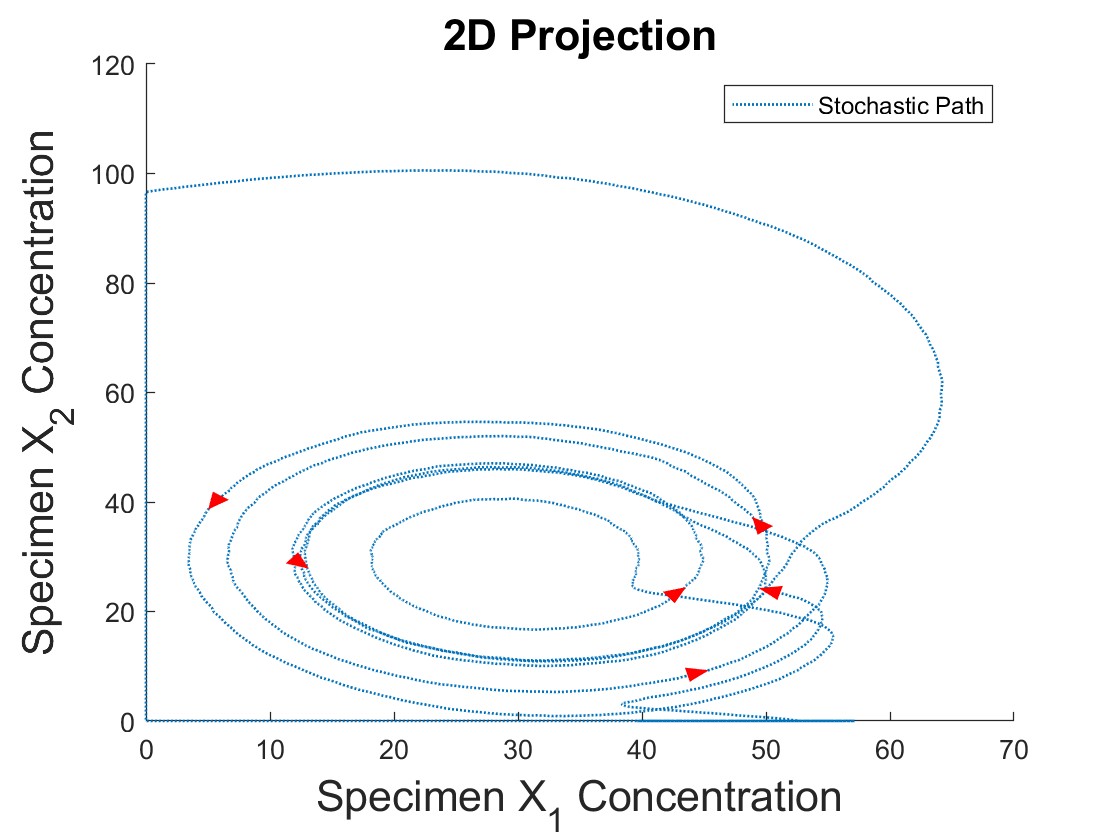}
    \caption{ }
  \end{subfigure}
  \caption[Successful and failed recovery of the R\"{o}ssler system~\eqref{rosslerode}]{Successful (a-b) and failed (c-d) recovery of the R\"{o}ssler system~\eqref{rosslerode} for $a=b=0.1$, $c=18$ (chaotic regime). Once specimen $\mathrm{Y}_i$ introduced via $\Psi_{QSST}$ is perturbed away from partial equilibrium, which is near-extinction for this example, the preservation of dynamics is broken and the chemical system behaves irregularly in (c-d). Arrows point in direction of increasing time in the $2$D projection, where molecule counts have been normalized to concentrations for all plots to enable scaled comparison. See Chapter~\ref{oldchapter6} for further simulations.} 
  \label{quasiillustrate}
\end{figure}

We have varied $\mu$ and observed the convergence of the stochastic trajectories for $\Psi_{QSST}$ in Figure~\ref{QSSAComparison} (a), after initializing the specimen molecule counts to $\mathrm{X}_1(0) = 10V, \mathrm{Y}_1(0) =30V$. Further observations may be made by choosing the initial condition $\mathrm{Y}_1(0) = 0$ as in (b), where no qualitative distinctions may be found for all (sufficiently small) values of $\mu$ being tested.

The accuracy of quasi-steady state approximations in general are largely impacted by the validity of their perturbation analysis, which often assumes an instantaneous convergence of the fast variable to a partial equilibrium. Noise or extinction of slow specimen may break the preservation of dynamics  by perturbing fast specimen away from their partial equilibrium, where minute $\mu$ indicates a longer observation timeframe of validity. Once perturbed, initial data is quickly forgotten during the relaxation to the quasi-steady state, stabilizing the trajectories to incorrect values and continuing the propagation (see Figures~\ref{QSSAComparison} (a),~\ref{quasiillustrate}). This deviance may be mitigated by decreasing the value of $\mu$ when detected. 

$\Psi_{Kow}$ is also affected by perturbation, due to imposing strict adherence to the solution manifold $y_1 = 1/x_1 \iff \mathrm{Y}_1 = V^2/\mathrm{X}_1$ for dynamical equivalence which is not guaranteed under stochastic settings. Furthermore, specimen concentrations must be initialized to the precise quantity $y_1(0) = 1/x_1(0)$. This is in stark contrast to $\Psi_{QSST}$, where the asymptotic equivalence holds for all realistic (non-negative) quantities of $\mathrm{Y}_1$ and is thus infinitely more tolerant of noise contamination in initialization, given that $\mu$ is sufficiently small.

At this point, it seems apt to preemptively present our insights from attempting Kowalski Transformations for more complex ODEs in Chapter~\ref{oldchapter6}, where the induced general system required exceptionally large volumes for successful encapsulation of the dynamics. In part, Figure~\ref{QSSAplot} (a) hints at this phenomenon by converging the general system induced by $\Psi_{Kow}$ to slightly incorrect values, even in the simple linear decay~\eqref{lineardecay}. More details about the impracticality of Kowalski Transformations (while using the Lorenz system as an example) are given in Appendix~\ref{chaoticsystemsinappendix}.

Although both $\Psi_{Kow}$ and $\Psi_{QSST}$ are prone to stochastic errors, it is possible to extend the observation timeframe of the general system induced by the latter technique by decreasing $\mu$. Generally speaking, $\Psi_{QSST}$ showed more accuracy and less variance (e.g. Figure~\ref{QSSAplot}), was significantly less reliant on initialization, and operated excellently for low reactor volumes in which $\Psi_{Kow}$ showed substantial deficiencies (Figure~\ref{finalfigure}). And perhaps most critically, the network formed by canonical inversion of the general system induced by $\Psi_{Kow}$ possesses no mechanism to regenerate $\mathrm{Y}_s$ after extinction has occurred--therefore permanently altering the dynamics of the network once stochasticity pulls $\mathrm{Y}_s$ molecule counts down to $0$ (Appendix~\ref{chaoticsystemsinappendix}). This is an overly restrictive limitation, especially when quantities of $\mathrm{X}_s$ and $\mathrm{Y}_s$ are required to vary inversely proportionally for preservation of the dynamics of the original system. We therefore reject Kowalski Transformations as a kineticization technique for the inversion framework.

\section{Selection of Bimolecularization Strategy}\label{selectionofbimolecularstrait}

\begin{figure}[h!]
  \begin{subfigure}[b]{0.5\textwidth}
    \includegraphics[width=\textwidth]{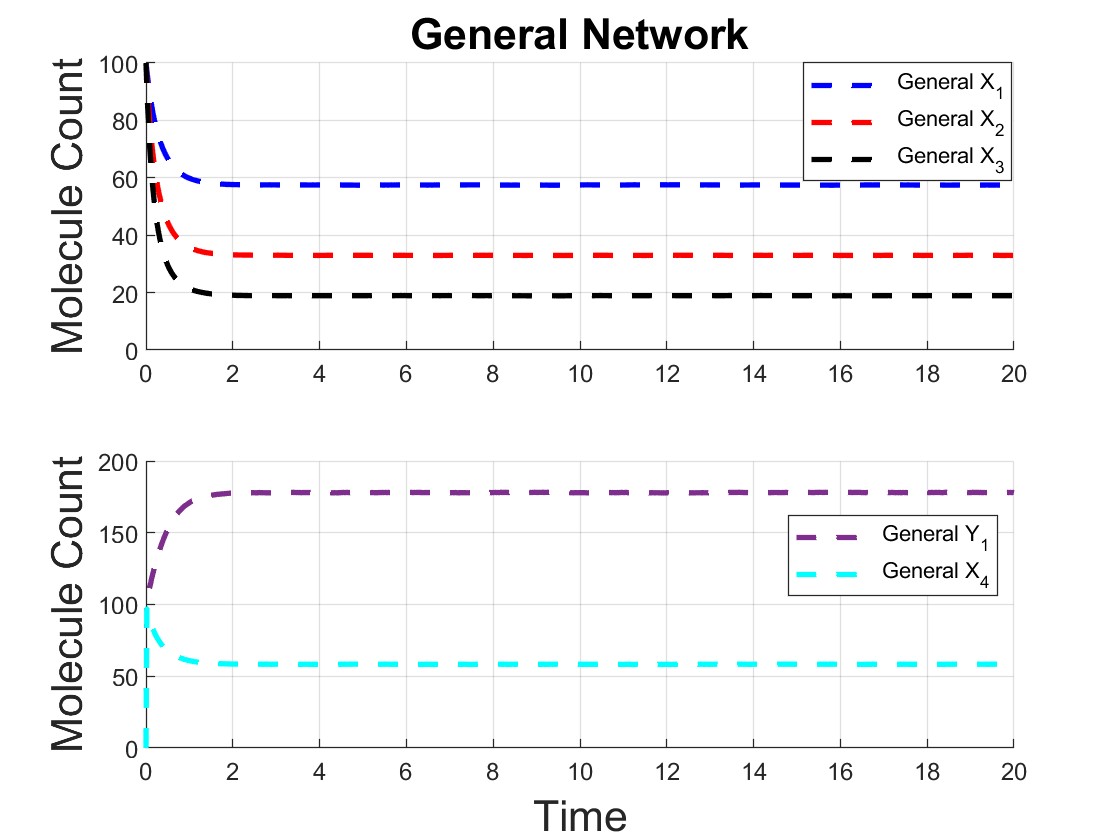}
    \caption{ }
  \end{subfigure}
          \hfill
  \begin{subfigure}[b]{0.5\textwidth}
    \includegraphics[width=\textwidth]{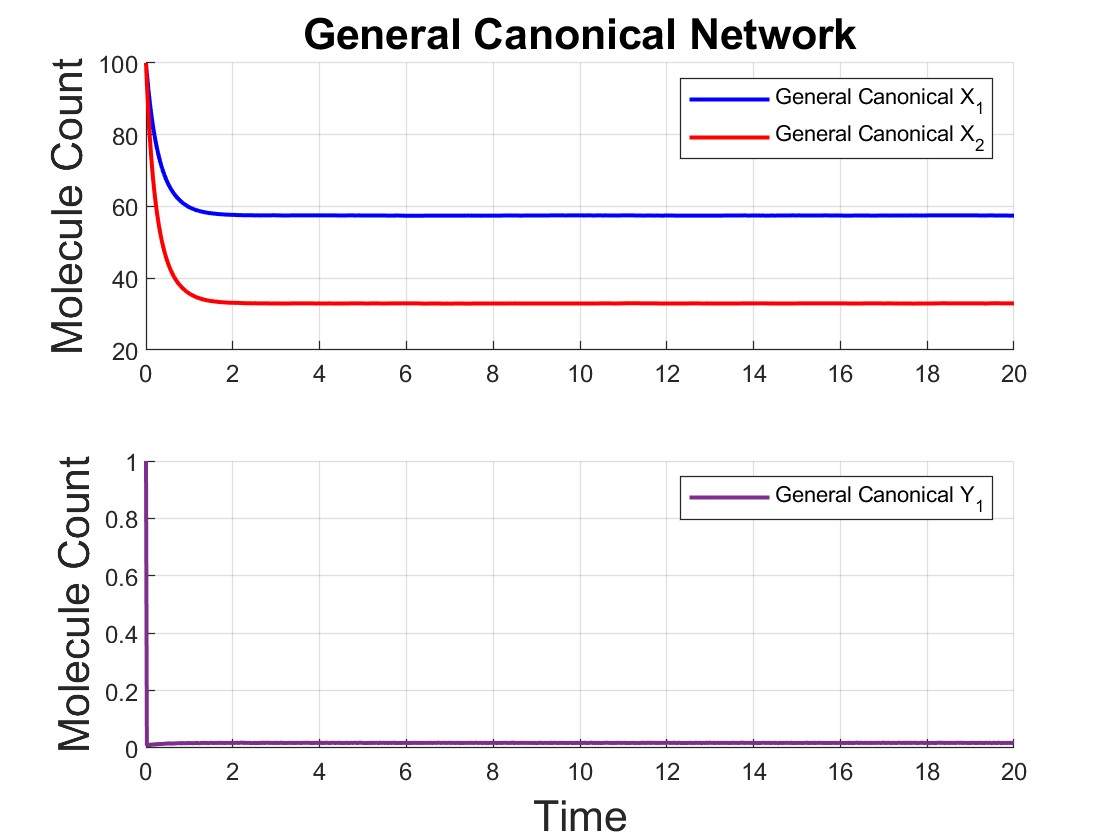}
    \caption{ }
  \end{subfigure}
        \hfill
  \begin{subfigure}[b]{0.5\textwidth}
    \includegraphics[width=\textwidth]{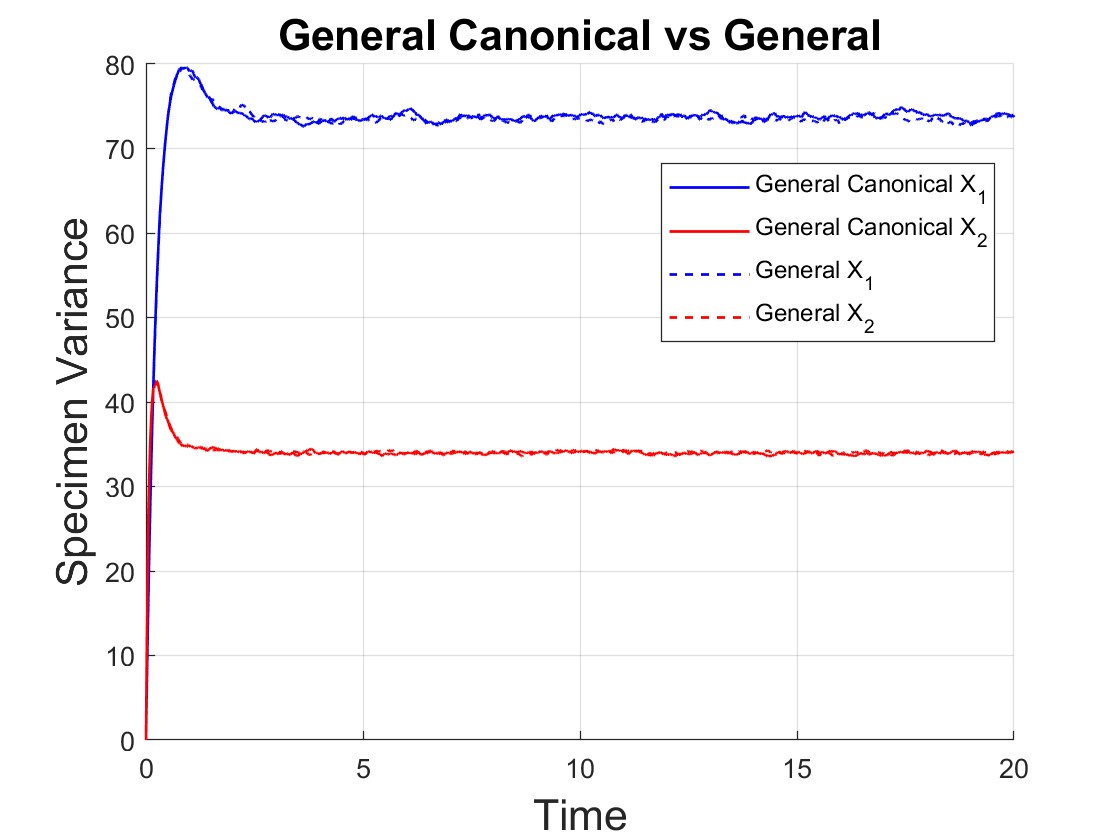}
    \caption{ }
  \end{subfigure}
  \hfill
    \begin{subfigure}[b]{0.5\textwidth}
    \includegraphics[width=\textwidth]{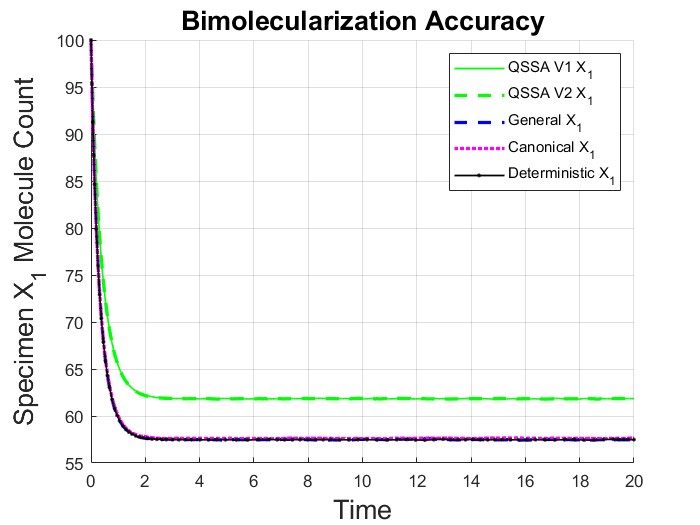}
    \caption{ }
  \end{subfigure}
  \hfill
  \begin{subfigure}[b]{0.5\textwidth}
    \includegraphics[width=\textwidth]{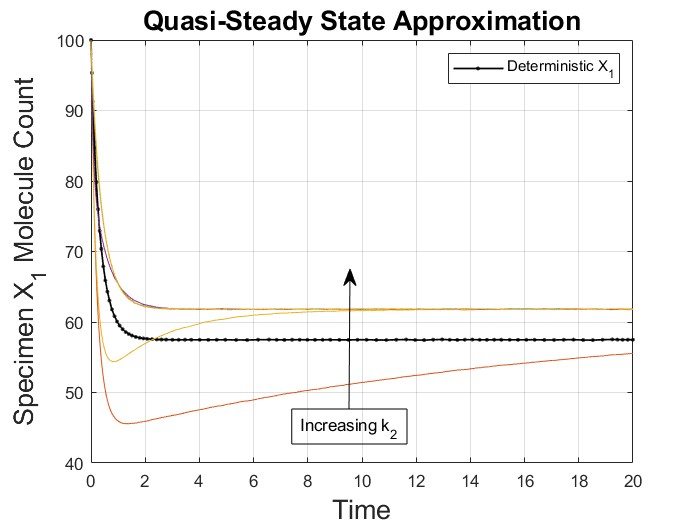}
    \caption{ }
  \end{subfigure}
    \hfill
  \begin{subfigure}[b]{0.5\textwidth}
    \includegraphics[width=\textwidth]{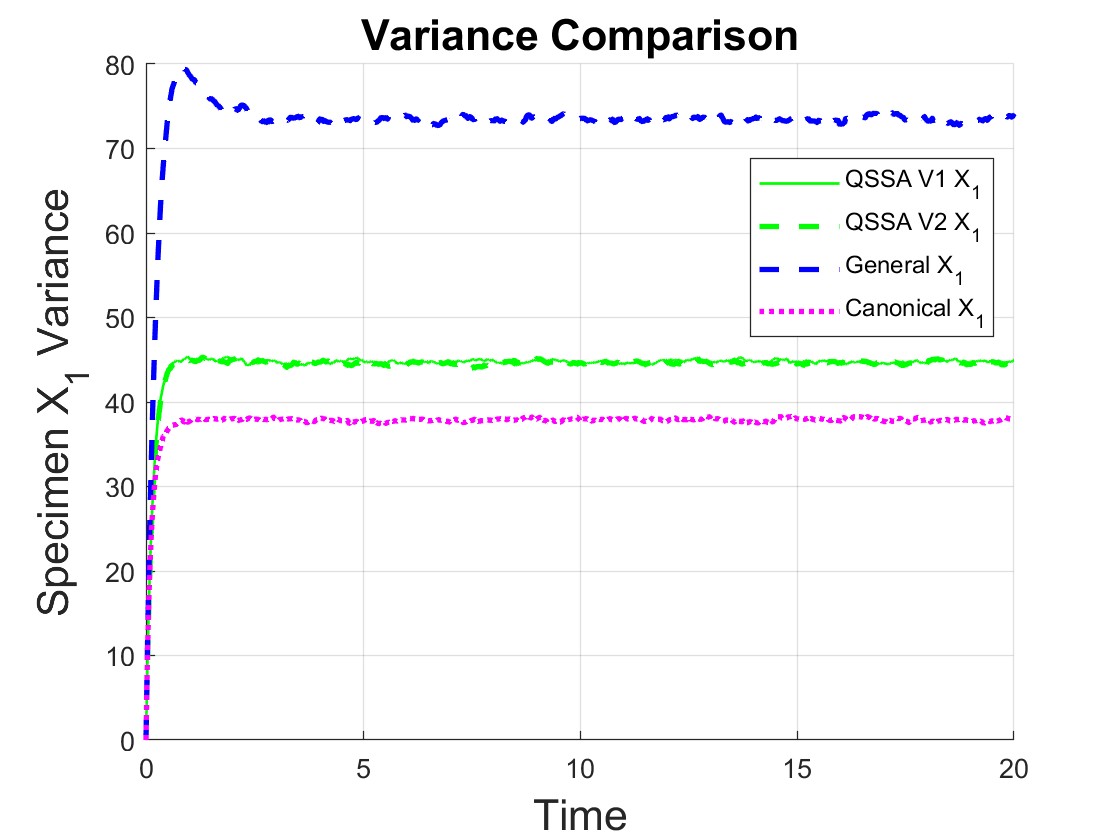}
    \caption{ }
  \end{subfigure}
  \caption[Summary of the data generated for bimolecularization techniques]{A summary of the data generated for bimolecularization technique selection. $k_2 = 10^{-1}, 10^{0}, \dots , 10^7$ have been used for (e) which plots (V1), where (V2) looks identical (plot not included). $k_2 = 10^7, k_1 = 1, k_3 = k_2/2$ were used to generate (d),(f), and the reactor volume $V = 100$ is used for all plots.} 
  \label{summarizedplot}
\end{figure}

As the Kerner Polynomialization for~\ref{model1} (i.e.~\ref{model2}) displayed the NEC phenomenon in which the equilibrium was easily removed, we instead choose the Taylor expansion~\ref{model3} as our test system. For simplicity, we analyze cubic representations for quadraticization:
\begin{equation}\label{taylorcubic}
    \Dot{x_1} = 1-x_1-x_1^2-\frac{x_1^3}{2} .
\end{equation}
An application of the General Quadraticization Algorithm (Section~\ref{generalquadraticizationalgorithm}) gives
\begin{equation}\label{kernerquadraticized}
    \begin{aligned}
        &\Dot{x_1} = 1-x_1-x_2-\frac{x_1x_2}{2}, \\
        &\Dot{x_2} = 2x_1-2x_2-2x_1x_2-x_2^2,
    \end{aligned}
\end{equation}
which is kineticized by $\Psi_{QSST}$ to form a \textit{cubic} polynomial, whose trimolecular canonical inversion may be identified as $r_1, \dots r_{10}$ of the network~\eqref{bimoleculargeneral} with $r^\prime_3$ substituted by
\begin{equation}\label{trimoleculargeneral}
    r_3: \quad \mathrm{X}_1 + \mathrm{X}_2 + \mathrm{Y}_1 \stackrel{1}{\longrightarrow} \mathrm{X}_2 + \mathrm{Y}_1.
\end{equation}
This network $r_1, \dots r_{10}$ is referenced in the following discussion as~\eqref{trimoleculargeneral} and denoted `General Canonical' in Figure~\ref{summarizedplot} (b-c), where it is compared with its bimolecularization~\eqref{bimoleculargeneral}. Another iteration of General Quadraticization to~\eqref{kernerquadraticized} while preserving kineticness\footnote{We note the usage of a modified version of the algorithm. An iteration of General Quadraticization reduces cubic terms to quadratic terms, and quadratic terms to monic terms. If the monic term presents to be cross-negative and the original quadratic term kinetic, we reject the degree reduction and leave the term unaltered.} gives
 \begin{equation}\label{bimoleculargeneralODE}
    \begin{aligned}
        \Dot{x_1} &= 1-x_1-x_1x_4-\frac{x_1x_2}{2}, \\
        \Dot{x_2} &= 2x_1-2x_2-2x_1x_2-x_2^2,\\
        \Dot{y_1} &=\frac{1}{\mu} \left(1-x_1y_1\right),\\
        \Dot{x_3} &= x_2 - 3x_3 - x_3x_4 - \frac{3x_2x_3}{2} + 2x_1^2  - 2x_1x_3 , \\
        \Dot{x_4} &= \frac{1}{\mu}x_2 + 2x_1y_1 - 2x_4-\left(2+\frac{1}{\mu}\right)x_1x_4-x_2x_4,
    \end{aligned}
\end{equation}
for $x_3 = x_1x_2$ and $x_4 = x_2y_1$. We form the bimolecular canonical inversion
\begin{equation}\label{bimoleculargeneral}
\resizebox{\textwidth}{!}{$
\begin{array}{lll}%
r_{1}: \quad \varnothing \stackrel{1}{\longrightarrow} \mathrm{X}_1,& \quad r_{2}: \quad \mathrm{X}_1 \stackrel{1}{\longrightarrow} \varnothing, & \quad r^\prime_{3}: \quad \mathrm{X}_1 + \mathrm{X}_4 \stackrel{1}{\longrightarrow} \mathrm{X}_4,\\
r_{4}: \quad \mathrm{X}_1 + \mathrm{X}_2 \stackrel{1/2}{\longrightarrow} \mathrm{X}_2,& \quad r_{5}: \quad \mathrm{X}_1 \stackrel{2}{\longrightarrow} \mathrm{X}_1 + \mathrm{X}_2,& \quad r_{6}: \quad \mathrm{X}_2  \stackrel{2}{\longrightarrow} \varnothing,\\
r_{7}: \quad \mathrm{X}_1 + \mathrm{X}_2 \stackrel{2}{\longrightarrow} \mathrm{X}_1,& \quad r_{8}: \quad 2\mathrm{X}_2 \stackrel{1}{\longrightarrow}\mathrm{X}_2,& \quad r_{9}: \quad \varnothing \stackrel{1/\mu}{\longrightarrow} \mathrm{Y}_1,\\
r_{10}: \quad \mathrm{X}_1 + \mathrm{Y}_1 \stackrel{1/\mu}{\longrightarrow} \mathrm{X}_1,& \quad r_{11}: \quad 2 \mathrm{X}_1 \stackrel{2}{\longrightarrow} 2\mathrm{X}_1+\mathrm{X}_3,& \quad r_{12}: \quad \mathrm{X}_3  \stackrel{3}{\longrightarrow} \varnothing,\\
r_{13}: \quad \mathrm{X}_1+\mathrm{X}_3 \stackrel{2}{\longrightarrow} \mathrm{X}_1,& \quad r_{14}: \quad \mathrm{X}_2 + \mathrm{X}_3 \stackrel{3/2}{\longrightarrow} \mathrm{X}_2,& \quad r_{15}: \quad \mathrm{X}_2 \stackrel{1}{\longrightarrow} \mathrm{X}_2 + \mathrm{X}_3,\\
r_{16}: \quad \mathrm{X}_3 + \mathrm{X}_4 \stackrel{1}{\longrightarrow} \mathrm{X}_4,& \quad r_{17}: \quad \mathrm{X}_2  \stackrel{1/\mu}{\longrightarrow} \mathrm{X}_2 + \mathrm{X}_4,& \quad r_{18}: \quad \mathrm{X}_1+\mathrm{X}_4 \stackrel{k_{18}}{\longrightarrow} \mathrm{X}_1,\\
r_{19}:  \quad \mathrm{X}_1+\mathrm{Y}_1 \stackrel{2}{\longrightarrow} \mathrm{X}_1+\mathrm{Y}_1+\mathrm{X}_4, & \quad r_{20}: \quad \mathrm{X}_4 \stackrel{2}{\longrightarrow} \varnothing, & \quad r_{21}: \quad \mathrm{X}_2+\mathrm{X}_4 \stackrel{1}{\longrightarrow} \mathrm{X}_2,
\end{array}$}
\end{equation}
where $k_{18} = 2+1/\mu$. On the other hand, the canonical inversion of~\eqref{taylorcubic} gives
\begin{equation}\label{canonicaltay}
\begin{aligned}
&r_{1}: \quad \varnothing \stackrel{1}{\longrightarrow} \mathrm{X}_1, \quad \quad r_{3}: \quad 2\mathrm{X}_1 \stackrel{1}{\longrightarrow} \mathrm{X}_1,\\
&r_{2}: \quad \mathrm{X}_1 \stackrel{1}{\longrightarrow} \varnothing, \quad\quad r_{4}^\prime: \quad 3\mathrm{X}_1 \stackrel{1/2}{\longrightarrow} 2\mathrm{X}_1,\\
\end{aligned}
\end{equation}
where the trimolecular reaction $r_4^\prime$ may be bimolecularized via the Quasi-Steady State Approximation. Following the derivation in Section~\ref{quasisteadystatetransformation}, we get from~\eqref{thisintermediateequation} the bimolecular approximation
\begin{equation}\label{actualapplicationofQSST}
\begin{aligned}
&r_4: \quad 2\mathrm{X}_1 \stackrel{k_1}{\rightarrow} \mathrm{Z}, \quad r_5: \quad \mathrm{Z} \stackrel{k_2}{\rightarrow} 2\mathrm{X}_1, \\
&r_6: \quad \mathrm{X}_1+ \mathrm{Z} \stackrel{k_{3}}{\rightarrow} \left(\nu_{i3}^+  +\nu_{j3}^+  +\nu_{k3}^+ \right) \mathrm{X}_1+\nu_{z3}^+ \mathrm{Z},
\end{aligned}
\end{equation}
where choosing $n = 1$ in a relation identical to~\eqref{rateconstant} gives two possible scenarios
\begin{gather*}
    \nu_{z3}^+ = 0, \quad \nu_{i3}^+ +\nu_{j3}^+ +\nu_{k3}^+ = 2,\tag{V1}\\
        \nu_{z3}^+ = 1, \quad \nu_{i3}^+ +\nu_{j3}^+ +\nu_{k3}^+ = 0.\tag{V2}
\end{gather*}
Before considering purely bimolecularization strategies, we note some non-trivial observations on the `General Canonical' network~\eqref{trimoleculargeneral} and the `General' network~\eqref{bimoleculargeneral}. A further iteration of the quadraticization algorithm produces two additional specimen in the bimolecular network~\eqref{bimoleculargeneral}, introducing more degrees of freedom for stochasticity to penetrate the variation in stochastic simulations. It then stands to reason that the bimolecular network must have greater variance than the trimolecular network~\eqref{trimoleculargeneral}. Shockingly, Figure~\ref{summarizedplot} (c) shows no substantive difference between the molecule count variance of the two networks for specimen $\mathrm{X_1, X_2}$. Due to complexity, we have so far been unable to verify if this uncanny phenomenon is replicated for higher order models, for instance in a bimolecularization of the Taylor network induced by expanding~\ref{model3} to order $20$.

Both networks~\eqref{trimoleculargeneral} and~\eqref{bimoleculargeneral} stabilize under repeated sampling to the deterministic steady states for $\mathrm{X_1, X_2}$ (Figure~\ref{summarizedplot} (a-b), where deterministic trajectories are removed for clarity). But once again, the dynamics of the stabilization are far from trivial. The transformation $\Psi_{QSST}$ induces a moderately fast variable which we have termed $\mathrm{Y}_1$ in both networks, which quickly goes extinct in the trimolecular network~\eqref{trimoleculargeneral} (Figure~\ref{summarizedplot} (b)). In contrast, the bimolecular inversion~\eqref{bimoleculargeneral} converges the molecule count of $\mathrm{Y}_1$ to a non-zero value (Figure~\ref{summarizedplot} (a)). This does not appear to affect the validity of the Quasi-Steady State Transformation; indeed, the General Quadraticization Algorithm preserves deterministic dynamics, and does not cause extinction in $\mathrm{X}_1$. Thus as long as $\mu$ remains minute, it is natural to expect the validity of~\eqref{bimoleculargeneral}. However, we carefully note that the proof given in Appendix \ref{ProofImportant} of the asymptotic validity of $\Psi_{QSST}$ (Theorem~\ref{QSSTtheorem}) may not be applied directly to the system~\eqref{bimoleculargeneralODE} which is no longer in the general form required by Tikhonov’s theorem~\cite{QSSATomiTheoryTichnoff} due to an iteration of the General Quadraticization Algorithm.


We briefly remark that the agitation of $\mathrm{X}_4$ into a fast specimen is instigated by $\mathrm{Y}_1$. The number of channel firings impacting $\mathrm{X_1, X_2, X_3}$ in a single realization used to produce Figure~\ref{summarizedplot} (a-b) are of order $\mathcal{O}(10^4)$ for both~\eqref{trimoleculargeneral} and~\eqref{bimoleculargeneral}, where $\mathrm{X}_3$ is relevant only to the latter. In contrast, the channel firings impacting $\mathrm{Y_1}$  are of $\mathcal{O}(10^5)$ for~\eqref{trimoleculargeneral} and $\mathcal{O}(10^6)$ for~\eqref{bimoleculargeneral}. Channel firings impacting $\mathrm{X_4}$ are also of $\mathcal{O}(10^6)$, and $\mu = 10^{-4}$ has been used for both networks to reduce computational intensity, breaking with the general convention of $\mu = 10^{-6}$ used throughout this paper. Therefore, we further take note of the asymmetric behaviour of an iteration of the General Quadraticization Algorithm increasing the relevance of $\mathrm{Y}_1$.

Figure~\ref{summarizedplot} (d) plots averaged trajectories of the canonical network~\eqref{canonicaltay} induced by the cubic~\eqref{taylorcubic}, denoted `Canonical', as well as the deterministic solution of \ref{model1} to evaluate the performance of the bimolecularized networks~\eqref{bimoleculargeneral} and (V1--V2) in remaining faithful to their original model. A subtle difference exists between the steady states of ~\eqref{bimoleculargeneral} and (V1--V2), of which the former more authentically outlines deterministic dynamics. (e) verifies the convergence of the Quasi-Steady State Approximation~\eqref{actualapplicationofQSST} as $k_2$ is increased, revealing that the $k_2 = 10^7$ used for simulations is within the appropriate range to obtain bimolecular approximations of the highest resolution. In contrast, (f) reveals that variance is largest for the `General' network~\eqref{bimoleculargeneral}, requiring more experiments to be performed in order to collect enough samples for accurate averaging. 

We cannot straightforwardly conclude the superiority of one bimolecularization strategy over the other. Perhaps the most general take-away from our study of the test system~\eqref{taylorcubic} is the success of the General Quadraticization Algorithm\footnote{That is, introducing additional specimen $\mathrm{X}_3 , \mathrm{X}_4 $ did not increase the variance for the `General' network~\eqref{bimoleculargeneral} (Figure~\ref{summarizedplot} (c)), which was entirely unexpected. Unlike in Chapter~\ref{oldchapter4}, exploding solutions were not immediately detected, nor did the variance rapidly grow with time (Figure~\ref{summarizedplot} (d),(f)). Thus, we may conclude that short-time explosions are not a general feature necessarily induced by Kerner Polynomialization.}, which essentially iterates Kerner Polynomialization substitutions for degree reduction. Our conclusion for the inverse framework arising from investigation of the test system is that bimolecuarization should be implemented on a case-by-case basis while considering the complexity of the inversion, importance of the accuracy of the data, and the cost of Monte Carlo experiments resulting in the need to limit variance.

%% file: chapter7.tex
\chapter{Derivation and Application of Inversion Framework}\label{oldchapter6}
We coalesce  our observations into the following inversion framework for the chemical integration of ODE systems. 

\begin{mdframed}
(\textbf{Five-Step Inversion Framework for Chemical Integration of ODEs})

\noindent\rule{16.3cm}{0.5pt}
\vspace{1pt}

\noindent \textbf{(B1): Polynomialization.} Given an arbitrary ODE system, use either Kerner Polynomialzation or a suitable polynomial approximation such as Taylor expansions to obtain a polynomial system approximating the original system. If Kerner Polynomialzation yields a system displaying NEC, series expansions are recommended as an alternative.

\vspace{10pt}
\noindent \textbf{(B2): Affine Transformation.} Conventional chemical dynamics assume non-negative specimen concentrations. To encapsulate desired dynamics of the original system, make an affine transformation to push regions in which the dynamics occur to the positive cone $\mathbb{R}_{ > 0}^n$, that is, using the substitution $\mathbf{x} \leftarrow \mathbf{A} \mathbf{x} - \boldsymbol{\mathcal{T}}_\mathbf{x}$ for $\mathbf{A} \in \mathbb{R}^{n\times n}$ and $\boldsymbol{\mathcal{T}}_\mathbf{x} \in \mathbb{R}^{n}$. Note that the order of \textbf{(B1)} and \textbf{(B2)} may be swapped, not necessarily yielding identical results.

\vspace{10pt}
\noindent\textbf{(B3): Kineticization.} Apply the kinetic transformation $\Psi_{QSST}$ to eliminate all cross-negative terms, using $\omega_s, p_s(\mathbf{x}) \equiv 1$ for $\mathrm{X}_s \in \mathcal{X}_2$ (see Definition~\ref{QSSAnewdefinition}). The appropriate value of $\mu$ may be determined via computational simulations (following steps \textbf{(B1-5)}) before lab-based chemical implementation.

\vspace{10pt}
\noindent\textbf{(B4): Direct Canonical Inversion.} Perform a canonical inversion of the kinetic polynomial to obtain a chemical reaction network, or,

\vspace{10pt}
\noindent\textbf{(B4): Optional Bimolecularization.} Depending on the circumstance (e.g. for molecular computation practicable using current biotechnology), we bimolecuarize the system. The General Quadraticization Algorithm does not guarantee a kinetic output but may competently represent deterministic dynamics, especially if NEC is not detected. A Quasi-Steady State Transformation guarantees a mass-action kinetic implementation, but renders the output network stiff, potentially accumulating errors in chemical implementations. We first recommend General Quadraticization, unless:

\noindent (i) The algorithm is too costly to terminate, 

\noindent (ii) The output network obtained via canonical inversion is intractably labor-intensive to implement due to the number of added specimen, 

\noindent (iii) The NEC is detected, 

\noindent (iv) The output polynomial demonstrates non-kinetic terms which are continually generated via repeated applications of General Quadraticization following the Kineticization step \textbf{(B3)}. Note that persistent introduction of new variables from kineticization negates the guaranteed termination of the quadraticization algorithm.

\noindent In such instances, we instead recommend applying Quasi-Steady State Approximations for bimolecularization. Afterwards, canonically invert the resulting ODE system into a bimolecular reaction network.

\vspace{10pt}
\noindent\textbf{(B5): Optional Repeated Sampling.} Average over a sufficient number of realizations for convergence, predicted preemptively by computational stochastic simulations and depending on the accuracy of the measured output required in the experiment. 
\end{mdframed}

\begin{figure}[h]
  \begin{subfigure}[b]{1\textwidth}
    \includegraphics[width=\textwidth]{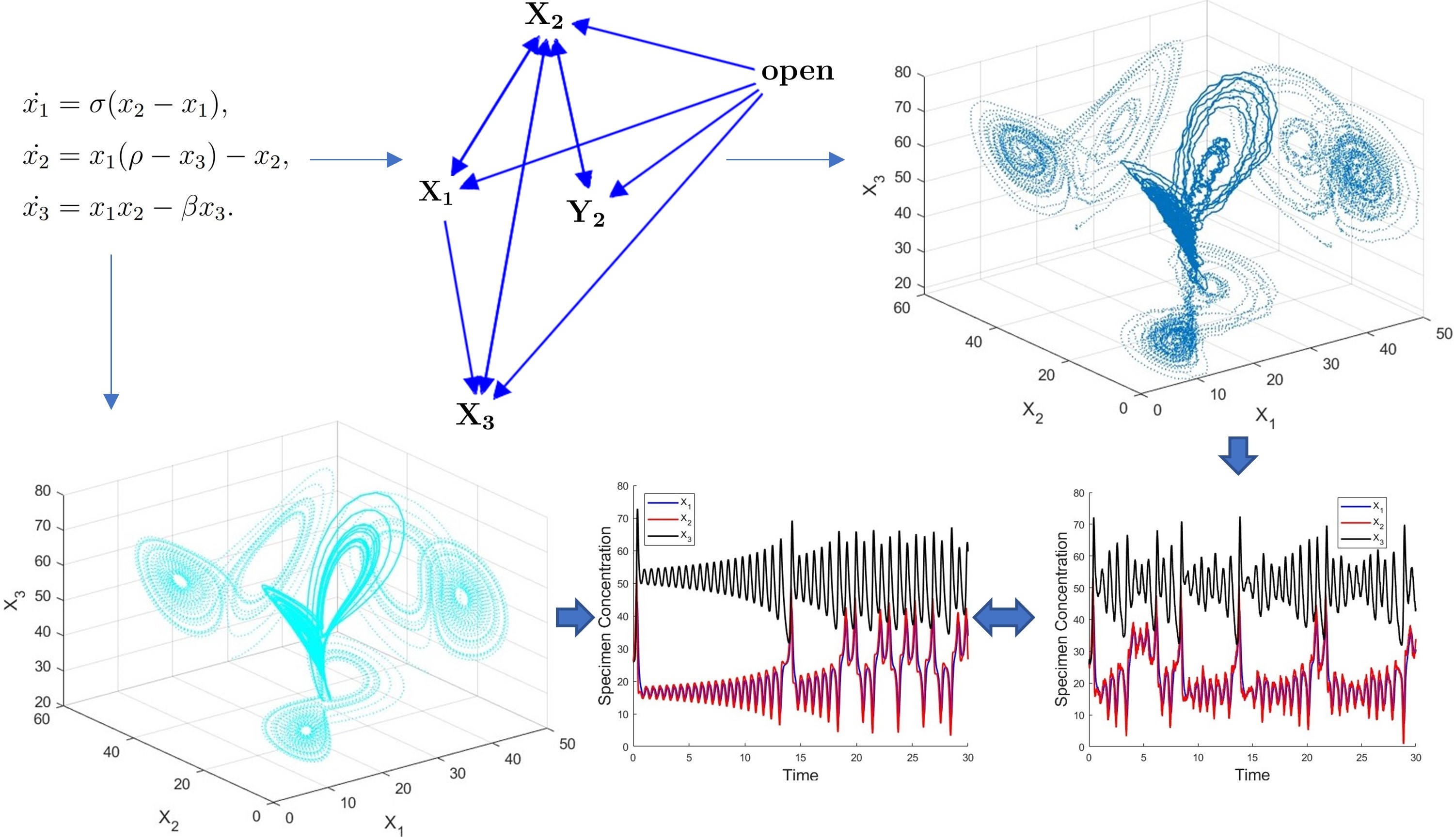}
  \end{subfigure}
  \caption[The inverse framework in action]{A concise pictorial summary of the inverse framework in action. The non-kinetic Lorenz system induces an open system in its chemical inversion.}
  \label{inverseframeworkpicture}
\end{figure}

\noindent Now that we have solidified the inversion framework, we seek to demonstrate its utility by chemically inverting and stochastically simulating meaningful ODE systems. In our view, an ultimate confirmation of its capabilities may come from either the validation or rejection of an open problem hypothesized in the literature, which should make use of the inversion framework during the study. For the rest of this chapter, we put our focus on the application of the proposed framework, where a  visual illustration of the framework in action is included as Figure~\ref{inverseframeworkpicture}. Four disjoint experiments simulating non-kinetic ODE systems are carried out and their results are concisely summarized.

The first test system is taken from~\cite{NonlinearSystems} and given in Cartesian and polar form:

\begin{minipage}{.55\textwidth}
\begin{equation}\label{hopfbifuraction}
\begin{aligned}
&\Dot{x_1}=\xi x_1-\zeta x_2-x_1\left(x_1^{2}+x_2^{2}\right), \\
&\Dot{x_2}=\zeta x_1+\xi x_2-x_2\left(x_1^{2}+x_2^{2}\right),
\end{aligned}
\end{equation}
\end{minipage}
\begin{minipage}{.35\textwidth}
\begin{equation}\label{polarform}
\begin{aligned}
&\dot{r}=\xi r-r^{3}, \\
&\dot{\theta}=\zeta .
\end{aligned}
\end{equation}
\end{minipage}

\vspace{2pt}

\noindent The system~\eqref{hopfbifuraction}--\eqref{polarform} undergoes a Hopf bifurcation as $\xi$ increases, birthing a stable limit cycle from a stable focus as $\xi$ crosses the origin. After a suitable affine transformation to translate the limit cycle to the positive quadrant and kineticization via $\Psi_{QSST}$, we set $\zeta = 1$ and $V = 500$ to amplify oscillatory dynamics. We will use the network~\eqref{hopfnetwork} induced by this system to validate an open problem hypothesized by by Plesa et al~\cite{TomiStatInferencePaper} that several statistical methods proposed in~\cite{AutoCorrAndDFT} to classify oscillations in noisy time series data are not sensitive enough to be used for time series born from deterministic systems undergoing a bifurcation.

\begin{figure}[h!]
  \begin{subfigure}[b]{0.325\textwidth}
    \includegraphics[width=\textwidth]{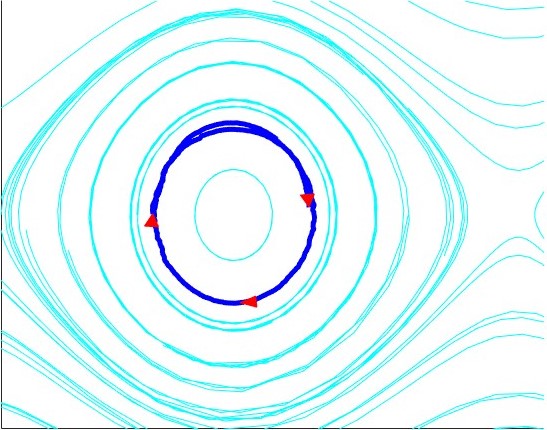}
    \caption{ }
  \end{subfigure}
  \hfill
  \begin{subfigure}[b]{0.325\textwidth}
    \includegraphics[width=\textwidth]{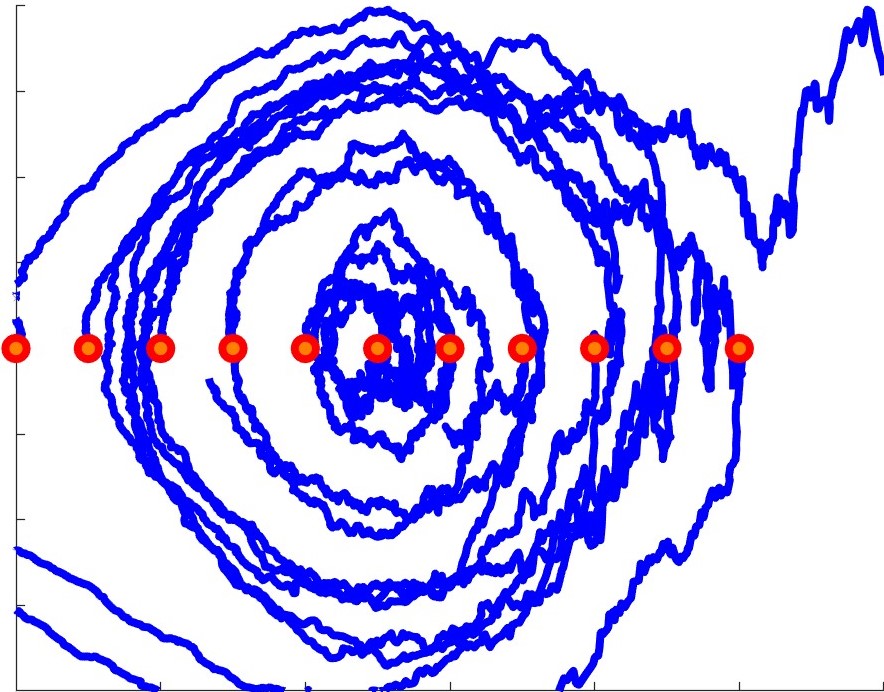}
    \caption{ }
  \end{subfigure}
    \hfill
    \begin{subfigure}[b]{0.325\textwidth}
    \includegraphics[width=\textwidth]{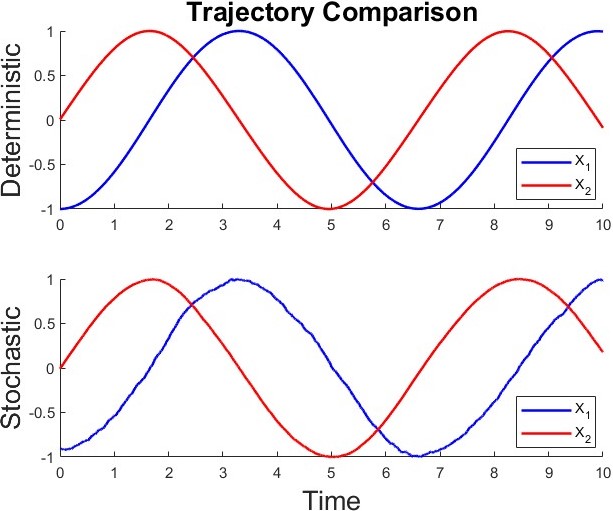}
    \caption{ }
  \end{subfigure}
  \hfill
  \begin{subfigure}[b]{0.325\textwidth}
    \includegraphics[width=\textwidth]{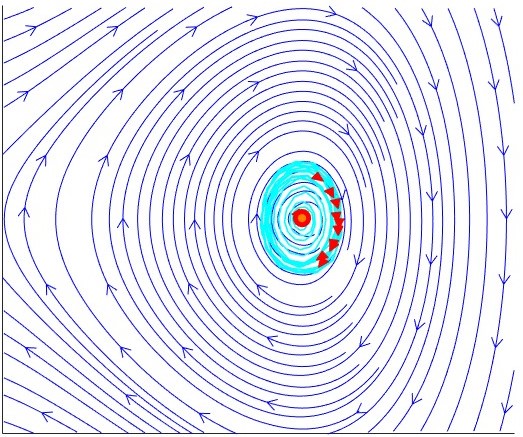}
    \caption{ }
  \end{subfigure}
    \hfill
      \begin{subfigure}[b]{0.325\textwidth}
    \includegraphics[width=\textwidth]{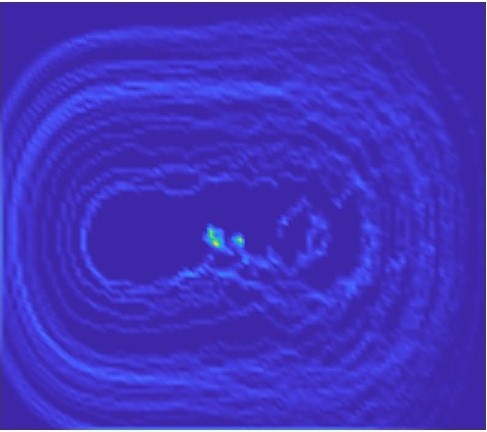}
    \caption{ }
  \end{subfigure}
  \hfill
  \begin{subfigure}[b]{0.325\textwidth}
    \includegraphics[width=\textwidth]{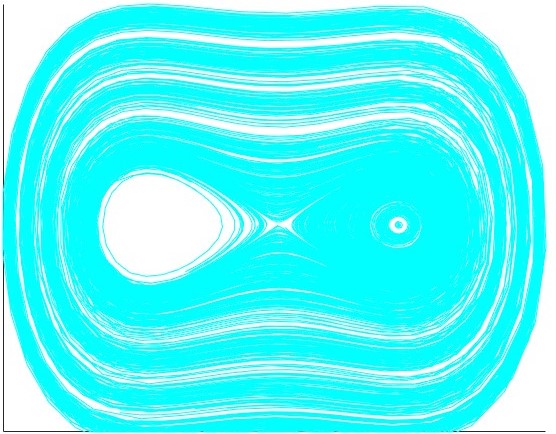}
    \caption{ }
  \end{subfigure}
    \hfill
  \caption[On the chemical simulation of a shifted pendulum]{A ``chemical'' phase plane traced using individual stochastic trajectories is given in (b). Starting points have been highlighted using coloured dots as in (d), and red arrows point in direction of increasing time. In (c), deterministic and stochastic trajectories (the latter averaged over $20$ realizations) have been renormalized  (identically translated and scaled) to live in the target range $[-1,1]$.}
  \label{thisisthependulumplot}
\end{figure}

The second test system is given by the so-called shifted ``pendulum'' 
\begin{equation}\label{pendulum}
    \Dot{x_1} = x_2 - \mathcal{T}_{x_2}, \quad \Dot{x_2} = \cos(x_1 - \mathcal{T}_{x_1}),
\end{equation}
where we take $V = 2000$ to sample exotic behaviours within the eye region (Figure~\ref{thisisthependulumplot} (a)) at a higher resolution. After making a few comments on observed characteristics of chemical trajectories (stochastic simulations of the induced network), we will put forward several exemplary computationally generated figures whose replication in biochemical laboratories may be used as a part of a roadmap to the successful chemical integration of ODE systems.

The third and fourth experiments are done by chemically simulating the chaotic R\"{o}ssler~\eqref{rosslerode} and Lorenz~\eqref{lorenzode} systems, given by 

\begin{minipage}{.45\textwidth}
\begin{equation}\label{rosslerode}
\begin{aligned}
&\Dot{x_1}=-x_2-x_3, \\
&\Dot{x_2}=x_1+a x_2, \\
&\Dot{x_3}=b+x_3(x_1-c).
\end{aligned}
\end{equation}
\end{minipage}
\begin{minipage}{.45\textwidth}
\begin{equation}\label{lorenzode}
\begin{aligned}
&\Dot{x_1}=\sigma(x_2-x_1), \\
&\Dot{x_2}=x_1(\rho-x_3)-x_2, \\
&\Dot{x_3}=x_1 x_2-\beta x_3 .
\end{aligned}
\end{equation}
\end{minipage}

\vspace{5pt}

\noindent Deterministic trajectories obtained by numerically solving the original non-kinetic ODEs~\eqref{rosslerode},~\eqref{lorenzode} are contrasted against their chemical realizations. In particular, several chemical trajectories are observed to display the signature dynamics of the R\"{o}ssler and Lorenz attractors more quickly than their deterministic counterparts initialized identically due to motion induced by stochasticity, and Poincare maps are drawn for additional comparison. Further discussions highlighting the rationale behind the selection of $\Psi_{QSST}$ as the kineticization strategy instead of $\Psi_{Kow}$, using the chemical simulation of the Lorenz Attractor, are included in Appendix~\ref{chaoticsystemsinappendix}, and the reaction network forms of~\eqref{hopfbifuraction},~\eqref{rosslerode},~\eqref{lorenzode} are written out in Appendix~\ref{thenetworksappendix}.

\begin{figure}[h]
  \begin{subfigure}[b]{0.5\textwidth}
    \includegraphics[width=\textwidth]{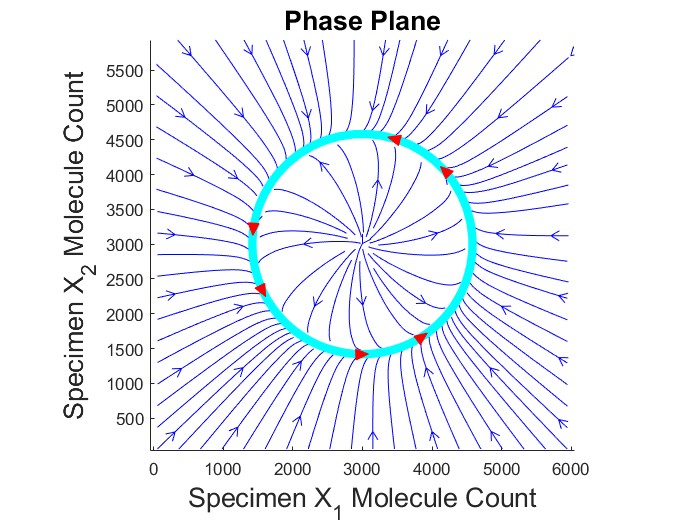}
    \caption{Deterministic Solution}
  \end{subfigure}
  \hfill
  \begin{subfigure}[b]{0.5\textwidth}
    \includegraphics[width=\textwidth]{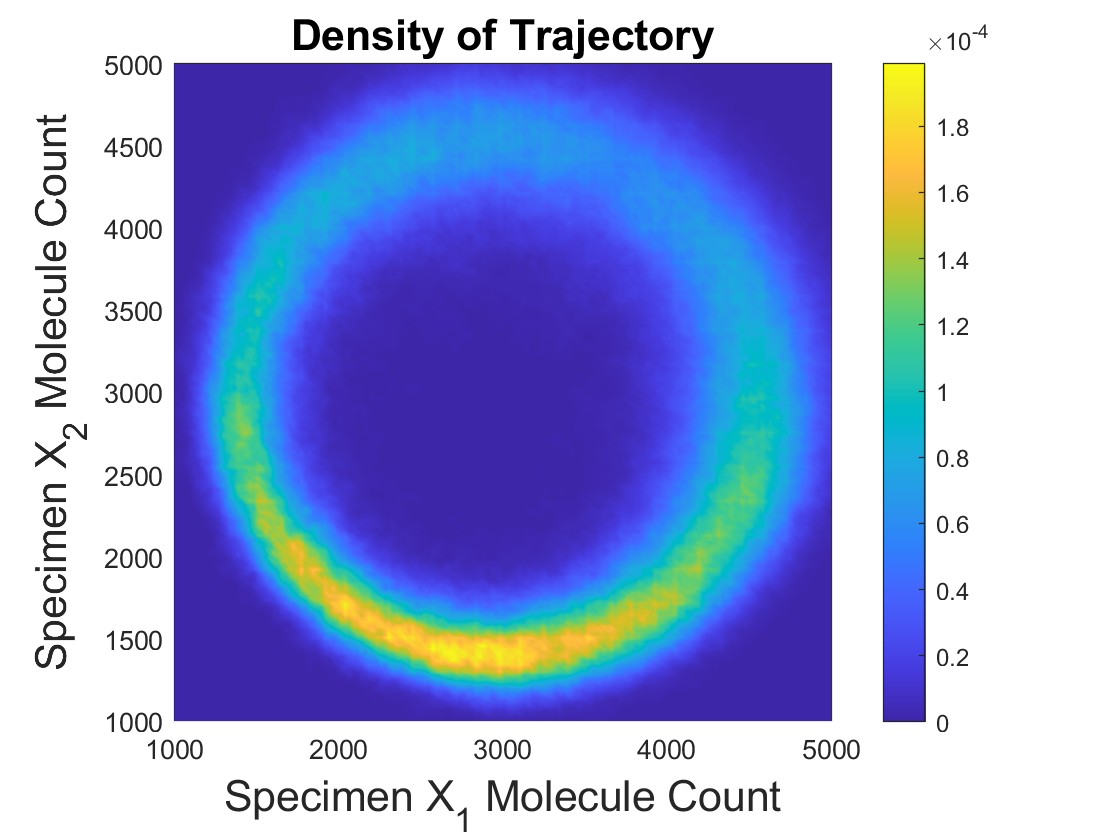}
    \caption{Chemical Realization}
  \end{subfigure}
  \caption[Joint density of Hopf system derived from single long-time trajectory]{The inversion framework has been used to invert the non-kinetic system~\eqref{hopfbifuraction} for $\xi = 10$. In the chemical setting, the stable limit cycle is detected by monitoring a single long-time trajectory, and the joint density is derived in (b). In Appendix~\ref{discreteappendix} we argue that the downward bias is caused by discrete size forces using a novel technique
which we have termed \textit{pseudo-propensity analysis}.}
  \label{starterplot}
\end{figure}
For the first experiment, we stochastically simulate the chemical reaction network~\eqref{hopfnetwork} found by applying the inverse framework to the non-kinetic ODE~\eqref{hopfbifuraction}. Figure~\ref{starterplot} (b) gives the joint density measured from monitoring a long-time ($t=2000$) trajectory, which successfully manifests exotic (limit cycle) dynamics shown in the deterministic phase plane (a). Therefore, we may now use this model to investigate an open problem proposed by Plesa et al~\cite{TomiStatInferencePaper}. It is hypothesized that several statistical methods used in~\cite{AutoCorrAndDFT} to distinguish quasi-cycles\footnote{In stochastic time series data, quasi-cycles are induced by stable nodes/foci, whereas limit cycles are induced by stable (deterministic) limit cycles.} and limit cycles in time series are in fact not powerful enough to detect and classify   limit cycles in noisy time series data born from bifurcated systems. The methods in question may be summarized as Discrete Fourier Transforms (DFT) to compute power spectra, Autocorrelation Functions (ACF), and probability density function (PDF) shape comparison. 

In order to examine their viability in classifying oscillations in noisy data, two networks with cubic RREs undergoing a homoclinic and a multiple limit cycle bifurcation are suggested in~\cite{TomiStatInferencePaper} as models to generate the time series before and after bifurcation has occurred. In the original work, a 
consideration of the shape of the density is given to argue the concealment of any deterministic cycles underlying the time series near bifurcation points, but other proposed statistical avenues (Autocorrelations and power spectra) are not explored. Using the reaction network~\eqref{hopfnetwork} derived from~\eqref{hopfbifuraction}, we may probe this further and fully validate or reject Plesa's hypothesis.

\begin{figure}[h!]
  \begin{subfigure}[b]{0.5\textwidth}
    \includegraphics[width=\textwidth]{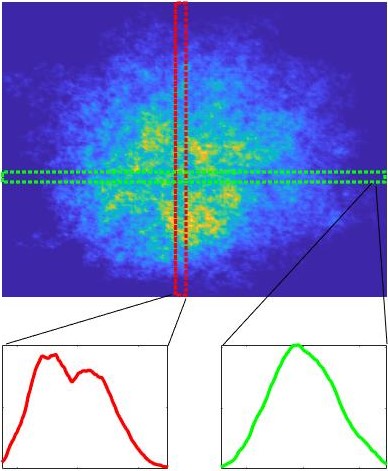}
    \caption{$\xi = 0.1$}
  \end{subfigure}
  \hfill
  \begin{subfigure}[b]{0.5\textwidth}
    \includegraphics[width=\textwidth]{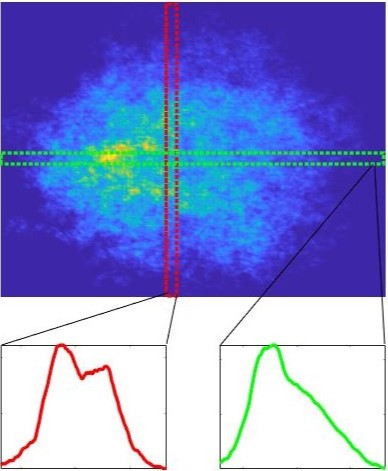}
    \caption{$\xi = -0.1$}
  \end{subfigure}
    \begin{subfigure}[b]{1\textwidth}
    \includegraphics[width=.33\textwidth]{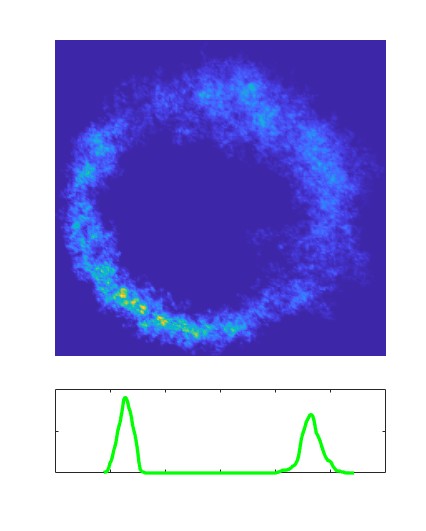}\hfill
    \includegraphics[width=.33\textwidth]{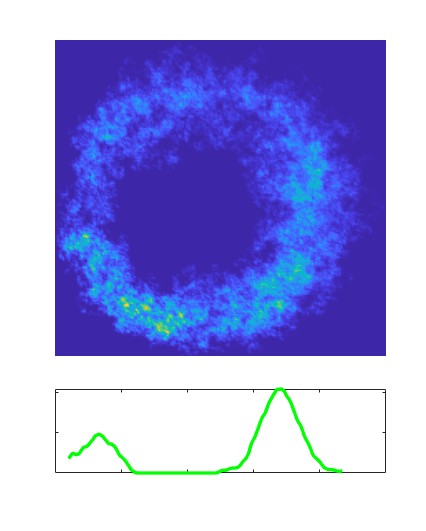}\hfill
    \includegraphics[width=.33\textwidth]{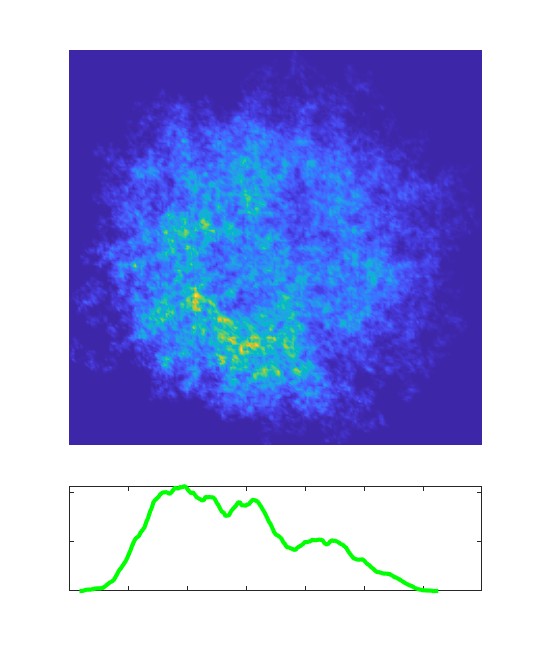}\hfill
    \caption{$\xi = 12, 8, 1,$ from left to right, where marginal density has been estimated horizontally. $\xi \downarrow 0$ obfuscates the crater ridge induced by the limit cycle for higher values of $\xi > 0$.}
  \end{subfigure}
  \caption[Deriving joint density from individual short-time trajectories]{Joint density are derived from individual short-time time series trajectories. Histogram data have been smoothed to approximate continuous marginal densities. The two-sided tails of the densities where no trajectory presence is detected have been truncated in order to render the shape or outline of marginal distribution clearly visible. Sharply near the bifurcation point $\xi = 0$, the stochastic trajectories manifest minimal distinctions and the two deterministically distinct systems (e.g. \eqref{hopfbifuraction} with $\xi = \pm \varepsilon$ for $0<\varepsilon \ll 1$) may not be distinguished by comparing marginal densities.}
  \label{goodfigure}
\end{figure}

\begin{figure}[h]
  \begin{subfigure}[b]{0.325\textwidth}
    \includegraphics[width=\textwidth]{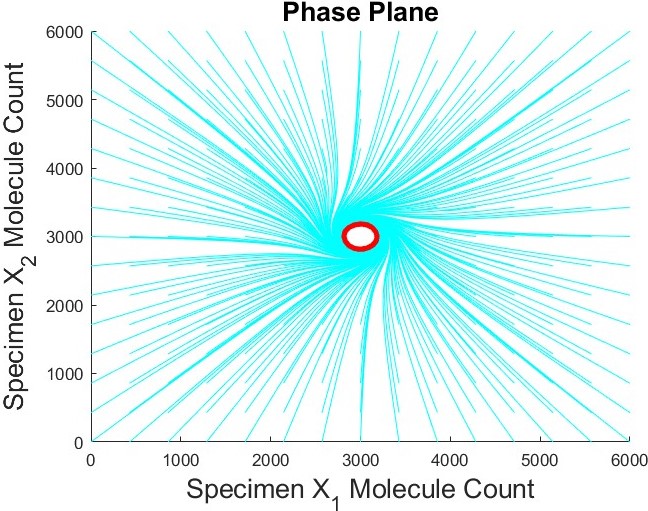}
    \caption{ }
  \end{subfigure}
  \hfill
  \begin{subfigure}[b]{0.325\textwidth}
    \includegraphics[width=\textwidth]{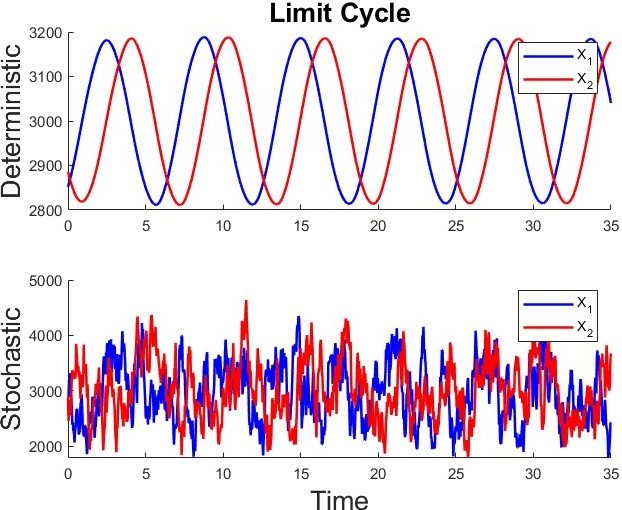}
    \caption{ }
  \end{subfigure}
    \hfill
    \begin{subfigure}[b]{0.325\textwidth}
    \includegraphics[width=\textwidth]{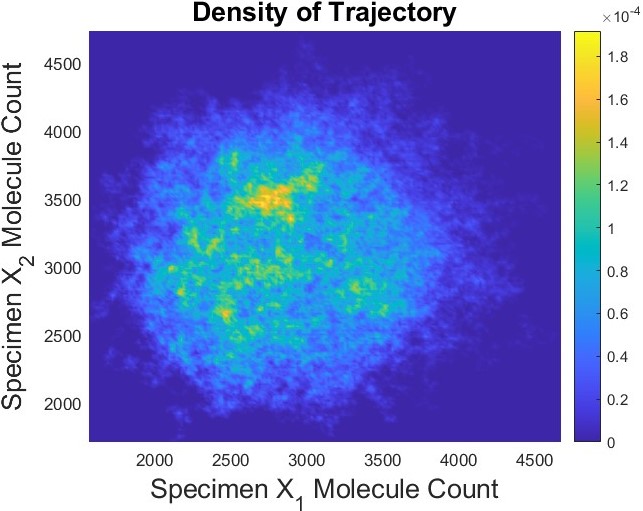}
    \caption{ }
  \end{subfigure}
  \hfill
  \begin{subfigure}[b]{0.325\textwidth}
    \includegraphics[width=\textwidth]{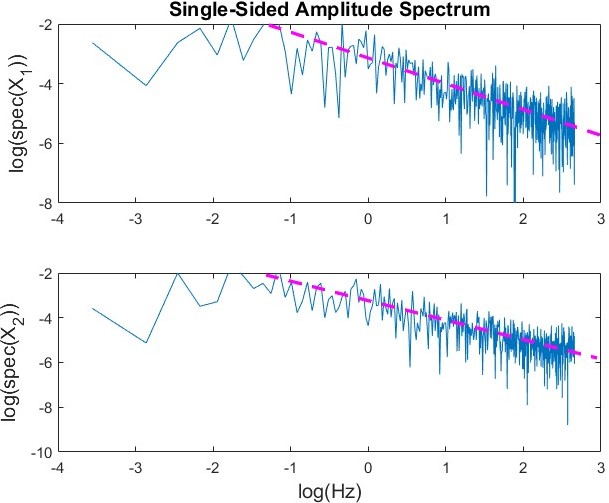}
    \caption{ }
  \end{subfigure}
    \hfill
      \begin{subfigure}[b]{0.325\textwidth}
    \includegraphics[width=\textwidth]{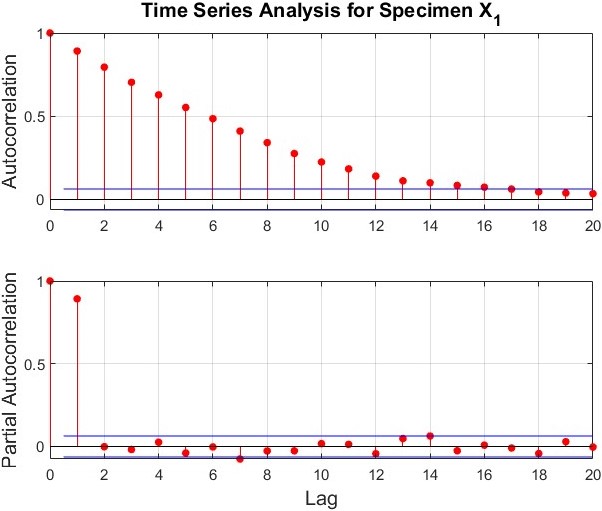}
    \caption{ }
  \end{subfigure}
  \hfill
  \begin{subfigure}[b]{0.325\textwidth}
    \includegraphics[width=\textwidth]{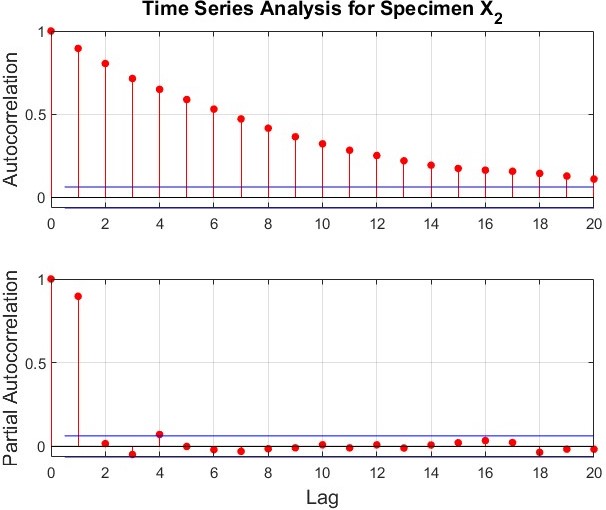}
    \caption{ }
  \end{subfigure}
    \hfill
      \begin{subfigure}[b]{0.325\textwidth}
    \includegraphics[width=\textwidth]{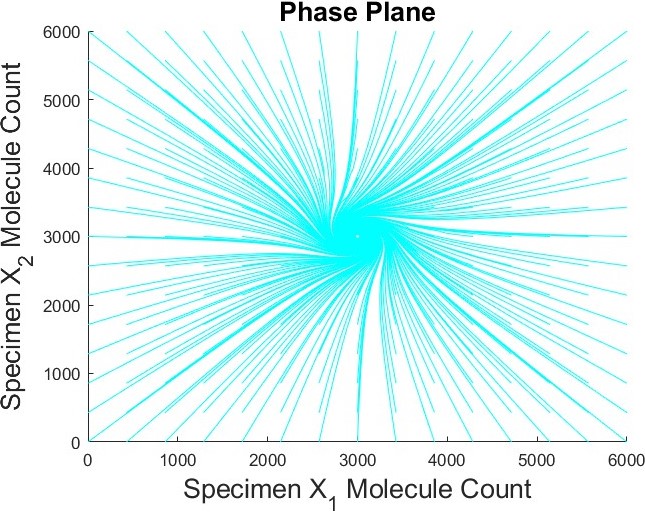}
    \caption{ }
  \end{subfigure}
  \hfill
  \begin{subfigure}[b]{0.325\textwidth}
    \includegraphics[width=\textwidth]{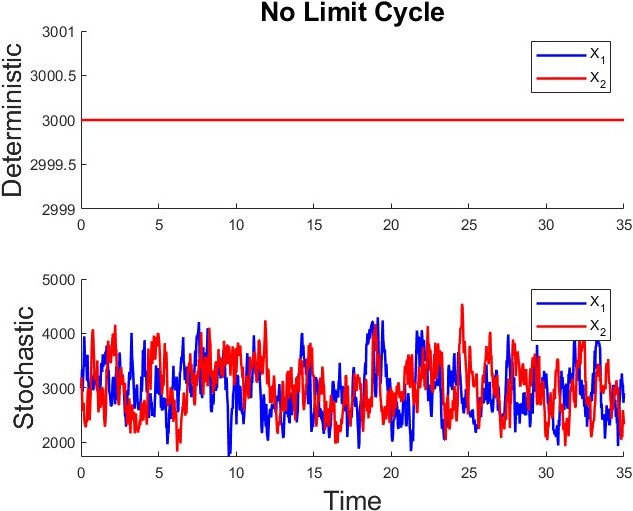}
    \caption{ }
  \end{subfigure}
    \hfill
    \begin{subfigure}[b]{0.325\textwidth}
    \includegraphics[width=\textwidth]{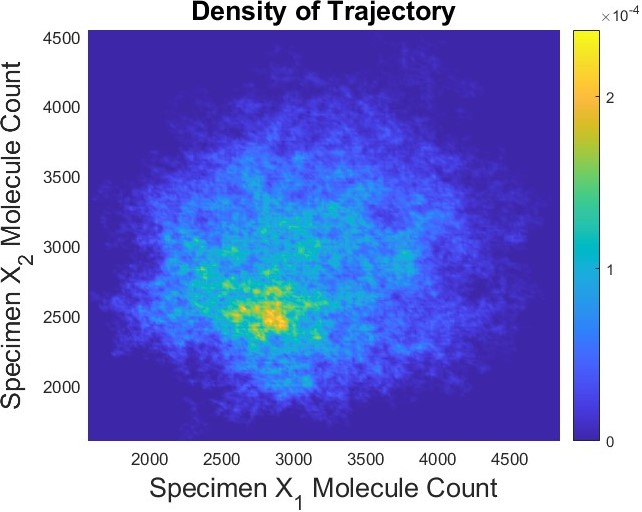}
    \caption{ }
  \end{subfigure}
  \hfill
  \begin{subfigure}[b]{0.325\textwidth}
    \includegraphics[width=\textwidth]{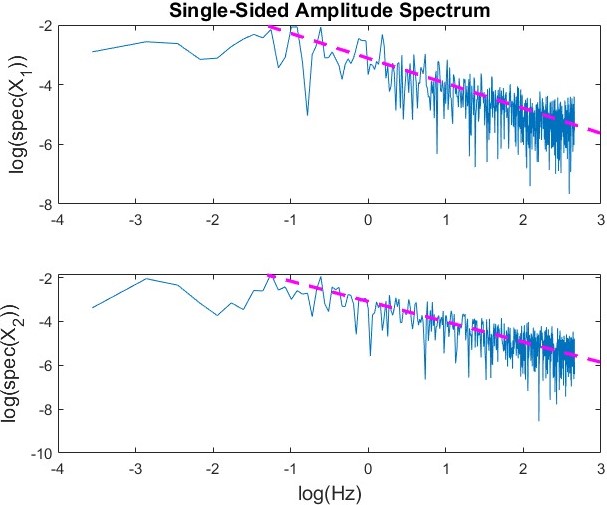}
    \caption{ }
  \end{subfigure}
    \hfill
      \begin{subfigure}[b]{0.325\textwidth}
    \includegraphics[width=\textwidth]{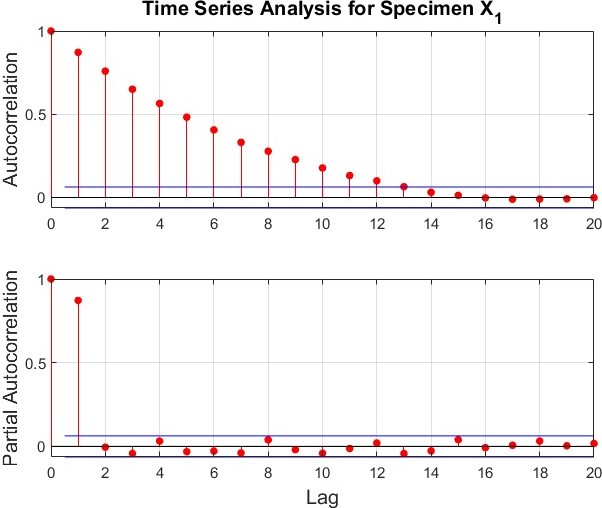}
    \caption{ }
  \end{subfigure}
  \hfill
  \begin{subfigure}[b]{0.325\textwidth}
    \includegraphics[width=\textwidth]{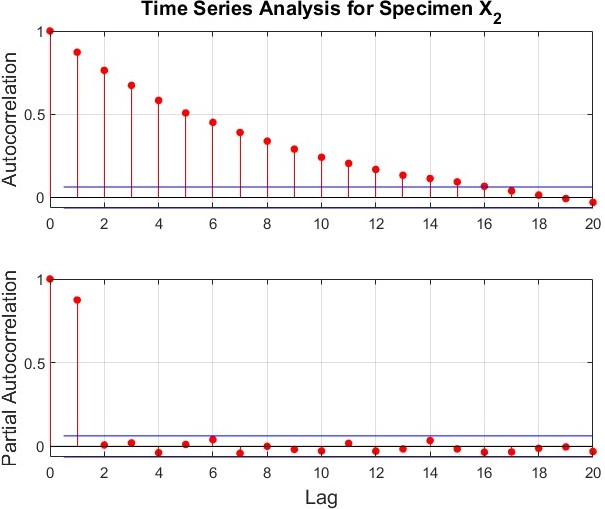}
    \caption{ }
  \end{subfigure}
    \hfill
  \caption[Autocorrelation and power spectra of bifurcated system]{The upper half (a-f) represents $\xi = 0.1$ and the lower half (g-l) represents $\xi = -0.1$. The limit cycle is outlined in pink in (a), which collapses into a stable focus for $\xi < 0$ as shown in (g). Partial Autocorrelation (e-f), (k-l) reveals that all correlations derived from shifted time series data is explainable by the interdependence of the data with its immediately preceding state ($1$ time lag). Time series analysis using Fast Fourier Transforms and Autocorrelations are done using pre-existing \mcode{Matlab} packages~\cite{fftMatlabProper,AutocorrelationMatlabProper}.}
  \label{largeplot}
\end{figure}

Firstly, the methods utilized in~\cite{AutoCorrAndDFT} presupposes the existence of a `large' limit cycle in the underlying deterministic system, which paves way to a clearly visible crater ridge in the stochastically observed joint density. It is intuitively evident that as parameter values are chosen to obfuscate the crater (Figure~\ref{goodfigure} (c)), the proposed methods need not be sensitive enough to determine the existence of limit cycles based on individual time series realizations which are stochastically similar, but have been born from topologically nonequivalent deterministic systems. Indeed, we note that the marginal densities estimated by measuring the data along a vertical or horizontal sliver of the observation domain disproves the hypothesis of non-normality induced exclusively by limit cycles in the case of bifurcated systems, as the approximated density in Figure~\ref{goodfigure} (b) portrays a marginal crater as well as a very skewed bell-curve. We contrast this with (a), where parameters have been chosen to develop a subtle limit cycle. See also Figure~\ref{largeplot} (a),(c),(g),(i) for a further illustration. 

More generally as $|\xi| \downarrow 0$, the damping of Autocorrelation signatures and evolution of power spectra showed no discernible difference whatsoever between quasi-cycles and limit cycles. Any subtle differences observed during individual realizations are consistently unreplicated in repeated experiments. Furthermore, the joint density measured from monitoring a single stochastic trajectory frequently displayed strong asymmetry for both quasi-cycles and limit cycles alike. The previous concerns raised by Plesa et al~\cite{TomiStatInferencePaper} that the methods suggested in~\cite{AutoCorrAndDFT} cannot possibly discriminate between deterministic topological differences near bifurcation points via studying individual time series realizations could experimentally be entirely validated, using different chemical test systems to that proposed in the original work. 

We therefore sought to determine if even for bifurcated systems, there were ideal conditions that could enhance our probability of success in distinguishing minute topological differences within stochastic time series data. We chose to initialize simulations of~\eqref{hopfnetwork} at different points; the stochastic quasi-cycle precisely on the deterministic stable focus and the stochastic limit cycle precisely on the deterministic limit cycle, in order to compel instantaneous information preservation for limit cycles and information decay for quasi-cycles in the time series data.

We have successfully observed a trend of a slower corrosion of Autocorrelations, where limit cycles more reliably maintain correlation values above the statistically insignificant threshold (Figure~\ref{largeplot} (e-f),(k-l)). The power spectrum in (d),(j) displays an identical decline, and could not be used to make inferences based on their general shape or configuration. Further statistical tools are desired to make a categorical determination unreliant on visual observation. Reinitializing the simulations from identical starting points confounded our data analysis due to involving trajectories during the relaxation process to the stable states, which only further weighted the inherent similarities between our two stochastic systems. This in turn frequently rendered any distinctions made from the damping of Autocorrelations irreplicable in repeated experiments. For the best case scenario, see Figure~\ref{hiddenlimitcycle}, where both systems were propagated from the focus. The limit cycle data appears to demonstrate a very slightly slower damping in Autocorrelations, but it is unclear if this result is statistically significant.

However, it is evident that contrary to the results in~\cite{AutoCorrAndDFT}, Autocorrelation signatures of time series data obtained by chemically simulating~\eqref{hopfbifuraction} did not succeed in inheriting the periodicity of the limit cycle in the low-rotation dynamical regime $\zeta = 1$, implying a failure in cycle detection. We further note that unless simulations are run for an exceptionally elongated timescale sufficient to produce stabilized joint densities (e.g. $t = 2000$ as in Figure~\ref{starterplot} instead of the $t = 35$ used in Figure~\ref{largeplot}), the derived densities from quasi-cycles are scarcely normally distributed. Therefore, without prior information about the parameter values chosen to produce the time series, we were unable to devise a categorical methodology to classify and distinguish the data based on the proposed statistical techniques. Although our results indicate that minute differences in bifurcated time series data are accentuated by truncating early timescales, or after determining when sufficient relaxation has occurred, we emphasize that decreasing $|\xi|$ further by a factor of $10$ destroys any observations made for the parameter values of Figure~\ref{largeplot}.

Moving on to the second experiment to demonstrate the capabilities of our framework, we simulate the shifted pendulum~\eqref{pendulum} as a chemical reaction network, where expansions to the $20$th order are taken in Figure~\ref{thisisthependulumplot} (a-c). (c) gives a comparison of deterministic expectations and stochastic trajectories which are averaged over $20$ realizations and plotted in (a), where trajectories have been renormalized to live in the target range $[-1,1]$ without affecting frequency. This particular experiment illustrates the utility of series expansions in allowing for the chemical translation of information contained in the original system such as frequency and amplitude. 
\clearpage
\newgeometry{top=8mm, bottom=10mm, left = 10mm, right = 10 mm}     

\cfoot{}
\clearpage
\cfoot{\thepage}
\restoregeometry

We have found that averaging over chemical (i.e. stochastically simulated) trajectories may not always provide satisfactory approximations of deterministic dynamics. For example, a Taylor expansion around $(x_1,x_2)=(-1,0)$ has been taken to the fourth order and the deterministic trajectory monitored until no further change could be observed ($t = 10^3$ suffices) in Figure~\ref{thisisthependulumplot} (d). However a single chemical realization of the system identically initialized and measured until a shorter time $t = 10^2$ in (e) outlines a completely deviant trajectory, which contributes to nonsensical averages. In (f), we plot all regions accessible by deterministic trajectories starting within the domain shown in (d), and verify that the chemical path in (e) was tracing a wider range of deterministic dynamics due to motion induced by stochasticity.

\begin{figure}[h]
  \begin{subfigure}[b]{0.325\textwidth}
    \includegraphics[width=\textwidth]{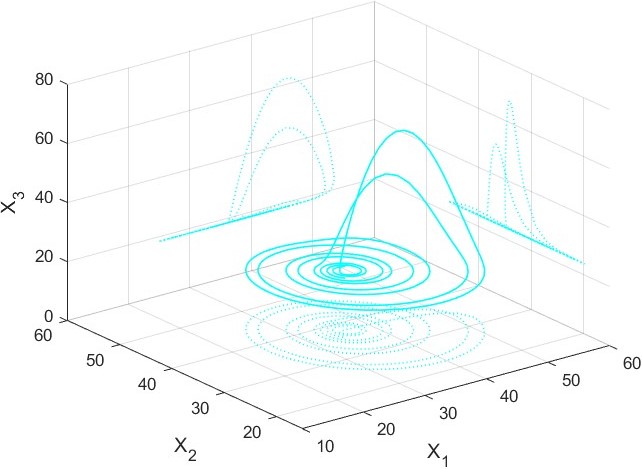}
    \caption{ }
  \end{subfigure}
  \hfill
  \begin{subfigure}[b]{0.325\textwidth}
    \includegraphics[width=\textwidth]{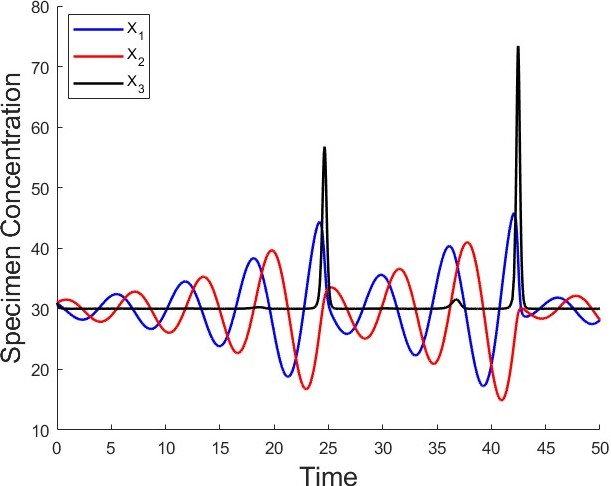}
    \caption{ }
  \end{subfigure}
    \hfill
    \begin{subfigure}[b]{0.325\textwidth}
    \includegraphics[width=\textwidth]{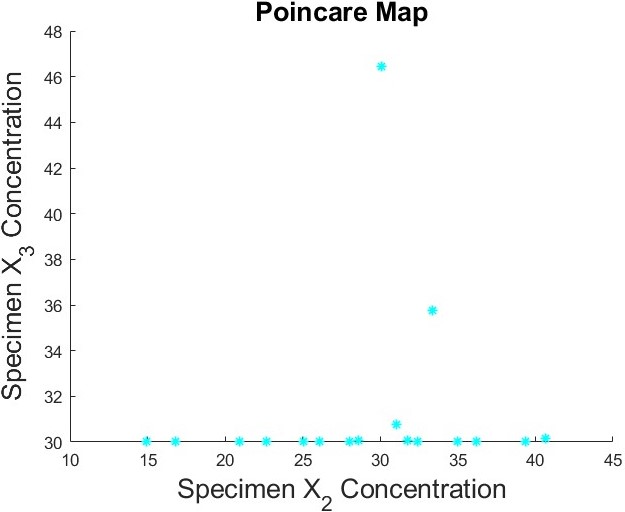}
    \caption{ }
  \end{subfigure}
  \hfill
  \begin{subfigure}[b]{0.325\textwidth}
    \includegraphics[width=\textwidth]{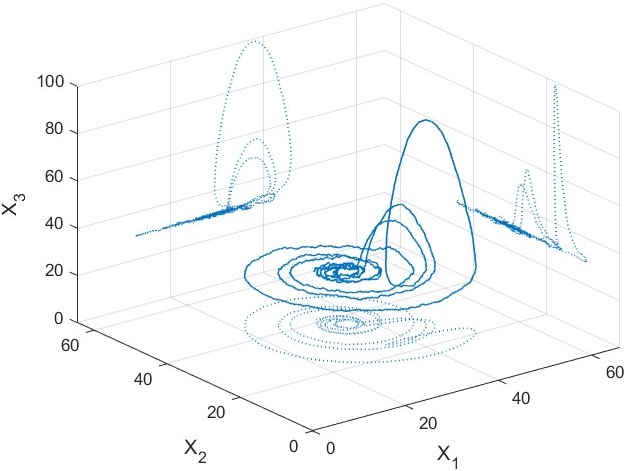}
    \caption{ }
  \end{subfigure}
    \hfill
      \begin{subfigure}[b]{0.325\textwidth}
    \includegraphics[width=\textwidth]{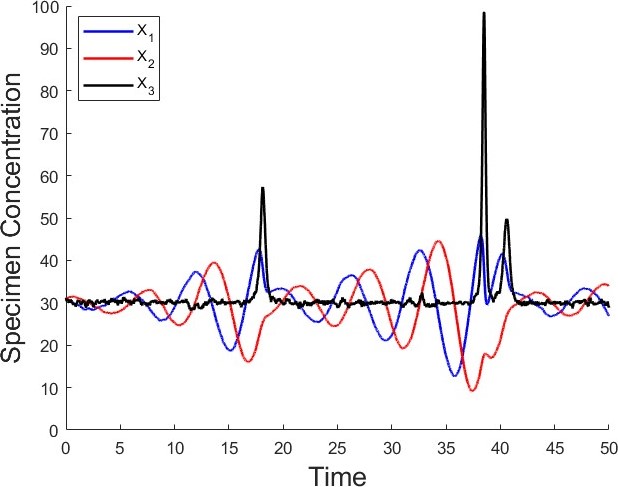}
    \caption{ }
  \end{subfigure}
  \hfill
  \begin{subfigure}[b]{0.325\textwidth}
    \includegraphics[width=\textwidth]{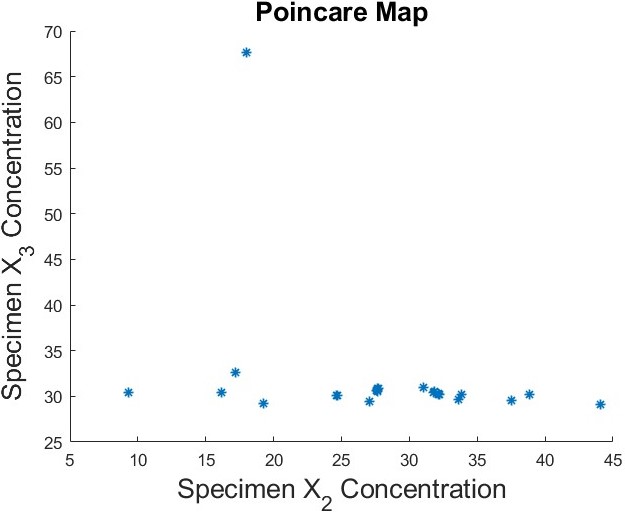}
    \caption{ }
  \end{subfigure}
    \hfill
      \begin{subfigure}[b]{0.325\textwidth}
    \includegraphics[width=\textwidth]{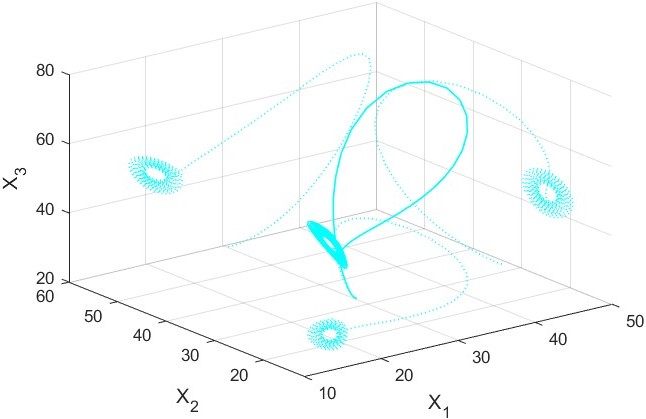}
    \caption{ }
  \end{subfigure}
  \hfill
  \begin{subfigure}[b]{0.325\textwidth}
    \includegraphics[width=\textwidth]{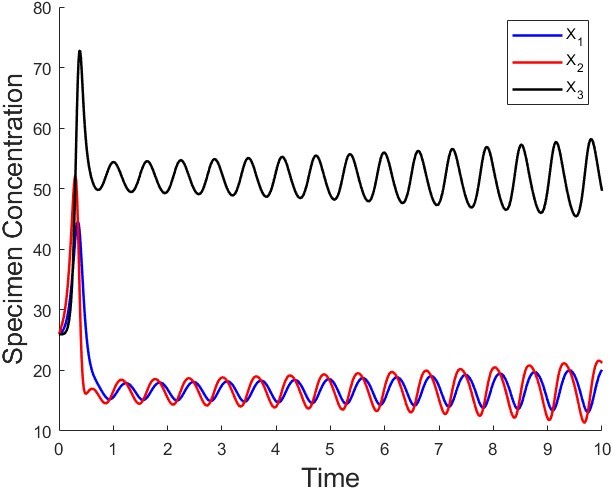}
    \caption{ }
  \end{subfigure}
    \hfill
    \begin{subfigure}[b]{0.325\textwidth}
    \includegraphics[width=\textwidth]{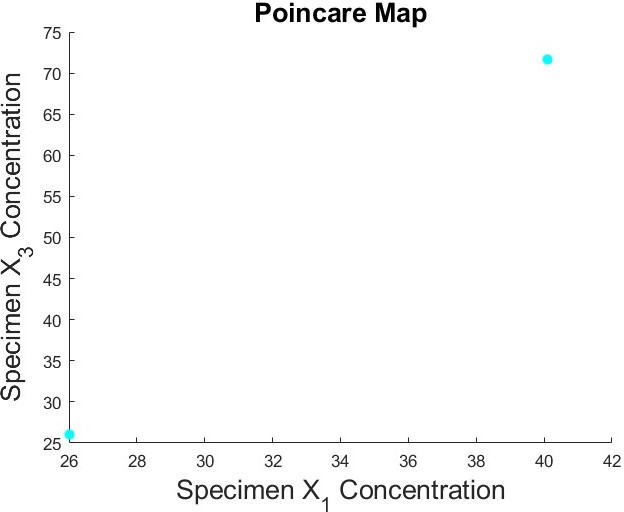}
    \caption{ }
  \end{subfigure}
  \hfill
  \begin{subfigure}[b]{0.325\textwidth}
    \includegraphics[width=\textwidth]{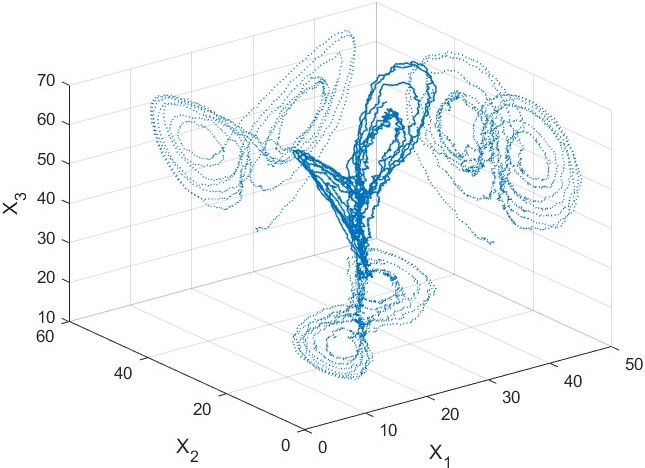}
    \caption{ }
  \end{subfigure}
    \hfill
      \begin{subfigure}[b]{0.325\textwidth}
    \includegraphics[width=\textwidth]{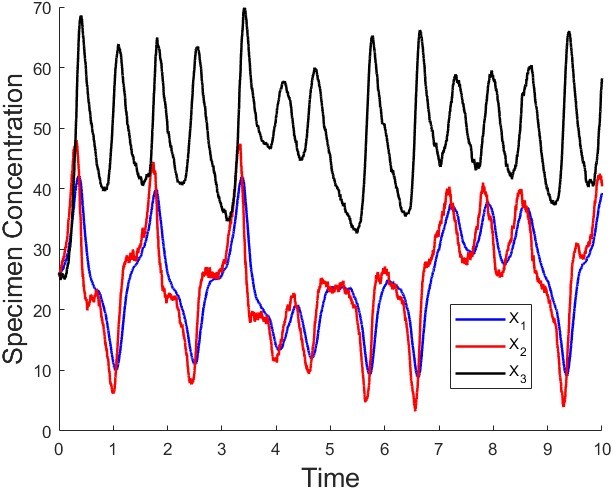}
    \caption{ }
  \end{subfigure}
  \hfill
  \begin{subfigure}[b]{0.325\textwidth}
    \includegraphics[width=\textwidth]{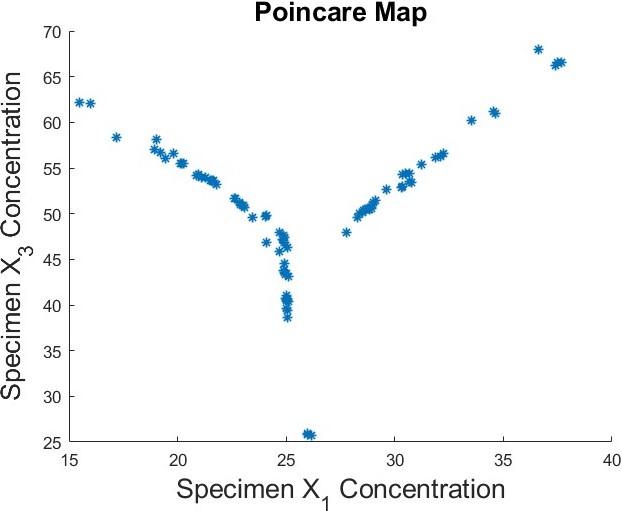}
    \caption{ }
  \end{subfigure}
    \hfill
  \caption[Chemical simulation of strange attractors]{The upper half (a-f) represents the R\"{o}ssler system ($a=b=0.2,$ $c=8,$ $V=1500$), and the lower half (g-l) represents the Lorenz system ($\sigma = 10,$ $\rho = 28,$ $\beta = 8/3$, $V=500$). The first rows (a-c),(g-i) give the solutions by numerically solving the non-kinetic systems~\eqref{rosslerode} and~\eqref{lorenzode}, and the second rows (d-f),(j-l) give an individual realization from chemical simulation using the inverse framework.}
  \label{chaos}
\end{figure}

These results piqued our interest in studying simulated chemical dynamics within the context of deterministic systems which are very sensitive to initial conditions. The R\"{o}ssler~\eqref{rosslerode} and Lorenz~\eqref{lorenzode} systems have been treated with AutoGillespie and simulated in Figure~\ref{chaos}. Indeed, the chemical realization picks up on characteristic dynamics deterministically unobserved within the simulation timeframe. For example, (g-i) shows only very early time oscillatory behaviours for the Lorenz system, while (j-l) already depicts the famous butterfly dynamics. Noise can also be discovered within Poincare maps--see (c),(f),(i),(l). It is not a surprise that Monte Carlo simulations do not converge stochastic trajectories to their deterministic expectation due to chaos.

Let us note that a similar figure to Figure~\ref{chaos} (j) is contained in Wilhelm's work~\cite{Wilhelm}, but it depicts a purely deterministic trajectory. Furthermore, kineticization is performed via $\Psi_{Kow}$ which is deterministically excellent but chemically undesirable, due to excessive limitations imposed on reactor volumes $V$. See Appendix~\ref{chaoticsystemsinappendix} for a further discussion.

To our knowledge, the chemical reaction network forms of these chaotic systems discovered by the inversion framework have not previously been reported within the literature.  

\clearpage
\newgeometry{top=8mm, bottom=10mm, left = 10mm, right = 10 mm}  

\cfoot{}
\clearpage
\cfoot{\thepage}
\restoregeometry

%% file: conclusions.tex
\chapter{Conclusion}

In this work, we formalized chemical notions into a coherent mathematical framework which highlighted the issue of cross-negative terms and their incompatible interpretation into mass-action kinetic networks. The concept of non-kinetic canonical inversion (Definition~\ref{noncanonicalinversiondefinition}) previously did not exist within the literature, but is helpful in establishing a clear description of kineticization strategies ($\Psi_{Kow}, \Psi_{QSST}$). 

Multiple methods were newly developed to design a general inversion framework for the chemical integration of ODEs. As polynomialization strategies, we proposed series expansions or more generally, polynomial approximations guided to be free from cross-negative terms by imposing penalties. Among other techniques, we also defined a bimolecularization strategy based on degree reduction via repeated substitutions, termed the General Quadraticization Algorithm which was introduced originally in~\cite{KernerSleightofHand} but refined substantially for clarity. Faced with the need to chemically realize arbitrary ODE systems, we designed and implemented the novel AutoGillespie Algorithm, capable of autonomously chemically inverting and simulating an input polynomial ODE system. 

We then cycled through various test systems to select appropriate techniques for the inversion framework. The simulation results for adequate polynomialization raised qualitative distinctions (e.g. NEC) in the dynamics effectuated by varying models, but rather than highlighting the superiority of one strategy over another, simulations informed the propriety of all strategies being considered. The selection of bimolecularization techniques followed a similar trend, but raised unresolved questions--namely, a further iteration of the quadraticization algorithm after $\Psi_{QSST}$ appeared to shift the steady state of the induced specimen $\mathrm{Y}$, seemingly without impacting the validity of the quasi-steady state assumption.

Possibly due to several existing methods (e.g. Kerner Polynomialization and $\Psi_{Kow}$) having been proven to fully preserve deterministic dynamics, a thorough investigation of series expansions in the chemical inversion of ODE systems is noticeably lacking in the literature. However, we have established through extensive simulations that attempting to perfectly preserve original dynamics may diminish returns. $\Psi_{Kow}$ was discovered to be largely undesirable in stochastic settings, and consequently superseded by $\Psi_{QSST}$ in the AutoGillespie Algorithm. Note that $\Psi_{QSST}$ may be considered to be a polynomial approximation strategy to non-kinetic polynomials, in contrast to $\Psi_{Kow}$ which assumes no approximation at all. We further showed that even low-order series expansions as done in Chapter $4$ may be more amenable to Monte Carlo methods in chemical simulations due to limited variance, providing a proof of principle to the hypothesis that approximate Taylor networks can possess qualitatively distinct dynamics that may be more desirable than exact networks (Section~\ref{proofofconcept}).

In Chapter $6$, we solidified a general inversion framework for the chemical integration of ODEs based on our simulations. To demonstrate its utility, a system unconsidered in~\cite{TomiStatInferencePaper} was kineticized and simulated to substantiate a key concern hypothesized in the previous work that statistical methods of~\cite{AutoCorrAndDFT} will be insufficient to distinguish time series realizations undergoing a deterministic bifurcation. The framework was further applied to chemically simulate the shifted pendulum, where a phase plane was drawn solely using individual realizations of stochastic trajectories. We propose the replication of Figure~\ref{thisisthependulumplot} (b),(e) in the lab setting as a challenge--in particular (b) requires complicated bimolecularization prior to synthetic implementation, and is therefore much more difficult. Certain individual trajectories impacted by stochasticity-induced motion were shown to outline the general contours of the phase plane, which piqued our interest in the simulation of the chaotic Lorenz and R\"{o}ssler systems, done promptly courtesy of AutoGillespie. To our knowledge, the inverted networks used to simulate the two strange attractors have not previously been uncovered in the literature, and are non-equivalent with existing chemical interpretations~\cite{LorenzChemical}. The proposed inversion framework can mass generate chemical reaction networks approximating a very wide range of exotic dynamics, as well as instantaneously producing multiple novel test systems for scientific and mathematical analysis. 

The simulations for the two chaotic attractors in video format, scaled to time, and a sample AutoGillespie code redesigned for ease of reading, are freely available to the general public at:  \\
1. https://youtu.be/CCplQMosFLI \quad 2. https://github.com/leesh-1/AutoGillespie 

\vspace{8 pt}

\noindent {\Large\textbf{Acknowledgements}}

\vspace{5 pt}

\noindent Su Hyeong Lee would like to thank Professor Radek Erban and Dr.Tomislav Plesa for their supervision during this project.


%% file: appendix1.tex
\chapter{Supplement to Chapter 1}\label{appendixA}

In 1864, Cato Guldberg and Peter Waage published a seminal paper proposing a method to determine the quantitative behaviours of chemical specimen, which has come to be known as the Law of Mass Action~\cite{LawMassActionHistory}. Unfortunately, their original paper was completely ignored by the wider academic community due to being written in Norwegian. Unsatisfied, Guldberg and Waage sought to publish their work in French, culminating in a paper titled ``Etudes sur les affinités chimiques" in 1867. After further refinement and validation of the theory via thermochemical techniques~\cite{LawMassActionThermochemicalHistory}, a Dutch chemist named Jacobus van’t Hoff independently rederived their work purely from the foundational principles of thermodynamics, where their work finally received widespread recognition. 

It may be noted that other researchers were simultaneously laying the foundations of the Law of Mass Action within this timeframe. For instance, it appears as though William Esson (1838-1916), an academically brilliant fellow at Merton College, Oxford, deduced a similar rudimentary theory on the basis of experimental data and intuitive reasoning~\cite{Merton}. This theory was the result of his collaborations with Augustus Harcourt~\cite{NatureHarcourt}, a chemist at Christ Church, Oxford, and coincides with the timeframe of the work by Gulberg and Waage formalizing the Law of Mass Action~\cite{HarcourtAndEsson}.

The Law of Mass Action asserts that under a ``suitable environment\footnote{In many references, a ``suitable environment'' does not appear to be rigorously defined. Usually, an environment in which temperature is unchanging qualifies, as the reaction rate coefficients are likely preserved.}'', the speed at which a reaction occurs is proportional to the active masses within the system, usually represented by chemical concentrations of the reactants taken to the power of their respective stoichiometric
coefficients~\cite{LawMassAction}. For example,  the rate of the forward (first) reaction in the model system 
$$
r_1: \quad \nu_{11}^- \mathrm{X}_1 + \nu_{21}^- \mathrm{X}_2 \stackrel{k_1}{\longrightarrow} \nu_{31}^+\mathrm{X}_3 + \nu_{41}^+ \mathrm{X}_4, \quad r_2 : \quad \nu_{32}^- \mathrm{X}_3 + \nu_{42}^- \mathrm{X}_4 \stackrel{k_2}{\longrightarrow} \nu_{12}^+\mathrm{X}_1 + \nu_{22}^+ \mathrm{X}_2
$$
is identical to the product of the rate coefficient and active masses of specimen $\mathrm{X}_1$, $\mathrm{X}_2$, which gives that
$$
\text{forward reaction rate}=k_{1} [\mathrm{X}_1]^{\nu_{11}^-}[\mathrm{X}_2]^{\nu_{21}^-}=k_{1} x_1^{\nu_{11}^-}x_2^{\nu_{21}^-}. 
$$
Similarly, the rate of the backward (second) reaction is given by 
$$
\text{backward reaction rate}=k_{2} [\mathrm{X}_3]^{\nu_{32}^-}[\mathrm{X}_4]^{\nu_{42}^-}=k_{2} x_3^{\nu_{32}^-}x_4^{\nu_{42}^-}. 
$$
At equilibrium, we expect both rates to be identical in value, and this induces
\begin{equation}\label{equilibriumconstant}
k_{1} x_1^{\nu_{11}^-}x_2^{\nu_{21}^-} = k_{2} x_3^{\nu_{32}^-}x_4^{\nu_{42}^-} \iff K:= \frac{k_1}{k_2} = \frac{x_3^{\nu_{32}^-}x_4^{\nu_{42}^-}}{x_1^{\nu_{11}^-}x_2^{\nu_{21}^-}}
\end{equation}
to be the equilibrium constant under the assumption of a stable environment, usually satisfied by controlling the temperature to be static under laboratory settings. In other words, the equilibrium constant is defined by~\eqref{equilibriumconstant} and numerically quantifies a dynamical system which displays, at least on a macroscophic level, no further tendency for change.

\begin{figure}[h!]
  \begin{subfigure}[b]{0.5\textwidth}
    \includegraphics[width=\textwidth]{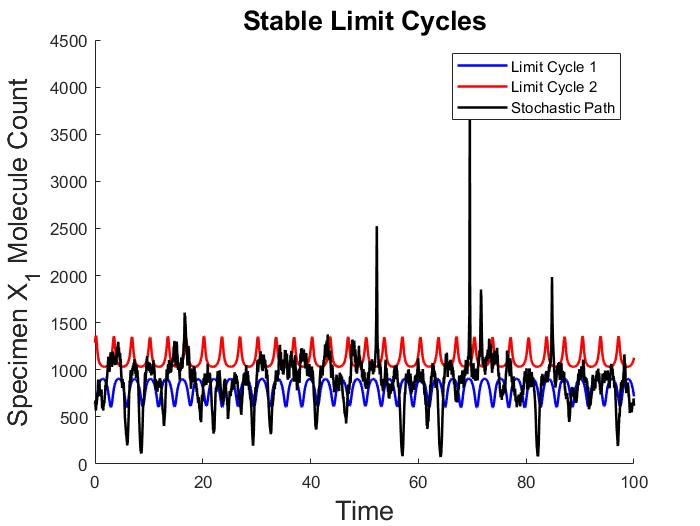}
    \caption{ }
  \end{subfigure}
  \hfill
  \begin{subfigure}[b]{0.5\textwidth}
    \includegraphics[width=\textwidth]{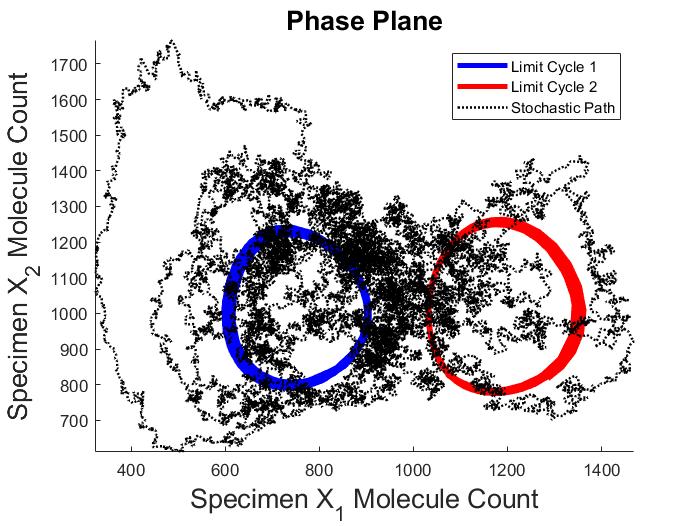}
    \caption{ }
  \end{subfigure}
  \caption[Examples of stochastic simulations of chemical reaction networks]{A simulation of the sample system composed of specimen $\mathrm{X}_1$, $\mathrm{X}_2$  given in Appendix B of~\cite{TomiStatInferencePaper} is provided. This system displays two stable limit cycles, which have been plotted deterministically in blue and red. Stochastic paths have been drawn in black, which shows a random switching between the two stable limit cycles. The bistable system has been non-dimensionalized, and abstractly models exotic stochastic switching between two stable states unobserved in the deterministic setting. This highlights the differences between deterministic and stochastic trajectories. }\label{topfigure}
\end{figure}

The discovery of the Law of Mass Action caused a revolution in the study of chemical systems, and led to the advent of chemical reaction network theory~\cite{ChemicalReactionNetworkTheory}. In turn, the formalization of chemical reactions into a coherent mathematical framework by pure and applied mathematicians alike has catalyzed rapid progress in the elucidation of previously intractable complex biochemical systems from the molecular level to the systems level\footnote{A Decay-Dimerization reaction network (see~\cite{implicitTauLeap} Section V, example A) is an example of molecular level analysis. A circadian oscillator (see equation~\eqref{stochasticnetworkcircadian}) is an example of a systems level analysis which elucidates periodicity of living systems.}. Chemical reaction networks have been shown to display a wide range of elegant dynamics, from demonstrating properties of memory~\cite{CRNmemory} and pattern formation~\cite{Murray2} to reproducing downright exotic behaviours such as multistability and limit cycles (Figure \ref{topfigure}) observed in deterministic systems~\cite{TomiStatInferencePaper}. Using the language of reaction network theory, design principles such as positive/negative feedback loops and biomedical controllers have been proposed in Systems and Synthetic Biology~\cite{TomiPraiseBiMolecularController}.

\section{Application-Oriented Regulatory Network}
\begin{figure}[h!]
    \begin{subfigure}[b]{0.5\textwidth}
    \includegraphics[width=\textwidth]{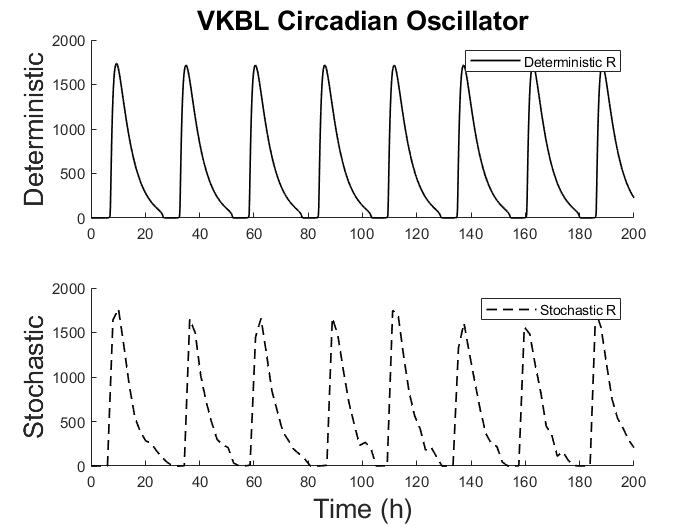}
    \caption{ }
  \end{subfigure}
  \hfill
  \begin{subfigure}[b]{0.5\textwidth}
    \includegraphics[width=\textwidth]{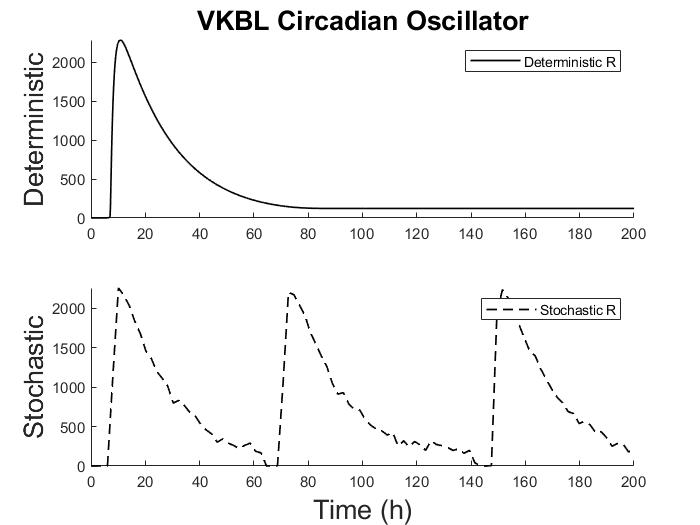}
    \caption{ }
  \end{subfigure}
  \caption[VKBL circadian oscillator]{A more application-oriented example may be found in the biomedical VKBL oscillator. In (a), we observe oscillations of approximately $24$ hours in both deterministic and stochastic settings. But as the degradation rate $\delta_R$ of the Repressor $\mathrm{R}$ is decreased (corresponding to $k_{14}$ and $\mathrm{X}_{8}$ in~\eqref{stochasticnetworkcircadian}), the system undergoes a supercritical Hopf bifurcation~\cite{GeneRegNetworkeasypaper}. Near the bifurcation point, deterministic oscillations are completely damped whereas periodic stochastic oscillations remain in (b).}
  \label{preprojectfigure}
\end{figure}

An application-driven example of a biomedically important chemical reaction network is the Vilar–Kueh–Barkai–Leibler (VKBL) circadian oscillator~\cite{VKBL}, which has been proposed as a minimal construction encapsulating the behaviour of periodic cycles observed experimentally \textit{in vitro}~\cite{VKBLexperimentdata}. Exact stochastic simulation algorithms such as the Gillespie Algorithm~\cite{GillespieFirstandDirectReactions} or the Modified Next Reaction Algorithm~\cite{modifiednextreactions} may be used to scrutinize the oscillator at the systems level, where it displays exotic dynamics which have been shown to be robust to inherent cellular noise (Figure~\ref{preprojectfigure} (b)). 

The VKBL reaction network is given by
\begin{equation}\label{stochasticnetworkcircadian}
\begin{aligned}
&r_{1}: \quad \mathrm{X}_6 + \mathrm{X}_1 \stackrel{k_1}{\longrightarrow} \mathrm{X}_3, \quad \quad r_{9}: \quad \mathrm{X}_5 \stackrel{k_9}{\longrightarrow} \varnothing,\\
&r_{2}: \quad \mathrm{X}_3 \stackrel{k_2}{\longrightarrow} \mathrm{X}_6 + \mathrm{X}_1, \quad\quad r_{10}: \quad \mathrm{X}_7 \stackrel{k_{10}}{\longrightarrow} \varnothing,\\
&r_{3}: \quad \mathrm{X}_6 + \mathrm{X}_2 \stackrel{k_3}{\longrightarrow} \mathrm{X}_4, \quad\quad r_{11}: \quad \mathrm{X}_5 \stackrel{k_{11}}{\longrightarrow} \mathrm{X}_6+\mathrm{X}_5, \\
&r_{4}: \quad \mathrm{X}_4 \stackrel{k_4}{\longrightarrow} \mathrm{X}_6 + \mathrm{X}_2, \quad\quad r_{12}: \quad \mathrm{X}_7 \stackrel{k_{12}}{\longrightarrow} \mathrm{X}_8 + \mathrm{X}_7,\\
&r_{5}: \quad \mathrm{X}_1 \stackrel{k_{5}}{\longrightarrow} \mathrm{X}_5 + \mathrm{X}_1, \quad\quad r_{13}: \quad \mathrm{X}_6 \stackrel{k_{13}}{\longrightarrow} \varnothing,\\
&r_{6}: \quad \mathrm{X}_3 \stackrel{k_{6}}{\longrightarrow} \mathrm{X}_5+\mathrm{X}_3, \quad\quad r_{14}: \quad \mathrm{X}_8 \stackrel{k_{14}}{\longrightarrow} \varnothing,\\
&r_{7}: \quad \mathrm{X}_2 \stackrel{k_{7}}{\longrightarrow} \mathrm{X}_7 + \mathrm{X}_2, \quad\quad r_{15}: \quad \mathrm{X}_6+\mathrm{X}_8 \stackrel{k_{15}}{\longrightarrow} \mathrm{X}_9,\\
&r_{8}: \quad \mathrm{X}_4 \stackrel{k_{8}}{\longrightarrow} \mathrm{X}_7 + \mathrm{X}_4, \quad\quad r_{16}: \quad \mathrm{X}_9 \stackrel{k_{16}}{\longrightarrow} \mathrm{X}_8,\\
\end{aligned}
\end{equation}
with rate vector (units of measurement are $\text{molecule}^{-1} \text{h}^{-1}$ for $k_1$, $k_3$, $k_{15}$, and $\text{h}^{-1}$ otherwise where $\text{h}$ stands for hour)
\begin{equation}\label{standardparameters}
    \boldsymbol{k} = (1,50,1,100,50,500,0.01,50,10,0.5,50,5,1,0.2,2,1).
\end{equation}
The deterministic counterpart dictating concentration dynamics of~\eqref{stochasticnetworkcircadian} are given by the Law of Mass Action:
\begin{equation}\label{deterministicnetworkcircadian}
\begin{aligned}
\Dot{x_1}=& k_2  x_3-k_1 x_1 x_6, \\
\Dot{x_2}=& k_4 x_4-k_3 x_2 x_6, \\
\Dot{x_3}=& k_1 x_1 x_6-k_2 x_3, \\
\Dot{x_4}=& k_3 x_2 x_6-k_4 x_4, \\
\Dot{x_5}=& k_6 x_3+k_5 x_1-k_9 x_5, \\
\Dot{x_6}=& k_{11} x_5+k_2 x_3+k_4 x_4 \\
&-x_6\left(k_1 x_1+k_3 x_2+k_{15} x_8+k_{16}\right), \\
\Dot{x_7}=& k_{8} x_4+k_{7} x_2-k_{10}x_7, \\
\Dot{x_8}=& k_{12} x_7-k_{15} x_6 x_8+k_{13} x_9 -k_{14} x_8, \\
\Dot{x_9}=& k_{15} x_6 x_8-k_{13} x_9.
\end{aligned}
\end{equation}

The VKBL model highlights two important features of complex biological systems. Firstly, stochastic analysis is essential for a thorough study of many proposed systems level networks, such as intracellular networks whose specimen are often sparsely populated~\cite{TomiCatastrophePaper}. Previous models of circadian oscillators have been shown to display non-robustness in the presence of noise, implying that such mechanisms are highly unlikely to be found in nature as a by-product of evolutionary processes~\cite{VKBLexperimentdata,incorrectcircadianmodel}. Furthermore, when a model system undergoes deterministic bifurcation, it is frequently reported in the literature that no substantive difference is found in stochastic simulations sharply at the bifurcation point, whereas the topology of the dynamics of the deterministic system changes drastically~\cite{TomiStatInferencePaper} (Figure~\ref{preprojectfigure} (b)).

Secondly, the VKBL oscillator suggests that living organisms may have adapted to take advantage of molecular noise to induce advantageous oscillations in their regulatory networks. In the circadian context, this implies that cells can possess more accurate internal clocks in stochastic settings than in purely deterministic settings, allowing for a  versatile response (such as preemptive preparatory gene transcription) to periodic environmental stimuli such as light/dark cycles and changes in temperature. These qualitatively distinct properties are completely neglected under a purely deterministic analysis. This highlights the need to undertake a stochastic analysis of deterministic models in biochemical settings via simulation of the system as chemical reaction networks.

\section{Further Motivations}
The design and implementation of reaction networks continues to be of interest in Systems and Synthetic Biology, as molecular models for complex phenomena and blueprints for synthetic design, and in Numerical Analysis and Statistical Theory, as test systems for simulation algorithms and acute parameter inference~\cite{TomiStatInferencePaper}. Circuitry for representing primitive calculations such as addition, division, and $m$-th root finding in chemical matter have been proposed~\cite{primitivecomputationCRN}, and biomolecule-based programming languages for computing roots of a small range of algebraic functions (such as polynomials) and nonlinear equations (exponential, logarithmic, and a subset of trigonometric equations) have been developed. In particular, the paper~\cite{molecularRootcomputation} depicts the implementation of Newton iterations for root-finding in an idealized abstract solution.

In $2010$, Soloveichik et al~\cite{DNAComputing} showed that DNA molecules may be designed to implement a very large class of chemical reaction networks, and gave a proof of principle by compiling limit cycle oscillators. This raises the possibility that in the distant future, DNA computing, with its massive parallel computation capabilities, may enable the simulation of highly complex reaction networks composed of tens of thousands of molecules in real time, revolutionizing the study of any system found in nature that may be described through the lens of chemical reaction network theory. Therefore, it is of strong interest to design and implement chemical reaction networks encapsulating the behaviour of \textit{arbitrary} dynamics.

We note that this closely relates to the second part of Hilbert's $16$th problem~\cite{Hilbert16problemproperreference}, which seeks to classify the number of limit cycles and their relative positions for two dimensional polynomial ODEs of $m$ degrees
\begin{equation}
    \begin{aligned}
&\Dot{x_1}=\mathcal{P}_1(x_1,x_2), \\
&\Dot{x_2}=\mathcal{P}_2(x_1,x_2),
\end{aligned}
\end{equation}
which remains unsolved for any $m>1$. Therefore, the study of inverting arbitrary ODE systems into two dimensional mass-action kinetic polynomials to investigate their dynamics is likely very difficult, as a special case of Hilbert's $16$th problem.

%% file: appendix2.tex
\chapter{Supplement to Chapters~\ref{oldchapter2}--\ref{oldchapter3}}\label{chapter2appendix}

As the notation $\mathbb{Z}_{\ge 0}^\mathcal{X}$ may not necessarily be standard in other fields of mathematics, further clarification is given via an example. We define the set 
\begin{equation}
    \mathcal{X} = \{\mathrm{X}_1, \mathrm{X}_2, \mathrm{X}_3, \mathrm{X}_4, \mathrm{X}_5 \},
\end{equation}
which implies 
\begin{equation}
    \mathbb{Z}_{\ge 0}^\mathcal{X} = \left\{\sum_{i \in \{ 1,\dots 5\}} b_i\mathrm{X}_i: b_i \in  \mathbb{Z}, b_i \ge 0 \right\}.
\end{equation}
The superscript $\mathcal{X}$ emphasizes the dependence of the elements of $\mathbb{Z}_{\ge 0}^\mathcal{X}$ on the basis set $\mathcal{X}$. It is also common to write $\mathbb{Z}_{\ge 0}^{|\mathcal{X}|}$ or $\mathbb{Z}_{\ge 0}^N$ for $N = |\mathcal{X}|$. Finite cardinality of $\mathcal{X}$ means that there exists a bijective mapping between a finite subset of $\mathbb{Z}$ and $\mathcal{X}$.

\section{Quasi-Steady State Transformation}\label{ProofImportant}

To prove the validity of the Quasi-Steady State Transformation, we first start by introducing Tikhonov's theorem~\cite{QSSATomiTheoryTichnoff}. 

\begin{theorem}[Tikhonov's theorem]\label{TikhonovTheoreminAppendix}
For the continuous functions $f_i: \mathbb{R}^{n+m+1} \to \mathbb{R}$, $g_j: \mathbb{R}^{n+m+1} \to \mathbb{R}$ and $\mathbf{x} = (x_1, \dots, x_n)^\top\in \mathbb{R}^n$, $\mathbf{y} = (y_1, \dots, y_m)^\top\in \mathbb{R}^m$, we consider the \textit{general} ODE system initialized at $\mathbf{x}(t_0) =  \mathbf{x}_0$, $\mathbf{y}(t_0) = \mathbf{y}_0$,
\begin{equation}\label{generalsetup}
    \begin{aligned}
\Dot{x_i} &=f_i(\mathbf{x}, \mathbf{y},t), \quad \text{for } i \in \{1,\dots,n\},\\
\mu \Dot{y_j}&=g_j(\mathbf{x}, \mathbf{y}, t), \quad \text{for } j \in \{1,\dots,m\} .
\end{aligned}
\end{equation}
Here $\mu$ is meant to be a small parameter and $t$ may be viewed as time. Taking the limit $\mu \downarrow 0$ induces the \textit{degenerate system}
\begin{equation}\label{degeneratesystem}
    \begin{aligned}
\Dot{x_i} &=f_i(\mathbf{x}, \mathbf{y}, t),  \quad \text{for } i \in \{1,\dots,n\}, \\
y_j &=\varphi_j(\mathbf{x}, t), \quad \text{for } j \in \{1,\dots,m\},
\end{aligned}
\end{equation}
where the second equation gives a root $\varphi = (\varphi_1,\dots,\varphi_m)^\top$ of the steady state of the \textit{adjoined system} 
\begin{equation}\label{trueadjoinedsystem}
\Dot{y_j} =g_j(\mathbf{x}, \mathbf{y}, t), \quad \text{for } j \in \{1,\dots,m \}.
\end{equation}
If the following conditions hold:

\noindent (i) $\varphi(\mathbf{x}, t)$ gives a stable isolated root of the adjoined system,

\noindent (ii) The initialization $\mathbf{y}(t_0) = \mathbf{y}_0$ lies within the domain of influence of the root $\mathbf{y}=\varphi(\mathbf{x}, t)$ in the adjoined system,

\noindent Then the general system asymptotically tends to the degenerate system in the limit $\mu \downarrow 0$. 
\end{theorem}

\noindent Let us now justify Theorem~\ref{QSSTtheorem}. For the reader's convenience, we first concisely recall Definition~\ref{QSSAnewdefinition} (the full definition is given in Section~\ref{QSSA}):

\begin{mdframed}
\begin{definition*}
Consider an $|\mathcal{X}|$-dimensional polynomial ODE system~\eqref{polynomialrighthandside}
\begin{equation*}
    \Dot{\mathbf{x}} = \mathcal{P}(\mathbf{x}),
\end{equation*}
and its inversion network $\mathcal{R}$. Given $\mathrm{X}_s \in \mathcal{X}$, disjointly partition $\mathcal{R} = \mathcal{R}_1^s \cup \mathcal{R}_2^s$ such that $r \in \mathcal{R}_1^s$ are kinetic and $r \in \mathcal{R}_2^s$ are non-kinetic. Rewrite system~\eqref{polynomialrighthandside} in the form of equation~\eqref{QSSAoriginalequation}, given below:
\begin{equation*}
\begin{aligned}
\Dot{x_{s}} &=\sum_{j \in \mathcal{R}^s_{1}} k_j\nu_{s j} \mathbf{x}^{\boldsymbol{\nu}_j^-}-\sum_{j^\prime \in \mathcal{R}^s_{2}} \norm{k_{j^\prime}\nu_{s j^\prime}} \mathbf{x}^{\nu_{j^\prime}^-}, \quad \text {for } \mathrm{X}_s \in \mathcal{X}, \\
x_{s}&\left(t_{0}\right) \ge 0, \quad t_0 \text{ initial time}.
\end{aligned}
\end{equation*}
Disjointly partition $\mathcal{X} = \mathcal{X}_1 \cup \mathcal{X}_2$ where $\mathcal{X}_1$ contains kinetic species and $\mathcal{X}_2$ non-kinetic species, respectively. We further enforce the initial condition $x_s(t_0) > 0$ for $\mathrm{X}_s \in \mathcal{X}_2$. Then, the \textit{degenerate system} is given by 
\begin{equation}\label{degeneratesysteminappenedix}
\begin{aligned}
\Dot{x_{s}} =\sum_{j \in \mathcal{R}^s_{1}} k_j\nu_{s j} \mathbf{x}^{\boldsymbol{\nu}_j^-}-\sum_{j^\prime \in \mathcal{R}^s_{2}} \norm{k_{j^\prime}\nu_{s j^\prime}} \mathbf{x}^{\boldsymbol{\nu}_{j^\prime}^-}, & \quad \text {for } \mathrm{X}_s \in \mathcal{X}_{1}, \\
\Dot{x_{s}}=\sum_{j \in \mathcal{R}^s_{1}} k_j\nu_{s j} \mathbf{x}^{\boldsymbol{\nu}_j^-} -\omega_{s}^{-1} x_{s} p_{s}(\textbf{x}) y_{s} \left(\sum_{j^\prime \in \mathcal{R}^s_{2}} \norm{k_{j^\prime}\nu_{s j^\prime}} \mathbf{x}^{\boldsymbol{\nu}_{j^\prime}^-}\right) , & \quad \text {for } \mathrm{X}_s \in \mathcal{X}_{2},
\end{aligned}
\end{equation}
which satisfies the aforementioned initial conditions, with an \textit{adjoined system} for $\mathrm{X}_s \in \mathcal{X}_2$ given by
\begin{equation}\label{adjoinedsystem}
    \begin{aligned}
    \mu \Dot{y_{s}} &=\omega_{s}-x_{s} p_{s}(\mathbf{x}) y_{s}, \\
    y_{s}\left(t_{0}\right) &\geq 0, \quad t_0 \text{ initial time.}
    \end{aligned}
\end{equation}
Here $\mu, \omega_s \in \mathbb{R}_{>0}$, $y_s(t_0)$ may be any non-negative value, and the polynomial $p(\mathbf{x})$ must map the non-negative set $\mathbb{R}_{\ge 0}^\mathcal{X}$ into the positive region $\mathbb{R}_{>0}^{\mathcal{X}_2}$. The complete system composed of the degenerate and adjoined systems is called the \textit{general system}. Then, the map $\Psi_{QSST} :  \mathbb{P}_{m}\left(\mathbb{R}^{\mathcal{X}} ; \mathbb{R}^{\mathcal{X}}\right) \rightarrow$ $\mathbb{P}_{m^\prime}\left(\mathbb{R}^{\mathcal{X}\sqcup\mathcal{X}_{2}} ; \mathbb{R}^{\mathcal{X}\sqcup\mathcal{X}_{2}}\right)$ which maps the right hand side of~\eqref{polynomialrighthandside} to the right hand side of the general system is called the \textit{Quasi-Steady State Transformation}. 
\end{definition*}
\end{mdframed}

\noindent\begin{theorem*}
The general system induced by $\Psi_{QSST}$ is asymptotically equivalent to \eqref{polynomialrighthandside} in the limit $\mu \downarrow 0$, given the agreement of the initial conditions posed in Definition~\ref{QSSAnewdefinition}.
\end{theorem*}

\noindent \textit{Proof.} Tikhonov’s theorem is usually a dimension reduction strategy, translating a general system with the small parameter $\mu$ to a more analytically tractable form given by the degenerate system. But here, we shall follow the inverse direction. Namely, we start with the \textit{first degenerate system}
\begin{equation}\label{fds}
    \Dot{x_{s}} =\sum_{j \in \mathcal{R}^s_{1}} k_j \nu_{s j} \mathbf{x}^{\boldsymbol{\nu}_j^-}-\sum_{j^\prime \in \mathcal{R}^s_{2}} \norm{k_{j^\prime}\nu_{s j^\prime}} \mathbf{x}^{\boldsymbol{\nu}_{j^\prime}^-}, \quad \text {for } \mathrm{X}_s \in \mathcal{X}, \quad x_{s}\left(t_{0}\right) \ge 0,
\end{equation}
and attempt to induce kinetic terms which asymptotically approximate the summands induced by non-kinetic reaction channels $j^\prime \in \mathcal{R}^s_{2}$. This gives rise to a degenerate system, whose general system will form a kinetic polynomial.

The degenerate system takes the form~\eqref{degeneratesystem},
\begin{equation*}
    \begin{aligned}
\Dot{x_i} &=f_i(\mathbf{x}, \mathbf{y}, t),  \quad \text{for } i \in \{1,\dots,|\mathcal{X}|\}, \\
y_j &=\varphi_j(\mathbf{x}, t), \quad \text{for } j \in \{1,\dots,|\mathcal{X}_2|\}, 
\end{aligned}
\end{equation*}
where we consider a reordering of specimen so that $\mathrm{X}_i \in \mathcal{X}_2$ for $i \in \{1,\dots,|\mathcal{X}_2|\}$. We aim for the functions $f_i(\mathbf{x}, \varphi(\mathbf{x}, t), t)$ to recover the right hand sides of the first degenerate system~\eqref{fds} for $i \in \{1,\dots,|\mathcal{X}_2|\}$. Considering the simplest manipulations possible, a natural choice is the form $y_i=1/x_i$, i.e. $\varphi_i(\mathbf{x}, t) = 1/x_i$. Then, a degenerate system that describes~\eqref{fds} may be written in the kinetic form
\begin{equation}\label{thisthat}
\begin{aligned}
\Dot{x_{s}} =\sum_{j \in \mathcal{R}^s_{1}} k_j\nu_{s j} \mathbf{x}^{\boldsymbol{\nu}_j^-}-\sum_{j^\prime \in \mathcal{R}^s_{2}} \norm{k_{j^\prime}\nu_{s j^\prime}} \mathbf{x}^{\boldsymbol{\nu}_{j^\prime}^-}, & \quad \text {for } \mathrm{X}_s \in \mathcal{X}_{1}, \\
\Dot{x_{s}}=\sum_{j \in \mathcal{R}^s_{1}} k_j\nu_{s j} \mathbf{x}^{\boldsymbol{\nu}_j^-} -x_{s} y_{s} \left(\sum_{j^\prime \in \mathcal{R}^s_{2}} \norm{k_{j^\prime}\nu_{s j^\prime}} \mathbf{x}^{\boldsymbol{\nu}_{j^\prime}^-}\right) , & \quad \text {for } \mathrm{X}_s \in \mathcal{X}_{2},
\end{aligned}
\end{equation}
along with the roots \begin{equation}
    y_s = 1/x_s, \quad \text {for } \mathrm{X}_s \in \mathcal{X}_{2}.
\end{equation}
Note that substituting the roots into~\eqref{thisthat} fully recovers~\eqref{fds}. We must now find an adjoined system that takes the given $\varphi$ as a root. Assuming $x_s(t)>0$ for $t \ge t_0$, we have $y_s = 1/x_s \iff 0 = 1-x_s y_s$, which may be seen as a steady state of 
\begin{equation}
    \Dot{y_s} = 1-x_s y_s, \quad \text{for } \mathrm{X}_s \in \mathcal{X}_2.
\end{equation}
Then the general system is given by the equations~\eqref{thisthat} with the ``adjoined system'' (note that the naming convention given in~\cite{TomiInversePaper} is slightly inconsistent with~\eqref{trueadjoinedsystem})
\begin{equation}
    \mu\Dot{y_s} = 1-x_s y_s, \quad \text{for } \mathrm{X}_s \in \mathcal{X}_2.
\end{equation}
This fully intuits Definition~\ref{QSSAnewdefinition}; the form given there only takes a slightly more general structure to the roots $\varphi$ (i.e. $\varphi_s(\mathbf{x},t) := \omega_{s}(x_{s} p_{s}(\mathbf{x}))^{-1}$).

Now the proof of the theorem is fairly straightforward. Let us remark that as noted below Theorem~\ref{QSSTtheorem}, in order for the dynamical equivalence to be preserved, $x_s(t)>0$ is enforced for $t \ge t_0$ for $\mathrm{X}_s \in \mathcal{X}_2$ in Definition~\ref{QSSAnewdefinition}. We assume this in the proof. Under the conditions of Tikhonov’s theorem, it is sufficient for asymptotic equivalence to show:

\noindent (i) $\omega_{s}\left(x_{s} p_{s}(\mathbf{x})\right)^{-1}$ is an isolated steady state of the adjoined system,

\noindent (ii) The steady state in (i) is globally stable within the non-negative cone $\mathbf{y}_0 \in \mathbb{R}^{\mathcal{X}_2}_{\ge 0}$.

As for (ii), we impose in the theorem that $\omega_{s}, x_{s}, p_{s}(\mathbf{x}) > 0$ for $t > t_0$, so the stability of the adjoined system is immediately established. It is also clear from substitution that $\omega_{s}\left(x_{s} p_{s}(\mathbf{x})\right)^{-1}$ is a steady state of the adjoined system as it nullifies the right hand side. Finally, noting that the root $\varphi_s$ is unique (thus isolated), the proof is complete. Note further that substituting $y_s = \omega_{s}\left(x_{s} p_{s}(\mathbf{x})\right)^{-1}$ into~\eqref{degeneratesysteminappenedix} recovers the system~\eqref{QSSAoriginalequation}.

\section{General Quadraticization}
We provide a step-by-step application of the General Quadratidization Algorithm to establish an intuitive grasp of the methodology. 

\begin{example}\label{easyexamplequadraticize}
We consider a $3$-dimensional cubic polynomial system~\eqref{typicalvariable} denoted in the natural variables $x$, $y$, and $z$ and its translation to the formal notation used in~\eqref{generalpolynomialsystem} and throughout this paper (equations~\eqref{translatedvariable1} and~\eqref{translatedvariable2}), where abstract indices $c_j$ have been swapped with $i$,$j$,$k$ for readability:

\begin{minipage}{.4\textwidth}
\begin{equation}\label{typicalvariable}
\begin{aligned}
&\Dot{x}=xy^2, \\
&\Dot{y}=z^2, \\
&\Dot{z}=1.
\end{aligned}
\end{equation}
\end{minipage}
\begin{minipage}{.45\textwidth}
\begin{equation}\label{translatedvariable1}
\begin{aligned}
&\Dot{x_1}=x_1x_2^2 = x_1x_2x_2, \\
&\Dot{x_2}=x_3^2 = x_3x_3, \\
&\Dot{x_3}=1.
\end{aligned}
\end{equation}
\end{minipage}
\vspace{5pt}
\begin{equation}\label{translatedvariable2}
\begin{aligned}
&\Dot{x_1}=0 + \sum_{i \in \{1,2,3\}} 0 \cdot x_i + \sum_{i,j \in \{1,2,3\}} 0 \cdot x_ix_j +  x_1x_2x_2 + \sum_{\substack{i,j,k \in \{1,2,3\} \\ \{ i,j,k\} \neq \{1,2,2 \}}} 0 \cdot x_ix_jx_k ,\\
&\Dot{x_2}=0 + \sum_{i \in \{1,2,3\}} 0 \cdot x_i + x_3x_3 + \sum_{\substack{i,j \in \{1,2,3\}\\ \{i,j\} \neq \{3,3 \}}} 0 \cdot x_ix_j +  \sum_{\substack{i,j,k \in \{1,2,3\} }} 0 \cdot x_ix_jx_k ,\\
&\Dot{x_3}=1 + \sum_{i \in \{1,2,3\}} 0 \cdot x_i + \sum_{i,j \in \{1,2,3\}} 0 \cdot x_ix_j + \sum_{\substack{i,j,k \in \{1,2,3\} }} 0 \cdot x_ix_jx_k .
\end{aligned}
\end{equation}
A single iteration is sufficient for termination of the quadraticization algorithm. We consider the first equation for $\dot{x}_1$, and $\dot{x}_2$, $\dot{x}_3$ are treated analogously. We introduce $\eta_{ij} := x_i x_j$ to obtain
\begin{equation}
    \Dot{x_1}=0 + \sum_{i \in \{1,2,3\}} 0 \cdot x_i + \sum_{i,j \in \{1,2,3\}} 0 \cdot \eta_{ij} +  \eta_{12}x_2 + \sum_{\substack{i,j,k \in \{1,2,3\} \\ \{ i,j,k\} \neq \{1,2,2 \}}} 0 \cdot \eta_{ij}x_k,
\end{equation}
a multivariate quadratic. Differentiation of $\eta_{12}$ gives
\begin{equation}
    \dot{\eta}_{12} = \dot{x}_1 x_2 + x_1 \dot{x}_2.
\end{equation}
Under the algorithm, the first summand is mapped to the form
\begin{equation}
     \dot{x}_1 x_2 = 0 \cdot x_2+ \sum_{i \in \{1,2,3\}} 0 \cdot \eta_{i2} + \sum_{i,j \in \{1,2,3\}} 0 \cdot \eta_{ij} +  \eta_{12}\eta_{22} + \sum_{\substack{i,j,k \in \{1,2,3\} \\ \{ i,j,k\} \neq \{1,2,2 \}}} 0 \cdot \eta_{ij}\eta_{k2},
\end{equation}
where the second summand becomes 
\begin{equation}\label{samplecubicexample}
     x_1 \dot{x}_2 = x_1 \cdot 0 + \sum_{i \in \{1,2,3\}} \eta_{1i} \cdot 0 + x_1\eta_{33} + \sum_{\substack{i,j \in \{1,2,3\}\\ \{i,j\} \neq \{3,3 \}}} x_1\eta_{ij} \cdot 0 +  \sum_{\substack{i,j,k \in \{1,2,3\} }} \eta_{1i}\eta_{jk} \cdot 0.
\end{equation}
As all summands are at most quadratic, the algorithm terminates. 
\end{example}

\section{Quasi-Steady State Approximation}\label{newexample}

\begin{example}
The case in which reactants consist of three distinct specimen is trivially managed by the presented derivation in Section~\ref{quasisteadystatetransformation}. Thus we instead apply this technique to the trimolecular reaction
\begin{equation}
    r_1:\quad \mathrm{X}_1 + 2\mathrm{X}_2 \stackrel{k}{\rightarrow} 3\mathrm{X}_2 + \sum_{\ell \in \{3,\dots,10\}} \nu_{\ell 1} \mathrm{X}_{\ell}.
\end{equation}
Note that the deterministic dynamics are given by the RREs 
\begin{equation}\label{compareexample}
    \begin{aligned}
    &\Dot{x_1}=- k x_1x_2^2, \\
&\Dot{x_2}= k x_1x_2^2, \\
&\Dot{x_\ell}=\nu_{\ell1} k x_1x_2^2,\quad \ell \in \{3,\dots,10\}.
\end{aligned}
\end{equation}
Revisiting~\eqref{thisintermediateequation}, there are ${}_3 C_2$ choices for $\mathrm{X}_i, \mathrm{X}_j$, where the remaining reactant is allocated to $\mathrm{X}_k$. For example, we may take 
\begin{equation}
    \mathrm{X}_i = \mathrm{X}_2,\quad \mathrm{X}_j = \mathrm{X}_2, \quad \mathrm{X}_k = \mathrm{X}_1,
\end{equation}
where the double identification of $\mathrm{X}_2$ implies the identities 
\begin{equation}\label{theidentity}
    \Dot{x_1} = \Dot{x_k},\quad \Dot{x_2} = \Dot{x_i} + \Dot{x_j}.
\end{equation}
This induces the bimolecular approximation~\eqref{thisintermediateequation}:
\begin{equation}
\begin{aligned}
&r_1: \quad \mathrm{X}_2+\mathrm{X}_2 \stackrel{k_1}{\rightarrow} \mathrm{Z}, \quad r_2: \quad \mathrm{Z} \stackrel{k_2}{\rightarrow} \mathrm{X}_2+\mathrm{X}_2, \\
&r_3: \quad \mathrm{X}_1 + \mathrm{Z} \stackrel{k_{3}}{\rightarrow} \nu_{i3} \mathrm{X}_2+\nu_{j3} \mathrm{X}_2+\nu_{k3} \mathrm{X}_1 +\sum_{\ell \in \{3,\dots,10\}} \nu_{\ell 3} \mathrm{X}_{\ell}+\nu_{z3} \mathrm{Z},
\end{aligned}
\end{equation}
where the rate coefficients $k_1,k_2,k_3$ and stoichiometric coefficients $\nu_{i3}, \nu_{j3}, \nu_{k3}, \nu_{z3} $ must be determined. Identifying~\eqref{goodapprox} and~\eqref{compareexample} via~\eqref{theidentity} gives
\begin{equation}
      \begin{aligned}
      &-k =\left(\nu_{k3}-1\right) \frac{k_{1} k_{3}}{k_{2}},  \\
& k =\left(\nu_{i3} + \nu_{j3} +2\nu_{z3}-2\right) \frac{k_{1} k_{3}}{k_{2}}, \\
& \nu_{\ell 1} k =\nu_{\ell3} \frac{k_{1} k_{3}}{k_{2}}, \quad \ell \in \{3,\dots,10\}.
\end{aligned}  
\end{equation}
We may impose for any $n \in \mathbb{Z}_{\ge 1}$ the constraint
\begin{equation}\label{rateconstant}
    k = \frac{n k_1 k_3}{k_2}, 
\end{equation}
where we choose $n = 1$ to simplify the example. Then, we must satisfy
\begin{equation}
    -1 = \nu_{k3}-1, \quad 1 = \nu_{i3} + \nu_{j3} +2\nu_{z3}-2, \quad \nu_{\ell 1} = \nu_{\ell 3}, \quad \ell \in \{3,\dots,10\}.
\end{equation}
The values $\nu_{k 3} = 0$ and $\nu_{\ell 1} = \nu_{\ell 3}$ are fixed. Non-uniqueness arises due to the degrees of freedom in the second identity, of which any valid choice presents an approximation. For instance by choosing $\nu_{i3} = 0$, $\nu_{j3}=\nu_{z3} = 1$, we derive the bimolecular network 
\begin{equation}
\begin{aligned}
&r_1: \quad \mathrm{X}_2+\mathrm{X}_2 \stackrel{k_1}{\rightarrow} \mathrm{Z}, \quad r_2: \quad \mathrm{Z} \stackrel{k_2}{\rightarrow} \mathrm{X}_2+\mathrm{X}_2, \\
&r_3: \quad \mathrm{X}_1 + \mathrm{Z} \stackrel{k_{3}}{\rightarrow}  \mathrm{X}_2  +\sum_{\ell \in \{3,\dots,10\}} \nu_{\ell 1} \mathrm{X}_{\ell}+ \mathrm{Z}.
\end{aligned}
\end{equation}
All that remains is to set the rate coefficients to be consistent with~\eqref{rateconstant} as $k_2 \to \infty$ while enforcing $k_1 \ll k_2$ to ensure an instantaneous and continued near-extinction of $\mathrm{Z}$. $k_1 = 1$, $k_3 = k k_2$ is one possible choice, as well as $k_1, k_3 = \sqrt{k k_2}$. Note that this approximation will inevitably render the system stiff due to only being valid in the limiting case $k_2 \to \infty$. 
\end{example}

\chapter{On Gillespie and AutoGillespie}\label{GillespieAutoGillespie}
In Section~\ref{AutoGillespieSection}, we proposed Taylor expansions as a general polynomialization strategy which may also be used as a remedy when Kerner Polynomialization fails to terminate or becomes intractably complex. When all arbitrary ODE systems arising in practice are considered, cross-negative terms will invariably appear in series expansions. It is furthermore difficult to estimate \textit{a priori} how high orders of expansion must be for sufficient encapsulation of the dynamics of the original system, thus an algorithm capable of autonomously stochastically simulating an arbitrary polynomial ODE system after inversion into a chemical reaction network must be developed.

The key to designing such an algorithm is the detection and tracking of cross-negative terms in input polynomials, which enables the application of nonlinear transformations for translation into a mass-action kinetic system. One possible implementation of this novel algorithm in \mcode{Matlab} is presented, via the \mcode{symbolic} \mcode{math} \mcode{toolbox}~\cite{SymbolicMathToolbox} maintaining the \mcode{children} and \mcode{symvar} packages. This AutoGillespie formulation implements the Kowalski Transformation, but can straightforwardly be modified to implement the Quasi-Steady State Transformation.

\begin{mdframed}
(\textbf{AutoGillespie Algorithm Conversational Pseudocode})

\noindent\rule{16.3cm}{0.5pt}
\vspace{1pt}

\noindent \textbf{(C1):} Take as input the right hand side of any general symbolic $|\mathcal{X}|$-dimensional polynomial ODE system of degree $m$~\eqref{polynomialrighthandside}. Recall that such systems can be rewritten in the form~\eqref{QSSAoriginalequation}, which is helpful in visualizing the algorithm:
\begin{equation}\tag{\ref{QSSAoriginalequation}}
    \frac{\mathrm{d} x_{s}}{\mathrm{~d} t} =\sum_{j \in \mathcal{R}^s_{1}} k_j\nu_{s j} x^{\boldsymbol{\nu}_j^-}-\sum_{j^\prime \in \mathcal{R}^s_{2}} \norm{k_{j^\prime}\nu_{s j^\prime}} x^{\boldsymbol{\nu}_{j^\prime}^-}, \quad \text {for } \mathrm{X}_s \in \mathcal{X}.
\end{equation}
Store all symbolic variables (specimen) that appear within the input polynomial system via the \mcode{symvar} package.
\vspace{10pt}

\noindent\textbf{(C2):} We will identify each term as kinetic or non-kinetic. Introduce the \textit{ghost particles} $g_1$, $g_2$ and \textit{ghost rates} $k_{g1}$, $k_{g2}$, which ensures proper implementation via \mcode{children}, a package utilizing internal \mcode{Matlab} classification rules. Call and expand the right hand side of $\Dot{x}_s$ after adding in $k_{g1}\cdot g_1 + k_{g2}\cdot g_2$.

\vspace{10pt}
\noindent\textbf{(C3):} Using \mcode{children}, count the number of terms that appear in the right hand side of $\Dot{x}_s$. Enter into a loop, which returns each existing monomial and the corresponding coefficient as ordered symbolic vectors. The coefficient vector is considered a segment of the rate vector $\mathbf{k}$.

\vspace{10pt}
\noindent\textbf{(C4):} Knock out the ghost particles by substituting $k_{g1}$, $g_1$, $k_{g2}$, $g_2 = 0$. Return the indices in which the rate vectors are negative, which identifies the candidates for cross-negative terms.

\vspace{10pt}
\noindent\textbf{(C5):} Recall Definition~\ref{crossnegative}, which was unconventionally given to clarify the AutoGillespie Algorithm. Substitute $x_s = 0$ and $x_{s^\prime} = 1$ for $\forall   \mathrm{X}_{s^\prime} \in \mathcal{X} \setminus \{\mathrm{X}_{s} \} $ into the vector of monomials in \textbf{(C4)}. Return the indices in which their entries survive (i.e. are non-zero).

\vspace{10pt}
\noindent\textbf{(C6):} Match the indices obtained in \textbf{(C4), (C5)}. If their intersection is non-empty, save this set as a pointer to the position of cross-negative terms, as well as adding the specimen $\mathrm{X}_s$ to $\mathcal{X}_2$. Introduce the specimen $\mathrm{Y}_s$, which will form the adjoined system. (See Definition~\ref{kowalskidefinition} for the definitions of $\mathrm{Y}_s,\mathcal{X}_2$.)

\vspace{10pt}
\noindent\textbf{(C7):} For $\forall \mathrm{X}_s \in \mathcal{X}_2$, we will represent the application of $\Psi_{Kow}$ to the right hand side of $\Dot{x}_s$. Save a copy of the monomial and coefficient vectors. Nullify locations in the vector of monomials and coefficients targeted by the pointers obtained in \textbf{(C6)} and return the results in a polynomial form, which represent all kinetic terms (i.e. dot product the two ordered vectors to return the polynomial form).

\vspace{10pt}
\noindent\textbf{(C8):} Now nullify locations in the original vectors not targeted by the pointers and return the results in a polynomial form, which represent the cross-negative terms. Multiply $y_s\cdot x_s$ to this polynomial and add to the returned kinetic polynomial in \textbf{(C7)}. This gives the degenerate system~\eqref{Kowalskidegenerate}.

\vspace{10pt}
\noindent\textbf{(C8):} Now we represent the adjoined system. For $\mathrm{X}_s \in \mathcal{X}_2$, call the right hand side of the original $\Dot{x}_s$ and multiply $-y_s^2$. Return this polynomial as the right hand side of $\Dot{y}_s$, which is kinetic. Formally add $\mathrm{Y}_s$ into the specimen set $\mathcal{X}$. 
\vspace{10pt}

\noindent\textbf{(C9):} We have fully represented the general system mapped to via $\Psi_{Kow}$. Using the methods given in \textbf{(C1)-(C4)}, we may canonically invert the general system into a mass-action kinetic reaction network. This is done by vectorizing the general system into a long concatenation of monomial vectors and the corresponding (concatenated) coefficient vector. The vector retained by applying the euclidean norm to each entry of the coefficient vector may now be identified as the rate vector given by the canonical inversion of the general system, which we stochastically simulate. 

\vspace{10pt}
\noindent\textbf{(C10):} Launch the Gillespie Algorithm with the vectorization in \textbf{(C9)}. The propensities may be computed symbolically and returned as a function handle by entering into a loop which recognizes the individual degrees of each variable that appears in a monomial, using the \mcode{symbolic} \mcode{math} \mcode{toolbox} and the tricks described in \textbf{(C1)-(C4)}.
\end{mdframed}


As we increase the order of Taylor expansions, the canonically inverted network grows vastly high dimensional. This version of AutoGillespie must therefore siphon through a  large number of reaction channels, and the computational cost incurred limits simulations to very early timescales. Therefore, it is desirable to seek a modified version of the algorithm that produces long-time stochastic trajectories even when the network has an excessive number of reaction channels. 

This may be accomplished by altering the structure of the Gillespie Algorithm in \textbf{(C10)}. Within the canonical inversion, any reaction channel adds or removes precisely one specimen from the reactants, which is determined by the sign of the concatenated coefficient vector in \textbf{(C9)}. In other words, for specimen $\mathrm{X}_s\in \mathcal{X}$, all terms that appear within the kineticized right hand side of $\Dot{x}_s$ are canonically inverted to a reaction channel that either adds or removes one $\mathrm{X}_s$ molecule. We apply a similar strategy to that detailed in \textbf{(C4-C6)} by identifying the negative real elements of the ordered coefficient vector and extracting a set of pointers which locate the terms inverted to remove one $\mathrm{X}_s$ molecule. After computing closed form propensities, their symbolic algebraic expressions are summed. Repeating this procedure for the terms which are inverted to add an $\mathrm{X}_s$ molecule essentially classifies all terms in the right hand side of $\Dot{x}_s$ into two pseudo-reactions, which add or remove one copy of $\mathrm{X}_s$. In this manner, we exploit the structure of canonical inversions to formalize any $|\mathcal{X}|$-dimensional \textit{kinetic} polynomial into a reaction network with at most $2|\mathcal{X}|$ reaction channels\footnote{If starting from an \textit{arbitrary} polynomial of $|\mathcal{X}|$ dimensions, kineticization via $\Psi_{Kow}$ or $\Psi_{QSST}$ may introduce a specimen with concentration $y_s$ per every specimen concentration $x_s$. In this sense, an arbitrary $|\mathcal{X}|$-dimensional polynomial can be considered to have been inverted into a network with at most  $4|\mathcal{X}|$ reaction channels.}.

In this approach, increasing the order of expansion drastically from $5$ to $20$ yields no substantive increase in run time in the simulation, although the symbolic computation of the expansion takes longer. However, this needs to be done \textit{precisely once}. After we have obtained the truncated polynomial series of the right hand side of an arbitrary ODE system, we may use them to compute the closed form propensities and return a function handle vector of length at most $4|\mathcal{X}|$, which again, need only be computed once. Given that the computing platform has enough resources to perform a symbolic computation for series expansions of high order, the Gillespie Algorithm is almost guaranteed to produce long time trajectories. This modified AutoGillespie Algorithm is used in Chapter~\ref{oldchapter6}. The generalization to using Quasi-Steady State Transformations instead of Kowalski Transformations is trivial. 

%% file: appendix3.tex
\chapter{Supplement to Chapters~\ref{oldchapter4}--\ref{oldchapter6}}\label{initialconcentration}

\begin{figure}[h]
  \begin{subfigure}[b]{0.5\textwidth}
    \includegraphics[width=\textwidth]{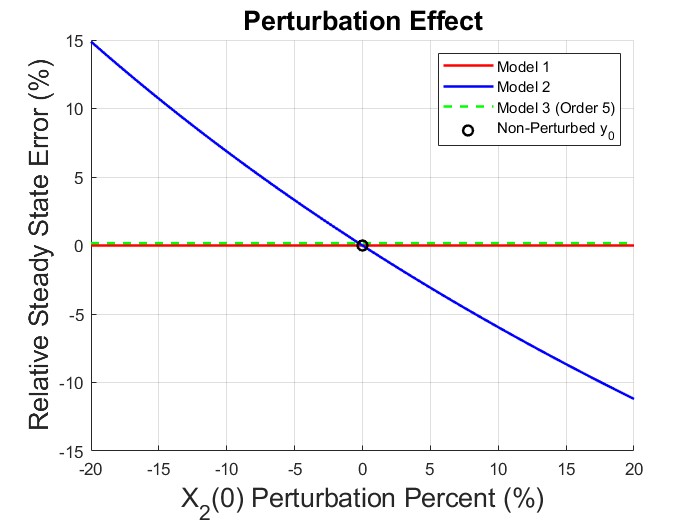}
    \caption{Relative Errors at Steady State}
  \end{subfigure}
  \hfill
  \begin{subfigure}[b]{0.5\textwidth}
    \includegraphics[width=\textwidth]{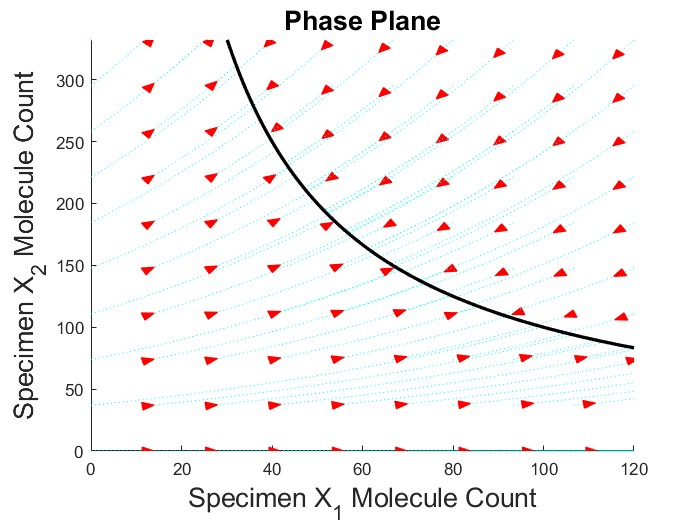}
    \caption{Phase Plane}
  \end{subfigure}
  \caption[Noise contamination in initial concentrations of~\ref{model2}.]{Noise contamination of initial data $x_2(0) = \mathrm{X}_2(0)/V$ for $V = 100$ stabilizes the steady state to a perturbed value. The phase plane is plotted in (b), where the steady state curve $x_2 = 1/x_1$ is shown in black and red arrows point in the direction of increasing time. Solution trajectories are plotted in light blue.}
  \label{perturbplot1}
\end{figure}
Fairly trivial analytical progress may be made for \ref{model2}, which immediately gives
\begin{equation}
    \frac{\mathrm{d}x_2}{\mathrm{d}x_1} = x_2 \implies x_2 = \frac{x_2(0)}{\exp{(x_1(0))}} \exp{(x_1)}.
\end{equation}
Substitution of $x_2$ into the steady state curve $x_2 = 1/x_1$ gives the equilibrium $x$ value
\begin{equation}\label{deterministicequilibrium1}
    1-\frac{x_2(0)}{\exp{(x_1(0))}} x_1\exp{(x_1)} = 0,
\end{equation}
which numerically generates Figure~\ref{perturbplot1} (a). Contamination in the initial solution concentrations is observed to shift the equilibrium from its unperturbed expectation, roughly in a one-to-one ratio with the perturbation percent of $x_2(0)$. In contrast, we see that the multi-molecular Taylor network is unaffected as of yet, due to being composed of a singular specimen.

\section{Generation of Figure~\ref{dynamicsofthefirstmodel}}\label{FigureGeneration}

\begin{figure}[h!]
  \begin{subfigure}[b]{0.5\textwidth}
    \includegraphics[width=\textwidth]{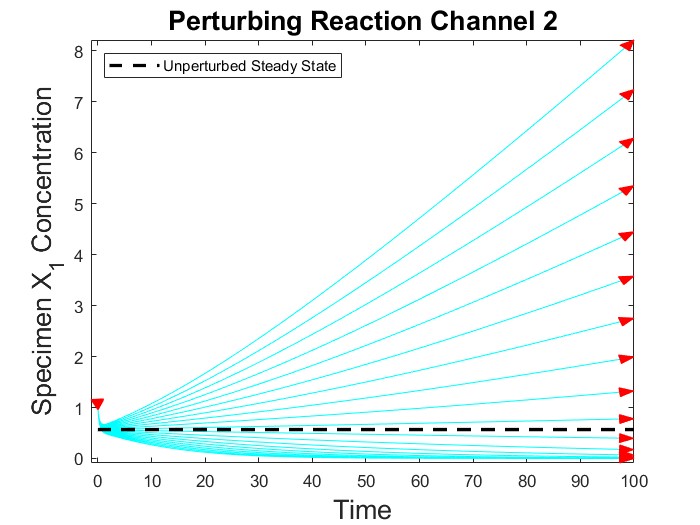}
  \end{subfigure}
  \hfill
  \begin{subfigure}[b]{0.5\textwidth}
    \includegraphics[width=\textwidth]{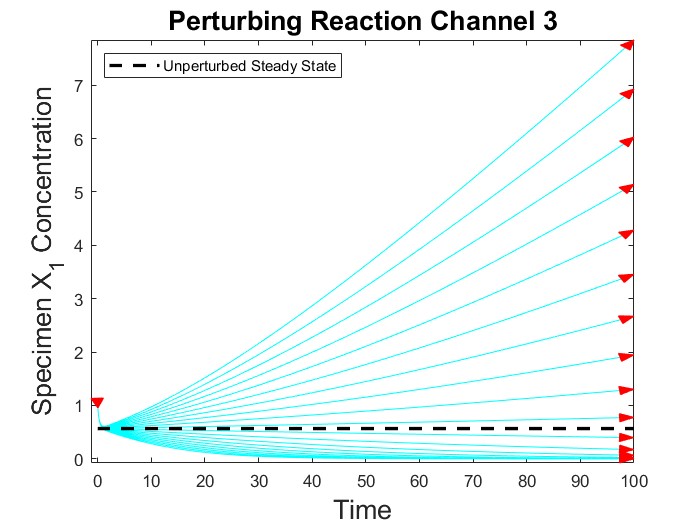}
  \end{subfigure} 
  \caption[Additional example of perturbation to reaction channel]{Perturbations ranging from $-10\%$ to $10\%$ have been applied singularly to a reaction channel. Perturbations of other channels leave identical results. Deterministic trajectories are drawn in cyan, and red arrows point in direction of increasing time.}
  \label{unchaged}
\end{figure}

Figure~\ref{dynamicsofthefirstmodel} (c),(d) gives  pseudocolour plots denoting concentrations at the end time $t = 80$. The corresponding trajectories for both specimen $\mathrm{X}_1$ and $\mathrm{X}_2$ are given in (a),(b). We have taken $k_1,k_2 = 1$ as a natural choice, and for justification it is helpful to consider how~\ref{model2} was derived and subsequently formalized into~\eqref{formalizedmodel2}. We began by denoting 
\begin{equation*}
    \Dot{x_1} = 1 - x_1\exp{(x_1)} = k_1 - k_2 x_1\exp{(x_1)},
\end{equation*}
and introduced $x_2 = \exp{(x_1)}$ to obtain
\begin{equation}\label{derivationcontroversial}
    \Dot{x_2} = \exp{(x_1)}\Dot{x_1} = x_2 \left( k_1 - k_2 x_1x_2\right) = k_1x_2 - k_2 x_1x_2^2.
\end{equation}
In the right hand side of $\Dot{x_2}$ in~\eqref{formalizedmodel2}, we have identified $k_1$, $k_2$ as independent rate coefficients $k_3$, $k_4$. To study the synthetic implementation of~\eqref{derivationcontroversial}, it is natural to consider the circumstance where noise deviates the rates $k_3$, $k_4$ away from their intended values, $k_1,k_2=1$. It is also possible to make different perturbations, e.g. by adding noise exclusively to $k_3$, but the general dynamics  remain unchanged (Figure~\ref{unchaged}).

\section{Elucidation of Upward Bias in Figure~\ref{onlythemostsnecessary}}\label{importantbitinappendix}

\begin{figure}[h!]
  \begin{subfigure}[b]{0.5\textwidth}
    \includegraphics[width=\textwidth]{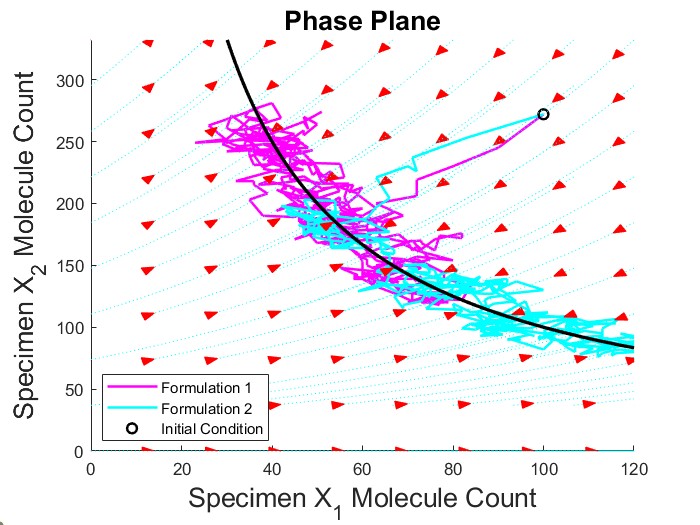}
    \caption{$V = 100$, single realization }
  \end{subfigure}
  \hfill
  \begin{subfigure}[b]{0.5\textwidth}
    \includegraphics[width=\textwidth]{agoodphaseplane1.jpg}
    \caption{$V = 100$}
  \end{subfigure}
    \begin{subfigure}[b]{0.5\textwidth}
    \includegraphics[width=\textwidth]{agoodx.jpg}
    \caption{$V = 100$}
  \end{subfigure}
  \hfill
  \begin{subfigure}[b]{0.5\textwidth}
    \includegraphics[width=\textwidth]{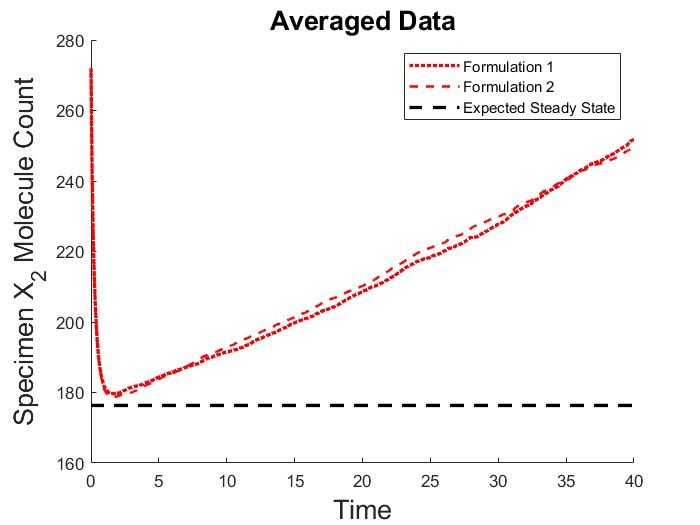}
    \caption{$V = 100$}
  \end{subfigure}
      \begin{subfigure}[b]{0.5\textwidth}
    \includegraphics[width=\textwidth]{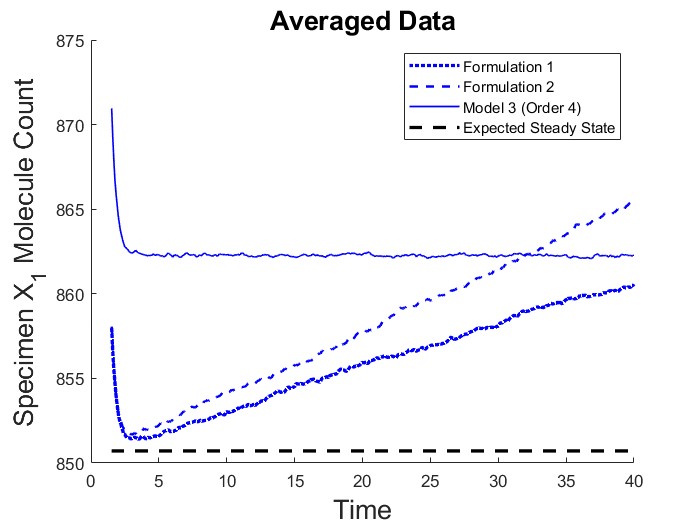}
    \caption{$V = 1500$}
  \end{subfigure}
  \hfill
  \begin{subfigure}[b]{0.5\textwidth}
    \includegraphics[width=\textwidth]{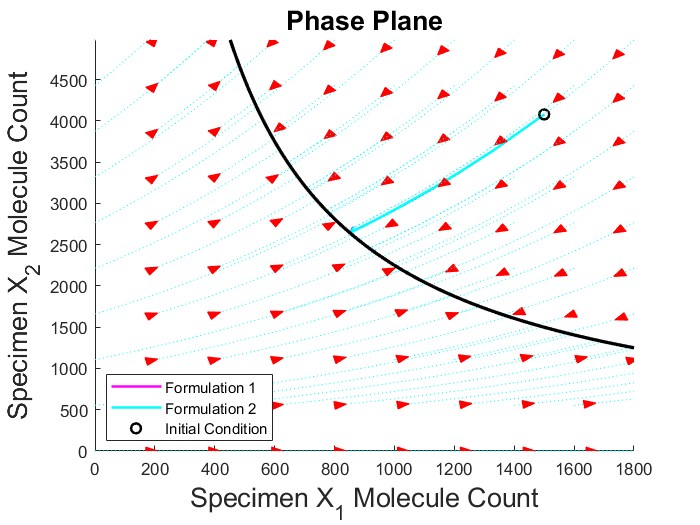}
    \caption{$V = 1500$}
  \end{subfigure}
  \caption[Investigation of upward bias in trajectories]{All plots with the exception of (a) are generated by averaging over $10^5$ realizations measured until $40$ seconds. The steady state wall $x_2 = 1/x_1$ is plotted in black in the phase plane. (a) illustrates that while the stochastic trajectory correctly predicts the wall, a single realization is inadequate to compute the root of~\ref{model1}. (c-e) confirms that the upward tilt exists for both phase planes (b),(f), though for larger volumes the deviance appears to be scaled downward. We note that the fourth-order Taylor network in (e) converges to a slightly error-prone steady state, as predicted deterministically in Figure~\ref{taylorinvestigation} (b-c), but this is mitigated by raising the order of expansion by $1$ (plot not included).}
  \label{whyisthisplotincluded}
\end{figure}

We aim to provide a brief account of the observed tilt in Figure~\ref{onlythemostsnecessary}, in both the phase plane and the averaged trajectories. In Figure~\ref{whyisthisplotincluded}, the trajectories of $\mathrm{X_1,X_2}$ in (c-d) are consistent with the deviance depicted in (b), demonstrated by an upward rise in averaged molecule counts. Increasing reactor volume $V$ appears to damp this behaviour as shown in (f), but a closer investigation in (e) by truncating early-time dynamics clarifies that the deviance is simply being scaled downward due to larger volume.

\begin{figure}[h]
  \begin{subfigure}[b]{0.5\textwidth}
    \includegraphics[width=\textwidth]{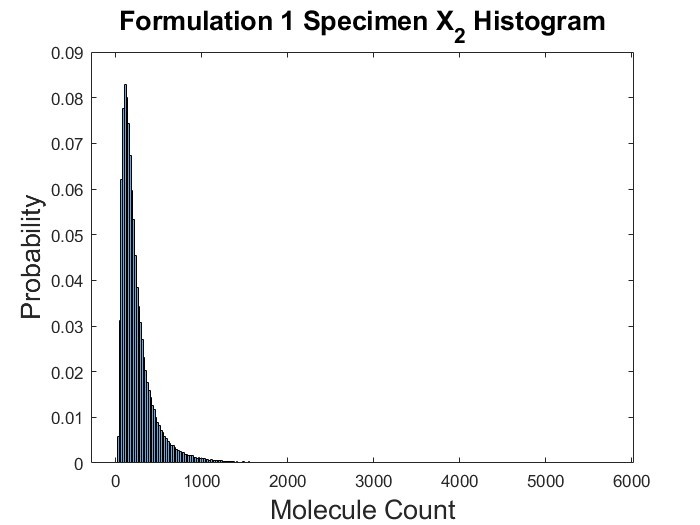}
    \caption{$V = 100$}
  \end{subfigure}
  \hfill
  \begin{subfigure}[b]{0.5\textwidth}
    \includegraphics[width=\textwidth]{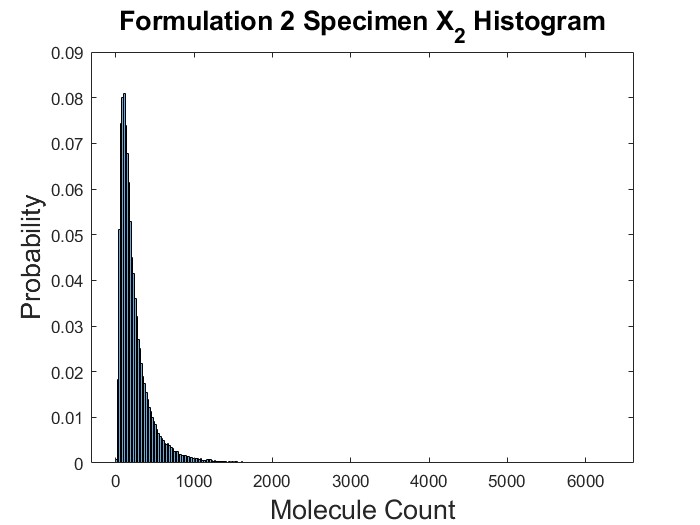}
    \caption{$V = 100$}
  \end{subfigure}
    \hfill  \begin{subfigure}[b]{0.5\textwidth}
    \includegraphics[width=\textwidth]{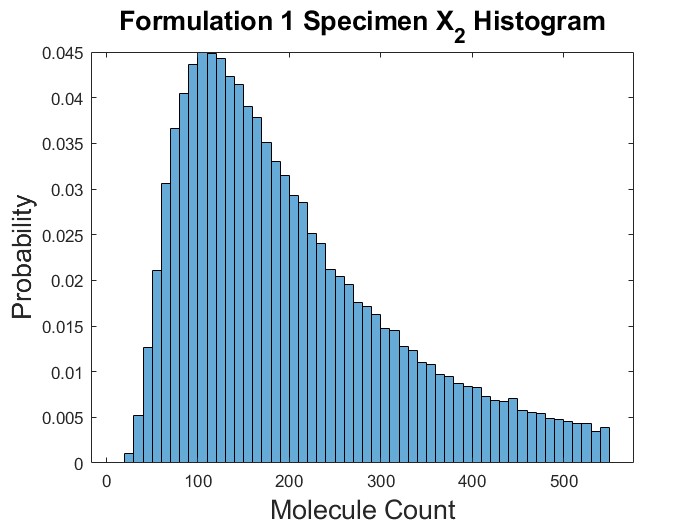}
    \caption{$V = 100,$ outliers removed}
  \end{subfigure}
  \hfill
  \begin{subfigure}[b]{0.5\textwidth}
    \includegraphics[width=\textwidth]{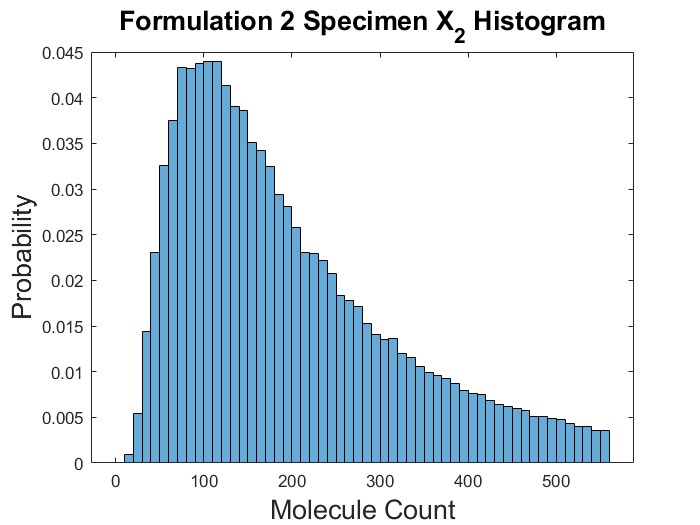}
    \caption{$V = 100$, outliers removed}
  \end{subfigure}
  \hfill
    \begin{subfigure}[b]{0.5\textwidth}
    \includegraphics[width=\textwidth]{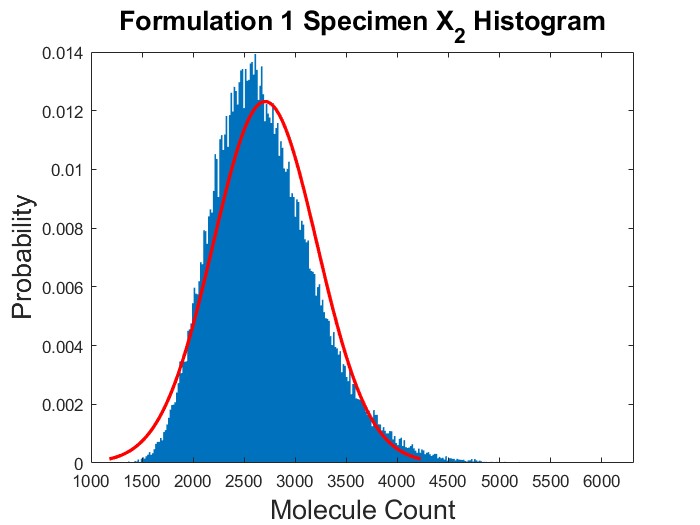}
    \caption{$V = 1500$ (Skewed Normal distribution)}
  \end{subfigure}
  \hfill
  \begin{subfigure}[b]{0.5\textwidth}
    \includegraphics[width=\textwidth]{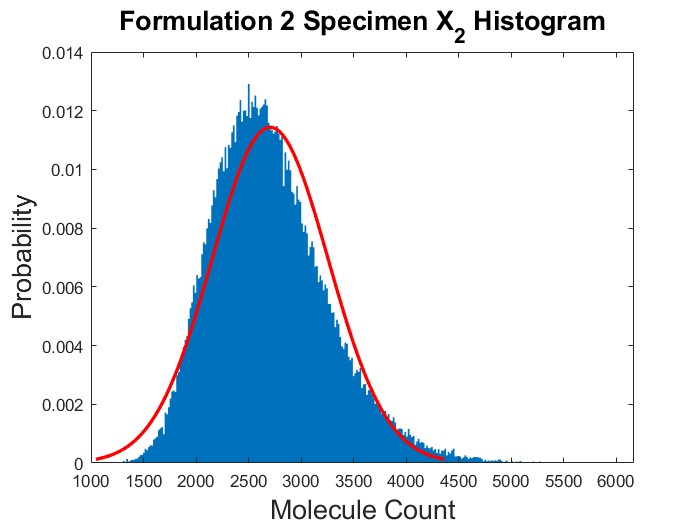}
    \caption{$V = 1500$ (Skewed Normal distribution)}
  \end{subfigure}
  \caption[Observing histograms of specimen $\mathrm{X}_2$]{All plots are generated by averaging over $10^5$ realizations measured at $t = 40$. Removing the outliers in (a-b) as is done in (c-d) correspond to elimination of exploding solutions, relieving the bias. The progression of the distribution given in (e) is shown in Figure~\ref{distributioninvestigationplot}. } 
  \label{basichistogram}
\end{figure}

For a further investigation, we measure the distribution of specimen molecule counts at specific points in time, as is done in Chapter~\ref{oldchapter4}. Figure~\ref{basichistogram} gives the distributions for $t = 40$. The observed upper bounds of specimen $\mathrm{X_2}$ given in (a-b) are detected to be approximately $6000$ for both Formulation $1$~\eqref{Formulation1} and Formulation $2$~\eqref{Formulation2} alike. A removal of outliers in the distribution reduces the bound for both formulations to around $500$ with the rightmost tail end cut off, lessening the bias. 

It is difficult to definitively determine the cause of this phenomenon, but our simulations offer some clues. Figure~\ref{whyisthisplotincluded} (a) plots an individual trajectory, which crashes appropriately into the steady state wall $x_2 = 1/x_1$ and then travels along the stable curve. Assuming that long-time trajectories stay on this curve, $x_2 = 1/x_1$ ($\mathrm{X}_2 = V^2/\mathrm{X}_1$ in molecule counts) forces the specimen quantities to vary inversely proportionally to each other. That is, if $\mathrm{X}_1$ grows smaller, $\mathrm{X}_2$ is agitated to become larger much faster than a linear relationship, and vice versa. Therefore an upward bias may not be surprising when molecule counts are averaged. This phenomenon is precisely depicted in Figure~\ref{dynamicsofthefirstmodel}, and more specifically (c-d), describing the Nonexistent Equilibrium Catastrophe (NEC) manifested by the deterministic perturbed system.

\begin{figure}[h!]
    \includegraphics[width=.24\textwidth]{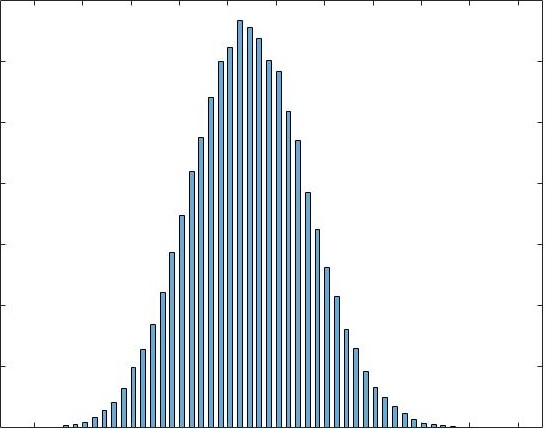}\hfill
    \includegraphics[width=.24\textwidth]{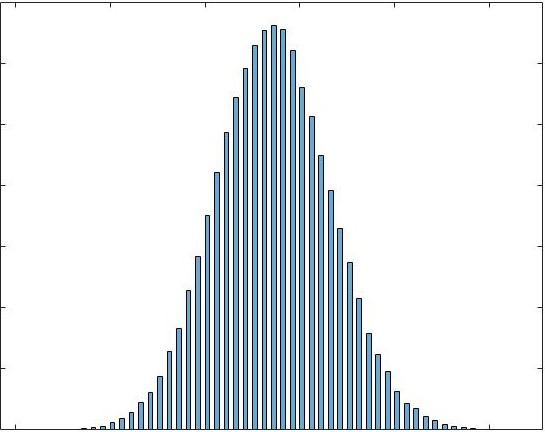}\hfill
    \includegraphics[width=.24\textwidth]{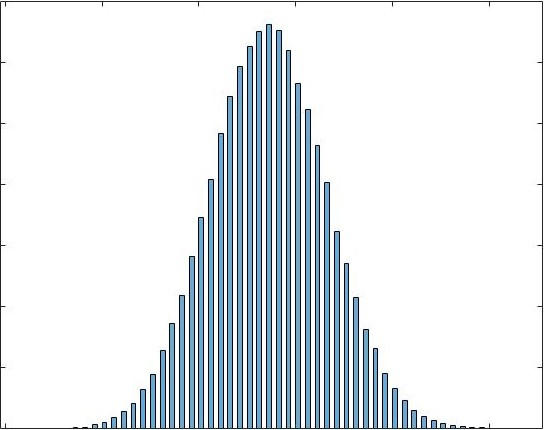}\hfill
    \includegraphics[width=.24\textwidth]{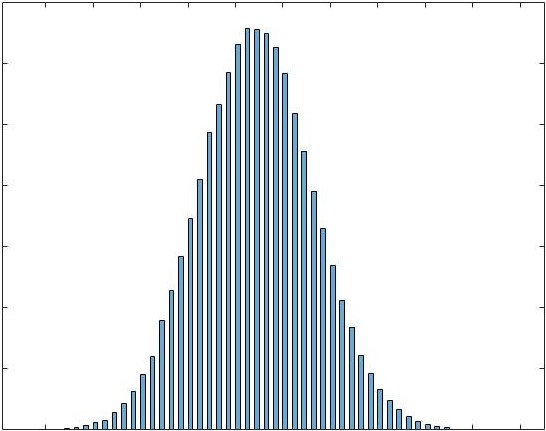}\hfill
    \\[\smallskipamount]
    \includegraphics[width=.24\textwidth]{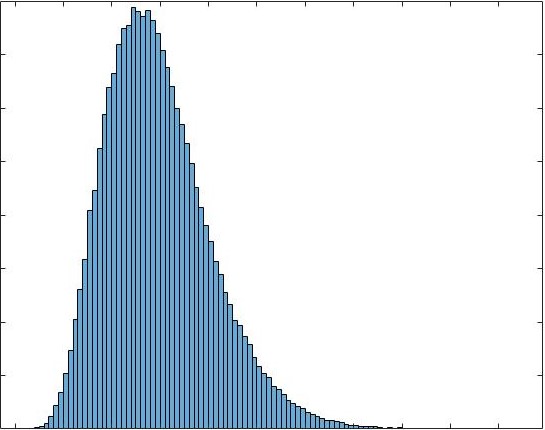}\hfill
    \includegraphics[width=.24\textwidth]{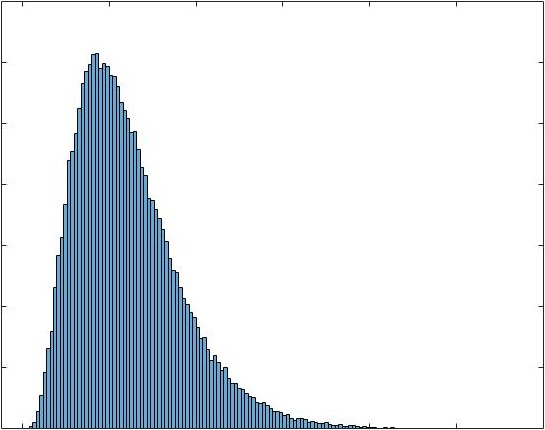}\hfill
    \includegraphics[width=.24\textwidth]{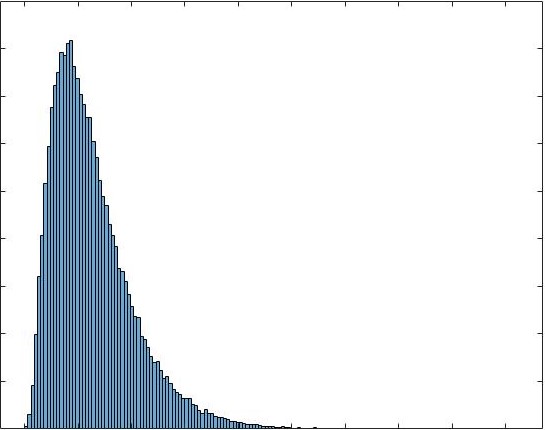}\hfill
    \includegraphics[width=.24\textwidth]{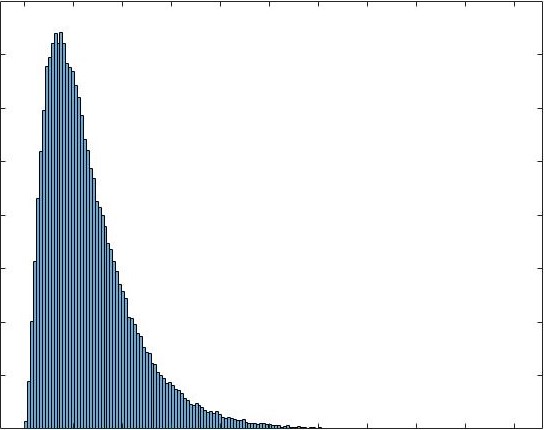}\hfill
    \\[\smallskipamount]
    \includegraphics[width=.24\textwidth]{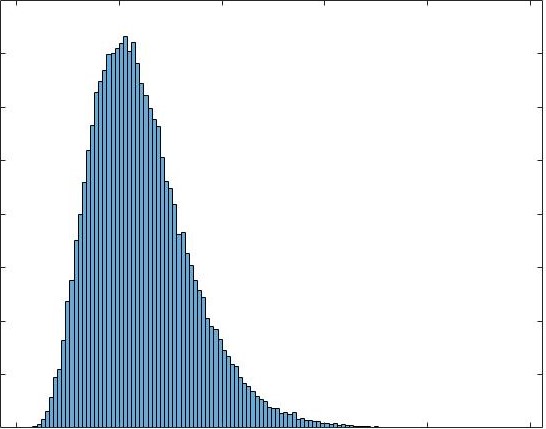}\hfill
    \includegraphics[width=.24\textwidth]{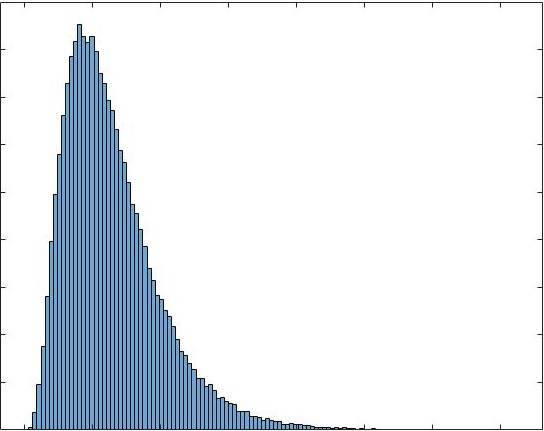}\hfill
    \includegraphics[width=.24\textwidth]{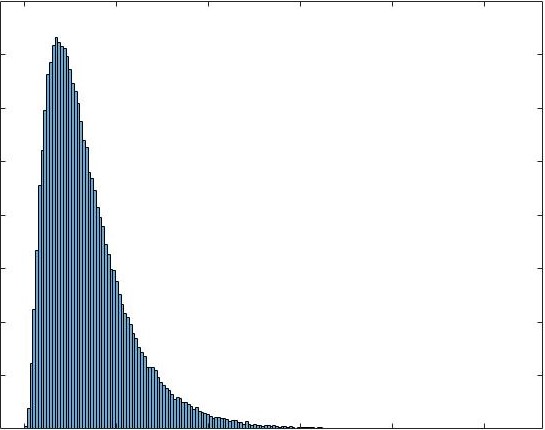}\hfill
    \includegraphics[width=.24\textwidth]{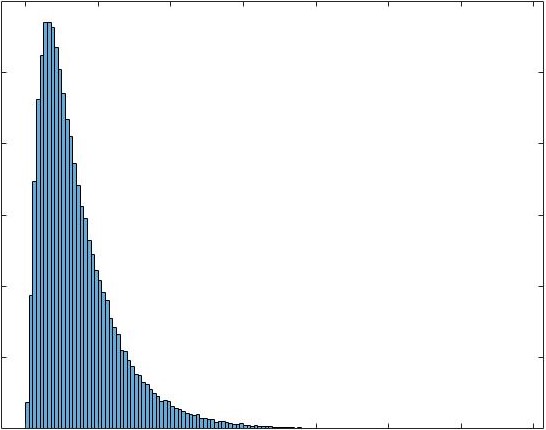}\hfill
  \caption[Sampled distributions for specimen $\mathrm{X}_1$ measured at $t = 10,20,30,40$]{Sampled distributions for specimen $\mathrm{X}_1$ measured at $t = 10,20,30,40$ for reactor volume $V = 100$. The top, middle, and bottom represent the Taylor network of order $4$~\eqref{model3}, Formulation $1$~\eqref{Formulation1}, and Formulation $2$~\eqref{Formulation2}, respectively. A progressively worsening skew is observed for the two formulations, consistent with the trajectories in Figure~\ref{onlythemostsnecessary}.}
  \label{skewinx}
\end{figure}

\cfoot{}
\clearpage
\cfoot{\thepage}

\clearpage
\cfoot{}
\clearpage
\cfoot{\thepage}

%% file: appendix4.tex
\section{Ideal Differentiation Scenario}\label{appendixD}
\begin{figure}[h!]
    \includegraphics[width=.33\textwidth]{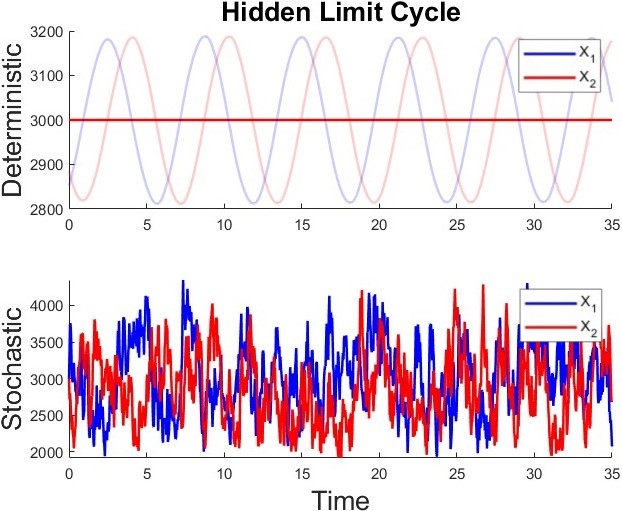}\hfill
    \includegraphics[width=.33\textwidth]{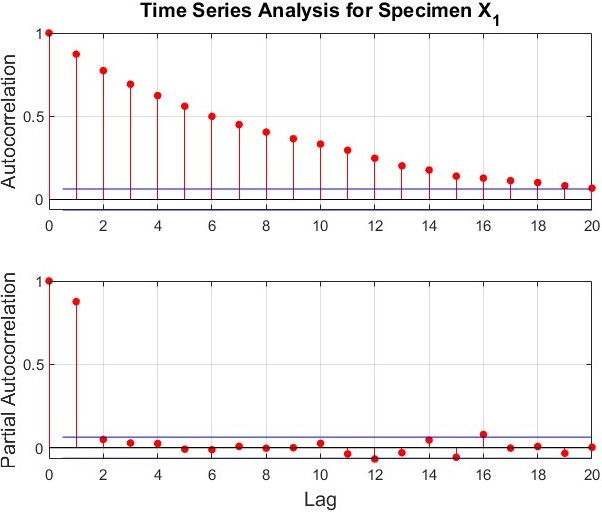}\hfill
    \includegraphics[width=.33\textwidth]{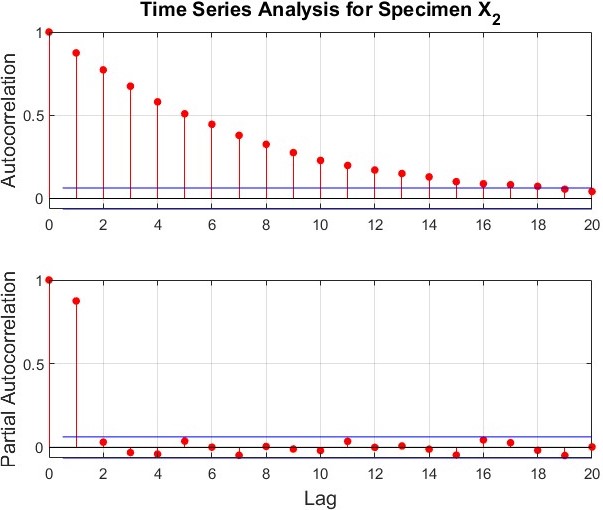}\hfill
    \\[\smallskipamount]
    \includegraphics[width=.33\textwidth]{actualnolimcyc1.jpg}\hfill
    \includegraphics[width=.33\textwidth]{actualnolimcyc2.jpg}\hfill
    \includegraphics[width=.33\textwidth]{actualnolimcyc3.jpg}
    \caption[Ideal case of time series differentiation]{Both systems (\eqref{hopfbifuraction} with $\xi = \pm 0.1$) were propagated from the focus, deterministically masking the limit cycle in the top row for $\xi = 0.1$. The Autocorrelation signatures preserve (\textit{very}) slightly more information for the top row, but it is unclear if this is a statistically significant result. }\label{hiddenlimitcycle}
\end{figure}

\section{R\"{o}ssler and Lorenz Attractors}\label{chaoticsystemsinappendix}

The R\"{o}ssler~\eqref{rosslerode} and Lorenz~\eqref{lorenzode} systems are given by 

\begin{minipage}{.45\textwidth}
\begin{equation}\tag{\ref{rosslerode}}
\begin{aligned}
&\Dot{x_1}=-x_2-x_3, \\
&\Dot{x_2}=x_1+a x_2, \\
&\Dot{x_3}=b+x_3(x_1-c).
\end{aligned}
\end{equation}
\end{minipage}
\begin{minipage}{.45\textwidth}
\begin{equation}\tag{\ref{lorenzode}}
\begin{aligned}
&\Dot{x_1}=\sigma(x_2-x_1), \\
&\Dot{x_2}=x_1(\rho-x_3)-x_2, \\
&\Dot{x_3}=x_1 x_2-\beta x_3 .
\end{aligned}
\end{equation}
\end{minipage}

\vspace{5pt}

\noindent The Lorenz system was constructed as a simplified model describing  atmospheric turbulence, with $\sigma=10, \rho = 28, \beta = 8/3$ originally proposed. Remarking on the aperiodicity of computed solutions, Lorenz comments~\cite{OriginalLorenz} on the impossibility of long-range weather forecasting without knowing infinitely precise data on atmospheric conditions, which are surely out of reach of human technology. Since its conception, the Lorenz model has found applications in a wide range of interdisciplinary areas such as lasers~\cite{LorenzLaser} and osmosis~\cite{LorenzOsmosis}. 

The induced flow can be seen to be composed of two spirals suspended and glued together in the state space $\mathbb{R}^3$, where trajectories switch back and forth between the two objects. At the time of the discovery of the R\"{o}ssler system, this behaviour was poorly understood, thus R\"{o}ssler sought to simplify the analysis by limiting the observations to a singular spiral~\cite{ChaosAndFractalsBook}. 

The R\"{o}ssler system was originally studied using the parameters $a=0.2, b=0.2,$ $c=5.7$, which generates only a single spiral instead of a dual one. Although R\"{o}ssler notes that his system has ``no longer an immediate physical interpretation~\cite{OriginalRossler}'', we have successfully provided an approximate chemical interpretation of his work in Chapter~\ref{oldchapter6}. The characteristic flow demonstrates the return of the outer portion of the spiral after an appropriate twist resembling the formation of the M\"{o}bius strip, which is simpler to analyze than the flow of the Lorenz system. 

\begin{figure}[h]
  \begin{subfigure}[b]{0.325\textwidth}
    \includegraphics[width=\textwidth]{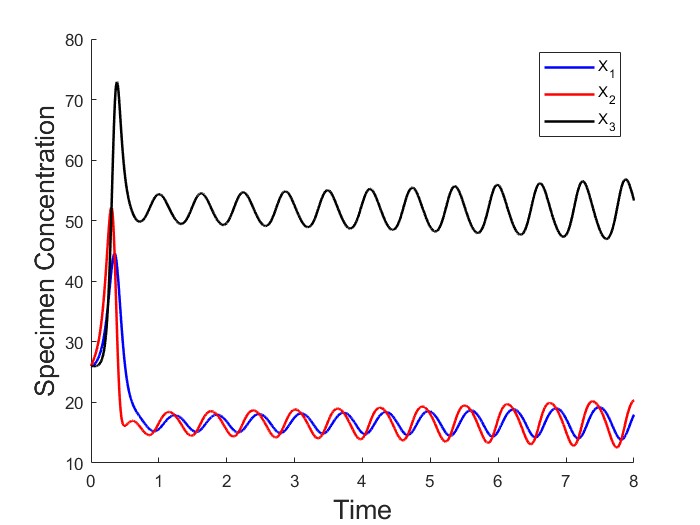}
    \caption{$V = 50000$ or $50$ }
  \end{subfigure}
  \hfill
  \begin{subfigure}[b]{0.325\textwidth}
    \includegraphics[width=\textwidth]{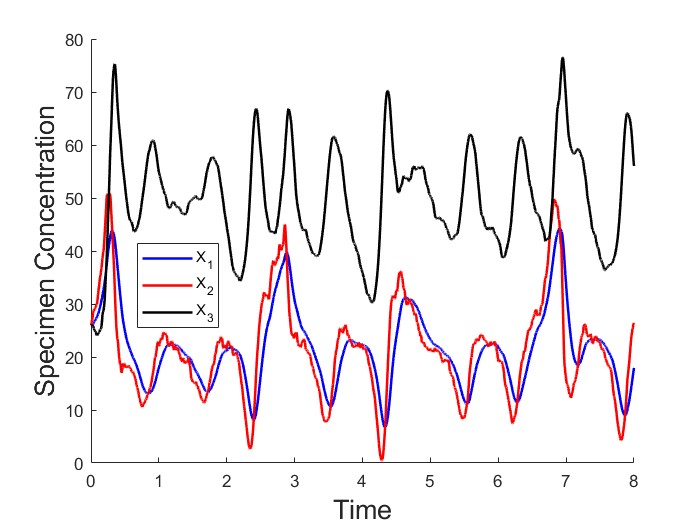}
    \caption{$\Psi_{Kow}$, $V=50000$ }
  \end{subfigure}
    \hfill
    \begin{subfigure}[b]{0.325\textwidth}
    \includegraphics[width=\textwidth]{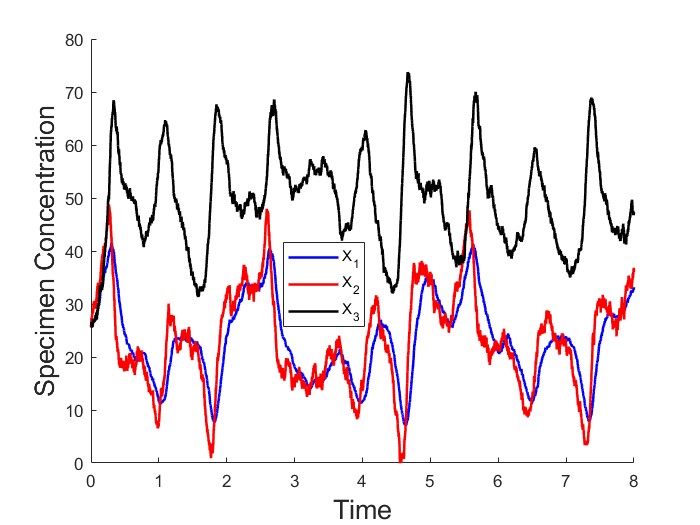}
    \caption{$\Psi_{QSST}$, $V=50$ }
  \end{subfigure}
  \hfill
  \begin{subfigure}[b]{0.325\textwidth}
    \includegraphics[width=\textwidth]{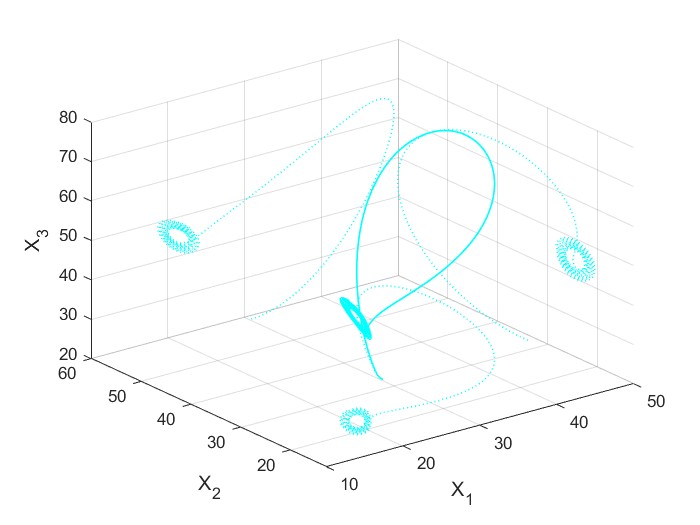}
    \caption{$V = 50000$ or $50$ }
  \end{subfigure}
    \hfill
      \begin{subfigure}[b]{0.325\textwidth}
    \includegraphics[width=\textwidth]{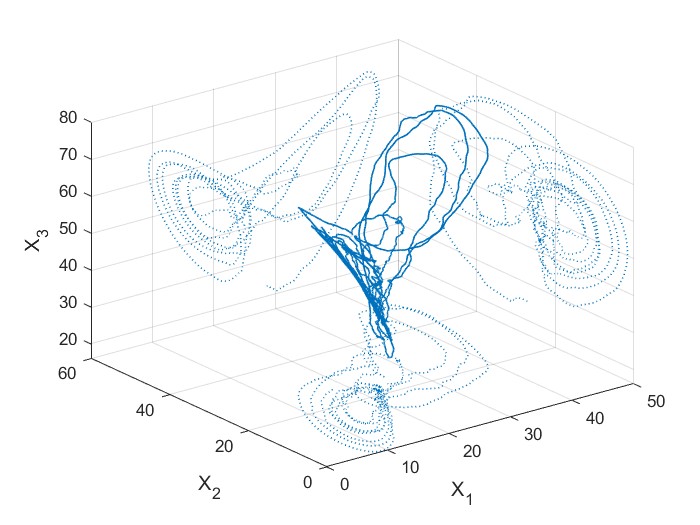}
    \caption{$\Psi_{Kow}$, $V=50000$ }
  \end{subfigure}
  \hfill
  \begin{subfigure}[b]{0.325\textwidth}
    \includegraphics[width=\textwidth]{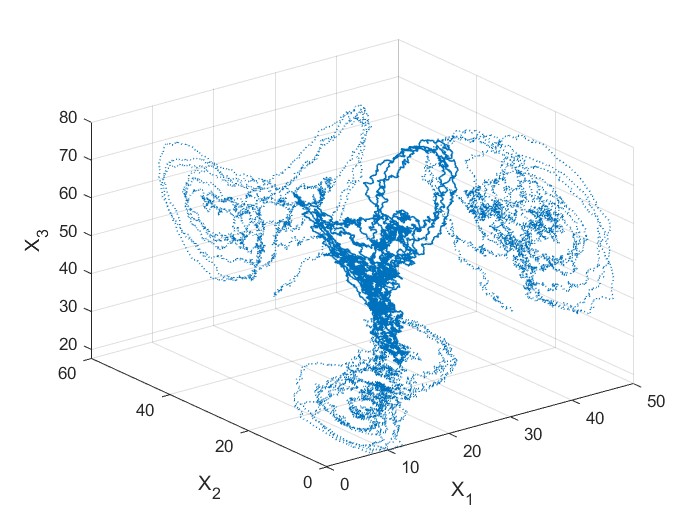}
    \caption{$\Psi_{QSST}$, $V=50$ }
  \end{subfigure}
    \hfill
      \begin{subfigure}[b]{0.45\textwidth}
    \includegraphics[width=\textwidth]{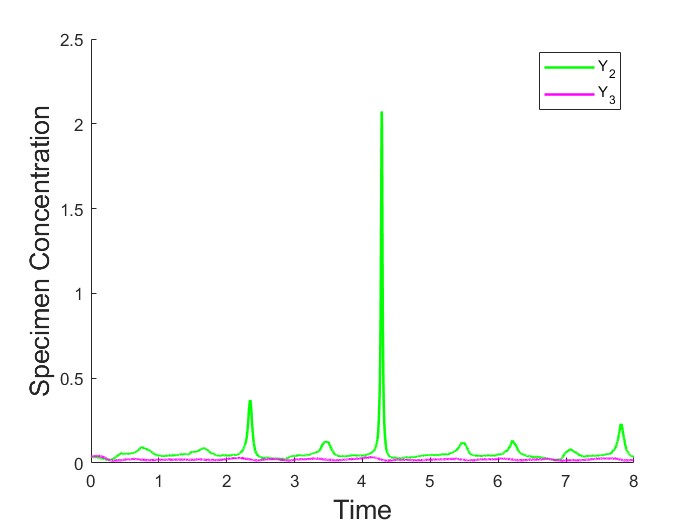}
    \caption{$\Psi_{Kow}$, $V=50000$}
  \end{subfigure}
  \hfill
  \begin{subfigure}[b]{0.45\textwidth}
    \includegraphics[width=\textwidth]{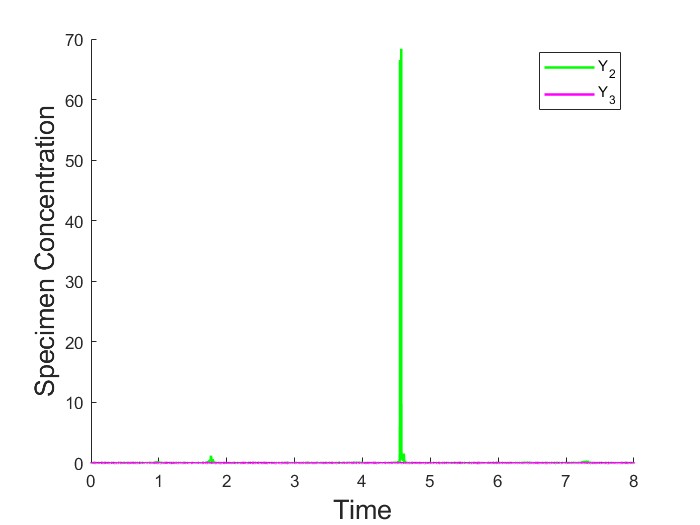}
    \caption{$\Psi_{QSST}$, $V=50$}
  \end{subfigure}
    \hfill
      \begin{subfigure}[b]{1\textwidth}
    \includegraphics[width=\textwidth]{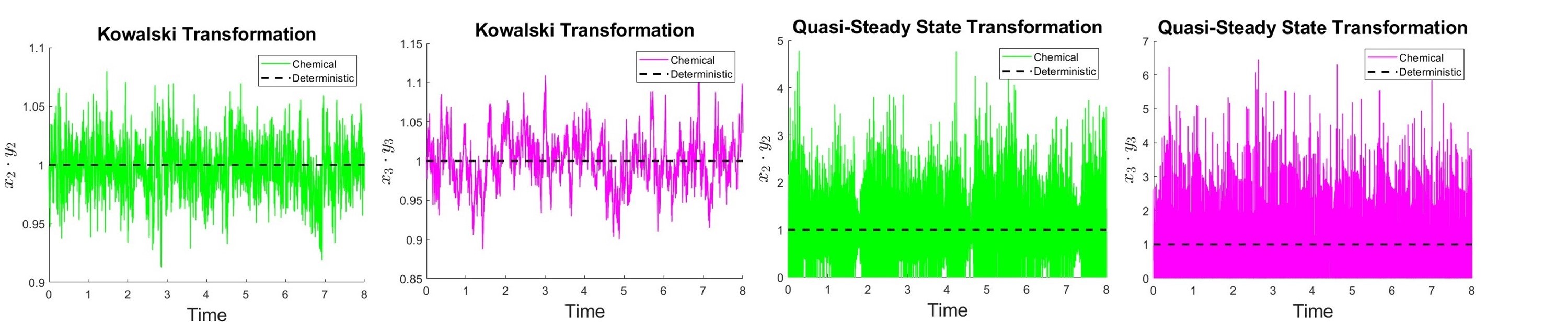}
    \caption{A confirmation of the deterministic expectation $x_2y_2=x_3y_3=1$ in the networks induced by $\Psi_{Kow}, \Psi_{QSST}$}
  \end{subfigure}
    \hfill
  \caption[Kineticization of Lorenz system]{ (a),(d) plots the deterministic solution of~\eqref{lorenzode} for comparison, where identical trajectories are traced for $V=50000,50$. Note that $\Psi_{QSST}$ has produced noisy trajectories in (c),(f),(i) due to low reactor volume $V$. This can be mitigated by increasing $V$, and this figure illustrates the limitations of $\Psi_{Kow}$ which fails to transcribe deterministic dynamics even for much larger volumes than what is sufficient for $\Psi_{QSST}$, such as $V = 8000$ (plot not included). The themes described in this section, suggesting that Quasi-Steady State Transformations produce more robust networks than Kowalski Transformations, are consistently reproduced across all test systems we have considered.}
  \label{finalfigure}
\end{figure}

In Figure~\ref{finalfigure} (b),(e), we plot a successful chemical inversion of the Lorenz system via $\Psi_{Kow}$, a kineticization technique we have avoided using in Chapter~\ref{oldchapter6}. In lower reactor volumes (e.g. $V \approx \mathcal{O}(10^2)$), the restriction $\mathrm{Y}_s = V^2/\mathrm{X}_s$ imposes too heavy a constraint to successfully encapsulate deterministic dynamics via simulation of the network induced by the general system. The network fails to produce sensible trajectories even for $V  \approx \mathcal{O}(10^3)$; further increasing the reactor volume to $V  \approx \mathcal{O}(10^4 \sim 10^5)$, however, gives successful chemical representations of the Lorenz system. This example shows that not all dynamical regions of a deterministic system are well-preserved in the chemical simulation, such as when $x_i\downarrow 0$ where $x_i$ appears as a variable in the original ODE system. Kowalski transformations induce systems in which desired dynamics live exclusively in poorly preserved regions, and deterministic information cannot wholly be transcribed to chemical reinterpretations except in solutions with extremely large volumes. 

It is worth noting another peculiarity in the analysis of kinetcization techniques. Namely, $\Psi_{QSST}$ also restricts solution concentrations to the manifold $1=x_sy_s$, for $\mathrm{X}_s \in \mathcal{X}_2$ (see Definition~\ref{QSSAnewdefinition}). This restriction is imposed by the adjoined system~\eqref{QSSAIntroducedSystem}
\begin{equation}
    0 \approx \mu \Dot{y_{s}} =\omega_{s}-x_{s} p_{s}(\mathbf{x}) y_{s}
\end{equation}
for $\omega_s,p_{s}(\mathbf{x}) = 1 $ as $\mu \downarrow 0$. Therefore, we expect $\Psi_{QSST}$ to show inadequacies for reactor volumes in which $\Psi_{Kow}$ fails. Contrary to our expectations, $\Psi_{QSST}$ is excellent for small $V$, evidenced by Figure~\ref{finalfigure} (c),(f) where the reactor volume $V$ has been shrunk by $\mathcal{O}(10^3)$ which successfully captures the butterfly dynamics of the Lorenz Attractor, though noisy. In contrast, $V = 10^4$ rarely produces any sensible trajectories when $\Psi_{Kow}$ is used as the kineticization technique.

Figure~\ref{finalfigure} (g-h) describes the solution concentrations of specimen $\mathrm{Y}_2, \mathrm{Y}_3$, where $\mathrm{Y}_1$ has not been introduced as no cross-negative terms have been found in the right hand side of $\Dot{x_1}$ in system~\eqref{lorenzode} after affine transformation. A spike shown in (h) corresponds to temporary extinction of $\mathrm{X}_2$ in (c). In (i), we verify the relation $x_2y_2=x_3y_3=1$, where more noise is detected in $\Psi_{QSST}$ due to \textit{much} lower reactor volume.

The themes that we have described here, suggesting that Quasi-Steady State Transformations are more applicable as a kineticization strategy than Kowalski Transformations, are consistently replicated across all test systems we have considered. In particular, we have observed that Quasi-Steady State Transformations are robust even for initial conditions and reactor volumes which force the restriction $\mathrm{Y}_s=V^2/\mathrm{X}_s < 1$ to be maintained at regular time intervals throughout simulation propagation, which appears to translate to a hyperactive switching of $\mathrm{Y}_s$ between $0$ (extinction) and $1$ to diminish the impact of discrete size forces.

We strongly suspect that the frequent failure of Kowalski Transformations comes from the fact that the network formed by canonical inversion of the general system induced by $\Psi_{Kow}$ possesses no mechanism to regenerate $\mathrm{Y}_s$ after extinction has occurred\footnote{That is, $y_s$ is multiplied to every term in the right hand side of the adjoined system~\eqref{Kowalskiadjoined}.}--therefore permanently altering the dynamics of the network once stochasticity pulls $\mathrm{Y}_s$ molecule counts down to $0$. This is an overly restrictive limitation, especially when quantities of $\mathrm{X}_s$ and $\mathrm{Y}_s$ are required to vary inversely proportionally for preservation of the dynamics of the original system. In contrast, Quasi-Steady State Transformations induce general systems whose canonical inversions allow for the fast regeneration of $\mathrm{Y}_s$ after extinction\footnote{That is, $\omega_s/\mu \gg 0$ for $\omega_s>0$, $\mu \downarrow 0$--see~\eqref{QSSAIntroducedSystem}.}. 

The irreversible extinction of $\mathrm{Y}_s$ deactivates entire branches of the network corresponding to the terms in the RREs which were responsible, before the kineticization, for driving molecule counts of $\mathrm{X}_s$ negative even when $x_s=0$. This in turn biases the firing of reaction channels towards the increase of $\mathrm{X}_s$, resulting in a frequent blow-up of solution concentrations in simulations. Diverging molecule counts translate to diverging propensities, driving down the expected next reaction time $\tau$ in the Gillespie Algorithm and making simulation termination unlikely.

\clearpage
\newgeometry{top=8mm, bottom=10mm, left = 10mm, right = 10 mm}     

\cfoot{}
\clearpage
\cfoot{\thepage}
\restoregeometry

%% file: appendix4andhalf.tex
\chapter{Pseudo-propensity Analysis}\label{discreteappendix}

In our long time simulation of the system~\eqref{hopfbifuraction} in Figure~\ref{starterplot},
\begin{equation}\tag{\ref{hopfbifuraction}}
    \begin{aligned}
&\Dot{x_1}=\xi x_1-\zeta x_2-x_1\left(x_1^{2}+x_2^{2}\right), \\
&\Dot{x_2}=\zeta x_1+\xi x_2-x_2\left(x_1^{2}+x_2^{2}\right),
\end{aligned}
\end{equation}
there was a perceptible bias in the derived joint density in the southward direction of the phase plane. Namely, the density appeared more concentrated in contrast to other regions of the limit cycle where they were more diffuse. Using a technique which we call pseudo-propensity analysis, we will argue that these are the result of discrete-size forces only visible in chemical simulations. 

\begin{figure}[h]
  \begin{subfigure}[b]{0.5\textwidth}
    \includegraphics[width=\textwidth]{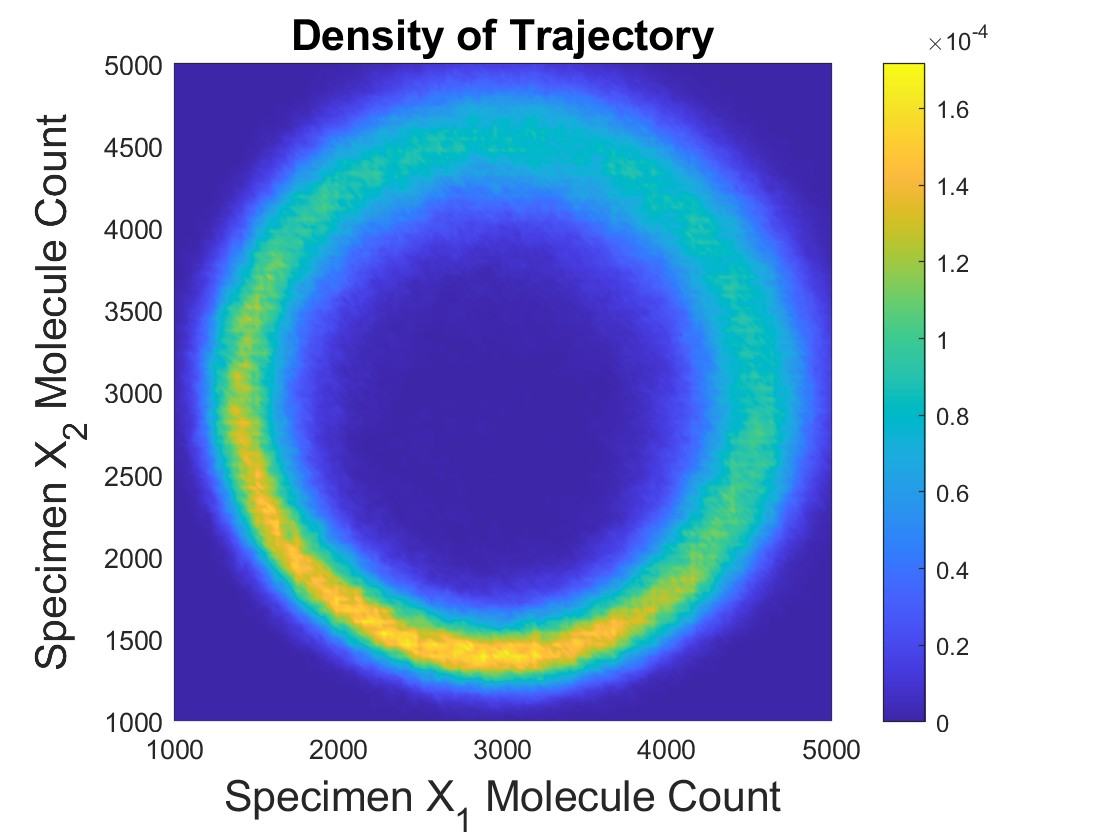}
    \caption{$\zeta = 10$ }
  \end{subfigure}
  \hfill
  \begin{subfigure}[b]{0.5\textwidth}
    \includegraphics[width=\textwidth]{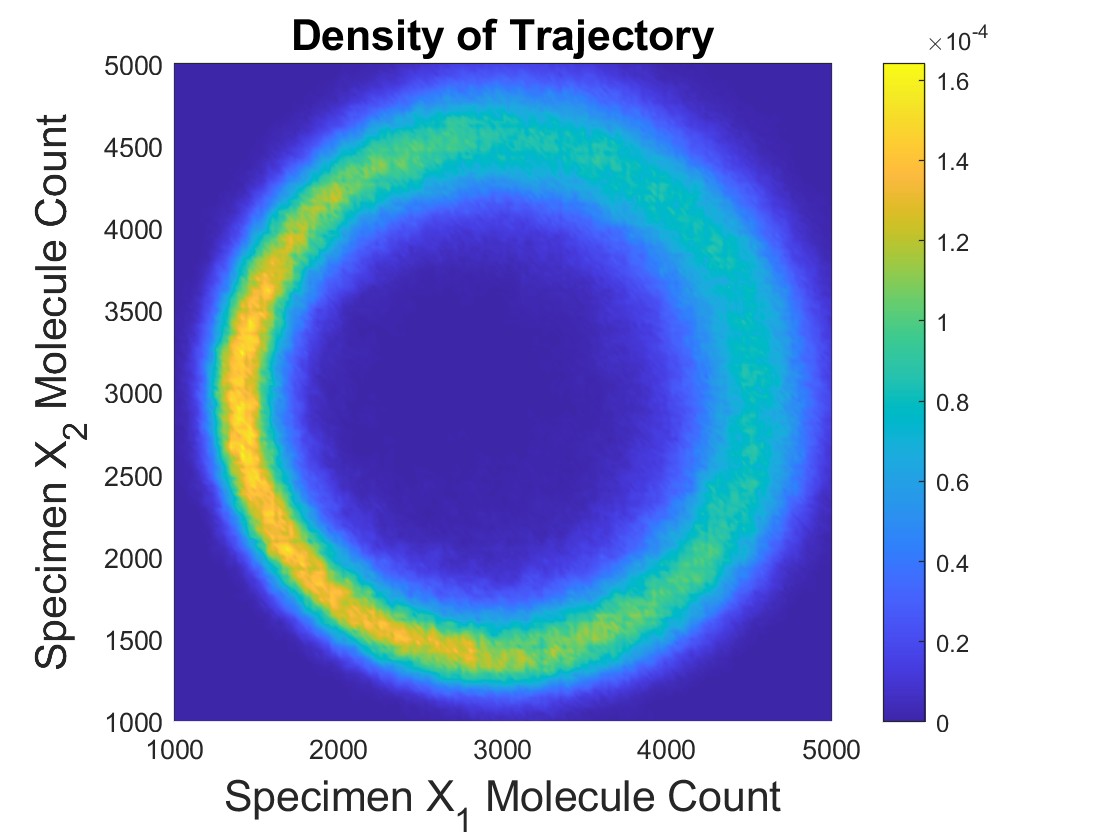}
    \caption{$\zeta = -10$}
  \end{subfigure}
  \caption[A depiction of bias in chemical simulation]{The bias concentrates trajectories in different locations, depending on the sign of $\zeta$. In both plots we represent~\eqref{firstaffine} chemically reacting for parameters $\mathcal{T}_{x_1}=\mathcal{T}_{x_2}=6,$ $\xi = 10,$ $V=500$ by preemptively scaling the ODE by reactor volume and propagating the simulation in a solution of volume $1$. Clockwise rotation ($\zeta < 0$) induces a bias in the west, whereas a counterclockwise rotation ($\zeta > 0$)  concentrates trajectories in the south.}
  \label{discreteargueplot}
\end{figure}

Firstly, it is helpful for our purposes to very briefly outline the operations that AutoGillespie performs, following the inversion framework. In this case, we first make the substitution $x_1 \leftarrow x_1 - \mathcal{T}_{x_1}$, $x_2 \leftarrow x_2 - \mathcal{T}_{x_2}$ to push limit cycle dynamics into the positive cone:
\begin{equation}\label{firstaffine}
\begin{aligned}
&\Dot{x_1}=\xi \left(x_1 - \mathcal{T}_{x_1}\right)-\zeta \left(x_2 - \mathcal{T}_{x_2} \right)-\left(x_1 - \mathcal{T}_{x_1}\right)\left(\left(x_1 - \mathcal{T}_{x_1}\right)^{2}+\left(x_2 - \mathcal{T}_{x_2} \right)^{2}\right), \\
&\Dot{x_2}=\zeta \left(x_1 - \mathcal{T}_{x_1}\right)+\xi \left(x_2 - \mathcal{T}_{x_2} \right)-\left(x_2 - \mathcal{T}_{x_2} \right)\left(\left(x_1 - \mathcal{T}_{x_1}\right)^{2}+\left(x_2 - \mathcal{T}_{x_2} \right)^{2}\right).
\end{aligned}
\end{equation}
Expanding the right hand side for parameter choices $\xi,\zeta \in [-10,10],$ $\mathcal{T}_{x_1}=\mathcal{T}_{x_2}=6$ reveals cross-negative terms, and an application of $\Psi_{QSST}$ multiplies all cross-negative terms with $x_1y_1$ or $x_2y_2$ and introduces the adjoined system
\begin{equation}\label{theadjoinedsystem}
\begin{aligned}
    &\mu \Dot{y_1} = 1 - x_1y_1,\\
&\mu \Dot{y_2} = 1 - x_2y_2.
\end{aligned}
\end{equation}
The latter steps of the AutoGillespie program canonically inverts all terms in the general system into a chemical reaction network and feeds vectorized information into the Gillespie Algorithm for simulation.

We make a few remarks before presenting numerical results that confirm our heuristics. Firstly, one distinction between the ODE system~\eqref{hopfbifuraction} and the induced chemical reaction network is that while the former may move through negative regions $\mathbf{x} \notin \mathbb{R}^{2}_{\ge 0}$, the latter system may not progress further into the negative direction after specimen extinction, for any dimension $x_1,x_2$. Therefore the non-negative  portions of the $x_1,x_2$-axis form an impenetrable boundary for the motion of the chemical trajectory to be confined by, in the south/west regions of the phase plane. An equivalent boundary does not exist in the north/east regions, thus there is expected to be more freedom of movement in that area. 

However, in order for this boundary to directly take effect, chemical trajectories must crash into this impenetrable wall and be physically prevented from moving into the negative region. A prerequisite is the extinction of chemical specimen, be it either $\mathrm{X}_1$ or $\mathrm{X}_2$. We observed in Definition~\ref{QSSAnewdefinition} that the adjoined system mapped to by $\Psi_{QSST}$ is quick to induce an instantaneous blow-up of fast specimen in finite time upon observing specimen extinction\footnote{For example, let $x_1 = 0$ in ~\eqref{theadjoinedsystem} and note that $\Psi_{QSST}$ is only valid in the limit $\mu \downarrow 0$.}, thus any physical prevention of the travel of the trajectory should be detectable by monitoring specimen concentrations of $\mathrm{Y}_1, \mathrm{Y}_2$. For $\mu = 10^{-6}$, the maximum molecule counts of both specimen were sampled to be $1$, which held frequently in repeated realizations (but occasionally rising to $2$). 

If the trajectory is not connecting with the axis walls, then another explanation is the existence of an additional force emanating from the boundary which dissuades the chemical trajectory from close contact. The derived densities in Figure~\ref{discreteargueplot} indeed support the hypothesis of an external repulsive force coming from the non-negative $x_1,x_2$-axis that cuts across the determinstically expected grain of rotation, constraining movement and concentrating trajectories (see Figure~\ref{guess} (a),(c) for their phase planes). We will now argue that these forces are caused by discrete size effects, or more precisely, due to a propensity flow in the respective regions. 

In order to analyze the chemical simulation in a deterministic format, we chose to use the propensities as an indication of the general increase/decrease of the chemical specimen counts, directing the motion of the trajectory in the phase plane. Following the techniques developed in Appendix~\ref{GillespieAutoGillespie} for the creation of AutoGillespie, we collected and summed the closed form propensities into a symbolic vector of length eight, the individual entries of which are responsible for adding or removing precisely a single copy of the specimen $\mathrm{X}_1$,$\mathrm{X_2}$,$\mathrm{Y}_1$,$\mathrm{Y}_2$, respectively, upon an instance of channel firing.

We follow the ordering
\begin{equation}
    \mathbf{p} = (p_1,p_2,p_3,p_4,p_5,p_6,p_7,p_8)^\top,
\end{equation}
where $p_{1},p_{2}$ represent the propensity of the addition and removal of specimen $\mathrm{X}_1$. Further entry pairs analogously represent propensities regarding specimen $\mathrm{X}_2$,$\mathrm{Y}_1$,$\mathrm{Y}_2$, in that order (e.g. $p_7$ represents addition of $\mathrm{Y}_2$).  For reasons which will become clear later, $\mathbf{p}$ has been normalized so that its entries sum to $1$.

The phase plane, with the two axis representing molecule counts of specimen $\mathrm{X}_1, \mathrm{X}_2$, is coarse-grained into smaller rectangular grids and the floor\footnote{The floor of $a \in \mathbb{R}$ refers to the maximal integer less than or equal to $a$.} of the molecule count in the vertex closest to the origin are used in the propensity computation for each rectangle. We must also consider the molecule counts of the specimen $\mathrm{Y}_1, \mathrm{Y}_2$, which are introduced in the system to model cross-negative terms that deterministically reduces $\mathrm{X}_1, \mathrm{X}_2$.

\begin{figure}[h!]
  \begin{subfigure}[b]{0.5\textwidth}
    \includegraphics[width=\textwidth]{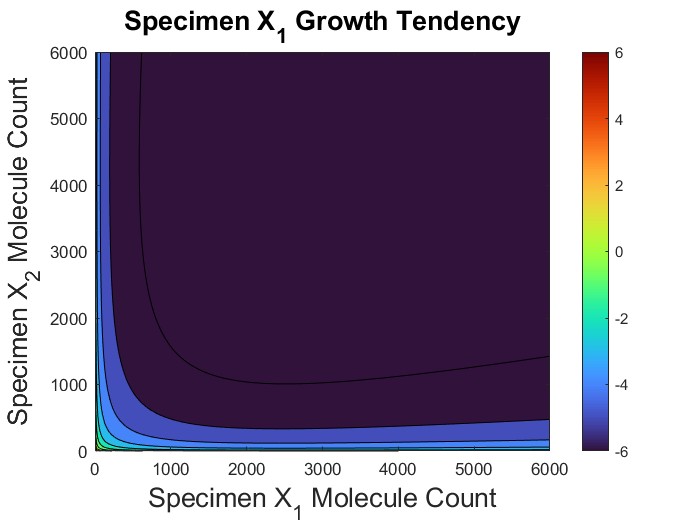}
    \caption{$(\mathrm{Y}_1, \mathrm{Y}_2 ) = (1,1)$}
  \end{subfigure}
  \hfill
  \begin{subfigure}[b]{0.5\textwidth}
    \includegraphics[width=\textwidth]{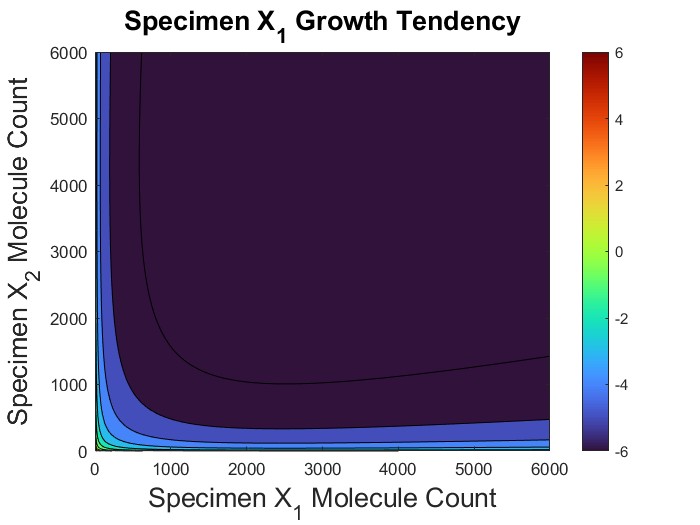}
    \caption{$(\mathrm{Y}_1, \mathrm{Y}_2 ) = (1,0)$}
  \end{subfigure}
    \hfill
  \begin{subfigure}[b]{0.5\textwidth}
    \includegraphics[width=\textwidth]{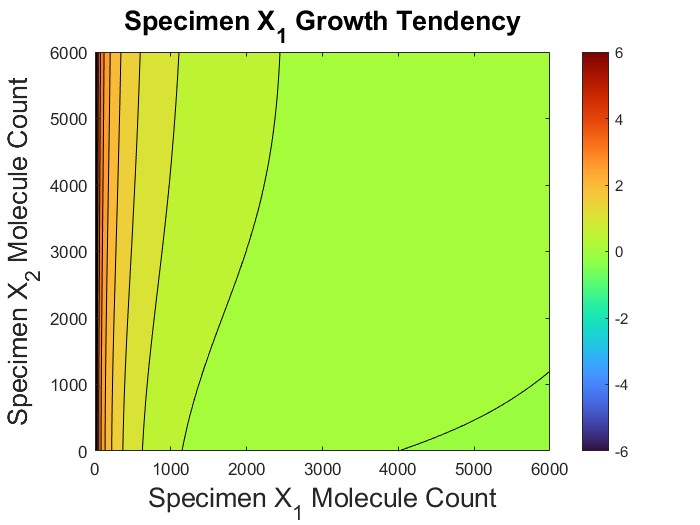}
    \caption{$(\mathrm{Y}_1, \mathrm{Y}_2 ) = (0,1)$}
  \end{subfigure}
  \hfill
  \begin{subfigure}[b]{0.5\textwidth}
    \includegraphics[width=\textwidth]{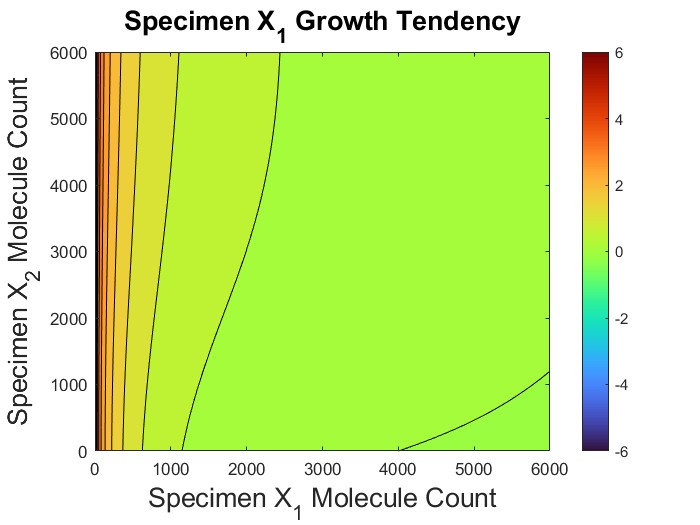}
    \caption{$(\mathrm{Y}_1, \mathrm{Y}_2 ) = (0,0)$}
  \end{subfigure}
  \caption[Contours of tendencies computed from $\mathrm{X}_1$ propensities]{Growth tendency of $\mathrm{X}_1$ for parameter values of Figure~\ref{discreteargueplot}, $\zeta = 10$. }
  \label{PropensityX1increase}
\end{figure}

\begin{figure}[h!]
  \begin{subfigure}[b]{0.5\textwidth}
    \includegraphics[width=\textwidth]{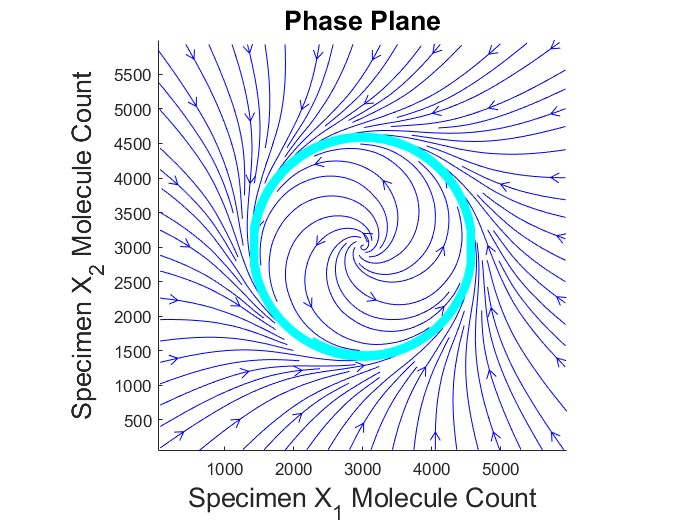}
    \caption{$\zeta = 10$}
  \end{subfigure}
  \hfill
  \begin{subfigure}[b]{0.5\textwidth}
    \includegraphics[width=\textwidth]{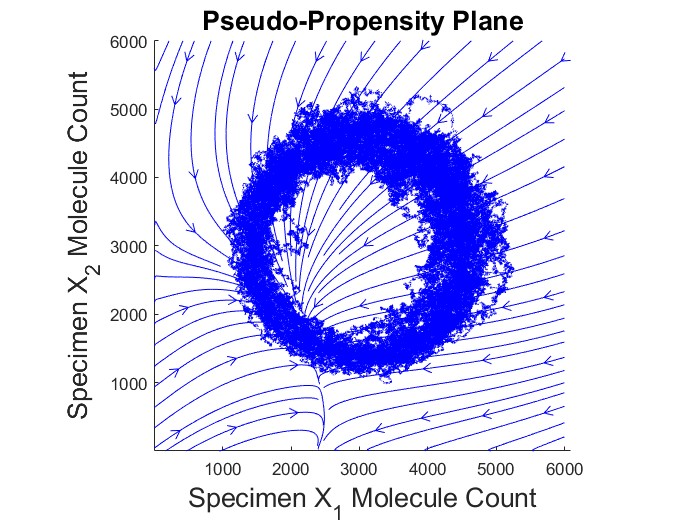}
    \caption{$\zeta = 10$}
  \end{subfigure}
    \hfill
  \begin{subfigure}[b]{0.5\textwidth}
    \includegraphics[width=\textwidth]{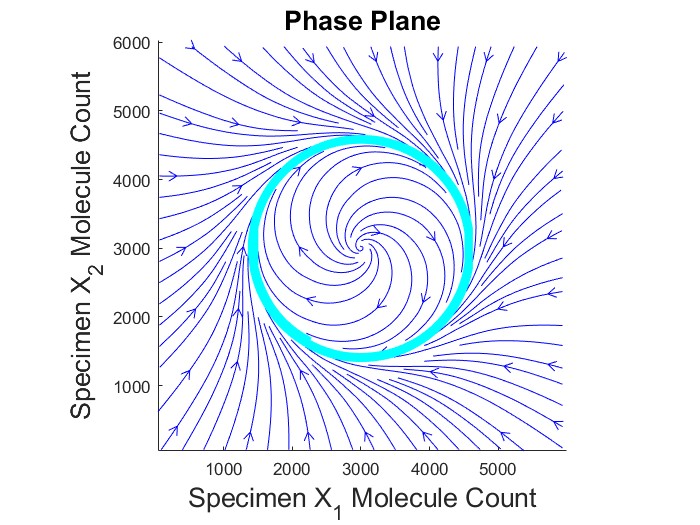}
    \caption{$\zeta = -10$}
  \end{subfigure}
  \hfill
  \begin{subfigure}[b]{0.5\textwidth}
    \includegraphics[width=\textwidth]{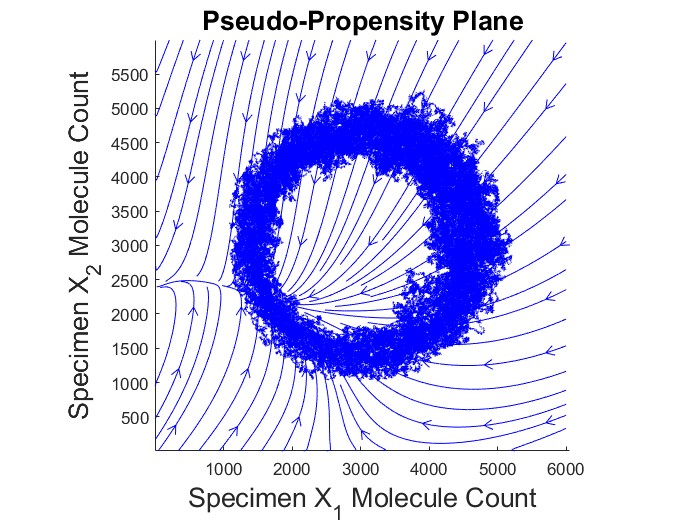}
    \caption{$\zeta = -10$}
  \end{subfigure}
  \hfill
    \begin{subfigure}[b]{0.5\textwidth}
    \includegraphics[width=\textwidth]{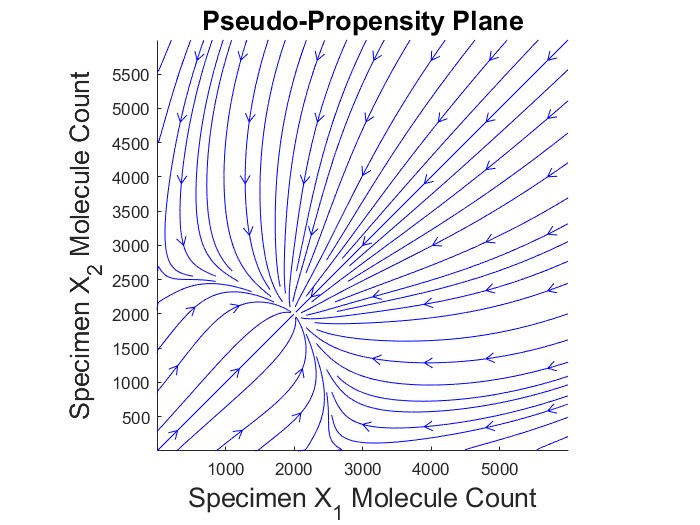}
    \caption{$\zeta = 0$ }
  \end{subfigure}
  \hfill
  \begin{subfigure}[b]{0.5\textwidth}
    \includegraphics[width=\textwidth]{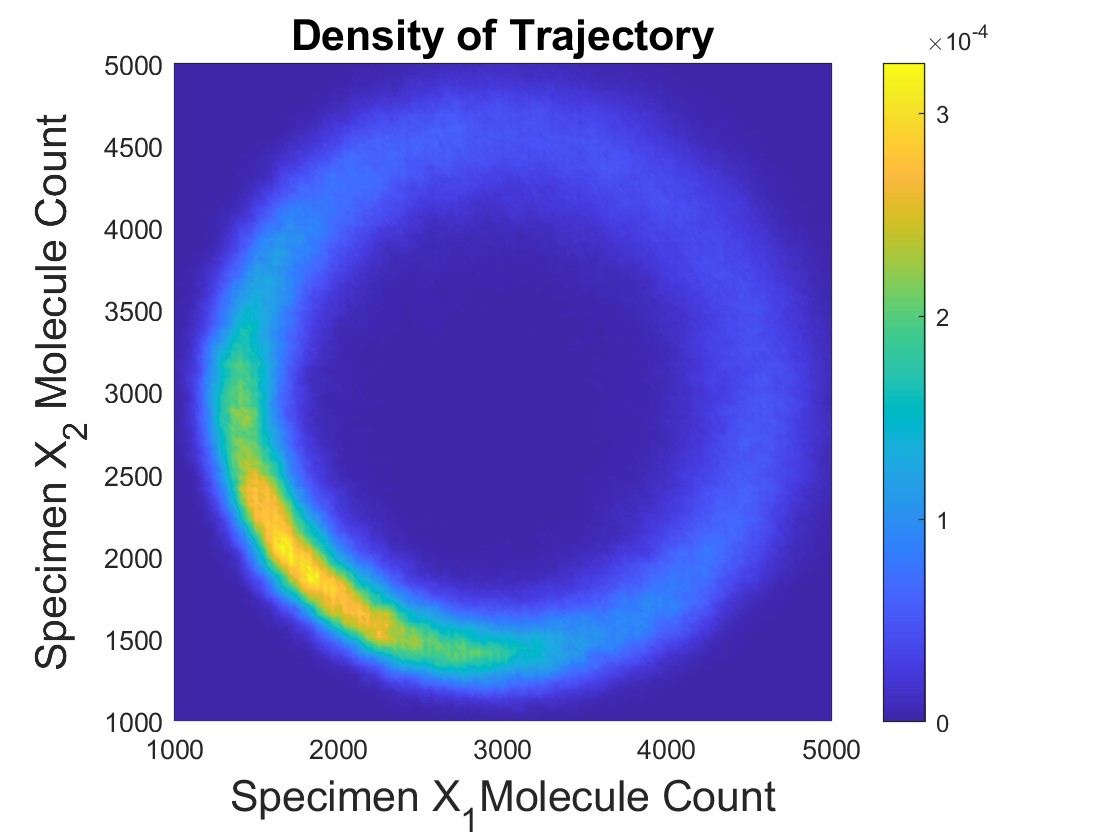}
    \caption{$\zeta = 0$ }
  \end{subfigure}
  \caption[A summary of the results of Appendix~\ref{discreteappendix}]{Trajectory flow predicted solely by propensity, for parameter values of Figure~\ref{PropensityX1increase}. Stochastic paths are pinched in the southwest region as the weighted propensity flow predicts. More research is desired to refine the accuracy and widen the applicability of this technique, which presents yet another potential extension to this project. See also Appendix~\ref{appendixE}.}
  \label{guess}
\end{figure}

The first approximation we make is that the molecule counts of $\mathrm{Y}_1, \mathrm{Y}_2$ are either $1$ or $0$, as repeated simulations of the system~\eqref{hopfbifuraction} at $\mu = 10^{-6}$ reliably shows the maximal specimen counts of both specimen limited to $1$, deviating only rarely from this trend. From this we obtain Figure~\ref{PropensityX1increase}.

In each plot, we have calculated the growth tendency to be $\log{(p_1/p_2)}$. Positive tendency therefore means that $\mathrm{X}_1$ is more likely to be added propensity-wise, and negative tendency the opposite. Note that $\mathrm{Y}_1 = 1$ in (a-b) heavily encourages the removal of $\mathrm{X}_1$. This is to be expected, as the existence of $\mathrm{Y}_1$ activates channels representing cross-negative terms which remove $\mathrm{X}_1$ that lay dormant during its extinction. In contrast, $\mathrm{X}_1$ shows a strong proclivity to be added when $\mathrm{Y}_1 = 0$ in (c-d). It is the alternating push and pull of these four different tendencies activated by molecule counts of $(\mathrm{Y}_1, \mathrm{Y}_2)$ that organizes joint densities into a circular shape depicted in Figure~\ref{discreteargueplot}.

We note in particular that Figure~\ref{PropensityX1increase} (c-d) shows a push for increasing $\mathrm{X}_1$ as its molecule count becomes smaller, which is intensified in the $1000-2000$ region. The plots for $\mathrm{X}_2$ are analogous. Although this may form the basis for an existence argument of the hypothesized ``external repulsive force'', we may further develop this idea by attempting to deterministically sketch the predicted path of the trajectory based on propensities.

\begin{figure}[h!]
  \begin{subfigure}[b]{0.5\textwidth}
    \includegraphics[width=\textwidth]{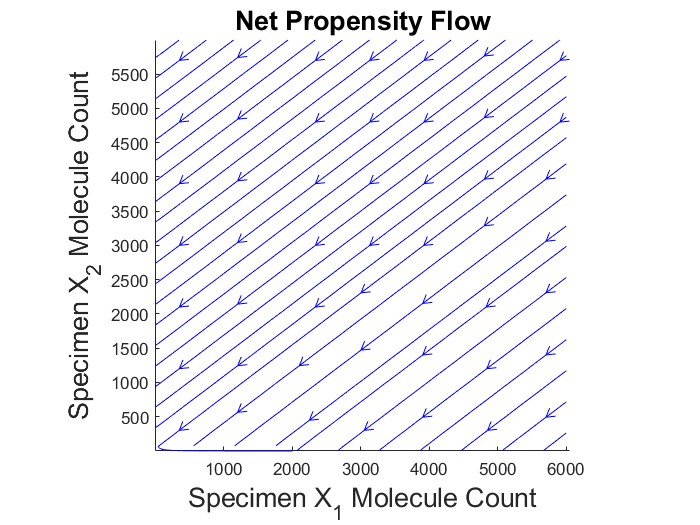}
    \caption{$(\mathrm{Y}_1, \mathrm{Y}_2 ) = (1,1)$}
  \end{subfigure}
  \hfill
  \begin{subfigure}[b]{0.5\textwidth}
    \includegraphics[width=\textwidth]{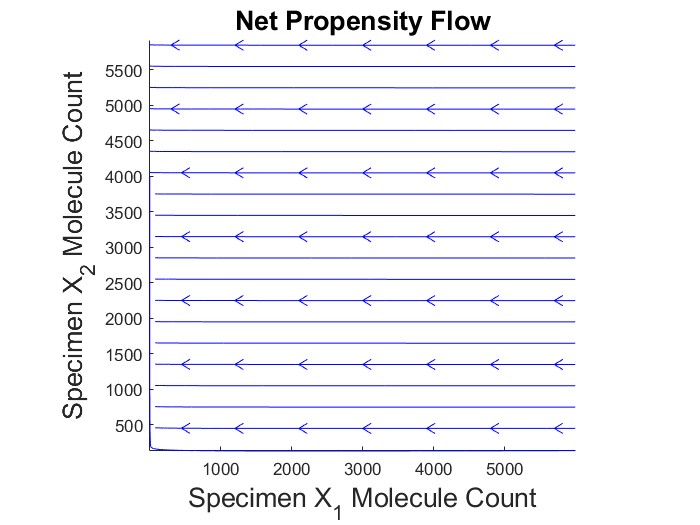}
    \caption{$(\mathrm{Y}_1, \mathrm{Y}_2 ) = (1,0)$}
  \end{subfigure}
    \hfill
  \begin{subfigure}[b]{0.5\textwidth}
    \includegraphics[width=\textwidth]{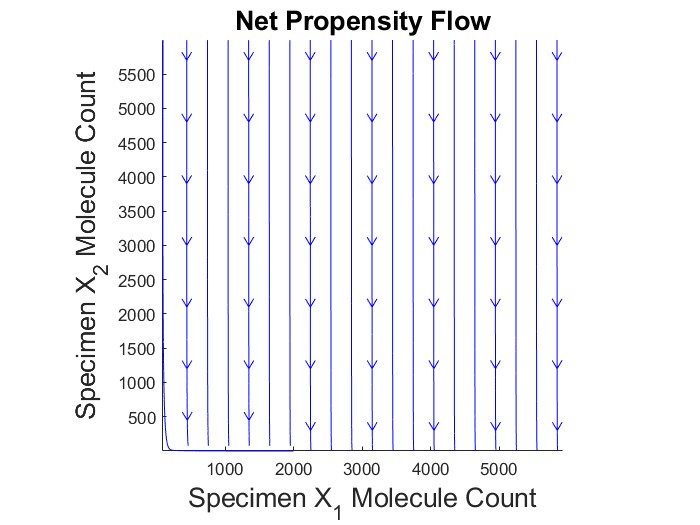}
    \caption{$(\mathrm{Y}_1, \mathrm{Y}_2 ) = (0,1)$}
  \end{subfigure}
  \hfill
  \begin{subfigure}[b]{0.5\textwidth}
    \includegraphics[width=\textwidth]{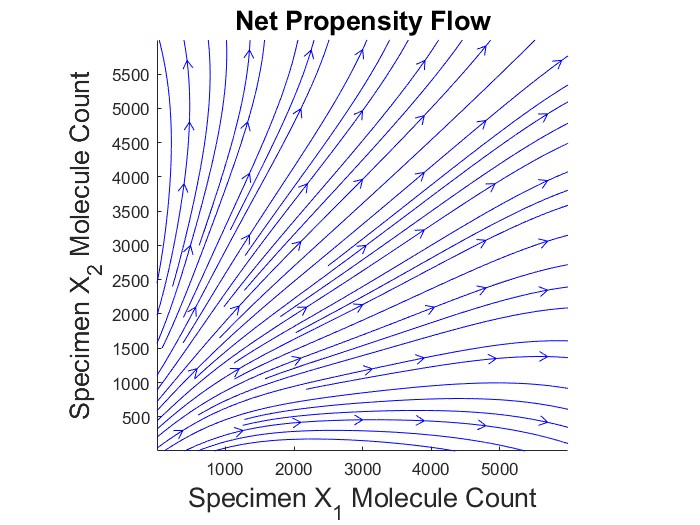}
    \caption{$(\mathrm{Y}_1, \mathrm{Y}_2 ) = (0,0)$}
  \end{subfigure}
  \caption[Rudimentary flow induced by net propensity vectors]{Trajectory flow predicted solely by propensity, for parameter values of Figure~\ref{PropensityX1increase}. }
  \label{rudimentaryflow}
\end{figure}

Doing so requires a different approach to computing tendencies. It is unlikely that simply adding the tendency data for the four scenarios considered in Figure~\ref{PropensityX1increase} will yield sensible results, because tendencies must be scaled by the likelihood that each state of $(\mathrm{Y}_1, \mathrm{Y}_2)$ is realized in the simulation. Furthermore, adding tendencies will result in terms of the form
\begin{equation}\label{confusing}
    \log{\left(\frac{p_1(1,1)p_1(1,0)p_1(0,1)p_1(0,0)}{p_2(1,1)p_2(1,0)p_2(0,1)p_2(0,0)} \right)},
\end{equation}
where $p_1(i,j)$ represents the value of $p_1$ for $(\mathrm{Y}_1,\mathrm{Y}_2) = (i,j)$. It is unclear what~\eqref{confusing} seeks to represent.

Therefore, we take an entirely different track and perform a \textit{pseudo-propensity analysis}. The \textit{net propensity increase} of $\mathrm{X}_1$ and $\mathrm{X}_2$ are defined to be $p_1 - p_2$ and $p_3 - p_4$. Identical coarse-graining leading up to Figure~\ref{PropensityX1increase} is repeated, and this time we attach to each vertex of the rectangles (formed in the coarse-graining) the net propensity increase for $\mathrm{X}_1$,$\mathrm{X}_2$. This allows us to define a net propensity vector at each vertex, which accumulate to give the flow in Figure~\ref{rudimentaryflow}. Similar observations to those stemming from Figure~\ref{PropensityX1increase} may be made.

\begin{figure}[h]
  \begin{subfigure}[b]{0.5\textwidth}
    \includegraphics[width=\textwidth]{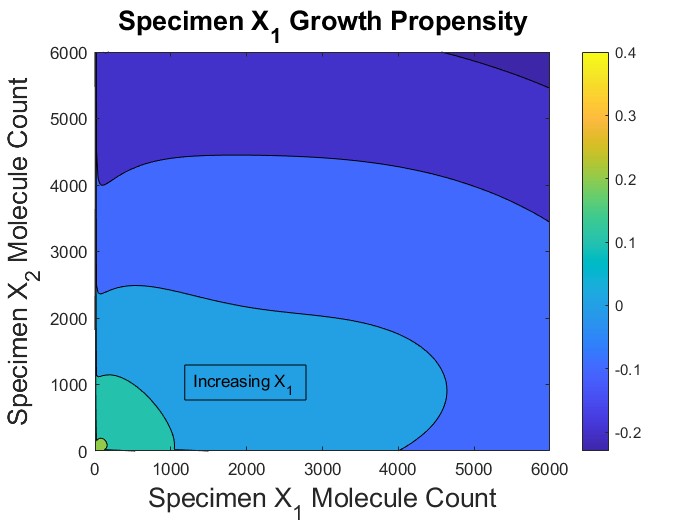}
    \caption{ }
  \end{subfigure}
  \hfill
  \begin{subfigure}[b]{0.5\textwidth}
    \includegraphics[width=\textwidth]{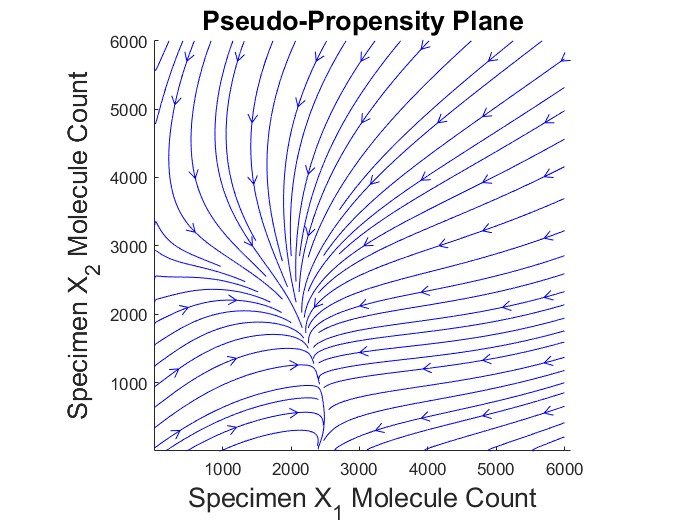}
    \caption{ }
  \end{subfigure}
  \caption[Flow induced by weighted propensity vectors]{Flow in the pseudo-propensity plane induced by weighting net propensity vectors in Figure~\ref{rudimentaryflow}. In particular (a) depicts the increase in $\mathrm{X}_1$ molecule counts in the southwest region. Analogous data is generated for $\mathrm{X}_2$ and used to compute (b) via automatic interpolation done by \mcode{Matlab}. We note that the interpolation overwrites small tendencies in the data to account for perceived errors; in particular (a) shows $\mathrm{X}_1$ increase (very slightly, when contour levels are checked) in the south region while (b) depicts it decreasing.}
  \label{PseudoPropensityFlow}
\end{figure}

To collect this data into a coherent plot, we must weigh (a-d) in Figure~\ref{rudimentaryflow} by the likelihood that each state $(\mathrm{Y}_1,\mathrm{Y}_2)$ exists, before adding the weighted propensities. This is an extremely complex task because the probability of a single state vector realization $(\mathrm{X}_1,\mathrm{X}_2,\mathrm{Y}_1,\mathrm{Y}_2)$ is impacted by its value in the previous state. To compute the likelihood of the previous state, we must look further up into the history axis, forming a very long chain of calculations to compute the correct weights. 

We have therefore used a series of rudimentary approximations to proceed with the analysis, starting by assuming a uniform probability of existence of all immediately preceding states that lead to the current state in order to compute the weights. We expect that the accuracy of the approximation will be enhanced as states further upward the history axis are considered, for example by making the uniform existence assumption for all states several reaction events prior to the current state.

To choose the timescale at which we travel down the history axis, we count the number of firings $F$ of the reaction channels that impact specimen $\mathrm{X}_1,\mathrm{X}_2,\mathrm{Y}_1, \mathrm{Y_2}$. For the parameter values of Figure~\ref{discreteargueplot}, the number of channel firings satisfy $F_{\mathrm{X}_1}/F_{\mathrm{X}_2} \approx F_{\mathrm{Y}_1}/F_{\mathrm{Y}_2} \approx 1$, and $F_{\mathrm{Y}_1}/F_{\mathrm{X}_1} \approx 2$. Denoting the reaction timescale of specimen $\mathrm{X}_1$ as $\tau_1$ and $\mathrm{Y}_2$ as $\tau_2$, we choose the unit time-stepping length of the history axis $\tau$ as $\tau_1 \gg \tau > \tau_2$. In this timescale, we need only consider singular reaction events altering specimen $\mathrm{Y}_1,\mathrm{Y}_2$ that may happen concurrently. The transitions of $(\mathrm{X}_1,\mathrm{X}_2)$ are neglected in the analysis due to the definition of $\tau$. Note that our intent here is to provide a rudimentary sketch of a technique that may be refined through further research.

Recall that after many simulations, we reliably observed that $\mathrm{Y}_1, \mathrm{Y_2} \in \{ 0,1\}$. Suppose that an immediately preceding state in the history axis has the coordinates $(\mathrm{Y}_1, \mathrm{Y_2}) = (1,0)$. Then, $\mathrm{Y}_1$ can only decrease, or stay constant for the next state; in contrast $\mathrm{Y}_2$ can only increase, or stay constant for the next state. For $\mathrm{Y}_1$ to decrease, we must fire a reaction channel whose propensity is summed in $p_6$, and for $\mathrm{Y}_2$ to increase, we must fire a reaction channel whose propensity is summed in $p_7$. For $\mathrm{Y}_1$ to remain constant in the next state, we must fire any other reaction channel whose propensity is not summed in $p_6$, so the propensity is $1-p_6$ (note that $p_5 \ll 1$ is irrelevant as we assume that the molecule count of $\mathrm{Y}_1$ cannot be greater than $1$). Analogously, for $\mathrm{Y}_2$ to remain constant, we must fire any other reaction channel whose propensity is not summed in $p_7$, so the propensity is $1-p_7$ (again, $p_8$ is irrelevant as $\mathrm{Y}_2$ cannot decrease). This induces the final approximation, which is to view such propensities as an estimate for the transition probabilities from one state to another, called \textit{pseudo-propensities}. 

For example, the propensity at state $(\mathrm{Y}_1, \mathrm{Y_2}) = (1,1)$ is given
\begin{equation}
    \mathbf{p} = (p_1(1,1),p_2(1,1),p_3(1,1),p_4(1,1),p_5(1,1),p_6(1,1),p_7(1,1),p_8(1,1))^\top.
\end{equation}
In the preceding state, $(\mathrm{Y}_1, \mathrm{Y_2})$ could have been in coordinates $(1,1), (1,0), (0,1),$ $(0,0)$. In the case $(\mathrm{Y}_1, \mathrm{Y_2}) = (1,1)$, neither $\mathrm{Y}_1$ nor $\mathrm{Y}_2$ have changed. We therefore compute the weight $(1-p_6(1,1))(1-p_8(1,1))$ as the probability of transition from state $(1,1)$ to $(1,1)$. In the case $(\mathrm{Y}_1, \mathrm{Y_2}) = (1,0)$, $\mathrm{Y}_1$ has not changed but $\mathrm{Y}_2$ has increased. The weight is therefore given as $(1-p_6(1,0))p_7(1,0)$. Considering the remaining cases $(\mathrm{Y}_1, \mathrm{Y_2}) = (0,1)$ or $(0,0)$, we obtain the final weighting as a sum of the weights
\begin{equation}
    (1-p_6(1,1))(1-p_8(1,1)) + (1-p_6(1,0))p_7(1,0) + p_5(0,1)(1-p_8(0,1)) + p_5(0,0)p_7(0,0).
\end{equation}
We take this to be the likelihood of existence of state $(\mathrm{Y}_1, \mathrm{Y_2}) = (1,1)$ and multiply the net propensities in Figure~\ref{rudimentaryflow} (a) with this value. Of course, this is done individually for each vertex formed in the coarse graining. Repeating this procedure for the net propensities in Figure~\ref{rudimentaryflow} (b-d) and adding the four weighted propensities together yields the flow induced by pseudo-propensities in Figure~\ref{PseudoPropensityFlow}.

We note that Figure~\ref{PseudoPropensityFlow} (b) does not appear very accurate in the northeast region, but gives sensible results in the southwest region. It depicts the existence of a repulsion force emanating from the west, and assuming that the travel of the trajectory is counterclockwise for $\zeta = 10$, the pesudo-propensity flow wraps around the stochastic trajectory in a way consistent with a restriction or concentration of movement as it passes through this region. However, this is where the similarities end; the flow in the east appears to cut across the grain of deterministically expected movement. Similar observations may be made for the case $\zeta = -10$ where the trajectory travels clockwise; the region of validity is now the south, while generally the north cuts across the grain of expected movement. The pseudo-propensity plane shows analogous trends; in both instances, we note the lack of $90$ degree rotational symmetry observed in the deterministic phase plane. See further Figure~\ref{guess}.

At the beginning of this discussion, we hypothesized about the existence of repulsive forces coming from the $x_1$,$x_2$-axis, which appear to be depicted in the $x_2$-axis of Figure~\ref{guess} (b) and $x_1$-axis of (d). Then it stands to reason that in the case of no rotation $\zeta = 0$, these forces will propagate the stochastic trajectory away from both axes, constraining the movement to the northeast region. However, our pseudo-propensity plane contradicts this assessment by plotting a concentration of movement toward the southwest region of the phase plane. Surprisingly, we have discovered that long-time stochastic simulations starting from different initial conditions agrees very well with the predictions of the pseudo-propensity plane, shown in Figure~\ref{guess} (e-f).

\chapter{Networks in Chapter~\ref{oldchapter6}}\label{thenetworksappendix}
In this section, we give chemical reaction network forms for the test systems \eqref{hopfbifuraction}, \eqref{rosslerode}, \eqref{lorenzode} considered in Chapter~\ref{oldchapter6}. The network for the shifted pendulum~\eqref{pendulum} may be derived in an identical manner, although as Taylor expansions are taken to order $20$ difficulties arise when attempting to explicitly write down the reaction network.

For~\eqref{hopfbifuraction}, the substitution  $\mathbf{x} \leftarrow \mathbf{x} - \boldsymbol{\mathcal{T}}_\mathbf{x}$, $\boldsymbol{\mathcal{T}}_\mathbf{x} \in \mathbb{R}^{2}_{>0}$ to push the dynamics of interest into the positive cone induces the system~\eqref{firstaffine}, which may be expanded as
\begin{equation}\label{f1}
\begin{aligned}
    \Dot{x_1} &= x_1\left(\xi - 3\mathcal{T}_{x_1}^2 - \mathcal{T}_{x_2}^2 \right) + 3x_1^2\mathcal{T}_{x_1}+x_2\left(-\zeta -2\mathcal{T}_{x_1}\mathcal{T}_{x_2} \right)-x_1x_2^2\\ &+2x_1x_2\mathcal{T}_{x_2}+x_2^2\mathcal{T}_{x_1}-x_1^3+\left(-\xi \mathcal{T}_{x_1} + \zeta \mathcal{T}_{x_2} + \mathcal{T}_{x_1}^3+\mathcal{T}_{x_1}\mathcal{T}_{x_2}^2 \right),
\end{aligned}
\end{equation}
\begin{equation}\label{f2}
    \begin{aligned}
        \Dot{x_2} &= x_2\left(\xi - 3\mathcal{T}_{x_2}^2 - \mathcal{T}_{x_1}^2 \right) + 3x_2^2\mathcal{T}_{x_2}+x_1\left(\zeta -2\mathcal{T}_{x_2}\mathcal{T}_{x_1} \right)-x_2x_1^2\\ &+2x_1x_2\mathcal{T}_{x_1}+x_1^2\mathcal{T}_{x_2}-x_2^3+\left(-\xi \mathcal{T}_{x_2} - \zeta \mathcal{T}_{x_1} + \mathcal{T}_{x_2}^3+\mathcal{T}_{x_2}\mathcal{T}_{x_1}^2 \right).
    \end{aligned}
\end{equation}
In the region $\zeta,\xi \in [-10,10],$ $\mathcal{T}_{x_1} = \mathcal{T}_{x_2} = 6$ considered in this paper, cross-negative terms are detected in the $x_2,x_1$ terms of the right hand sides in~\eqref{f1}, \eqref{f2} as $|\zeta|<2\mathcal{T}_{x_1}\mathcal{T}_{x_2}$. Under $\Psi_{QSST}$, the general system is given
\begin{equation}\label{f3}
\begin{aligned}
    \Dot{x_1} &= x_1\left(\xi - 3\mathcal{T}_{x_1}^2 - \mathcal{T}_{x_2}^2 \right) + 3x_1^2\mathcal{T}_{x_1}+x_1x_2y_1\left(-\zeta -2\mathcal{T}_{x_1}\mathcal{T}_{x_2} \right)-x_1x_2^2\\ &+2x_1x_2\mathcal{T}_{x_2}+x_2^2\mathcal{T}_{x_1}-x_1^3+\left(-\xi \mathcal{T}_{x_1} + \zeta \mathcal{T}_{x_2} + \mathcal{T}_{x_1}^3+\mathcal{T}_{x_1}\mathcal{T}_{x_2}^2 \right),
\end{aligned}
\end{equation}
\begin{equation}\label{f4}
    \begin{aligned}
        \Dot{x_2} &= x_2\left(\xi - 3\mathcal{T}_{x_2}^2 - \mathcal{T}_{x_1}^2 \right) + 3x_2^2\mathcal{T}_{x_2}+x_1x_2y_2\left(\zeta -2\mathcal{T}_{x_2}\mathcal{T}_{x_1} \right)-x_2x_1^2\\ &+2x_1x_2\mathcal{T}_{x_1}+x_1^2\mathcal{T}_{x_2}-x_2^3+\left(-\xi \mathcal{T}_{x_2} - \zeta \mathcal{T}_{x_1} + \mathcal{T}_{x_2}^3+\mathcal{T}_{x_2}\mathcal{T}_{x_1}^2 \right),
    \end{aligned}
\end{equation}
\begin{equation}\label{f5}
\begin{aligned}
    &\mu \Dot{y_1} = 1 - x_1y_1,\\
&\mu \Dot{y_2} = 1 - x_2y_2,
\end{aligned}
\end{equation}
which is canonically inverted into the reaction
\begin{equation}\label{hopfnetwork}
\begin{array}{ll}%
r_{1}: \quad \varnothing \stackrel{k_1}{\longrightarrow} \mathrm{X}_1,& \quad \quad r_{2}: \quad 2\mathrm{X}_1 \stackrel{k_2}{\longrightarrow} 3\mathrm{X}_1, \\  r_{3}: \quad \mathrm{X}_1 + \mathrm{X}_2+\mathrm{Y}_1 \stackrel{k_3}{\longrightarrow} \mathrm{X}_2+\mathrm{Y}_1, &\quad \quad
r_{4}: \quad \mathrm{X}_1 + 2\mathrm{X}_2 \stackrel{k_4}{\longrightarrow} 2\mathrm{X}_2,\\    r_{5}: \quad \mathrm{X}_1 + \mathrm{X}_2 \stackrel{k_5}{\longrightarrow} 2\mathrm{X}_1 + \mathrm{X}_2,&\quad \quad   r_{6}: \quad 2\mathrm{X}_2  \stackrel{k_6}{\longrightarrow} \mathrm{X}_1+2\mathrm{X}_2,\\
r_{7}: \quad 3\mathrm{X}_1  \stackrel{k_7}{\longrightarrow} 2\mathrm{X}_1,& \quad \quad  r_{8}: \quad \varnothing \stackrel{k_8}{\longrightarrow}\mathrm{X}_1,\\   r_{9}: \quad \mathrm{X}_2 \stackrel{k_9}{\longrightarrow} 2\mathrm{X}_2, &\quad \quad
r_{10}: \quad 2\mathrm{X}_2  \stackrel{k_{10}}{\longrightarrow} 3\mathrm{X}_2,\\   r_{11}: \quad \mathrm{X}_1 + \mathrm{X}_2 + \mathrm{Y}_2 \stackrel{k_{11}}{\longrightarrow} \mathrm{X}_1+\mathrm{Y}_2,&\quad \quad  r_{12}: \quad 2\mathrm{X}_1 +\mathrm{X}_2 \stackrel{k_{12}}{\longrightarrow} 2\mathrm{X}_1,\\
r_{13}: \quad \mathrm{X}_1+\mathrm{X}_2 \stackrel{k_{13}}{\longrightarrow} \mathrm{X}_1+2\mathrm{X}_2,& \quad \quad  r_{14}: \quad 2\mathrm{X}_1 \stackrel{k_{14}}{\longrightarrow}2\mathrm{X}_1+ \mathrm{X}_2,\\   r_{15}: \quad 3\mathrm{X}_2 \stackrel{k_{15}}{\longrightarrow} 2\mathrm{X}_2, &\quad \quad
r_{16}: \quad \varnothing \stackrel{k_{16}}{\longrightarrow} \mathrm{X}_2,\\    r_{17}: \quad \varnothing  \stackrel{k_{17}}{\longrightarrow} \mathrm{Y}_1 ,& \quad \quad  r_{18}: \quad \mathrm{X}_1+\mathrm{Y}_1 \stackrel{k_{18}}{\longrightarrow} \mathrm{X}_1,\\ 
r_{19}:  \quad \varnothing \stackrel{k_{19}}{\longrightarrow} \mathrm{Y}_2, &  \quad \quad r_{20}: \quad \mathrm{X}_2 + \mathrm{Y}_2\stackrel{k_{20}}{\longrightarrow} \mathrm{X}_2, 
\end{array}
\end{equation}
for the reaction coefficients
\begin{equation}
    \begin{aligned}
        &k_1 = \xi - 3\mathcal{T}_{x_1}^2 - \mathcal{T}_{x_2}^2,\quad k_2 = 3\mathcal{T}_{x_1},\quad k_3 = \zeta +2\mathcal{T}_{x_1}\mathcal{T}_{x_2}, \quad k_4 = 1,\\
        &k_5= 2\mathcal{T}_{x_2}, \quad k_6=\mathcal{T}_{x_1},\quad k_7=1,\quad k_8 = -\xi \mathcal{T}_{x_1} + \zeta \mathcal{T}_{x_2} + \mathcal{T}_{x_1}^3+\mathcal{T}_{x_1}\mathcal{T}_{x_2}^2, \\
        &k_9 = \xi - 3\mathcal{T}_{x_2}^2 - \mathcal{T}_{x_1}^2,\quad k_{10} = 3\mathcal{T}_{x_2},\quad k_{11}= -\zeta +2\mathcal{T}_{x_2}\mathcal{T}_{x_1}, \quad k_{12} = 1,\\
        &k_{13}= 2\mathcal{T}_{x_1}, \quad k_{14}=\mathcal{T}_{x_2},\quad k_{15}=1,\quad k_{16} = -\xi \mathcal{T}_{x_2} - \zeta \mathcal{T}_{x_1} + \mathcal{T}_{x_2}^3+\mathcal{T}_{x_2}\mathcal{T}_{x_1}^2, \\
        & k_{17} = \frac{1}{\mu}, \quad k_{18} = \frac{1}{\mu}, \quad k_{19} = \frac{1}{\mu}, \quad k_{20} = \frac{1}{\mu}.
    \end{aligned}
\end{equation}
Note that appropriate scalings to the coefficients must be taken for non-unit reactor volume $V$. The networks of the R\"{o}ssler system~\eqref{rosslerode} and the Lorenz system~\eqref{lorenzode} follow analogous derivations. For $a,b,c,\mathcal{T}_{x_1},\mathcal{T}_{x_2},\mathcal{T}_{x_3}>0$, the R\"{o}ssler network is given
\begin{equation}\label{rosslernetwork}
\begin{array}{ll}%
r_{1}: \quad \mathrm{X}_1+\mathrm{X}_2+\mathrm{Y_1} \stackrel{k_1}{\longrightarrow} \mathrm{X}_2+\mathrm{Y_1},& \quad\quad r_{2}: \quad \mathrm{X}_1+\mathrm{X}_3+\mathrm{Y_1} \stackrel{k_2}{\longrightarrow} \mathrm{X}_1+\mathrm{X}_3+\mathrm{Y_1}, \\ r_{3}: \quad \varnothing \stackrel{k_3}{\longrightarrow} \mathrm{X}_1, & \quad\quad
r_{4}: \quad \mathrm{X}_1  \stackrel{k_4}{\longrightarrow} \mathrm{X}_1 + \mathrm{X}_2, \\  r_{5}: \quad \mathrm{X}_2 \stackrel{k_5}{\longrightarrow} 2\mathrm{X}_2,&\quad \quad r_{6}: \quad \mathrm{X}_2 + \mathrm{Y}_2  \stackrel{k_6}{\longrightarrow} \mathrm{Y}_2,\\
r_{7}: \quad \varnothing  \stackrel{k_7}{\longrightarrow} \mathrm{X}_3,& \quad\quad r_{8}: \quad \mathrm{X}_1 + \mathrm{X}_3 \stackrel{k_8}{\longrightarrow}\mathrm{X}_1,\\ r_{9}: \quad \mathrm{X}_3 \stackrel{k_9}{\longrightarrow} 2\mathrm{X}_3, & \quad\quad
r_{10}: \quad \mathrm{X}_1  \stackrel{k_{10}}{\longrightarrow} \mathrm{X}_1+\mathrm{X}_3,\\ r_{11}: \quad \mathrm{X}_3 + \mathrm{Y}_3 \stackrel{k_{11}}{\longrightarrow} \mathrm{Y}_3,&\quad \quad r_{12}: \quad \varnothing \stackrel{k_{12}}{\longrightarrow} \mathrm{Y}_1,\\
r_{13}: \quad \mathrm{X}_1+\mathrm{Y}_1 \stackrel{k_{13}}{\longrightarrow}  \mathrm{X}_1,&\quad \quad r_{14}: \quad \varnothing \stackrel{k_{14}}{\longrightarrow} \mathrm{Y}_2,\\  r_{15}: \quad \mathrm{X}_2 + \mathrm{Y}_2 \stackrel{k_{15}}{\longrightarrow} \mathrm{X}_2, &\quad\quad r_{16}: \quad \varnothing \stackrel{k_{16}}{\longrightarrow} \mathrm{Y}_3,\\  r_{17}: \quad \mathrm{X}_3 + \mathrm{Y}_3 \stackrel{k_{17}}{\longrightarrow} \mathrm{X}_3 ,&\quad \quad 
\end{array}
\end{equation}
for the reaction coefficients
\begin{equation}
    \begin{aligned}
        &k_1 = 1,\quad k_2 = 1,\quad k_3 = \mathcal{T}_{x_1}+\mathcal{T}_{x_3}, \quad k_4 = 1, \quad k_5= a, \quad k_6=\mathcal{T}_{x_1}+a\mathcal{T}_{x_2},\\
        &k_7=b,\quad k_8 = 1, \quad k_9 = \mathcal{T}_{x_1} +c,\quad k_{10} = \mathcal{T}_{x_3},\quad k_{11}= \mathcal{T}_{x_3}\left(\mathcal{T}_{x_1}+c\right), \quad k_{12} = \frac{1}{\mu},\\
        &k_{13}= \frac{1}{\mu}, \quad k_{14}=\frac{1}{\mu},\quad k_{15}=\frac{1}{\mu},\quad k_{16} = \frac{1}{\mu}, \quad k_{17} = \frac{1}{\mu}. 
    \end{aligned}
\end{equation}
Similarly, the Lorenz system for $\sigma,\rho,\beta,\mathcal{T}_{x_1},\mathcal{T}_{x_2},\mathcal{T}_{x_3}>0$ gives  
\begin{equation}\label{lorenznetwork}
\begin{array}{ll}%
r_{1}: \quad \mathrm{X}_2 \stackrel{k_1}{\longrightarrow} \mathrm{X}_2+\mathrm{X_1},& \quad\quad r_{2}: \quad \mathrm{X}_1 \stackrel{k_2}{\longrightarrow} \varnothing,\\ r_{3}: \quad \mathrm{X}_1 \stackrel{k_3}{\longrightarrow} \mathrm{X}_1 + \mathrm{X}_2, & \quad\quad
r_{4}: \quad \mathrm{X}_3 \stackrel{k_4}{\longrightarrow} \mathrm{X}_2 + \mathrm{X}_3, \\  r_{5}: \quad \mathrm{X}_1 + \mathrm{X}_2 + \mathrm{X}_3 + \mathrm{Y}_2 \stackrel{k_5}{\longrightarrow} \mathrm{X}_1  + \mathrm{X}_3 + \mathrm{Y}_2,& \quad\quad r_{6}: \quad \mathrm{X}_2 + \mathrm{Y}_2  \stackrel{k_6}{\longrightarrow} \mathrm{Y}_2,\\
r_{7}: \quad \mathrm{X}_1 + \mathrm{X}_2  \stackrel{k_7}{\longrightarrow} \mathrm{X}_1 + \mathrm{X}_2 + \mathrm{X}_3,&\quad \quad r_{8}: \quad \mathrm{X}_1 + \mathrm{X}_3 + \mathrm{Y}_3 \stackrel{k_8}{\longrightarrow}\mathrm{X}_1 + \mathrm{Y}_3,\\  r_{9}: \quad \mathrm{X}_1 + \mathrm{X}_3 + \mathrm{Y}_3 \stackrel{k_9}{\longrightarrow} \mathrm{X}_2 + \mathrm{Y}_3, & \quad\quad
r_{10}: \quad \mathrm{X}_3  \stackrel{k_{10}}{\longrightarrow} \varnothing,\\ r_{11}: \quad \varnothing \stackrel{k_{11}}{\longrightarrow} \mathrm{X}_3,&\quad \quad r_{12}: \quad \varnothing \stackrel{k_{12}}{\longrightarrow} \mathrm{Y}_2,\\
r_{13}: \quad \mathrm{X}_2+\mathrm{Y}_2 \stackrel{k_{13}}{\longrightarrow}  \mathrm{X}_2,&\quad \quad r_{14}: \quad \varnothing \stackrel{k_{14}}{\longrightarrow} \mathrm{Y}_3,\\ r_{15}: \quad \mathrm{X}_3 + \mathrm{Y}_3 \stackrel{k_{15}}{\longrightarrow} \mathrm{X}_3, & 
\end{array}
\end{equation}
for the reaction coefficients
\begin{equation}
    \begin{aligned}
        &k_1 = \sigma,\quad k_2 = \sigma,\quad k_3 = \rho+\mathcal{T}_{x_3}, \quad k_4 = \mathcal{T}_{x_1}, \quad k_5 = 1, \quad k_6= \rho \mathcal{T}_{x_1}+\mathcal{T}_{x_1}\mathcal{T}_{x_3},\\
        &k_7 = 1,\quad k_8 = \mathcal{T}_{x_2}, \quad k_9 = \mathcal{T}_{x_1} ,\quad k_{10} = \beta ,\quad k_{11}=\mathcal{T}_{x_1}\mathcal{T}_{x_2} + \beta \mathcal{T}_{x_3}, \quad k_{12} = \frac{1}{\mu},\\
        &k_{13}= \frac{1}{\mu}, \quad k_{14}=\frac{1}{\mu},\quad k_{15}=\frac{1}{\mu}. 
    \end{aligned}
\end{equation}

%% file: appendix5.tex
\chapter{Possible Extensions}\label{appendixE}
Chemical reaction network theory is a very active field of research, and many extensions to this work are possible. Below we list a few.

1. \textbf{Chemical implementation in wet labs.}

\noindent A natural extension to our work may be to synthetically implement the obtained networks within a biochemical laboratory. We propose the plots in Chapter~\ref{oldchapter6} as roadmaps to the chemical integration of ODEs as a proof of principle--some systems will require further bimolecularization, but are readily handled by the techniques we have introduced. However, a few caveats should the mentioned that confound the implementation.

The Gillespie Algorithm assumes well mixing of a solution~\cite{GillespieFirstandDirectReactions,helpfulGillespie}, which is an idealized version of reality where chemical specimen undergo Fickian diffusion in the absence of external forces, such as stirring. We note that both of the Quasi-Steady State based techniques ($\Psi_{QSSA},$ QSST) renders the system stiff due to introducing a fast variable to approximate a non-kinetic or higher-order reaction. Under canonical inversions, volume scalings by an integer power of $V$ were also applied to the reaction coefficients, inducing multiscale reaction channels for larger volumes. 

Chemically, stiff kinetic terms correspond to explosively reactive channels, whose rate of firing are orders of magnitude higher than that of slower channels. Very fast reactants are unlikely to coexist in the presence of other slower reactants, and form small spherical pockets in which the fast reactants have become extinct after translation to product specimen~\cite{fastreactionsphericalextinct}. This voids the assumption of well-mixing sought in our idealized abstract solution simulated by the Gillespie Algorithm. Likewise in areas such as DNA computing, multiscale reaction channels have been reported to accumulate chemical errors during its manifestation~\cite{DNAComputingBook}. Research is ongoing into design principles to better achieve stability~\cite{DNAComputingConference}, however we have reason to believe that synthetic implementations may not be as straightforward as one would imagine. 

But the benefits of a chemical implementation are enormous. Many of the simulations contained in this work took hours or if not days to generate, despite simulating early time behaviour (e.g. $t = 30$). In the chemical setting, an integration of a time-dependent ODE system until $30$ seconds requires precisely $30$ seconds of monitoring, and  not days of waiting.

2. \textbf{Development of non-stiff polynomial approximations}. 

\noindent The question of the optimal polynomial approximation relates to the founding question of Numerical Analysis~\cite{ApproximationTheoryBook}. We note the existence of a cutting-edge numerical analysis platform in \mcode{Matlab}, called \textit{Chebfun}~\cite{Chebfun}. An extension of this project may involve further development of polynomial approximation techniques based on a closer investigation of modern numerical analysis theory. We note that when fitting to exotic data such as in Figure~\ref{circadianoptimalselect}, splines yield much better results. However piecewise polynomials do not have a chemical interpretation. 

We briefly mentioned the possibility of deep-learning strategies to select the coefficients of a polynomial structure penalized not to have cross-negative terms in Section~\ref{optimalpolyselection}. Naive supervised training may be carried out by feeding a multilayer perceptron network successfully optimized coefficients, obtained by using Lagrange interpolation~\cite{properlagrangeinterpolation} or by Chebyshev polynomial expansion in an appropriate function space~\cite{Chebyshev}. We may minimize over a loss function defined via the $\ell_2$-norm as done previously. However being a supervised algorithm, the output will be at best approximations or recapitulations of the input coefficient data~\cite{MultilayerPerceptronProperBook}. For better results, unsupervised learning algorithms may be desired which learns and samples from the inherent distribution of successfully optimized coefficients, if they exist~\cite{MLmorebook}. Many extensions are possible dealing with optimal polynomial selection strategies using deep learning or otherwise, and new avenues may be freely explored. 

3. \textbf{Advancing the AutoGillespie program.}

\noindent Many extensions are possible to the AutoGillespie Algorithm. When non-canonical inversions are considered, we may no longer apply the tricks used in Appendix~\ref{GillespieAutoGillespie} to speed up our simulations. Approximate Stochastic Simulation Algorithms (approximate SSA) such as implicit Tau-leaping~\cite{implicitTauLeap} may be further built into the the algorithm in order to lessen computational resource usage. 

Quadraticization capabilities may also be added. We previously noted that just as Kerner Polynomialization does not provide unique output, General Quadraticization shares this feature as it exploits the same substitution methods for degree reduction. Therefore, we may investigate if there exists a combinatorial algorithm that always returns the optimal reaction network with the least number of added variables after quadraticization (e.g. see~\cite{proceeding}).

Another related extension is using the concept of \textit{sparsity}~\cite{sparsityeasy} of reaction networks and designing AutoGillespie to return or simulate the sparsest realization of a deterministic polynomial ODE. Lessening the number of reactions and specimen in the synthetic implementation is desirable, as it results in reduced cost (human and otherwise). Alternatively, it should be possible to use integer programming to return all possible kinetic interpretations of a given kineticized polynomial ODE system~\cite{minmaxrealization}, which may then be offered to the synthetic biologist as model candidates to choose from. As we saw in our simulations, the question of which network preserves the deterministic dynamics `optimally' in the stochastic simulation is a different question altogether--a categorical study via additional research should be carried out, starting with providing a rigorous definition of `optimal' in this context. 

%% file: ociamthesismain.bbl
\begin{thebibliography}{10}

\bibitem{Wilhelm}
Thomas Wilhelm.
\newblock {Chemical systems consisting only of elementary steps – a paradigma
  for nonlinear behavior}.
\newblock {\em Journal of Mathematical Chemistry}, 27:71--88, 2000.

\bibitem{TomiInversePaper}
Tomislav Plesa, Tom{\'a}{\v{s}} Vejchodsk{\'y}, and Radek Erban.
\newblock {Chemical reaction systems with a homoclinic bifurcation: an inverse
  problem}.
\newblock {\em Journal of Mathematical Chemistry}, 54:1884--1915, 2016.

\bibitem{easySparsePaper}
Fadil Santosa and Benjamin Weitz.
\newblock {An inverse problem in reaction kinetics}.
\newblock {\em Journal of Mathematical Chemistry}, 49:1507--1520, 2011.

\bibitem{TomiStatInferencePaper}
Tomislav Plesa, Tom{\'a}{\v{s}} Vejchodsk{\'y}, and Radek Erban.
\newblock {Test Models for statistical inference: two-dimensional reaction
  systems displaying limit cycle bifurcations and bistability}.
\newblock {\em Springer International Publishing}, pages 3--27, 2017.

\bibitem{FFTBookTheory}
E.~Oran Brigham.
\newblock {\em The Fast Fourier Transform and its Applications}.
\newblock Prentice Hall Signal Processing Series. Prentice Hall, 1988.

\bibitem{AutoCorrelationBookTheory}
Stephen~J. Taylor.
\newblock {\em Modelling Financial Time Series (2nd Edition)}.
\newblock World Scientific Publishing Company, 2007.

\bibitem{AutoCorrAndDFT}
Mario Pineda-Krch, Hendrik~J. Blok, Ulf Dieckmann, and Michael Doebeli.
\newblock {A tale of two cycles: distinguishing quasi-cycles and limit cycles
  in finite predator-prey populations}.
\newblock {\em Oikos}, 116(1):53--64, 2007.

\bibitem{GillespieFirstandDirectReactions}
Daniel~T Gillespie.
\newblock {A general method for numerically simulating the stochastic time
  evolution of coupled chemical reactions}.
\newblock {\em Journal of Computational Physics}, 22(4):403--434, 1976.

\bibitem{implicitTauLeap}
Muruhan Rathinama, Linda~R. Petzold, Yang Cao, and Daniel~T. Gillespie.
\newblock {Stiffness in stochastic chemically reacting systems: The implicit
  tau-leaping method}.
\newblock {\em The Journal of Chemical Physics}, 119(24), 2003.

\bibitem{originaltauleap}
Daniel~T. Gillespie.
\newblock {Approximate accelerated stochastic simulation of chemically reacting
  systems}.
\newblock {\em The Journal of Chemical Physics}, 115, 2001.

\bibitem{KowalskiWeak}
Krzysztof Kowalski.
\newblock {Universal formats for nonlinear dynamical systems}.
\newblock {\em Chemical Physics Letters}, 209(2):167--170, 1993.

\bibitem{QSSATomiTheoryTichnoff}
Wlodzimierz Klonowski.
\newblock {Simplifying principles for chemical and enzyme reaction kinetics}.
\newblock {\em Biophysical Chemistry}, 18:{73--87}, 1983.

\bibitem{KernerSleightofHand}
Edward~H. Kerner.
\newblock {Universal formats for nonlinear ordinary differential systems}.
\newblock {\em Journal of Mathematical Physics}, 22(7):1366--1371, 1981.

\bibitem{WilhelmAdvancedSchneider}
Klaus~R. Schneider and Thomas Wilhelm.
\newblock {Model reduction by extended quasi-steady-state approximation}.
\newblock {\em Journal of Mathematical Biology}, 40:443 -- 450, 2000.

\bibitem{fminsearchtheory}
Jeffrey~C. Lagarias, James~A. Reeds, Margaret~H. Wright, and Paul~E. Wright.
\newblock {Convergence Properties of the Nelder--Mead Simplex Method in Low
  Dimensions}.
\newblock {\em SIAM Journal on Optimization}, 9(1):112 -- 147, 1998.

\bibitem{fminsearch}
The MathWorks~Inc. (2022).
\newblock fminsearch (optimization): User's guide (r2022a).

\bibitem{moleculecomputegeneralexplain}
Xin Liang, Wen Zhu, Zhibin Lv, and Quan Zou.
\newblock {Molecular Computing and Bioinformatics}.
\newblock {\em Molecules}, 24(13):{2358}, 2019.

\bibitem{primitivecomputationCRN}
H.~J. Buisman, H.~M. Eikelder, P.~A. Hilbers, and A.~M. Liekens.
\newblock {Computing algebraic functions with biochemical reaction networks}.
\newblock {\em Artificial Life}, 15(1):5--19, 2009.

\bibitem{molecularRootcomputation}
Ziwei Shang, Changjun Zhou, and Qiang Zhang.
\newblock Chemical reaction networks' programming for solving equations.
\newblock {\em Current Issues in Molecular Biology}, 44(4):1725--1739, 2022.

\bibitem{TomiCatastrophePaper}
Tomislav Plesa, Alex Dack, and Thomas~E. Ouldridge.
\newblock Integral feedback in synthetic biology: negative-equilibrium
  catastrophe (arxiv), 2021.

\bibitem{gausianproper}
The MathWorks~Inc. (2022).
\newblock fitdist (statistics and machine learning toolbox): User's guide
  (r2022a).

\bibitem{WilkinsonPolynomialProperReference}
Robert~M Corless and Leili~Rafiee Sevyeri.
\newblock {The Runge Example for Interpolation and Wilkinson's Examples for
  Rootfinding}.
\newblock {\em SIAM Review}, 62(1):{231--243}, 2020.

\bibitem{properlagrangeinterpolation}
Ram~Bakhsha Srivastava and Saurabh Shukla.
\newblock {\em Numerical accuracies of Lagrange's and Newton polynomial
  interpolation: Numerical accuracies of Interpolation formulas}.
\newblock LAP LAMBERT Academic Publishing, 2012.

\bibitem{MultilayerPerceptronProperBook}
Stephen Marsland.
\newblock {\em Machine Learning: An Algorithmic Perspective}.
\newblock Crc Machine Learning \& Pattern Recognition. Chapman and Hall, 2009.

\bibitem{NonlinearSystems}
Jon Chapman (OCIAM~Oxford University).
\newblock Nonlinear systems lecture notes.
\newblock Online, 2021.
\newblock Last Accessed: August 2022. Accessed at:
  https://courses-archive.maths.ox.ac.uk/node/48977.

\bibitem{fftMatlabProper}
The MathWorks~Inc. (2022).
\newblock fft (fourier analysis and filtering): User's guide (r2022a).

\bibitem{AutocorrelationMatlabProper}
The MathWorks~Inc. (2022).
\newblock autocorr (econometrics toolbox): User's guide (r2022a).

\bibitem{LorenzChemical}
Douglas Poland.
\newblock {Cooperative catalysis and chemical chaos: a chemical model for the
  Lorenz equations}.
\newblock {\em Physica D: Nonlinear Phenomena}, 65(2):{86--99}, 1993.

\bibitem{LawMassActionHistory}
Robin~E Ferner and Jeffrey~K Aronson.
\newblock {Cato Guldberg and Peter Waage, the history of the Law of Mass
  Action, and its relevance to clinical pharmacology}.
\newblock {\em British journal of clinical pharmacology}, 81(1):52--55, 2016.

\bibitem{LawMassActionThermochemicalHistory}
E.~Wang Lund.
\newblock {Guldberg and Waage and the law of mass action}.
\newblock {\em Journal of Chemical Education}, 42(10):548--550, 1965.

\bibitem{Merton}
Ton~Yeh of~Merton College University~of Oxford.
\newblock William esson.
\newblock Online.
\newblock Last Accessed: July 2022. Accessed at: https://www.merton.ox.ac.uk
  /sites/default/files/inline-files/William-Esson.pdf.

\bibitem{NatureHarcourt}
Nature Publishers.
\newblock {Augustus Geosrge Vernon-Harcourt, 1834–1919}.
\newblock {\em Nature}, 134(963), 1934.

\bibitem{HarcourtAndEsson}
Harcourt Vernon and Esson William.
\newblock {On the laws of connexion between the conditions of a chemical change
  and its amount}.
\newblock {\em Philosophical Transactions}, 156:{193--221}, 1866.

\bibitem{LawMassAction}
S.W. Hinkley and Chris~P. Tsokos.
\newblock {A stochastic model for chemical equilibrium}.
\newblock {\em Mathematical Biosciences}, 21(1):85--102, 1974.

\bibitem{ChemicalReactionNetworkTheory}
Martin Feinberg.
\newblock {\em Foundations of Chemical Reaction Network Theory}.
\newblock Applied Mathematical Sciences. Springer, 2019.

\bibitem{CRNmemory}
Naren Ramakrishnan and Upinder Bhalla.
\newblock {Memory switches in chemical reaction space}.
\newblock {\em PLOS Computational Biology}, 4(7):1--9, 2008.

\bibitem{Murray2}
J.D. Murray.
\newblock {\em Mathematical Biology II: Spatial Models and Biomedical
  Applications}.
\newblock Interdisciplinary Applied Mathematics. Springer, 2003.

\bibitem{TomiPraiseBiMolecularController}
Tomislav Plesa, Guy-Bart Stan, Thomas~E. Ouldridge, and Wooli Bae.
\newblock {Quasi-robust control of biochemical reaction networks via stochastic
  morphing}.
\newblock {\em Journal of The Royal Society Interface}, 2021.

\bibitem{VKBL}
José M.~G. Vilar, Hao~Yuan Kueh, Naama Barkai, and Stanislas Leibler.
\newblock Mechanisms of noise-resistance in genetic oscillators.
\newblock {\em Proceedings of the National Academy of Sciences},
  99(9):5988--5992, 2002.

\bibitem{VKBLexperimentdata}
Naama Barkai and Stanislas Leibler.
\newblock {Circadian clocks limited by noise}.
\newblock {\em Nature}, 403:267--268, 2000.

\bibitem{modifiednextreactions}
David Anderson.
\newblock {A modified next reaction method for simulating chemical systems with
  time dependent propensities and delays}.
\newblock {\em The Journal of Chemical Physics}, 127, 2007.

\bibitem{GeneRegNetworkeasypaper}
Hana~El Samad, Mustafa Khammash, Linda Petzold, and Dan Gillespie.
\newblock {Stochastic modelling of gene regulatory networks}.
\newblock {\em International Journal of Robust and Nonlinear Control},
  15:691--711, 2005.

\bibitem{incorrectcircadianmodel}
Jean-Christophe Leloup and Albert Goldbeter.
\newblock A model for circadian rhythms in drosophila incorporating the
  formation of a complex between the per and tim proteins.
\newblock {\em Journal of Biological Rhythms}, 13:70 -- 87, 1998.

\bibitem{DNAComputing}
David Soloveichik, Georg Seeliga, and Erik Winfree.
\newblock {DNA as a universal substrate for chemical kinetics}.
\newblock {\em Proceedings of the National Academy of Sciences (PNAS)},
  107(12):5393--5398, 2010.

\bibitem{Hilbert16problemproperreference}
Yu~Ilyashenko.
\newblock {Centennial history of Hilbert's 16th problem}.
\newblock {\em Bulletin of the American Mathematical Society},
  39(3):{301--354}, 2002.

\bibitem{SymbolicMathToolbox}
The MathWorks~Inc. (2022).
\newblock Symbolic math toolbox: User's guide (r2022a).

\bibitem{OriginalLorenz}
Edward~N. Lorenz.
\newblock {Deterministic Nonperiodic Flow}.
\newblock {\em Journal of Atmospheric Sciences}, 20(2):{130--148}, 1963.

\bibitem{LorenzLaser}
H.~Haken.
\newblock {Analogy between higher instabilities in fluids and lasers}.
\newblock {\em Physics Letters A}, 53(1):{77--78}, 1975.

\bibitem{LorenzOsmosis}
Stephan~I. Tzenov.
\newblock Strange attractors characterizing the osmotic instability, 2014.

\bibitem{ChaosAndFractalsBook}
Heinz-Otto Peitgen, Hartmut J\"{u}rgens, and Dietmar Saupe.
\newblock {\em Chaos and Fractals: New Frontiers of Science}.
\newblock Springer, 2004.

\bibitem{OriginalRossler}
Otto~Eberhard R{\"o}ssler.
\newblock {An equation for continuous chaos}.
\newblock {\em Physics Letters A}, 57(5):{397--398}, 1976.

\bibitem{helpfulGillespie}
Daniel~T. Gillespie.
\newblock The chemical langevin and fokker-planck equations for the reversible
  isomerization reaction.
\newblock {\em Journal of Physical Chemistry A}, 106:5063--5071, 2002.

\bibitem{fastreactionsphericalextinct}
Frank~E Marble.
\newblock {Mixing, Diffusion and Chemical Reaction of Liquids in a Vortex
  Field}.
\newblock {\em Biophysical Chemistry}, pages {581--596}, 1988.

\bibitem{DNAComputingBook}
Lila Kari, Elena Losseva, and Petr Sosik.
\newblock {\em DNA Computing and Errors: A Computer Science Perspective}.
\newblock Molecular Computational Models: Unconventional Approaches. IGI
  Global, 2005.

\bibitem{DNAComputingConference}
David Soloveichik and Bernard Yurke.
\newblock {\em DNA Computing and Molecular Programming (Proceedings of the 19th
  International Conference, DNA 19)}.
\newblock Lecture Notes in Computer Science. Springer, 2013.

\bibitem{ApproximationTheoryBook}
Lloyd~N. Trefethen.
\newblock {\em Approximation Theory and Approximation Practice, Extended
  Edition}.
\newblock Other Titles in Applied Mathematics. SIAM, 2019.

\bibitem{Chebfun}
Tobin~A. Driscoll, Nicholas Hale, and Lloyd~N. Trefethen.
\newblock Chebfun guide.
\newblock Online, 2014.
\newblock Last Accessed: August 2022. Accessed at:
  https://www.chebfun.org/docs/guide/chebfunguide.pdf.

\bibitem{Chebyshev}
Codruta Chis and F.~Cret.
\newblock {Approximating Functions with Chebyshev polynomials}.
\newblock {\em Scientifical Researches}, 11({2}):{481--484}, 2005.

\bibitem{MLmorebook}
David Barber.
\newblock {\em Bayesian Reasoning and Machine Learning}.
\newblock Lecture Notes in Computer Science. Cambridge University Press, 2012.

\bibitem{proceeding}
Andrey Bychkov and Gleb Pogudin.
\newblock Optimal monomial quadratization for ode systems.
\newblock In {\em Combinatorial Algorithms}, pages 122--136, Cham, 2021.
  Springer International Publishing.

\bibitem{sparsityeasy}
Fadil Santosa and Benjamin Weitz.
\newblock {An inverse problem in reaction kinetics}.
\newblock {\em Journal of Mathematical Chemistry}, 49:{1507--1520}, 2011.

\bibitem{minmaxrealization}
G{\'a}bor Szederk{\'e}nyi, Katalin~M. Hangos, and Tam{\'a}s P{\'e}ni.
\newblock {Maximal and minimal realizations of reaction kinetic systems:
  computation and properties}.
\newblock {\em MATCH Communications Mathematical Computer Chemistry},
  65(2):{309--332}, 2011.

\end{thebibliography}
